\documentclass[11pt,a4paper]{article}
\def\AmSTeX{\leavevmode\hbox{$\mathcal A\kern-.2em\lower.376ex%
        \hbox{$\mathcal M$}\kern-.2em\mathcal S$-\TeX}}
\usepackage{graphicx}
\usepackage[numbers,sort&compress]{natbib}
\usepackage{amsmath}
\usepackage{amssymb}
\usepackage{subfigure}
\usepackage{mathrsfs}
\usepackage{bm}
\usepackage[amsmath,thmmarks,hyperref]{ntheorem}
\ifx\pdfoutput\undefined \pdffalse \fi \ifx\pdfoutput\relax
\pdffalse \fi
\ifx\texonly\undefined\let\texonly\relax\fi
\ifx\endtexonly\undefined\let\endtexonly\relax\fi \texonly
  \let\htmlonly\iffalse
  \let\endhtmlonly\fi
\endtexonly

\htmlonly
  
\endhtmlonly
\texonly
  \usepackage{makeidx}

  \usepackage{multicol}
  
\usepackage[toc,page,title,titletoc,header]{appendix}
\endtexonly
\title{}
\author{\thanks{}}
\date{}
\begin{document}

\title{\textbf{The rare semi-leptonic $B_c$ decays involving orbitally excited final mesons }}

\author{Wan-Li~Ju\footnote{wl\_ju\_hit@163.com}~$^1$, Guo-Li Wang\footnote{gl\_wang@hit.edu.cn}~$^1$, Hui-Feng Fu\footnote{huifengfu@tsinghua.edu.cn}~$^2$,
  ~Zhi-Hui Wang\footnote{wzh19830606@163.com
}~$^3$, Ying Li\footnote{liying@ytu.edu.cn}~$^4$
  \\
{\it \small  $^1$Department of Physics, Harbin Institute of Technology, Harbin, 150001, China}\\
{ \small $^2$\it Department of Physics, Tsinghua University, Beijing, 100084, China} \\
{\it  \small $^3$ Department of Physics,~Beifang University of Nationalities,~Yinchuan,~750021,~China
}\\
{\it  \small $^4$ Department of Physics, Yantai University, Yantai 264-005, China} }


\maketitle

\baselineskip=20pt
\begin{abstract}

%
The rare processes $B_c\to D_{(s)J} ^{(*)}\mu\bar{\mu}$, where $D_{(s)J}^{(*)}$ stands for
the final meson $D_{s0}^*(2317)$, $D_{s1}(2460,2536)$,~$D_{s2}^*(2573)$,
$D_0^*(2400)$, $D_{1}(2420,2430)$
or~$D_{2}^*(2460)$, are studied within the Standard Model.
The  hadronic matrix elements are evaluated in the Bethe-Salpeter approach and furthermore a  discussion on the gauge-invariant condition of the annihilation hadronic currents is presented. Considering the penguin,  box,
annihilation, color-favored cascade and color-suppressed cascade
contributions, the  observables $\text{d}Br/\text{d}Q^2$,  $A_{LPL}$, $A_{FB}$ and  $P_L$ are calculated.

\end{abstract}

\clearpage
\section{Introduction}

The rare decays
$b\to s(d) l\bar{l}$ have particular features.
These transitions are of the single-quark flavor-changing neutral current~(FCNC) processes, which are forbidden at tree level in the Standard Model (SM)
but mediated by loop processes. Hence, within the SM, the $b\to s(d) l\bar{l}$ amplitudes are greatly suppressed.
The situation is different for the standard model extensions, where many new particles beyond the SM are predicted.
These new particles can virtually entry the loops relevant to FCNC processes or induce the transitions at tree level, which makes that the observables predicted in the standard model extensions may significantly deviate from the ones in the SM. This sensitive nature to the effects beyond the SM can be exploited as a tool for stringently testing the SM and indirectly hunting the New Physics (NP).

In literatures, the $b\to s l\bar{l}$ processes were extensively analyzed in the decays $B\to K^{(*)} l\bar{l}$. In  recent years, the decays $B\to K_1(1270,1400)l\bar{l}$ \cite{BtoKhigherstates1}, $B\to K^*_0(1430)l\bar{l}$~\cite{BtoKhigherstates2,BtoKhigherstates3,BtoKhigherstates4,BtoKhigherstates5,BtoKhigherstates7,BtoKhigherstates8,BtoKhigherstates9,BtoKhigherstates10} and $B\to K^*_2(1430)l\bar{l}$~\cite{BtoKhigherstates11,BtoKhigherstates12,BtoKhigherstates13,BtoKhigherstates14,BtoKhigherstates15,BtoKhigherstates16,BtoKhigherstates17,BtoKhigherstates18,BtoKhigherstates19,BtoKhigherstates20,BtoKhigherstates21,BtoKhigherstates9,BtoKhigherstates23} 
have also been emphasized. 
However, according to Ref.~\cite{pdg}, the mass differences among the $K^{(*)}_{J}$s, where  $K^{(*)}_{J}$s denote the mesons $K_1(1270)$, $K_1(1400)$, $K^*_0(1430)$ and $K^*_2(1430)$, are small and their widths are rather wide.
This leads to the problem that the observables in a certain kinematic region may receive contributions from several different channels and it is not easy to separate them confidently.  For instance, as estimated in Ref.~\cite{BtoKhigherstates9}, at $m_{K\pi}\sim1.4~\text{GeV}$, the longitudinal differential branching fraction $\text{d}Br_L(B\to K\pi l\bar{l})/\text{d} m^2_{K\pi} $ is affected by the channels $B\to K^*_0(1430)l\bar{l}$, $B\to K^*_2(1430)l\bar{l}$,  $B\to K^*(1680) l\bar{l}$ and $B\to K^*(1410)l\bar{l}$  un-negligibly. 
But this situation will be ameliorated, if the decays $B_c\to D_{sJ} ^{(*)}l\bar{l}$ are investigated. Compared with the $K^{(*)}_{J}$s, the mass differences among the $D_{sJ} ^{(*)}$ mesons are bigger and their widths are much narrower~\cite{pdg}. These features are helpful in reducing the interferences among the different channels. Hence in this paper, we are motivated to investigate the processes $B_c\to D_{sJ} ^{(*)}l\bar{l}$.

\begin{figure}[htbp]
\centering
\subfigure[~box diagram]{\includegraphics[width =
0.31\textwidth,height=0.13\textheight]{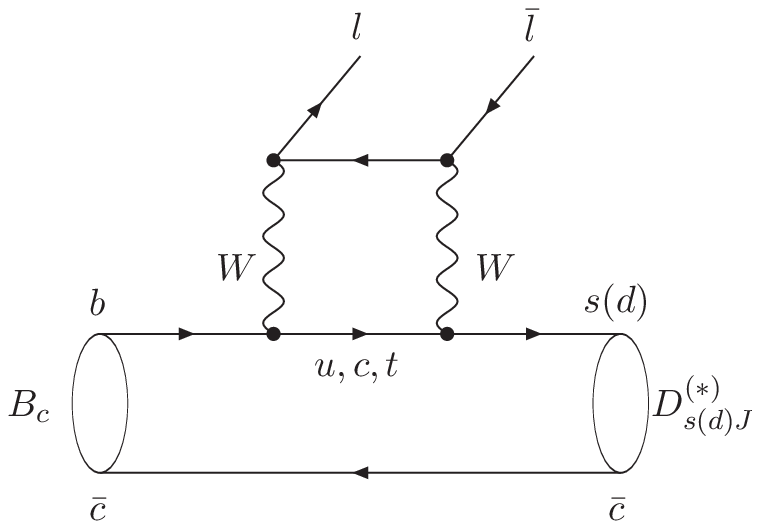}}
\subfigure[~$Z^0~(\gamma)$~penguin diagram]{\includegraphics[width =
0.31\textwidth,height=0.15\textheight]{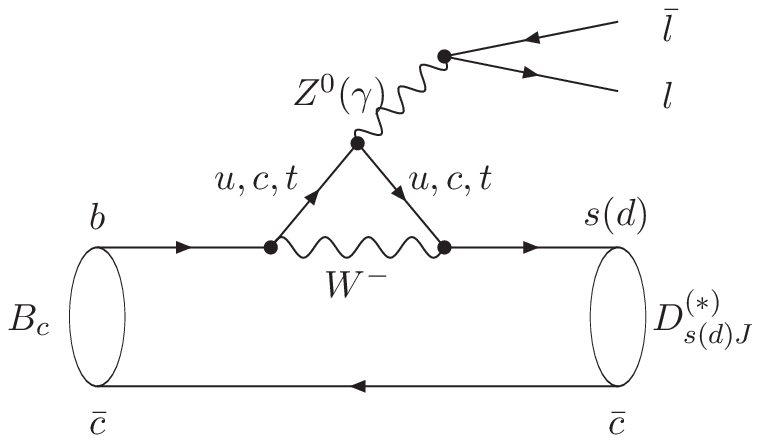}}
\subfigure[~annihilation diagram]{\includegraphics[width =
0.31\textwidth,height=0.11\textheight]{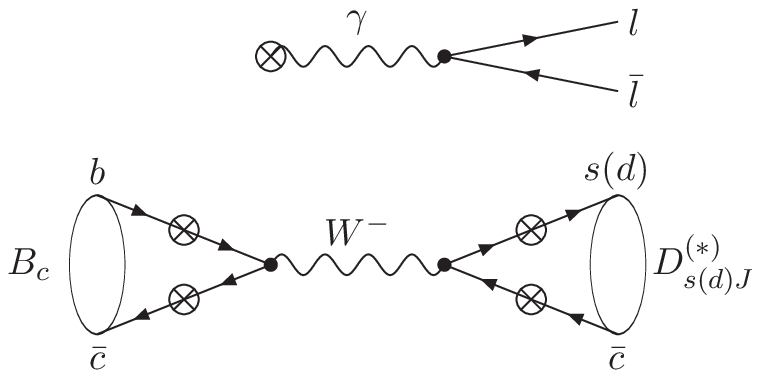}}
\\
\subfigure[~color-suppressed cascade diagram ]{\includegraphics[width =
0.35\textwidth,height=0.15\textheight]{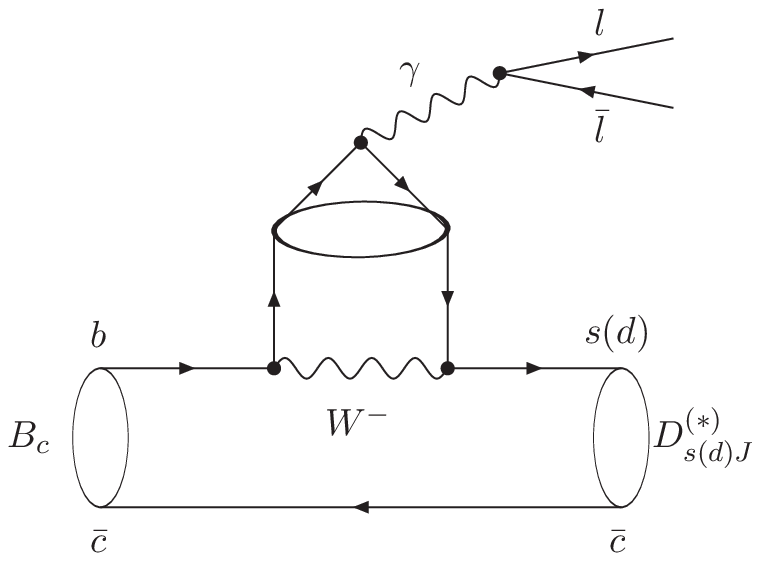}}
\subfigure[~color-facored cascade diagram]{\includegraphics[width =
0.35\textwidth,height=0.15\textheight]{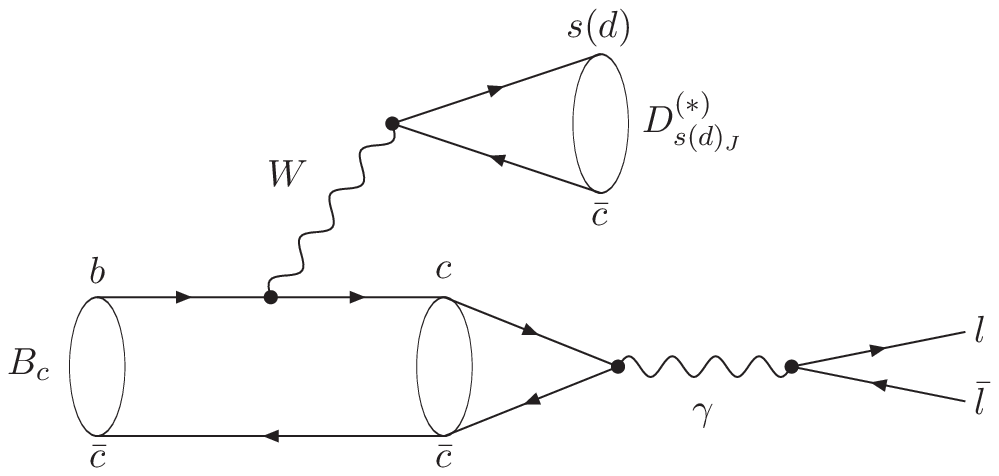}}\caption{Typical diagrams of $B_c\rightarrow D^{(*)}_{s(d)J} l\bar{l}$ process. In annihilation diagrams (c) the photon can be emitted from each quark, denoted by $\bigotimes$, and decays to the lepton pair.}
\label{figureSD1}
\end{figure}

In the previous works~\cite{BctoDs2317QCDsunrule,BctoDs2317lightcone}, the process $B_c\to D_{s0}^*(2317)l\bar{l} $ was calculated including only the $b\to s l\bar{l}$ effects, whose typical Feynman diagrams are Box and Penguin (BP) diagrams, as plotted in Figs.~\eqref{figureSD1} (a,~b). However, besides the BP effects, the Annihilation (Ann) diagrams, as shown in Figs.~\eqref{figureSD1} (c), also make un-negligible contributions. On one hand, both BP and Ann diagrams are of order $\mathcal{O}(\alpha_{em}G_f)$ and the ratio of their CKM matrix elements is $|V^*_{cb}V_{cs(d)}|/|V^*_{ts(d)}V_{tb}|\sim 1$. On the other hand, from Fig.~\eqref{figureSD1} (c), we see that the color factors of Ann diagrams are 3 times larger than those of  BP diagrams. Thus, when the decay $B_c\to D_{s0}^*(2317)l\bar{l} $ is analyzed, it is necessary to include the Ann effects.


In addition to the BP and Ann effects, the process $B_c\to D_{s0}^*(2317)l\bar{l} $ is also influenced by resonance cascade processes, such as $B_c\to D_{s0}^*(2317)J/\psi~(\psi(2S))\to D_{s0}^*(2317)l\bar{l} $. Their typical Feynman diagrams are illustrated in Figs.~\eqref{figureSD1} (d,~e). Transition amplitudes of these diagrams in the area $m^2_{l\bar{l}}\sim m^2_{J/\psi~(\psi(2S))}$ always become much larger than the BP and Ann ones. Hence, to avoid overwhelming the BP and Ann contributions, the regions around $m^2_{l\bar{l}}\sim m^2_{J/\psi~(\psi(2S))}$ should be experimentally removed.
In Ref.~\cite{BctoDs2317QCDsunrule}, the regions~\cite{gengBctoDs}, which are defined through comparing the BP and color-suppressed (CS) cascade contributions, are employed. However, in the $B_c\to D_{s0}^*(2317)l\bar{l} $ process, both the color-favored (CF) and CS diagrams exist. Furthermore, the CF transition amplitudes are expected to be larger than the CS ones by a 3
times larger color factor approximately. Thus, it is necessary to redefine these regions with both CF and CS cascade influences.

So in this paper, we  investigate $B_c\to D_{s0}^{*}(2317)l\bar{l}$ transition including BP, Ann, CS and CF contributions. In addition, in order to give a more comprehensive discussion on the semi-leptonic rare decays of $B_c$, the processes $B_c\to$$D_{s1}(2460,2536)l\bar{l}$, $B_c\to D_{s2}^*(2573)$$l\bar{l} $ and
$B_c\to D_{J}^{(*)}l\bar{l}$ are also analyzed.

In our calculations, the low-energy effective theory is employed~\cite{WILLSON}. Within this method, the short distance information of transition amplitude is factorized into the Wilson coefficients, while the long distance effects are described by the matrix element which is an operator sandwiched by the initial and
the final states. The Wilson coefficients in the SM can be attained perturbatively. But the matrix elements are of non-perturbative nature and in this paper we calculate them with
the Bethe-Salpeter (BS) method~\cite{chao-hsi chang1}.
In this method, the BS equation \cite{BS1,BS2} is employed to solve the wave functions for mesons, while the Mandelstam Formalism \cite{SMandelstam} is used to evaluate hadronic matrix elements. With such method, the hadronic matrix elements keep the relativistic effects from both the wave functions and the kinematics.
In our previous paper \cite{Juwangpaper1}, within the BS method, we calculated the $B_c\to D_{s,d}^{(*)} l\bar{l}$ rare transitions, whose final mesons are of S-wave states, and checked the gauge-invariance condition of the annihilation hadronic currents. In this paper, we investigate the processes $B_c\to D_{(s)J}^{(*)} l\bar{l}$, whose final mesons are of P-wave states, and furthermore, we give a more generalized conclusion: the annihilation hadronic currents obtained within the BS method satisfy the gauge-invariance condition, no matter what the $J^P$s of initial and final mesons are.

This paper is organized as follows.
In Section 2, we introduce the transition amplitudes corresponding to BP, Ann, CS and CF contributions and specify the involved hadronic matrix elements. Within Section 3, we calculate these hadronic matrix elements through the Bethe-Salpeter method and express the results in terms of form factors.
In Section 4, using these form factors, we compute the observables, including $\text{d}Br/\text{d}Q^2$,  $A_{LPL}$, $A_{FB}$ and  $P_L$. Section 5 is devoted to the discussions on the theoretical uncertainties. Finally, we summarize and conclude in Section 6.

%
%

\section{Transition Amplitudes of BP, Ann, CS and  CF Contributions}\label{123} 

In this section, we briefly review the transition amplitudes corresponding to BP, Ann, CS and  CF  effects. A more detailed introduction  can be found in our previous paper~\cite{Juwangpaper1}.

According to low-energy effective theory~\cite{WILLSON}, the transition amplitude describing the $b\to s(d) l\bar{l}$ (or equivalently, BP) contribution is,
\begin{equation}
\mathcal{M}_{BP}=i\frac{G_{F}\alpha_{em}}{2\sqrt{2}\pi}V_{tb}V_{ts(d)}^{*}\left\{\left[C^{eff}_{9}W_{\mu}-\frac{2m_{b}}{Q^2}C^{eff}_{7}W_{\mu}^{T}\right]\bar{l}\gamma^{\mu}l+C_{10}W_{\mu}\bar{l}\gamma^{\mu}\gamma_5l\right\},
\label{eq:amplitudePB}\end{equation}
where $Q=P_i-P_f$ and $P_{i(f)}$ stands for the momentum of the initial~(finial) meson. $V_{tb}$ and $V_{ts(d)}$ denote the CKM
matrix elements. $C_{10}$ is the Wilson coefficient.  $C^{eff}_{7,9}$ are the combinations of the Wilson coefficients which are multiplied by the same hadronic matrix elements. The numerical value of $C_{10}$ and the explicit expressions of $C^{eff}_{7,9}$ can be found in Ref.~\cite{AFaessler}. The hadronic matrix elements $W_{\mu}$ and $W_{\mu}^{T}$ are defined as \begin{equation}W_{\mu}=\langle f|
\bar{s}(\bar{d})\gamma_{\mu}(1-\gamma_{5})b| i\rangle ,
~~~~~W_{\mu}^{T}=\langle
f|\bar{s}(\bar{d})i\sigma_{\mu\nu
}(P_i-P_f)^{\nu}(1+\gamma_5) b| i\rangle, \label{eq:formfactorBP}\end{equation}
where the definition $\sigma^{\mu\nu} =(i/2)[\gamma^{\mu},\gamma^{\nu}]$ is used.

Based on the effective theory~\cite{WILLSON} and the factorization hypothesis~\cite{Factorization}, the transition amplitude describing the Ann effects is \cite{Juwangpaper1}
\begin{equation}
\begin{split}
\mathcal{M}_{Ann}
=&V_{cb}V^{*}_{cs(d)}\frac{ i
\alpha_{em}}{Q^2}\frac{G_{F}}{2\sqrt{2}\pi}\left(\frac{C_1}{N_c}+C_2\right)W_{ann}^{\mu}\bar{l}\gamma_{\mu}l,
\end{split}\label{eq:definitionMAnn}
\end{equation}
where $C_{1,2}$ are the Wilson coefficients, whose values can  be found in Ref.~\cite{AFaessler}.
The annihilation hadronic current $W_{ann}^{\mu}$ is defined as $W^{\mu}_{ann}=W^{\mu}_{1ann}+W^{\mu}_{2ann}+W^{\mu}_{3ann}+W^{\mu}_{4ann}$, where \begin{equation}
\begin{split}
W_{1ann}^{\mu}=&(-8\pi^2)\langle f|\bar{s}(\bar{d})
\gamma_{\alpha}(1-\gamma_5)c|0\rangle\langle0|\bar{c}\gamma^{\alpha}(1-\gamma_5)
\frac{1}{\not\!{p}_{q_1}-m_{q_1}+i \epsilon}(-\frac{1}{3})\gamma^{\mu}b| i\rangle,\\
W_{2ann}^{\mu}=&(-8\pi^2)\langle f|\bar{s}(\bar{d})
\gamma_{\alpha}(1-\gamma_5)c|0\rangle\langle0|\bar{c}(\frac{2}{3})\gamma^{\mu}
\frac{1}{\not\!{p}_{q_2}-m_{q_2}+i \epsilon}\gamma^{\alpha}(1-\gamma_5)b| i\rangle,\\
W_{3ann}^{\mu}=&(-8\pi^2)\langle f|\bar{s}(\bar{d}) (-\frac{1}{3})\gamma^{\mu}\frac{1}{\not\!{p}_{q_3}-m_{q_3}+i  \epsilon}\gamma_{\alpha}(1-\gamma_5)c|0\rangle\langle0|\bar{c}\gamma^{\alpha}(1-\gamma_5)b| i\rangle,\\
W_{4ann}^{\mu}=&(-8\pi^2)\langle f|\bar{s}(\bar{d})\gamma_{\alpha}(1-\gamma_5)\frac{1}{\not\!{p}_{q_4}-m_{q_4}+i  \epsilon}(\frac{2}{3})\gamma^{\mu}c|0\rangle\langle0|\bar{c}\gamma^{\alpha}(1-\gamma_5)b| i\rangle.\\
\end{split}\label{eq:formfactorAnn}
\end{equation}
$p_{q_{1-4}}$ and $m_{q_{1-4}}$ are momenta and
masses of the propagated quarks, respectively.

For the CS and CF cascade resonance effects, the transition amplitudes are~\cite{Juwangpaper1}
\begin{equation}
\begin{split}
\mathcal{M}_{CS}
=&i\frac{9 G_{F}}{2\sqrt{2}\alpha_{em}}V_{cb}V_{cs(d)}^{*}\left(C_1+\frac{C_2}{N_c}\right)\left[\underset{V=J/\psi,\psi(2S)}{\sum}\frac{\Gamma(V\rightarrow\bar{l}l)M_V}{Q^2-M_V^2+i\Gamma_VM_V}\right ]W_{\mu}\bar{l}\gamma^{\mu}l ,\\
\mathcal{M}_{CF}=&i
\frac{G_{F}\alpha_{em}}{2\sqrt{2}\pi}V_{cb}V^{*}_{cs(d)} \left(C_2+\frac{C_1}{N_c}\right)W_{CF}^{\mu}\bar{l}\gamma_{\mu}l,
\end{split}\label{eq:definitionMCS}
\end{equation}
where $M_V$ and $\Gamma_V$ are the mass and full width of the resonance meson, respectively.
$\Gamma(V\rightarrow\bar{l}l)$ denotes the branching width of the transition $V\rightarrow\bar{l}l$. The resonance meson $V$  stands for the particle $ J/\psi$ or $\psi(2S)$.
The CF hadronic current $W_{CF}^{\mu}$ is defined as
\begin{equation}
\begin{split}
W_{CF}^{\mu}=&\underset{V=J/\psi,\psi(2S)}{\sum}\frac{-16\pi^2}{3M_V^2}\langle 0|\bar{c}\gamma^{\mu}c|V\rangle\frac{i}{Q^2-M_V^2+i\Gamma_VM_V}\langle
V|\bar{c}\gamma^{\nu}(1-\gamma_5)b|i\rangle\langle
f|\bar{s}(\bar{d})\gamma_{\nu}(1- \gamma_5)c|0\rangle.
\end{split}
\label{eq:formfactorCF}
\end{equation}

Consequently, the total transition amplitude is \begin{equation}
\begin{split}\mathcal{M}_{Total}=\mathcal{M}_{BP}+\mathcal{M}_{Ann}+\mathcal{M}_{CS}+\mathcal{M}_{CF}.\end{split}\label{Mtotal}
\end{equation}

\section{Hadronic Transition Matrix Elements in the BS Method}

In Sec.~\ref{123}, the transition amplitudes of the $B_c\to D_{(s)J}^{(*)}\l\bar{l}$ processes are introduced and the hadronic matrix elements $W_{(T)}$, $W_{ann}$ and $W_{CF}$ are defined.
In this section, within the BS method, we show how to calculate these hadronic matrix elements.
In Sec.~\ref{subsec:1}, we express the  hadronic currents  as the integrals of the wave functions. Sec.~\ref{subsec:2} is devoted to showing the wave functions of the mesons which are involved in this paper. Using these  wave functions, we calculate the hadronic currents in Sec.~\ref{SecHadronicCurrentsin BS} and parameterize the results  in terms of form factors in Sec.~\ref{SecDefinFormFacotors}. In Sec.~\ref{Sec3-5}, we present the numerical results of the form factors.

\subsection{ General Arguments on Hadronic Currents }\label{subsec:1}
In this part, we  rewrite the hadronic currents as the integrals of the wave functions and present some general arguments.

According to the
Mandelstam formalism \cite{SMandelstam}, $W_{(T)}$ can be expressed as the integrals of the 4-dimensional BS wave functions.
In the spirit of the instantaneous approximation~\cite{ChangPRD49 3399}, the integrations with respect to $q^0_i$, where $q_i$ represents the relative momentum between the quark and anti-quark of the initial meson, can be performed first. And then we have~\cite{Juwangpaper1,chao-hsi chang1}
\begin{equation}\begin{split}
W&_{\mu}=-\int\frac{d^3\vec{q_i}}{(2\pi)^3}\mathrm{Tr}\left\{\frac{\not\!{P_i}}{M_i}\bar{\varphi}^{++}_{f}\gamma_{\mu}\left(1-\gamma_5\right)\varphi^{++}_{i}\right\},\\
W&^{\mu}_T=-\frac{1}{2}(P_i-P_f)_{\nu}\left(\mathcal{Y}^{\mu\nu}_{V}+\mathcal{Y}^{\nu\mu}_{A}\right),
\end{split}\label{eq:Mandelstam Formalism}\end{equation}
where  the hadronic tensors $\mathcal{Y}^{\mu\nu}_{V,A}$  are defined as
\begin{equation}
\begin{split}
\mathcal{Y}^{\mu\nu}&_{V}=-\int\frac{d^3\vec{q_i}}{(2\pi)^3}\mathrm{Tr}\left\{\frac{\not\!{P_i}}{M_i}\bar{\varphi}^{++}_{f}\gamma^{\mu}\gamma^{\nu}\varphi^{++}_{i}\right\},\\
\mathcal{Y}^{\mu\nu}&_{A} =-\int\frac{d^3\vec{q_i}}{(2\pi)^3}\mathrm{Tr}\left\{\frac{\not\!{P_i}}{M_i}\bar{\varphi}^{++}_{f}\gamma^{\mu}\gamma^{\nu}\gamma_5\varphi^{++}_{i}\right\}.
\end{split}\label{eq:YYT}
\end{equation}
The term $\varphi^{++}_{i(f)}$ in Eqs.~(\ref{eq:Mandelstam Formalism}-\ref{eq:YYT}) denotes the positive energy part of the initial~(finial) wave function \cite{ChangPRD49 3399} and will be specified in the next subsection. In this paper we ignore the negative-energy parts since they give negligible contributions.

For $W_{ann}$, similar to the derivations of Eq.~\eqref{eq:Mandelstam Formalism}, we have\footnote{While deriving Eqs.~(\ref{eq:w1w2w3w4ann1}-\ref{eq:w1w2w3w4ann4}), we employ the weak binding hypothesis~\cite{ChangPRD49 3399}. In this manner, the expansion $\omega_{1,2}\equiv\sqrt{m^2_{1,2}-q^2_{a,c}}=m_{1,2}+\frac{-q^2_{a,c}}{2m_{1,2}}+\cdots\cdots$ can be performed~\cite{ChangPRD49 3399} and in this paper only the leading term is kept. Under this approximation, we have the  relationships $(\alpha_1 \not\!{P}+\not\!{q}_{_{P\bot}}-m_1)\varphi^{++}_{i,f}\sim0$ and $\varphi^{++}_{i,f}(\alpha_2 \not\!{P}-\not\!{q}_{_{P\bot}}+m_2)\sim0$, which are quite useful to simplify $W_{ann}$.
},
\begin{equation}
\begin{split}
W&^{\mu}_{1ann}(i\rightarrow f )=\frac{8}{\pi^2}\left\{\int\frac{d^3\vec{q}_i}{(2\pi)^3}\frac{2\mathcal{F}^{\nu}_{i0}(i)(\alpha^{i}_1{P}^{\mu}_i+{q}^{\mu}_{a})
-\mathcal{F}^{\mu\nu}_{i+}(i)-\mathcal{F}^{\mu\nu}_{i-}(i)}{M_i(Q^2-2Q\cdot(\alpha^{i}_1P_i+q_{a})+i\epsilon)}\right\}
\\
&\times\left\{\int\frac{d^3\vec{q}_f}{(2\pi)^3}\frac{\mathcal{F}_{\nu}^{f0}(f)}{M_f}\right\},\\
\end{split}\label{eq:w1w2w3w4ann1}
\end{equation}
\begin{equation}
\begin{split}
W&^{\mu}_{2ann}(i\rightarrow f )=\frac{-16}{\pi^2}\left\{\int\frac{d^3\vec{q}_i}{(2\pi)^3}\frac{2\mathcal{F}^{\nu}_{i0}(i)(-\alpha^{i}_2{P}^{\mu}_i+{q}^{\mu}_{a})
+\mathcal{F}^{\mu\nu}_{i+}(i)-\mathcal{F}^{\mu\nu}_{i-}(i)}{M_i(Q^2+2Q\cdot(-\alpha^{i}_2P_i+q_{a})+i\epsilon)}\right\}
\\
&\times\left\{\int\frac{d^3\vec{q}_f}{(2\pi)^3}\frac{\mathcal{F}_{\nu}^{f0}(f)}{M_f}\right\},\\
\end{split}
\end{equation}
\begin{equation}
\begin{split}
W&^{\mu}_{3ann}(i\rightarrow f )=\frac{8}{\pi^2} \left\{\int\frac{d^3\vec{q}_f}{(2\pi)^3}\frac{2\mathcal{F}^{\nu}_{f0}(f)(\alpha^{f}_1{P}^{\mu}_f+{q}^{\mu}_{c}
)+\mathcal{F}^{\mu\nu}_{f+}(f)+\mathcal{F}^{\mu\nu}_{f-}(f) }{M_f(Q^2+2Q\cdot(\alpha^{f}_1P_f+q_{c})+i\epsilon)}\right\}\\
&\times \int\frac{d^3\vec{q}_i}{(2\pi)^3}  \frac{\mathcal{F}_{\nu}^{i0}(i)}{M_i},\\
\end{split}
\end{equation}
\begin{equation}
\begin{split}
W&^{\mu}_{4ann}(i\rightarrow f )=\frac{-16}{\pi^2} \left\{\int\frac{d^3\vec{q}_f}{(2\pi)^3}\frac{-2\mathcal{F}^{\nu}_{f0}(f)(\alpha^{f}_2{P}^{\mu}_f-{q}^{\mu}_{c}
)-\mathcal{F}^{\mu\nu}_{f+}(f)+\mathcal{F}^{\mu\nu}_{f-}(f) }{M_f(Q^2+2Q\cdot(\alpha^{f}_2P_f-q_{c})+i\epsilon)}\right\}\\
&\times \int\frac{d^3\vec{q}_i}{(2\pi)^3}  \frac{\mathcal{F}_{\nu}^{i0}(i)}{M_i},
\end{split}\label{eq:w1w2w3w4ann4}
\end{equation}
where $q_{a}$ is defined as $q_i-(P_i\cdot q_i/M_i^2)P_i$, while   $q_{c}=q_f-(P_f\cdot q_f/M_f^2)P_f$. The coefficients $\alpha^{i,f}_{1,2}$ are given as $\alpha^{i}_{1}=m_b/(m_b+m_c),~\alpha^{i}_{2}=m_c/(m_b+m_c),~\alpha^{f}_{1}=m_{s(d)}/(m_{s(d)}+m_c),~\alpha^{f}_{2}=m_c/(m_{s(d)}+m_c)$, where $m_{b,c,s,d}$ are masses of the constituent quarks.
The parameters $\mathcal{F}_{i0,i\pm}(i\to f)$ and $\mathcal{F}_{f0,f\pm}(i\to f)$ are defined as
\begin{equation}
\begin{split}
\mathcal{F}&^{\nu}_{i0}(i\to f)=\mathrm{Tr}\left\{\varphi_i^{++}\gamma^{\nu}(1-\gamma_5)\right\},~~~~~~~~~~
\mathcal{F}^{\mu\nu}_{i\pm}(i\to f)=\frac{1}{2}\mathrm{Tr}\left\{\varphi_i^{++}\gamma^{\nu}(1-\gamma_5)(\not\!{Q}\gamma^{\mu}\pm\gamma^{\mu}{\not\!{Q}})\right\},\\
\mathcal{F}&^{\nu}_{f0}(i\to f)=\mathrm{Tr}\left\{\bar{\varphi_f}^{++}\gamma^{\nu}(1-\gamma_5)\right\},~~~~~~~
\mathcal{F}^{\mu\nu}_{f\pm}(i\to f)=\frac{1}{2}\mathrm{Tr}\left\{\bar{\varphi_f}^{++}\gamma^{\nu}(1-\gamma_5)(\gamma^{\mu}{\not\!{Q}}\pm\not\!{Q}\gamma^{\mu})\right\}.
\end{split}\label{eq:w1w2w3w4ann4defi}
\end{equation}

Using Eqs.~(\ref{eq:w1w2w3w4ann1}-\ref{eq:w1w2w3w4ann4defi}),  we now discuss the gauge invariant condition of the Ann hadronic currents calculated in BS method. One may note that examining whether $W_{ann}$ satisfies the gauge invariant condition is equivalent to checking whether $W_{ann}\cdot Q$ is zero. If we multiply Eqs.~(\ref{eq:w1w2w3w4ann1}-\ref{eq:w1w2w3w4ann4}) by $Q^{\mu}$, it is obvious that  $\left(W_{1ann}\cdot Q\right)+\left(W_{2ann}\cdot Q\right)$ cancels $\left(W_{4ann}\cdot Q\right)+\left(W_{3ann}\cdot Q\right)$. Hence, we have $W_{ann}\cdot Q=0$. This implies that the Ann hadronic currents in BS method indeed satisfy the gauge invariant condition.
We stress that there is no need to specify the initial or final state in the process of obtaining $W_{ann}\cdot Q=0$. Thus, our conclusion is quite general.

For $W_{CF}$, in this paper, we do not go into any details of their calculations, because $W_{CF}$s involved in the $B_c\to D_{(s)J}^{(*)}\mu\bar{\mu}$ transitions can be  obtained from $W_{CF}(B_c\to D_{(s)}^{(*)}\mu\bar{\mu})$s by properly replacing the final decay constants. (We refer to Ref.~\cite{Juwangpaper1} for more details on $W_{CF}(B_c\to D_{(s)}^{(*)}\mu\bar{\mu})$ calculation.)
The decay constants of the scalar  and axial-vector mesons can be found in Ref.~\cite{BSwavefuctionp-wave1}. But due to the angular momentum conservation condition, the longitudinal decay constants of the tensor mesons are zero. Hence, we have $W_{CF}(B_c\to D_{s2}^*(2573)(D_{2}^*(2460))\mu\bar{\mu})=0$.


\subsection{ Wave Functions  in BS Method}\label{subsec:2}

In BS method, the meson
is considered to be a bound state of two constituent quarks and can be described by the
BS wave functions~\cite{BS1}. In the framework of instantaneous approximation~\cite{ChangPRD49 3399}, the time component of the BS wave functions' arguments can be integrated out and the BS equations are reduced to the Salpeter equations. By means of solving the Salpeter equations, we obtain the wave function~\cite{BSwavefuctions-wave1,BSwavefuctions-wave2,BSwavefuctionp-wave1,BSwavefuctionp-wave2} for each meson.

In the present work, the mesons  $D^*_{s0}(2317)$, $D^*_{0}(2400)$, $D^*_{s2}(2573)$, $D^*_{2}(2460)$, $D_{s1}(2460,2536)$, $D_{1}(2420,2430)$ and $B_c$ are relevant. In the following paragraphs, their wave functions are introduced.

\begin{flushleft}
\textbf{(1)Wave Functions of $D^*_{s0}(2317)$ and $D^*_{0}(2400)$}
\end{flushleft}

Based on Ref.~\cite{pdg}, $J^{P}$s of $D^*_{s0}(2317)$ and $D^*_{0}(2400)$ mesons are $0^+$. In this paper, we consider them as  $^3P_0$ states. In the BS approach, the positive energy wave function
 for $^3P_0$ state can be expressed as \cite{wangzhihuipwave}
\begin{equation}
\begin{split}
\varphi_{^3P_0}^{++}=a_{1}\left(\not\!q_{_{P_{\bot}}}+a_2\frac{\not\!P  \not\!q_{_{P_{\bot}}}}{M}+a_3+a_4\frac{\not\!P  }{M}\right),
\end{split}\label{eq:wavefunction3p0}
\end{equation}
where the parameters $a_{1-4}$ can be found in Ref.~\cite{wangzhihuipwave}. 

\begin{flushleft}
\textbf{(2)Wave Functions of $D^*_{s2}(2573)$ and $D^*_{2}(2460)$}
\end{flushleft}

From Ref.~\cite{pdg}, $J^{P}$s of $D^*_{s2}(2536)$ and $D^*_{2}(2460)$ mesons are $2^+$. In this paper, they are described as $^3P_2$ states.
The positive energy wave function for $^3P_2$ state is \cite{wangzhihuipwave}
\begin{equation}
\begin{split}
\varphi_{^3P_2}^{++}=\epsilon^{T}_{\mu\nu}q_{_{P_{\bot}}}^{\nu}\left\{q_{_{P_{\bot}}}^{\mu}\left[d_1+d_2\frac{\not\!P}{M}+d_3\frac{\not\!q_{_{P_{\bot}}}
}{M}-d_4\frac{\not\!P \not\! q_{_{P_{\bot}}}}{M^2}\right]+\gamma^{\mu}\left[d_5+d_6\frac{\not\!P}{M}+d_7\frac{\not\!q_{_{P_{\bot}}}}{M}+d_8\frac{\not\!P \not\! q_{_{P_{\bot}}}}{M^2}\right]\right\},
\end{split}\label{eq:wavefunction3p2}\end{equation}
where $\epsilon^{T}_{\mu\nu}$ is the polarization tensor. The parameters $d_{1-8}$ can be found in Refs.~\cite{wangzhihuipwave,BSwavefuctionp-wave2}.

\begin{flushleft}
\textbf{(3)Wave Functions of $D_{s1}(2460,2536)$ and $D_{1}(2420,2430)$}
\end{flushleft}

Unlike the mesons introduced above, $D_{s1}(2460,2536)$ and $D_{1}(2420,2430)$ can not be described by the pure $^{(2S+1)}L_J$ states.
Based on~\cite{Isgur:1990jf,Cheng2006dm}, we consider them as the mixtures of the $^1P_1$ and $^3P_1$ states, namely,
\begin{equation}
\begin{split}
\left(
\begin{array}{c}
 |D_1(2430)\rangle \\
|D_1(2420)\rangle
\end{array}
\right)=\mathcal{A}
\left(
\begin{array}{c}
 |D_{^1P_1}\rangle\\
|D_{^3P_1}\rangle
\end{array}\right)\equiv
\left(
\begin{array}{cc}
 \sin\alpha&\cos\alpha\\
\cos\alpha&-\sin\alpha
\end{array}\right)
\left(
\begin{array}{c}
 |D_{^1P_1}\rangle\\
|D_{^3P_1}\rangle
\end{array}\right),\\
\left(
\begin{array}{c}
 |D_{s1}(2460)\rangle \\
|D_{s1}(2536)\rangle
\end{array}
\right)=\mathcal{B}
\left(
\begin{array}{c}
 |D_{s^1P_1}\rangle\\
|D_{s^3P_1}\rangle
\end{array}\right)\equiv
\left(
\begin{array}{cc}
 \sin\beta&\cos\beta\\
\cos\beta&-\sin\beta
\end{array}\right)
\left(
\begin{array}{c}
 |D_{s^1P_1}\rangle\\
|D_{s^3P_1}\rangle
\end{array}\right),
\end{split}
\label{eq:mixDs}
\end{equation}
where $\alpha=\theta-\arctan (\sqrt{1/2})$ and $\beta=\theta_s-\arctan (\sqrt{1/2})$.
Based on the experimental observation \cite{Abe2003zm} and the  discussions in Ref.~\cite{Cheng2006dm}, the mixing angle $\theta=5.7^{\circ}$ is used in this paper. Besides, according to the analysis in the quark potential model \cite{Cheng2003id}, $\theta_s=7^{\circ}$ is employed.

From Eq.~(\ref{eq:mixDs}), the wave functions of  $D_{s1}(2460,2536)$ and $D_{1}(2420,2430)$  can be constructed from the ones of $^1P_1$ and $^3P_1$ states.
 In the BS method, the positive energy wave functions of  $^1P_1$ and $^3P_1$ states \cite{wangzhihuipwave} are
\begin{equation}
\begin{split}
\varphi_{^1P_1}^{++}=&b_{1}\left(\epsilon_A\cdot q_{_{P_{\bot}}}\right)\left(1+b_2\frac{\not\!P}{M}+b_3\not\!q_{_{P_{\bot}}}-b_4\frac{\not\!P\not\!q_{_{P_{\bot}}}}{M}\right)\gamma_5,\\
\varphi_{^3P_1}^{++}=&ic_1\epsilon_{\mu\nu\alpha\beta}P^{\nu}q_{_{P_{\bot}}}^{\alpha}\epsilon_A^{\beta}\left(M\gamma^{\mu}+c_2\gamma^{\mu}\not\!P+
c_3\gamma^{\mu} \not\!q_{_{P_{\bot}}}+c_4\gamma^{\mu} \not\!P \not\!q_{_{P_{\bot}}} \right)/M^2,
\end{split}\label{eq:wavefunction1p13p1}
\end{equation}where $\epsilon^{A}_{\mu}$ is the polarization vector of the axial-vector meson.
The explicit expressions of $b_{1-4}$ and $c_{1-4}$ can be found in Ref.~\cite{wangzhihuipwave} and their numerical values can be obtained by solving the Salpeter equations~\cite{BSwavefuctionp-wave1}. In the processes of solving the Salpeter equations, the masses of ${^1P_1}$ and ${^3P_1}$ states, namely, $M_{D_{(s)^1P_1}}$ and $M_{D_{(s)^3P_1}}$, are required.  In analogy to the case of $\eta_1-\eta_8$ mixing~\cite{Ecker:1988te}, we determine them from the following relationships~\cite{Verma:2011yw,Cheng:2003sm},
\begin{equation}
\begin{split}
 \mathcal{A}^{\dag}\left(
\begin{array}{cc}
 M^2_{D_1(2430)} & 0 \\
0 & M^2_{D_1(2420)}
\end{array}
\right)\mathcal{A}
=
\left(
\begin{array}{cc}
 M^2_{D_{^1P_1}} & \delta \\
\delta & M{}^2_{D_{^3P_1}}
\end{array}
\right),\\
 \mathcal{B}^{\dag}\left(
\begin{array}{cc}
 M^2_{D_{s1}(2460)} & 0 \\
0 & M^2_{D_{s1}(2536)}
\end{array}
\right)\mathcal{B}
=
\left(
\begin{array}{cc}
 M^2_{D_{s^1P_1}} & \delta_s \\
\delta_s & M{}^2_{D_{s^3P_1}}
\end{array}
\right),
\end{split}\label{eq:massmix}
\end{equation}
where $M_{D_1(2420,2430)}$ and $M_{D_{s1}(2460,2536)}$ stand for the physical masses and   we take them from  Ref.~\cite{pdg}.%

\begin{flushleft}
\textbf{(4)Wave Function of $B_c$}
\end{flushleft}


 The $B_c$ meson is considered as a $^1S_0$ state, whose the positive energy wave function can be written as~\cite{BSwavefuctions-wave1},
\begin{equation}
\varphi^{++}_{^1S_0}=e_1\left[e_2+\frac{\not\!P}{M}+\not\!q_{_{P_{\bot}}}e_3+\frac{\not\!q_{_{P_{\bot}}}\not\!P-\not\!P\not\!q_{_{P_{\bot}}}}{2M}e_4\right]\gamma_5.
\label{eq:wavefunction1s0}\end{equation}
where the parameters $e_{1-4}$ can be found in Ref.~\cite{BSwavefuctions-wave1}.

\subsection{ Calculations of Hadronic Matrix Elements }\label{SecHadronicCurrentsin BS}

In this part, we calculate the hadronic currents through the formalism introduced above.
Since  $W^{\mu}$s have been investigated extensively 
in our previous papers \cite{wangzhihuipwave,Yue:2013gxa,Wang:2011jt,Fu:2011zzo,JIANG:2013ufa,huifengfuJHEP}, here we do not introduce the $W^{\mu}$ calculations  but pay more attentions to $W^{\mu}_{T,ann}$s. Please recall that
 $W^{\mu}_{T}$s have been expressed in combinations of $\mathcal{Y}^{\mu\nu}_{V,A}$s within Eq.~\eqref{eq:Mandelstam Formalism}, while in Eqs.~(\ref{eq:w1w2w3w4ann1}-\ref{eq:w1w2w3w4ann4}),  $W^{\mu}_{ann}$s are  written in terms of $\mathcal{F}_{i,f0(\pm)}$s. Hence, in order to obtain $W^{\mu}_{T,ann}$, it is convenient to compute $\mathcal{Y}^{\mu\nu}_{V,A}$s and $\mathcal{F}_{i,f0(\pm)}$s first of all. From their definitions in Eq.~\eqref{eq:YYT} and Eq.~\eqref{eq:w1w2w3w4ann4defi}, we see that the calculations of $\mathcal{Y}^{\mu\nu}_{V,A}$s and $\mathcal{F}_{i,f0(\pm)}$s are channel-dependent and the channels under our consideration include $P\to S,T,A$ transitions, where $P,~S,~T,~A$ are the abbreviations for pseudo-scalar, scalar, tensor, axial-vector mesons, respectively.

\subsubsection{Hadronic Matrix Elements of $P\to S$ processes}\label{SecpStransitions}

First, we introduce the details of the $\mathcal{Y}^{\mu\nu}_{V,A}(P\to S)$ estimations.
We have expressed $\mathcal{Y}^{\mu\nu}_{V,A}$s  as the overlapping integrals of $\varphi^{++}_{i,f}$s in Eq.~\eqref{eq:YYT}.
In the $P\to S$ processes,  the initial wave function $\varphi^{++}_{i}$  corresponds to $\varphi^{++}_{^1S_0}$, while $\varphi^{++}_{f}$ should be $\varphi^{++}_{^3P_0}$. The expressions of $\varphi^{++}_{^1S_0}$ and $\varphi^{++}_{^3P_0}$ are given in Eq.~\eqref{eq:wavefunction1s0} and Eq.~\eqref{eq:wavefunction3p0}, respectively. Substituting Eqs.~(\ref{eq:wavefunction3p0}, \ref{eq:wavefunction1s0}) into Eq.~\eqref{eq:YYT},  the hadronic matrix elements $\mathcal{Y}^{\mu\nu}_{V,A}$s can be obtained.
In light of  the forbidden parity, we have $\mathcal{Y}^{\mu\nu}_{V}(P\rightarrow S)=0$, while for $\mathcal{Y}^{\mu\nu}_{A}(P\rightarrow S)$, it reads
\begin{equation}
\begin{split}
\mathcal{Y}^{\mu\nu}&_{A}(P\rightarrow S)=\int\frac{d^3\vec{q}}{(2\pi)^3}\frac{-4  a_1 e_1}{M_f M_i}\left\{M_i \left[g^{\mu \nu } \left(q_a\cdot q_b a_2 e_3 e_f+e_4 M_f q_a\cdot q_b +a_4 e_4 P_f\cdot q_a+a_4 e_2 e_f\right.\right.\right.\\
&\left.\left.-a_3 M_f\right)+q_b^{\mu } \left(q_a^{\nu } a_2 e_3 e_f+q_a^{\nu }e_4 M_f+a_2 P_f^{\nu }\right)-q_a^{\mu } \left(q_b^{\nu } a_2 e_3 e_f+q_b^{\nu }e_4 M_f+a_4 e_4 P_f^{\nu }\right)\right]\\
&-a_2 e_3 g^{\mu \nu } P_f\cdot q_a P_i\cdot q_b-P_i^{\mu } \left[q_b^{\nu } \left(e_2 M_f-a_2 e_3 P_f\cdot q_a\right)+P_f^{\nu } \left(a_2 e_3 q_a\cdot q_b+a_4 e_2\right)\right.\\
&\left.+a_3 e_3 M_f q_a^{\nu }\right]+P_f^{\mu } \left[q_a^{\nu } \left(a_4 e_4 M_i-a_2 e_3 P_i\cdot q_b\right)+P_i^{\nu } \left(a_2 e_3 q_a\cdot q_b+a_4 e_2\right)-a_2 M_i q_b^{\nu }\right]\\
&\left.-a_2 e_3 q_b^{\mu } P_i^{\nu } P_f\cdot q_a+a_2 e_3 q_a^{\mu } P_f^{\nu } P_i\cdot q_b+a_3 e_3 M_f q_a^{\mu } P_i^{\nu }+e_2 M_f g^{\mu \nu } P_i\cdot q_b+e_2 M_f q_b^{\mu } P_i^{\nu }\right\},
\end{split}
\end{equation}
where the definition of $q_a$ has been given in Sec.~\ref{subsec:1}, while $q_{b}$ is the relative momentum of the final meson.  Due to the spectator approximation, the retarded relationship between $q_{a}$ and $q_{b}$ reads \cite{chao-hsi chang1} \begin{equation}q_b^{\mu}= q_a+\alpha^{f}_{2}P_f^{\mu}-\alpha^{f}_{2}E_f P_i /M_i.\label{eq:retarded relationships}\end{equation}


Now we turn to the discussions of $\mathcal{F}_{i,f0(\pm)}(P\to S)$s.
In Eq.~\eqref{eq:w1w2w3w4ann4defi}, $\mathcal{F}_{i0(\pm)}$s are written in terms of $\varphi^{++}_{i}$s, while $\mathcal{F}_{f0(\pm)}$s are shown in the integrals of $\varphi^{++}_{f}$s. Similar to the calculations of $\mathcal{Y}^{\mu\nu}_{V,A}(P\to S)$s,  $\varphi^{++}_{i(f)}$  corresponds to $\varphi^{++}_{^1S_0(^3P_0)}$. So we have
\begin{equation}
\begin{split}
\mathcal{F}&^{\nu}_{i0}(P\to S)=4 e_1 \left(e_3 M_i q_a+P_i\right)^{\nu},\\
\mathcal{F}&^{\mu\nu}_{i+}(P\to S)={4 e_1 \left[-g^{\nu \mu } \left(e_3 M_i Q\cdot q_a+Q\cdot P_i\right)+Q^{\nu } \left(e_3 M_i q_a^{\mu }+P_i^{\mu }\right)+e_3 M_i Q^{\mu } q_a^{\nu }+Q^{\mu } P_i^{\nu }\right]},\\
\mathcal{F}&^{\mu\nu}_{i-}(P\to S)={4 i e_1 \left(e_3 M_i \epsilon ^{\nu \mu Qq_a}+\epsilon ^{\nu \mu QP_i}\right)},\\
\mathcal{F}&^{\nu}_{f0}(P\to S)=4a_1(a_4 P_{f}+{M_f}q_{c })^{\nu},\\
\mathcal{F}&^{\mu\nu}_{f+}(P\to S)={4 a_1 \left\{-g^{\nu \mu } \left(a_4 Q\cdot P_f+M_f Q\cdot q_c\right)+Q^{\nu } \left(a_4 P_f^{\mu }+M_f q_c^{\mu }\right)+a_4 Q^{\mu } P_f^{\nu }+M_f Q^{\mu } q_c^{\nu }\right\}},\\
\mathcal{F}&^{\mu\nu}_{f-}(P\to S)=-{4 i a_1 \left(a_4 \epsilon ^{\nu \mu QP_f}+M_f \epsilon ^{\nu \mu Qq_c}\right)}.\\
\end{split}\label{eq:p-sAnnhadroniccurents1}
\end{equation}

\clearpage
\subsubsection{Hadronic Matrix Elements of $P\to T$ processes}

Here we deal with $\mathcal{Y}^{\mu\nu}_{V,A}$ in the $P\to T$ precesses.
The calculations of $\mathcal{Y}^{\mu\nu}_{V,A}(P\to T)$ are similar to the ones of $\mathcal{Y}^{\mu\nu}_{V,A}(P\to S)$, except replacing the final wave function $\varphi^{++}_{^3P_0}$ by $\varphi^{++}_{^3P_2}$. The expression of $\varphi^{++}_{^3P_2}$ can be found in Eq.~\eqref{eq:wavefunction3p2}.
Hence, we have
\begin{equation}
\begin{split}
\mathcal{Y}^{\mu\nu}&_{V}(P\rightarrow T)=\int\frac{d^3\vec{q}}{(2\pi)^3}\frac{- 4 i e_1}{M_f^2 M_i^2}\epsilon^{T}_{\alpha\beta}q_{b}^{\beta}\left\{\mathcal{F}^{\alpha\mu\nu}_{V1}+\mathcal{F}^{\alpha\mu\nu}_{V2}+\mathcal{F}^{\alpha\mu\nu}_{V3}+\mathcal{F}^{\alpha\mu\nu}_{V4
}+\mathcal{F}^{\alpha\mu\nu}_{V5}+\mathcal{F}^{\alpha\mu\nu}_{V6}+\mathcal{F}^{\alpha\mu\nu}_{V7}
\right\},\\
\mathcal{Y}^{\mu\nu}&_{A}(P\rightarrow T)=\int\frac{d^3\vec{q}}{(2\pi)^3}\frac{-4 e_1}{M_f^2 M_i}\epsilon^{T}_{\alpha\beta}q_{b}^{\beta}\left\{-e_3\mathcal{F}^{\alpha\mu\nu}_{A1}-e_2M_i\mathcal{F}^{\alpha\mu\nu}_{A2}
-e_4M_i\mathcal{F}^{\alpha\mu\nu}_{A3}\right.\}.
\end{split}
\end{equation}
The expressions of $\mathcal{F}^{\alpha\mu\nu}_{Vl}$ and $\mathcal{F}^{\alpha\mu\nu}_{Ak}$, where $l=1,\dots ,7$ and $k=1,2,3$, are presented in Appendix.~A.

 Next, we pay attentions   to $\mathcal{F}_{i0(\pm)}(P\to T)$s.
 From Eq.~\eqref{eq:w1w2w3w4ann4defi}, we see that
 $\mathcal{F}_{i0(\pm)}(P\to T)$s are the same as  $\mathcal{F}_{i0(\pm)}(P\to S)$s,  due to the identical initial meson $B_c$ in the decays $P\to S,T$. The discussions of $\mathcal{F}_{i0(\pm)}(P\to S)$s have been performed in Sec.~\ref{SecpStransitions}.
  But for $\mathcal{F}_{f0(\pm)}(P\to T)$s, the situations are different. They should be calculated through Eq.~\eqref{eq:w1w2w3w4ann4defi}, with the final wave functions $\varphi^{++}_{f}$ being $\varphi^{++}_{^3P_2}$.  After factoring the polarization tensor out,  we have
\begin{equation}
\begin{split}
\mathcal{F}&_{f0}^{\nu}(P\to T)=\mathcal{E}_{f0}^{\nu\delta}(^3P_2)\epsilon^{T}_{\delta\sigma}q_{c}^{\sigma}/M_f,~~~\mathcal{F}^{\mu\nu}_{f+}(P\to T)=\mathcal{E}^{\mu\nu\delta}_{f+}(^3P_2)\epsilon^{T}_{\delta\sigma}q_{c}^{\sigma}/M_f,\\
\mathcal{F}&^{\mu\nu}_{f-}(P\to T)=\mathcal{E}^{\mu\nu\delta}_{f-}(^3P_2)\epsilon^{T}_{\delta\sigma}q_{c}^{\sigma}/M_f,
\end{split}
\end{equation}
where $\mathcal{E}_{f0,f\pm}(^3P_2)$ are defined as
\begin{equation}
\begin{split}
\mathcal{E}&_{f0}^{\nu\delta}(^3P_2)=8 \left\{\left(d_2 M_f-d_8\right) P_f^{\nu }q^{\delta}_c+M_f \left(d_3 q_c^{\nu }  q^{\delta}_c+d_5 g^{\delta\nu } M_f\right)+i d_8 \epsilon ^{\nu \delta P_f q_c}\right\},\\
\mathcal{E}&^{\mu\nu\delta}_{f+}(^3P_2)=4 M_f q_c^{\delta } \left[-g^{\nu \mu } \left(d_3 Q\cdot q_c+d_2 Q\cdot P_f\right)+Q^{\nu } \left(d_3 q_c^{\mu }+d_2 P_f^{\mu }\right)+d_3 Q^{\mu } q_c^{\nu }+d_2 Q^{\mu } P_f^{\nu }\right]\\
&-2 i d_8 \left[-2 i q_c^{\delta } \left(-g^{\nu \mu } Q\cdot P_f+Q^{\mu } P_f^{\nu }+Q^{\nu } P_f^{\mu }\right)-2 g^{\nu \mu } \epsilon ^{\delta QP_fq_c}+2 Q^{\nu } \epsilon ^{\delta \mu P_fq_c}+Q^{\delta } \epsilon ^{\nu \mu P_fq_c}\right.\\
&\left.-2 P_f^{\mu } \epsilon ^{\nu \delta Qq_c}+2 q_c^{\mu } \epsilon ^{\nu \delta QP_f}\right]+4 d_5 M_f^2 \left(Q^{\nu } g^{\delta \mu }-Q^{\delta } g^{\nu \mu }+Q^{\mu } g^{\nu \delta }\right),\\
\mathcal{E}&^{\mu\nu\delta}_{f-}(^3P_2)=2 i d_8 \left\{-2 g^{\delta \mu } \epsilon ^{\nu QP_fq_c}+2 q_c^{\nu } \left[\epsilon ^{\delta \mu QP_f}+i \left(Q^{\delta } P_f^{\mu }-g^{\delta \mu } Q\cdot P_f\right)\right]\right.\\
&-2 P_f^{\nu } \left[\epsilon ^{\delta \mu Qq_c}+i \left(Q^{\delta } q_c^{\mu }-g^{\delta \mu } Q\cdot q_c\right)\right]+2 g^{\nu \delta } \left[\epsilon ^{\mu QP_fq_c}+i \left(q_c^{\mu } Q\cdot P_f-P_f^{\mu } Q\cdot q_c\right)\right]\\
&\left.+Q^{\delta } \epsilon ^{\nu \mu P_fq_c}+2 \left(q_c^{\delta } \epsilon ^{\nu \mu QP_f}+Q^{\mu } \epsilon ^{\nu \delta P_fq_c}+Q\cdot P_f \epsilon ^{\nu \delta \mu q_c}-Q\cdot q_c \epsilon ^{\nu \delta \mu P_f}\right)\right\}\\
&-4 i M_f q_c^{\delta } \left(d_3 \epsilon ^{\nu \mu Qq_c}+d_2 \epsilon ^{\nu \mu QP_f}\right)-4 i d_5 M_f^2 \epsilon ^{\nu \delta \mu Q}.
\end{split}
\end{equation}

\clearpage
\subsubsection{
Hadronic Matrix Elements of $P\to A$ processes}

Due to the mixing nature of the final mesons as formulated in Eq.~\eqref{eq:mixDs}, the calculations of $\mathcal{Y}^{\mu\nu}_{V,A}(P\to A)$s and $\mathcal{F}_{i,f0(\pm)}(P\to A)$s are different from the cases of $P\to S $ and $P\to T$.
In order to obtain $\mathcal{Y}^{\mu\nu}_{V,A}(P\to A)$s and $\mathcal{F}_{i,f0(\pm)}(P\to A)$s, first of all, we  compute $\mathcal{Y}^{\mu\nu}_{V,A}(P\to A_{^3P_1,^1P_1})$s and $\mathcal{F}_{i,f0(\pm)}(P\to A_{^3P_1,^1P_1})$s. And then, based on the mixing relationships in Eq.~\eqref{eq:mixDs}, we  combine the results of $P\rightarrow A_{^3P_1}$ and $P\rightarrow A_{^1P_1}$.

For $\mathcal{Y}^{\mu\nu}_{V,A}(P\to A_{^3P_1,^1P_1})$s, we calculate them from Eq.~\eqref{eq:YYT}, with the initial wave function $\varphi^{++}_{i}$  being $\varphi^{++}_{^1S_0}$ and the final one $\varphi^{++}_{f}$ being $\varphi^{++}_{^3P_1,^1P_1}$. The expressions of $\varphi^{++}_{^3P_1,^1P_1}$ are given in Eq.~\eqref{eq:wavefunction1p13p1}, while the initial ones $\varphi^{++}_{^1S_0}$ is shown in Eq.~\eqref {eq:wavefunction1s0}. The results of $\mathcal{Y}^{\mu\nu}_{V,A}(P\to A_{^3P_1,^1P_1})$s read
\begin{equation}
\begin{split}
\mathcal{Y}_V^{\mu\nu}&(P\rightarrow A_{^3P_1})=\int\frac{d^3\vec{q}}{(2\pi)^3}\frac{- 8 c_1 c_4 e_1}{M_f^2 M_i^2} \epsilon ^{\nu P_fq_b\epsilon _A} \left[e_4 \left(M_i^2 \epsilon ^{\mu P_fq_aq_b}+2 P_i^{\mu } \epsilon ^{P_fP_iq_aq_b}-2 P_f\cdot P_i \epsilon ^{\mu P_iq_aq_b}\right.\right.\\
&\left.\left.+2 P_i\cdot q_b \epsilon ^{\mu P_fP_iq_a}\right)-e_2 M_i \epsilon ^{\mu P_fP_iq_b}\right],\\
\mathcal{Y}^{\mu\nu}_{A}&(P\rightarrow A_{^3P_1})=\int\frac{d^3\vec{q}}{(2\pi)^3}\frac{-8 i c_1 e_1 }{M_f^2 M_i}\epsilon ^{\nu P_fq_b\epsilon _A} \left\{q_b^{\mu } \left[M_i \left(c_4 e_4 P_f\cdot q_a+2 c_3 M_f^2\right)+c_4 e_2 P_f\cdot P_i\right]\right.\\
&\left.-P_f^{\mu } \left[c_4 \left(e_4 M_i q_a\cdot q_b+e_2 P_i\cdot q_b\right)-2 c_2 M_i\right]+M_f \left(e_4 M_i q_a^{\mu }+e_2 P_i^{\mu }\right)\right\},\\
\mathcal{Y}_V^{\mu\nu}&(P\rightarrow A_{^1P_1})=\int\frac{d^3\vec{q}}{(2\pi)^3}\frac{- 4 b_1 e_1 q_b\cdot \epsilon _A }{M_f M_i}\left\{M_i \left[g^{\mu \nu } \left(e_4 b_3 M_f q_a\cdot q_b+e_4 b_2 P_f\cdot q_a+M_f\right)-e_4q_a^{\mu } \right.\right.\\
&\left.\left(b_3 M_f q_b^{\nu }+b_2 P_f^{\nu }\right)+q_b^{\mu } \left(b_3 e_4 M_f q_a^{\nu }+b_4 P_f^{\nu }\right)\right]-b_4 e_3 g^{\mu \nu } P_f\cdot q_a P_i\cdot q_b+b_4 e_3 g^{\mu \nu } q_a\cdot q_b P_f\cdot P_i\\
&-P_i^{\mu } \left[q_b^{\nu } \left(b_3 e_2 M_f-b_4 e_3 P_f\cdot q_a\right)+P_f^{\nu } \left(b_4 e_3 q_a\cdot q_b+b_2 e_2\right)-e_3 M_f q_a^{\nu }\right]+P_f^{\mu } \left[q_a^{\nu } \left(b_2 e_4 M_i\right.\right.\\
&\left.\left.-b_4 e_3 P_i\cdot q_b\right)+P_i^{\nu } \left(b_4 e_3 q_a\cdot q_b+b_2 e_2\right)-b_4 M_i q_b^{\nu }\right]+b_4 e_3 q_a^{\nu } q_b^{\mu } P_f\cdot P_i-b_4 e_3 q_a^{\mu } q_b^{\nu } P_f\cdot P_i\\
&-b_4 e_3 q_b^{\mu } P_i^{\nu } P_f\cdot q_a+b_4 e_3 q_a^{\mu } P_f^{\nu } P_i\cdot q_b-e_3 M_f q_a^{\mu } P_i^{\nu }+b_3 e_2 M_f g^{\mu \nu } P_i\cdot q_b+b_2 e_2 g^{\mu \nu } P_f\cdot P_i\\
&\left.+b_3 e_2 M_f q_b^{\mu } P_i^{\nu }\right\},\\
\mathcal{Y}^{\mu\nu}_{A}&(P\rightarrow A_{^1P_1})=\int\frac{d^3\vec{q}}{(2\pi)^3}\frac{ 4 i b_1 e_1 q_b\cdot \epsilon _A}{M_f M_i^2} \left\{M_i \left[b_4 e_3 \left(-g^{\mu \nu } \epsilon ^{P_fP_iq_aq_b}-P_f^{\mu } \epsilon ^{\nu P_iq_aq_b}\right.\right.\right.\\
&+P_i^{\mu } \epsilon ^{\nu P_fq_aq_b}+q_b^{\mu } \epsilon ^{\nu P_fP_iq_a}+P_f^{\nu } \epsilon ^{\mu P_iq_aq_b}-P_i^{\nu } \epsilon ^{\mu P_fq_aq_b}-q_b^{\nu } \epsilon ^{\mu P_fP_iq_a}+P_f\cdot P_i \epsilon ^{\mu \nu q_aq_b}\\
&\left.+P_f\cdot q_a \epsilon ^{\mu \nu P_iq_b}+P_i\cdot q_b \epsilon ^{\mu \nu P_fq_a}\right)+M_f \left(b_3 e_2 \epsilon ^{\mu \nu P_iq_b}-e_3 \epsilon ^{\mu \nu P_iq_a}\right)+\left(b_4 e_3 q_a\cdot q_b\right.\\
&\left.\left.-b_2 e_2\right) \epsilon ^{\mu \nu P_fP_i}\right]-M_i^2 \left(e_4 b_3 M_f \epsilon ^{\mu \nu q_aq_b}-e_4 b_2 \epsilon ^{\mu \nu P_fq_a}+b_4 \epsilon ^{\mu \nu P_fq_b}\right)+b_4 \left(e_3 M_i q_a^{\mu }\right.\\
&\left.+2 P_i^{\mu }\right) \epsilon ^{\nu P_fP_iq_b}-b_4 \left(e_3 M_i q_a^{\nu }+2 P_i^{\nu }\right) \epsilon ^{\mu P_fP_iq_b}+2 \left[e_4 \left(b_3 M_f P_i^{\nu } \epsilon ^{\mu P_iq_aq_b}-b_3 M_f P_i^{\mu } \epsilon ^{\nu P_iq_aq_b}\right.\right.\\
&\left.-\epsilon ^{\mu \nu P_iq_a} \left(b_3 M_f P_i\cdot q_b+b_2 P_f\cdot P_i\right)+b_2 P_i^{\mu } \left(-\epsilon ^{\nu P_fP_iq_a}\right)+b_2 P_i^{\nu } \epsilon ^{\mu P_fP_iq_a}\right)\\
&\left.\left.+b_4 P_f\cdot P_i \epsilon ^{\mu \nu P_iq_b}+b_4 P_i\cdot q_b \epsilon ^{\mu \nu P_fP_i}\right]\right\}.
\end{split}\label{eq:P-A2}
\end{equation}

For $\mathcal{F}_{i0(\pm)}(P\to A_{^3P_1,^1P_1})$s, we see that they are identical to $\mathcal{F}_{i0(\pm)}(P\to S)$s. But as to $\mathcal{F}_{f0(\pm)}(P\to A_{^3P_1,^1P_1})$s, we need to compute them by substituting $\varphi^{++}_{^3P_1,^1P_1}$ into Eq.~\eqref{eq:w1w2w3w4ann4defi}. The results read
\begin{equation}
\begin{split}
\mathcal{F}&^{\alpha}_{f0}(A_{^1P_1})=4 b_1 q_c\cdot \epsilon _A \left(b_3 M_f q_c^{\alpha }+b_2 P_f^{\alpha }\right),\\
\mathcal{F}&^{\mu\alpha}_{f+}(A_{^1P_1})=\frac{4 b_1 q_c\cdot \epsilon _A }{M_f}\left[b_3 M_f \left(-g^{\alpha \mu } Q\cdot q_c+Q^{\mu } q_c^{\alpha }+Q^{\alpha } q_c^{\mu }\right)+b_2 \left(-g^{\alpha \mu } Q\cdot P_f+Q^{\mu } P_f^{\alpha }+Q^{\alpha } P_f^{\mu }\right)\right],\\
\mathcal{F}&^{\mu\alpha}_{f-}(A_{^1P_1})=-\frac{4 i b_1 q_c\cdot \epsilon _A \left(b_3 M_f \epsilon ^{\alpha \mu Qq_c}+b_2 \epsilon ^{\alpha \mu QP_f}\right)}{M_f},\\
\mathcal{F}&^{\alpha}_{f0}(A_{^3P_1})=4 c_1 \left[c_4 \left(q_c^{\alpha } M_f q_c\cdot \epsilon _A-q_c^2 \epsilon _A^{\alpha }\right)-{i \epsilon ^{\alpha P_fq_c\epsilon _A}}\right],\\
\mathcal{F}&^{\mu\alpha}_{f+}(A_{^3P_1})=\frac{1}{M_f}4 c_1 \left\{c_4 M_f \left[q_c\cdot \epsilon _A \left(-g^{\alpha \mu } Q\cdot q_c+Q^{\mu } q_c^{\alpha }+Q^{\alpha } q_c^{\mu }\right)-q_c^2 \left(-g^{\alpha \mu } Q\cdot \epsilon _A\right.\right.\right.\\
&\left.\left.\left.+Q^{\mu } \epsilon _A^{\alpha }+Q^{\alpha } \epsilon _A^{\mu }\right)\right]-i \left[g^{\alpha \mu } \left(-\epsilon ^{QP_fq_c\epsilon _A}\right)+Q^{\alpha } \epsilon ^{\mu P_fq_c\epsilon _A}+Q^{\mu } \epsilon ^{\alpha P_fq_c\epsilon _A}\right]\right\},\\
\mathcal{F}&^{\mu\alpha}_{f-}(A_{^3P_1})=\frac{1}{M_f^2}4 c_1 \left\{M_f \left[\epsilon _A^{\alpha } \left(q_c^{\mu } Q\cdot P_f-P_f^{\mu } Q\cdot q_c\right)+q_c^{\alpha } \left(P_f^{\mu } Q\cdot \epsilon _A-\epsilon _A^{\mu } Q\cdot P_f\right)\right.\right.\\
&\left.+P_f^{\alpha } \left(\epsilon _A^{\mu } Q\cdot q_c-q_c^{\mu } Q\cdot \epsilon _A\right)\right]+i c_4 \left[\left(q_c^{\mu } P_f^{\alpha }-q_c^{\alpha } P_f^{\mu }\right) \epsilon ^{QP_fq_c\epsilon _A}+\epsilon ^{\mu P_fq_c\epsilon _A} \left(q_c^{\alpha } Q\cdot P_f\right.\right.\\
&\left.\left.\left.-P_f^{\alpha } Q\cdot q_c\right)+\epsilon ^{\alpha P_fq_c\epsilon _A} \left(P_f^{\mu } Q\cdot q_c-q_c^{\mu } Q\cdot P_f\right)\right]\right\}.
\end{split}\label{eq:1p13p1annparas}
\end{equation}
Finally, with the results above and the mixing relationship in Eq.~\eqref{eq:mixDs}, we can calculate the hadronic matrix elements of the physical processes from
\begin{equation}
\begin{split}
&\left(
\begin{array}{c}
 \mathcal{Y}^{\mu\nu}_{V,A~~}(B_c\to D_1(2430)) \\
\mathcal{Y}^{\mu\nu}_{V,A~~}(B_c\to D_1(2420))
\end{array}
\right)=\mathcal{A}
\left(
\begin{array}{c}
  \mathcal{Y}^{\mu\nu}_{V,A~~}(B_c\to D_{^1P_1})\\
\mathcal{Y}^{\mu\nu}_{V,A~~}(B_c\to D_{^3P_1})
\end{array}\right),\\
&\left(
\begin{array}{c}
 \mathcal{Y}^{\mu\nu}_{V,A~~}(B_c\to D_{s1}(2460))\\
\mathcal{Y}^{\mu\nu}_{V,A~~}(B_c\to D_{s1}(2536))
\end{array}
\right)=\mathcal{B}
\left(
\begin{array}{c}
\mathcal{Y}^{\mu\nu}_{V,A~~}(B_c\to D_{s^1P_1})\\
\mathcal{Y}^{\mu\nu}_{V,A~~}(B_c\to D_{s^3P_1})
\end{array}\right),\\
&\left(
\begin{array}{c}
 \mathcal{F}_{f0(\pm)}(B_c\to D_1(2430)) \\
\mathcal{F}_{f0(\pm)}(B_c\to D_1(2420))
\end{array}
\right)=\mathcal{A}
\left(
\begin{array}{c}
  \mathcal{F}_{f0(\pm)}(B_c\to D_{^1P_1})\\
\mathcal{F}_{f0(\pm)}(B_c\to D_{^3P_1})
\end{array}\right),\\
&\left(
\begin{array}{c}
 \mathcal{F}_{f0(\pm)}(B_c\to D_{s1}(2460))\\
\mathcal{F}_{f0(\pm)}(B_c\to D_{s1}(2536))
\end{array}
\right)=\mathcal{B}
\left(
\begin{array}{c}
\mathcal{F}_{f0(\pm)}(B_c\to D_{s^1P_1})\\
\mathcal{F}_{f0(\pm)}(B_c\to D_{s^3P_1})
\end{array}\right).
\end{split}\label{eq:P-A3}
\end{equation}
During our calculations of Eq.~\eqref{eq:P-A3}, to avoid the kinematic confusion, we consider $M_f$ in Eqs.~(\ref{eq:P-A2}-\ref{eq:1p13p1annparas}) as the physical mass of the finial meson. (In this paper, the masses of $^1P_1$ and $^3P_1$ states introduced in Eq~\eqref{eq:massmix} are used only in solving the BS equations.) This approximation can also be found in the investigations of $B\to K_1(1270,1400)l\bar{l}$~\cite{Hatanaka:2008guK,Bashiry:2009wqK2,Bashiry:2009whK1,Ahmed:2011vrK,Li:2011nfK}.

\subsection{The Definitions of Form Factors}\label{SecDefinFormFacotors}
In the previous parts, we show how to calculate the hadronic currents. In order to show their results conveniently, here  we parameterize the hadronic matrix elements in terms of the form factors.
In this paper, we do not define the form factors of $W_{CF}$s, because as introduced in Sec.~\ref{subsec:1}, $W^{\mu}_{CF}(P\rightarrow S,A)$ can be obtained from $W^{\mu}_{CF}(P\rightarrow P,V)$ by some trivial replacements, while $W^{\mu}_{CF}(P\rightarrow T)=0$. Hence, in the following paragraphs, we pay more attentions to the form factors of $W_{(T)}$ and $W_{ann}$s.

In the case of the $P\to Sl\bar{l}$ transitions, according to the Lorentz symmetry and the gauge invariant condition of the Ann currents discussed in  Sec.~\ref{subsec:1}, we have \begin{equation}
\begin{split}
W&^{\mu}(P\rightarrow S)=F^S_z \left(P_+^{\mu}-\frac{P_+\cdot Q}{Q^2}Q^{\mu}\right)+F^S_0 \frac{P_+\cdot Q}{Q^2}Q^{\mu},\\
W&^{\mu}_{T}(P\rightarrow S)=\frac{-F^S_T }{M_i+M_f}\left\{Q^2P_+^{\mu}-(P_+\cdot Q)Q^{\mu}\right\},\\
W&^{\mu}_{ann}(P\rightarrow S)= B^S_{z}\left\{Q^2 P_+^{\mu}-{(P_+\cdot Q)}Q^{\mu}\right\},
\end{split}\label{eq:formfactorsPS}
\end{equation}where $P_+\equiv P_i+P_f$ and $F^S_{z}$,
$F^S_0$, $F_T^S$, $B^S_z$ are  form factors.

Similarly, for $P\to T l\bar{l}$ transitions, the definitions are shown as
\begin{equation}
\begin{split}
W^{\mu}(P\rightarrow T)=&\frac{iV^T }{(M_i+M_f)M_f}\epsilon^T_{\alpha\beta}Q^{\beta}\epsilon^{\mu\alpha Q P_+}-2A_0^T \frac{\epsilon_T^{\alpha\beta}Q_{\beta} Q_{\alpha}}{Q^2}Q^{\mu}-\frac{M_i+M_f}{M_f}A_1^T \left(\epsilon_T^{\mu\alpha}Q_{\alpha}\right.\\
&\left.-\frac{\epsilon_T^{\alpha\beta}Q_{\beta} Q_{\alpha}}{Q^2}Q^{\mu}\right)
+A_2^T \frac{\epsilon_T^{\alpha\beta}Q_{\beta} Q_{\alpha}}{M_{f}(M_i+M_f)}\left\{P_+^{\mu}-\frac{P_+\cdot Q}{Q^2}Q^{\mu}\right\},\\
W^{\mu}_T(P\to T)=&-i\frac{T_1^T}{M_f} \epsilon^T_{\alpha\beta}Q^{\beta}\epsilon^{\mu\alpha Q P_+}+\frac{T_2^T}{M_f} \left\{P_+\cdot Q \epsilon_T^{\mu\beta}Q_{\beta}-(\epsilon_T^{\alpha\beta}Q_{\beta} Q_{\alpha})P_+^{\mu}\right\}\\
&+\frac{T_3^T}{M_f} \left(\epsilon_T^{\alpha\beta}Q_{\beta} Q_{\alpha}\right)\left\{Q^{\mu}-\frac{Q^2}{P_+\cdot Q}P_+^{\mu}\right\},\\
W^{\mu}_{ann}(P\to T)=&(M_i-M_f)\left\{T_{1ann}^T  \frac{M_i^2}{M_f}\left(\epsilon_T^{\mu \alpha }Q_{\alpha}-\frac{Q^{\alpha}Q^{\beta}\epsilon^T_{\alpha\beta}}{Q^2}Q^{\mu}\right)+\frac{T^T_{zann}}{M_f}\epsilon_T^{\alpha\beta}Q_{\alpha}Q_{\beta} \left(P_+^{\mu}\right.\right.\\
&\left.\left.-\frac{P_+\cdot Q}{Q^2}Q^{\mu}\right)+\frac{1}{2} i \frac{V_{ann}^T}{M_f}\epsilon^T_{\alpha\beta}Q^{\beta} ~ \epsilon ^{\mu \alpha
QP_+}\right\},
\end{split}\label{eq:formfactorsPT}
\end{equation}where $V^T$, $A_1^T$, $A_2^T$, $A_0^T$, $T_1^T$, $T_2^T$, $T_3^T$, $T_{1ann}^T$, $T_{zann}^T$ and $V_{ann}^T$ are the
form factors.

As to $P\to Al\bar{l}$ decays, the definitions take the following forms,
\begin{equation}
\begin{split}
W^{\mu}(P\rightarrow A)=&\frac{iV^A }{M_i+M_f}\epsilon^{\mu\epsilon_{A} Q P_+}-2M_fA_0^A \frac{\epsilon_A \cdot Q}{Q^2}Q^{\mu}-(M_i+M_f)A_1^A \left(\epsilon_A^{\mu}-\frac{\epsilon_A\cdot Q}{Q^2}Q^{\mu}\right)\\
&+A_2^A \frac{\epsilon_A\cdot Q}{M_i+M_f}\left\{P_+^{\mu}-\frac{P_+\cdot Q}{Q^2}Q^{\mu}\right\},\\
W^{\mu}_T(P\to A)=&-iT_1^A \epsilon^{\mu\epsilon_A Q P_+}+T_2^A \left\{P_+\cdot Q \epsilon_A^{\mu}-(\epsilon_A \cdot Q)P_+^{\mu}\right\}+T_3^A \left(\epsilon_A\cdot Q\right)\left\{Q^{\mu}-\frac{Q^2}{P_+\cdot Q}P_+^{\mu}\right\},\\
W^{\mu}_{ann}(P\to A)=&(M_i-M_f)\left\{T_{1ann}^A  ~M_i^2\left(\epsilon_A^{\mu  }-\frac{Q\cdot\epsilon_A}{Q^2}Q^{\mu}\right)+T^A_{zann}Q\cdot\epsilon_A \left(P_+^{\mu}-\frac{P_+\cdot Q}{Q^2}Q^{\mu}\right)\right.\\&\left.+\frac{1}{2} i V_{ann}^A ~ \epsilon ^{\mu \epsilon_A
QP_+}\right\},
\end{split}\label{eq:formfactorsPA}
\end{equation} where $V^A$, $A_1^A$, $A_2^A$, $A_0^A$, $T_1^A$, $T_2^A$, $T_3^A$, $T_{1ann}^A$, $T_{zann}^A$ and $V_{ann}^A$ are the
form factors.

\subsection{ Numerical Results of Form Factors}\label{Sec3-5}
In this part, we present the numerical results of form factors and the according discussions.

\subsubsection{Parameters in the Calculations}\label{secBSparameter}

Here we specify the  involved parameters. First, the masses and the lifetimes of
$B_c$ and $D_{(s)J}^{(*)}$ are  required in our calculations and we take their values from
Ref.~\cite{pdg}. Second, the BS-inputs are also needed, which include the Cornell-Potential-Parameters (CPPs) and the masses of the constituent quarks.
 The CPPs can be found in Ref.~\cite{chaoHsichang_BS}. The masses of the constituent quarks are taken as $m_b=4.96$ GeV, $m_c=1.62$ GeV, $m_s=0.5$ GeV and $m_d=0.311$ GeV  \cite{huifengfuJHEP}.


\subsubsection{Results and Discussions on Form Factors}\label{Sec:Results and Discussions on Form Factors}

From the aforementioned parameters and the derivations in Sec.~\ref{SecHadronicCurrentsin BS},
the form factors  can be evaluated. In the following paragraphs, we will show and discuss them. 

In Fig.~3~(a),  the form factors of $W^{\mu}_{(T)}(B_c\to D^*_{s0}(2317))$ are presented.
These form factors are all positively related to $Q^2$. This behavior can be understood from the facts that 1) as shown in Eqs.~(\ref{eq:Mandelstam Formalism}-\ref{eq:YYT}), our hadronic currents $W^{\mu}_{(T)}$s are obtained from the integrals over the overlapping regions of the initial and final wave functions and  2)  due to the retarded relationship~in Eq.~\eqref{eq:retarded relationships}, the overlapping regions grow with increase in the variable $Q^2$.

In  recent years, $W^{\mu}_{(T)}(B_c\to D^*_{s0}(2317))$ have also been  calculated in the three-point QCD sum rules~\cite{BctoDs2317QCDsunrule} and light-cone quark model \cite{BctoDs2317lightcone}. The definitions of the $W^{\mu}_{(T)}$ form factors in Refs.~\cite{BctoDs2317lightcone,BctoDs2317QCDsunrule} are  different from the ones in this paper. But if the same definitions  are taken, the absolute values of our form factors are comparable with theirs.

Fig.~3~(b) shows the form factors of $W^{\mu}_{ann}(B_c\to D^*_{s0}(2317))$.  We see that $B^S_z$  are complex. The reason is that  in the calculations of the $W_{ann}$, the quark propagators are involved, as shown in Eqs.~(\ref{eq:w1w2w3w4ann1}-\ref{eq:w1w2w3w4ann4}). In order to deal with these propagators, we separate them into two parts: the principal value terms and $\delta$ function ones. The real part of $B^S_z$ comes from the principal value terms, while its imaginary part is caused by $\delta$ function terms.\footnote{The  monotonicity of the BP form factors and complexity of the Ann form factors can also be found in the case of  $B_c\to D_{(s)}^{(*)}\mu\bar{\mu}$ processes~\cite{Juwangpaper1}. And in Ref~\cite{Juwangpaper1}, there is a more detailed  discussion on them. }

Figs.~4~(a,~b) display the results of $W_{(T)}^{\mu}(B_c\to  D^*_{s2}(2573))$.  Similar to  $W^{\mu}_{(T)}(B_c\to D^*_{s0}(2317))$,  the form factors of $W^{\mu}_{(T)}(B_c\to  D^*_{s2}(2573))$  also increase monotonically as $Q^2$ grows.
This similarity comes from the facts that both $W^{\mu}_{(T)}(B_c\to D^*_{s0}(2317))$ and $W^{\mu}_{(T)}(B_c\to  D^*_{s2}(2573))$ are evaluated by Eqs.~(\ref{eq:Mandelstam Formalism}-\ref{eq:YYT}).

In Figs.~4~(c,~d), the Ann form factors of $B_c\to  D^*_{s2}(2573)l\bar{l}$ process are plotted.
One may note that the absolute values of these form factors are quite smaller than the ones of
$W^{\mu}_{ann}(B_c\to D^*_{s0}(2317))$. To see how this happens, one should recall that the Ann currents $W_{ann}$ are the sums of the terms $W_{ann1,\dots,ann4}$s. In the case of $W^{\mu}_{ann}(B_c\to D^*_{s0}(2317))$, the four terms all contribute. But as to $W^{\mu}_{ann}(B_c\to  D^*_{s2}(2573))$, the vanishing decay constant of the final meson forbids the $W_{ann1,ann2}$ contributions and leaves only $W_{ann3,ann4}$ terms. Compared with the sums of $W_{ann1}$ and $W_{ann2}$, the contributions of $W_{ann3}$ and $W_{ann4}$ are fairly suppressed. \footnote{The reason of this suppression is that $W_{ann3}$ and $W_{ann4}$ correspond to the diagrams where the virtual photons are emitted from the final quarks. Under the non-relativistic limit, the propagated quarks of these diagrams are highly off-shell and therefore  when calculating the amplitudes of these diagrams, the denominators are considerably large. Even though the relativistic effects are included, this kind of suppression is still not obviously ameliorated.} Thus, we see the smaller $W^{\mu}_{ann}(B_c\to  D^*_{s2}(2573))$ form factors in Figs.~4~(c,~d).

In Figs.~5~(a,~b) and Figs.~6~(a,~b), we plot the BP form factors of $B_c\to D_{s1}({2460,2536})l\bar{l}$.
First, we see that  the form factors of $W_{(T)}^{\mu}(B_c\to D_{s1}(2460,2536))$ are  not of the same sign. To
understand this feature,  recall that in order to calculate $W_{(T)}^{\mu}(B_c\to D_{s1}(2460,2536))$, the hadronic currents $W_{(T)}(B_c\to D_{s^1P_1,^3P_1})$ are first evaluated  and then we mix the results according to the mixing relationship in Eq.~\eqref{eq:P-A3}.
The form factors of $W_{(T)}(B_c\to D_{s^1P_1,^3P_1})$ are all of the same sign. But in the mixing step, we need to evaluate the sums and differences of the $W_{(T)}(B_c\to D_{s^1P_1,^3P_1})$ form factors. Hence, as illustrated in Figs.~5~(a,~b) and Figs.~6~(a,~b), the form factors with the different signs emerge.

Second, from  Figs.~6~(a,~b), one may note that the absolute values of $V^A$, $A^A_{1}$, $T_1^A$ and $T_2^A$ are much smaller than those of $A^A_{0,2}$ and $T_3^A$. This feature implies that the hadronic matrix element $W_{(T)}(B_c\to D_{s1}(2536)_{\bot})$ obtained in the BS method is suppressed significantly compared with  $W_{(T)}(B_c\to D_{s1}(2536)_{\|})$. Here $D_{s1}(2536)_{\bot(\|)}$ stands for the final meson $D_{s1}(2536)$ which is  transversely (longitudinally) polarized. 

Figs.~5~(c,~d) and Figs.~6~(c,~d) present the Ann form factors of $B_c\to D_{s1}({2460,2536})l\bar{l}$.
Due to the suppressions from the small decay constant of $D_{s1}(2536)$~\cite{BSwavefuctionp-wave1}, we see that  the form factors corresponding to $W_{ann}(B_c\to D_{s1}(2536))$ are much smaller than those of $W_{ann}(B_c\to D_{s1}(2460))$. 

In Figs.~7-10, we illustrate the form factors of $B_c\to D^{(*)}_{J}l
\bar{l}$ decays. 
%
%
The form factors of $W_{(T),ann}(B_c\to D^{(*)}_{J})$ behave similarly to the  $W_{(T),ann}(B_c\to D^{(*)}_{sJ})$ ones.
This is because 1) as discussed in Sec.~\ref{SecHadronicCurrentsin BS}, $W_{(T),ann}(B_c\to D^{(*)}_{J})$ and $W_{(T),ann}(B_c\to D^{(*)}_{sJ})$ are calculated within the same formalism and 2) in the BS method, due to the constituent mass relationship $m_s\sim m_d \ll m_c$, the wave functions of $D^{(*)}_{J}$ are quite comparable with the $D^{(*)}_{sJ}$ ones.
%
%
\section{The  Observables}
In the previous  section, we calculate the hadronic matrix elements within the BS method and express the results in terms of the form factors.
Using these form factors, the total amplitude $\mathcal{M}_{Total}$ in Eq.~\eqref{Mtotal} can be estimated.
From the obtained total amplitude, in this section, we  evaluate the physical observables.

\subsection{The Calculations of Observables}

In this part, we employ the helicity amplitude method~\cite{AFaessler} to calculate observables.

First of all, we need to split the total transition amplitudes as \begin{equation}\mathcal{M}_{Total}\equiv \mathcal{M}_{1}^{\mu}\bar{l}\gamma_{\mu}l+\mathcal{M}_{2}^{\mu}\bar{l}\gamma_{\mu}\gamma_{5}l, \label{eq:dedinitionM1M2}\end{equation} where $\mathcal{M}_{1(2)}^{\mu}$ can be determined by matching Eq.~\eqref{Mtotal} to the equation above.

And then by projecting $M_{1(2)}^{\mu}$ to the  helicity components $\epsilon^{\mu}_ {H} (t,0,\pm1)$, the helicity
amplitudes can be obtained, that is~\cite{AFaessler},
\begin{equation}\begin{split}
H^{1(2)}_{t,~\pm,~0}=\epsilon_H (t,~\pm,~0)\cdot M_{1(2)}.
\end{split}\label{eq:definitionHzft0}\end{equation} The explicit expressions of $\epsilon^{\mu}_ {H} (t,0,\pm1)$ are specified in Appendix. B.

Finally, according to the derivations in Ref.~\cite{AFaessler},
the  differential
branching fractions $\mathrm{d}Br / \mathrm{d}Q^2$, the forward-backward asymmetries
$A_{FB}$, the longitudinal polarizations of the final mesons $P_L$ and the leptonic longitudinal polarization
asymmetries $A_{LPL}$ can be expressed in terms of helicity
amplitudes, which are
\begin{equation}\begin{split}
\frac{\mathrm{d}Br}{\mathrm{d}Q^2}=&\frac{1}{(2\pi)^3\Gamma_{B_c}}\frac{\lambda^{1/2}Q^2}{24M^3_{B_c}}\sqrt{1-\frac{4m_l^2}{Q^2}}\mathcal{M}^2_H,\\
A_{FB}=&\frac{3}{4}\sqrt{1-\frac{4m^2_l}{Q^2}}\frac{2}{\mathcal{M}^2_H}\left\{\mathrm{Re}\left(H_+^{(1)}H_+^{\dag(2)}\right)-\mathrm{Re}\left(H_-^{(1)}H_-^{\dag(2)}\right)\right\},\\
P_L=&\frac{1}{\mathcal{M}^2_H} \left\{
H^{(1)}_0H^{\dag(1)}_0\left(1+\frac{2m^2_l}{Q^2}\right)+H^{(2)}_0H^{\dag(2)}_0
\left(1-\frac{4m^2_l}{Q^2}\right)+\frac{2m_l^2}{Q^2}3H^{(2)}_tH^{\dag(2)}_t
\right\},\\
A_{LPL}&\equiv\frac{dBr_{h=-1/2}/dQ^2-dBr_{h=1/2}/dQ^2}{dBr_{h=-1/2}/dQ^2+dBr_{h=1/2}/dQ^2}\\
=&\sqrt{1-\frac{4m^2_l}{Q^2}}\frac{2}{\mathcal{M}^2_H}\left\{\mathrm{Re}\left(H_+^{(1)}H_+^{\dag(2)}\right)+\mathrm{Re}\left(H_-^{(1)}H_-^{\dag(2)}\right)+\mathrm{Re}\left(H_0^{(1)}H_0^{\dag(2)}\right)\right\},
\end{split}\label{eq:observables}\end{equation}
where $h$ denotes the  helicity of  $l^-$, while the denotation $\lambda=(M_i^2-M_f^2){}^2+Q^2(Q^2-2M_i^2-2M_f^2)$ is employed. And the definition of  $\mathcal{M}_H$ is
\begin{equation}
\begin{split}
\mathcal{M}^2_H=& \left(H^{(1)}_+H^{\dag(1)}_+
+H^{(1)}_-H^{\dag(1)}_-+H^{(1)}_0H^{(1)\dag}_0\right)\left(1+\frac{2m^2_l}{Q^2}\right)+\\
&\left(H^{(2)}_+H^{\dag(2)}_++H^{(2)}_-H^{\dag(2)}_-
+H^{(2)}_0H^{(2)\dag}_0 \right)
\left(1-\frac{4m^2_l}{Q^2}\right)+\frac{2m_l^2}{Q^2}3H^{(2)}_tH^{\dag(2)}_t.
\end{split}
\end{equation}
Plugging the  helicity
amplitudes $H^{1(2)}_{t,~\pm,~0}$ into Eq.~\eqref{eq:observables}, the observables are obtained.

\subsection{Numerical Results of the Observables}
Within Figs.~11-18, the numerical values of the observables are presented in the solid (or dash-dot) lines, while their theoretical uncertainties are illustrated in the pale green (or pink) areas. In this part, we lay stress on the introductions of numerical results of the observables. And in next section, the systematic discussions on the theoretical uncertainties will be shown.

 When the numerical values of observables are calculated in this paper, we have considered  the BP, Ann, CS and CF diagrams. In order to show their influences clearly, for each channel, we plot 1) the observables where only BP contributions are considered, 2) the ones where  BP and CS effects are contained, 3) the ones with  BP  and Ann influences and 4) the ones including the BP, Ann, CS and CF diagrams. In the following paragraphes, their comparisons and discussions will be presented.

\subsubsection{The Observables of $B_c\to D^*_{s0}(2317)\mu\bar{\mu}$ decays}

In Figs.~11 (a,~b), the differential branching fractions of $B_c\to D^*_{s0}(2317)\mu\bar{\mu}$ process are illustrated. 

For  $\text{d}Br/\text{d}Q^2$ which includes only BP contributions, as shown in the dash-dot line of Fig.~11 (a), we see that $\text{d}Br/\text{d}Q^2$ is biggest around $Q^2\sim 10.5~\text{GeV}^2$ and suppressed considerably at the end points. This  is similar to the result in Ref.~\cite{BctoDs2317lightcone} but quite different from the one in Ref.~\cite{BctoDs2317QCDsunrule}. If the Ann effects are added, as plotted in the dash-dot line of Fig.~11 (b), $\text{d}Br/\text{d}Q^2$ is enhanced un-negligibly  around $Q\sim12.5~\text{GeV}^2$.

For $\text{d}Br/\text{d}Q^2$ which contains BP and CS effects, as plotted in the solid line of Fig.~11 (a), because of the Breit-Wigner propagators in $C_{9}^{CS}$, the significant enlargements emerge around the resonance regions. If the Ann and CF diagrams are included, as displayed in solid line of Fig.~11 (b), $\text{d}Br/\text{d}Q^2$ around $Q^2\sim M_{J/\psi}^2$ continues enlarging. But in light of the node structure of the $\psi(2S)$ wave function, which leads to the cancelations in the $W_{CF}(B_c\to D^*_{s0}(2317)\psi(2S) \to D^*_{s0}(2317)\mu\bar{\mu})$ calculation, $\text{d}Br/\text{d}Q^2$ around $Q^2\sim M_{\psi(2S)}^2$ changes imperceptibly. This feature can also be found in the processes $B_c\to D_{(s)}\mu\bar{\mu}$ \cite{Juwangpaper1}.

In Figs.~11~(c,~d), we illustrate $A_{LPL}$s of the $B_c\to D_{s0}^*(2317)\mu\bar{\mu}$ process.

 For $A_{LPL}$ which includes only BP diagrams, as shown in dash-dot line of Fig.~11.~(c), we note that $A_{LPL}\sim -1$ in the region $Q^2\in[2,15]~\text{GeV}^2$. In order to see how this happens, note that due to the relationship $C^{\text{eff}}_9\sim C_{10} \gg 2 m_b C^{\text{eff}}_{7}/(M_i+M_f)$, $\mathcal{M}_{BP}^{L}$ contributes to $\mathcal{M}_{BP}$ dominantly. (Hereafter, $\mathcal{M}_{BP(ann)}^{L(R)}$s stand for the BP (or Ann) amplitudes whose final leptons are all left (or right) handed.)
 This makes that for the relativistically boosted $\mu^{\pm}$, $dBr_{h=+1/2}/dQ^2$s are much bigger than $dBr_{h=-1/2}/dQ^2$s over the domain $Q^2\in[2,15]~\text{GeV}^2$.
 Hence, from the definition of $A_{LPL}$ in Eq.~\eqref{eq:observables}, we have $A_{LPL}\sim -1$. This feature can also be found in the decays $B_c\to D_{(s)}^{(*)}\mu\bar{\mu}$ \cite{Juwangpaper1}.

If the Ann effects are added, as given in dash-dot line of Fig.~11.~(d),  $A_{LPL}$ deviates from $-1$ strongly over the low $Q^2$ area, while in the high $Q^2$ region, this kind of deviation becomes weaker. To understand this feature,  recall that the real part of Ann form factor $\Re[B_{zann}^{S}]$ is positive within  the low $Q^2$ domain but turns negative when $Q^2\geq 12~\text{GeV}^2$, as shown in Fig.~3~(b). When $\Re[B_{zann}^{S}]>0$,
$\mathcal{M}_{ann}^{L}$ interferes destructively with $\mathcal{M}_{BP}^{L}$, making $\text{d}Br_{h=+1/2}/\text{d}Q^2$ suppressed. But if $\Re[B_{zann}^{S}]<0$,
there are constructive interferences between $\mathcal{M}_{ann}^{L}$ and  $\mathcal{M}_{BP}^{L}$, leading to the enhanced $\text{d}Br_{h=+1/2}/\text{d}Q^2$.
Hence, based on Eq.~\eqref{eq:observables}, $A_{LPL}$ should be quite larger than $-1$ in the low $Q^2$ domain but become smaller with the increase in $Q^2$.

Once the BP, Ann, CS and CF contributions are all considered, as seen in solid line of Fig.~11.~(d), one may find that $A_{LPL}\sim-1$ in the low $Q^2$ region.
This is
 due to  the cancelations   between Ann and CF transition amplitudes.

\subsubsection{The Observables of $B_c\to D^*_{s2}(2573)\mu\bar{\mu}$ decays}

Figs.~12 (a-h) depict observables of the $B_c\to D^*_{s2}(2573)\mu\bar{\mu}$ transition. Considering $W_{CF}(B_c\to D^*_{s2}(2573))=0$ as discussed in Sec.~\ref{subsec:1}, the $B_c\to D^*_{s2}(2573)\mu\bar{\mu}$ process does not receive any contributions from the CF diagrams. Hence,  in Figs.~12 (a-h),  we do not illustrate the observables which include  CF effects.

Within Figs.~12 (a,~b), we plot $\text{d}Br/\text{d}Q^2$s as the functions of $Q^2$. First, we see that $\text{d}Br/\text{d}Q^2(B_c\to D^*_{s2}(2573)\mu\bar{\mu})$s are much bigger than $\text{d}Br/\text{d}Q^2(B_c\to D^*_{s0}(2317)\mu\bar{\mu})$s around the $Q^2\sim0~\text{GeV}^2$ point. To understand this behavior,  note that 1) from Eq.~\eqref{eq:observables}, $\text{d}Br/\text{d}Q^2$s are almost proportional to the sum of $H^{(1,2)}_{\pm,0}H^{\dag(1,2)}_{\pm,0}$s and 2) in the low $Q^2$ area, the transverse contributions $H^{(1,2)}_{\pm}H^{\dag(1,2)}_{\pm}$s can be enhanced significantly by the $\gamma$ propagators. For $B_c\to D^*_{s2}(2573)\mu\bar{\mu}$ decay, both $H^{(1,2)}_{0}H^{\dag(1,2)}_{0}$s and $H^{(1,2)}_{\pm}H^{\dag(1,2)}_{\pm}$s contribute. But in $B_c\to D^*_{s0}(2317)\mu\bar{\mu}$ process, only
$H^{(1,2)}_{0}H^{\dag(1,2)}_{0}$s participate. Hence, around the $Q^2\sim0~\text{GeV}^2$ point, there are  enhancements in $\text{d}Br/\text{d}Q^2(B_c\to D^*_{s2}(2573)\mu\bar{\mu})$ but not in $\text{d}Br/\text{d}Q^2(B_c\to D^*_{s0}(2317)\mu\bar{\mu}) $.
Second, from Figs.~12 (a,~b), one may note that  $\text{d}Br/\text{d}Q^2$ including the BP and Ann effects
deviates imperceptibly  from the one with only BP contribution. This is because that  as plotted in Figs.~4 (c,~d), the Ann form factors are quite small, which  suppresses $\mathcal{M}_{ann}$ considerably so
that the Ann contributions are much less than the BP ones. Hence, as illustrated in Figs.~12 (a,~b), $\text{d}Br/\text{d}Q^2$s show the insensitivities to the Ann diagrams.

Figs.~12~(c,~d) are devoted to presenting the results of $A_{LPL}(B_c\to D_{s2}^*(2573)\mu\bar{\mu})$.
When the BP (and CS) effects are included, we see the similarities between $A_{LPL}(B_c\to D_{s2}^*(2573)\mu\bar{\mu})$s and $A_{LPL}(B_c\to D_{s0}^*(2317)\mu\bar{\mu})$s. If the Ann contributions are added, in analogy to the case of $\text{d}Br/\text{d}Q^2(B_c\to D^*_{s2}(2573)\mu\bar{\mu})$s,
 $A_{LPL}(B_c\to D_{s2}^*(2573)\mu\bar{\mu})$s also change slightly.

 In Figs.~12~(e,~f), we display $A_{FB}$s of the $B_c\to D_{s2}^*(2573)\mu\bar{\mu}$ process. In Fig.~12~(e), we see that $A_{FB}$s are positive  over the high $Q^2$ domain (except the resonance regions), while due to suppressions from the $\gamma $ penguin diagrams,  $A_{FB}$s turn negative in the low $Q^2$ region. Once the Ann influences are take into account, likewise for $\text{d}Br(B_c\to D_{s2}^*(2573)\mu\bar{\mu})/\text{d}Q^2$s and $A_{LPL}(B_c\to D_{s2}^*(2573)\mu\bar{\mu})$s, $A_{FB}$s   behave insensitively to Ann effects.

Figs.~12~(g,~h) show the results of $P_{L}(B_c\to D_{s2}^*(2573)\mu\bar{\mu})$s. When only the BP diagrams are contained,  $P_{L}$ is positively related to $Q^2$ in the low $Q^2$ region but inversely to $Q^2$ in the high $Q^2$ domain.
If the Ann effects are added, $P_{L}$s change negligibly.

\subsubsection{The Observables of $B_c\to D_{s1}(2460)\mu\bar{\mu}$ decays}

  Figs.~13 (a-h) present the observables of $B_c\to D_{s1}(2460)\mu\bar{\mu}$ process. When the BP (and CS) contributions are  under consideration, the $B_c\to D_{s1}(2460)\mu\bar{\mu}$ observables are similar to those of $B_c\to D_{s2}^*(2573)\mu\bar{\mu}$ decays.

  But once the CF and Ann effects are included, the $B_c\to D_{s1}(2460)\mu\bar{\mu}$ observables behave quite sensitively.
  More specifically, we see that 1) in  Figs.~13 (c,~d), $A_{LPL}$ which includes the BP (and CS) diagrams is negative in the low $Q^2$ region. But if the
  CF and Ann contributions are taken account of, $A_{LPL}$ turns positive; 2) in Figs.~13 (a,~b),  $\text{d}Br/\text{d}Q^2(B_c\to D_{s1}(2460)\mu\bar{\mu})$s around $Q^2=M_{J/\psi}^2$ are enlarged considerably by the CF contributions;
  3) in Figs.~13 (e-h), $P_L$s and $A_{FB}$s are suppressed fairly after the Ann and CF effects are added.

  These sensitive behaviors imply that  the CF and Ann contributions play important roles in the $B_c\to D_{s1}(2460)\mu\bar{\mu}$ process. Therefore, when the observables of $B_c\to D_{s1}(2460)\mu\bar{\mu}$ transition are calculated, besides the BP and CS Feynman diagrams, it is necessary to include the CF and Ann diagrams.

\clearpage

\subsubsection{The Observables of $B_c\to D_{s1}(2536)\mu\bar{\mu}$ decays}

In  Figs.~14~(a-h),  the observables of the decay $B_c\to D_{s1}(2536)\mu\bar{\mu}$ are illustrated. The behaviors of these observables are very different from those
in the $B_c\to D_{s1}(2460)\mu\bar{\mu}$ process.

 First, we see that if only the BP contribution is considered, $\text{d}Br/\text{d}Q^2(B_c\to D_{s1}(2536)\mu\bar{\mu})$ is much smaller than $\text{d}Br/\text{d}Q^2(B_c\to D_{s1}(2460)\mu\bar{\mu})$. To understand this smallness,  note that, as discussed in Sec.~\ref{Sec:Results and Discussions on Form Factors}, the BP form factors of the $B_c\to D_{s1}(2536)\mu\bar{\mu}$ process have  different signs. This makes that when $\mathcal{M}_{BP}(B_c\to D_{s1}(2536)\mu\bar{\mu})$ is calculated, the cancelations emerge between the positive BP form factors and the negative ones.  Hence, as shown in  Figs.~13,~14~(a), $\text{d}Br/\text{d}Q^2(B_c\to D_{s1}(2536)\mu\bar{\mu})\ll\text{d}Br/\text{d}Q^2(B_c\to D_{s1}(2460)\mu\bar{\mu})$.

 Second, we see that when only BP Feynman diagrams are included, $A_{FB}\sim 0$ and $ P_L\sim 1$ within the area $Q^2\in [1,6]~\text{GeV}^2$.
 To see how this happens, we note that  as concluded in Sec.~\ref{Sec:Results and Discussions on Form Factors},
 the hadronic current $W_{(T)}(B_c\to D_{s1}(2536)_{\bot})$ obtained in BS method is much smaller than  $W_{(T)}(B_c\to D_{s1}(2536)_{\|})$. This implies that, if only BP effects are considered, the transverse helicity  amplitudes in the $B_c\to D_{s1}(2460)\mu\bar{\mu}$ decay are considerably suppressed compared with the longitudinal  ones, namely, $H^{(1,2)}_{\pm}\ll H^{(1,2)}_{0}$.
 Hence, according to the expressions of $A_{FB}$ and $P_L$ in Eq.~\eqref{eq:observables}, over the domain $Q^2\in [1,6]~\text{GeV}^2$, $|A_{FB}|$ has a quite small value, while $ P_L$  almost equals one.

 Third, if the Ann and CF influences are contained, the $B_c\to D_{s1}(2536)\mu\bar{\mu}$ observables show the insensitivities. This is because  the decay constant of $D_{s1}(2536)$ is fairly small, which suppresses $\mathcal{M}_{ann}$ and $\mathcal{M}_{CF}$ strongly so that the BP contributions are quite bigger than the others. Hence, as illustrated in Figs.~14~(a-h), when the Ann and CF diagrams are added, there are no obvious deviations in the $B_c\to D_{s1}(2536)\mu\bar{\mu}$ observables outside the resonance regions.

 \clearpage

\subsubsection{The Observables of $B_c\to D_{J}^{(*)}\mu\bar{\mu}$ decays}

In Figs. 15-18 (a,~b), the differential branching fractions of $B_c\to D_{J}^{(*)}\mu\bar{\mu}$ are displayed. One may note that  $\text{d}Br(B_c\to D_{J}^{(*)}\mu\bar{\mu})/\text{d}Q^2$s are much smaller than $\text{d}Br(B_c\to D_{sJ}^{(*)}\mu\bar{\mu})/\text{d}Q^2$s. We attribute this smallness  to their suppressed CKM matrix elements. More specifically, for $B_c\to D_{sJ}^{(*)}\mu\bar{\mu}$, the CKM matrix element of BP diagrams is $V_{tb}V^*_{ts}\sim -A\lambda^2$~\cite{pdg}, while the one corresponding to Ann, CS and CF effects is $V_{cb}V^*_{cs}\sim A\lambda^2$~\cite{pdg}. But as to $B_c\to D_{J}^{(*)}\mu\bar{\mu}$, the CKM matrix element for BP diagrams is $V_{tb}V^*_{td}\sim A\lambda^3$~\cite{pdg}, while the one of Ann, CS and CF contributions is $V_{cb}V^*_{cd}\sim -A\lambda^3$~\cite{pdg}. Hence, when   $\text{d}Br/\text{d}Q^2(B_c\to D_{J}^{(*)}\mu\bar{\mu})$s are calculated, the small parameter $\lambda$ suppresses their numerical values.

In Figs. 15 (c,~d) and  Figs. 16-18 (c-h), the $A_{LPL}$s, $A_{FB}$s and $P_L$s of $B_c\to D_{J}^{(*)}\mu\bar{\mu}$  are shown. We see that these observables behave similarly to those in $B_c\to D_{sJ}^{(*)}\mu\bar{\mu}$ decays. The reasons are 1) in the present work, the Feynman diagrams corresponding to $B_c\to D_{J}^{(*)}\mu\bar{\mu}$ are analogous to those of the $B_c\to D_{sJ}^{(*)}\mu\bar{\mu}$ processes; 2) as shown in Sec.~\ref{Sec:Results and Discussions on Form Factors}, the $B_c\to D_{J}^{(*)}\mu\bar{\mu}$  form factors are quite similar to the $B_c\to D_{sJ}^{(*)}\mu\bar{\mu}$ ones.

\subsection{The Experimentally Excluded Regions and Integrated Branching Fractions}

Using the results of $\text{d}Br/\text{d}Q^2$s, as shown in Figs. 11-18 (a,~b), now we define the experimentally excluded regions.
According to the sensitivities to the CF effects, the decays $B_c\to D_{(s)J}^{(*)}\mu\bar{\mu}$ fall into two categories. The first category includes
$B_c\to D_0^*(2400)(D_{s0}^*(2317))\mu\bar{\mu}$, $B_c\to D_{s1}(2460)\mu\bar{\mu}$ and $B_c\to D_1(2430)\mu\bar{\mu}$ channels, which
are quite sensitive to the CF contributions. Through comparing $\text{d}Br/\text{d}Q^2$s which contain only BP and Ann effects with the ones which include BP, Ann, CS and CF contributions, we define their experimentally excluded region as
\begin{equation}\begin{split}
\text{Region}:Q^2>5~\text{GeV}^2.\label{eq:cuttingregion1}
\end{split}\end{equation}
The second category contains
$B_c\to D_2^*(2460)(D_{s2}^*(2573))\mu\bar{\mu}$, $B_c\to D_{s1}(2536)\mu\bar{\mu}$ and $B_c\to D_1(2420)\mu\bar{\mu}$ transitions, which
are not sensitive to the CF contributions. So their  experimentally excluded area is defined as
\begin{equation}\begin{split}
\text{Region}:Q^2>7~\text{GeV}^2.\label{eq:cuttingregion2}
\end{split}\end{equation}

 Based on the experimentally excluded regions introduced above, the integrated branching fractions are calculated and shown in Table. 1.
As seen in Table. 1, the branching fractions including BP and Ann effects are comparable with the ones containing both BP, Ann, CF and CS contributions. This implies that our experimentally excluded regions defined in Eqs.~(\ref{eq:cuttingregion1},~\ref{eq:cuttingregion2}) are workable.

\begin{table}[ph]
\caption{Branching ratio for each channel.}
\begin{center}
{\begin{tabular}{|c|c|c|c|c|c|}
\hline Modes&$Br^{BP+Ann}$&$Br^{BP+Ann+CS+CF}$\\
\hline $B_c\rightarrow D^*_{0}(2400)\mu \bar{\mu})$&$8.9^{+2.8}_{-2.3}\times10^{-11}$&$1.1^{+0.5}_{-0.4}\times10^{-10}$\\
\hline $B_c\rightarrow D^*_{s0}(2317)\mu \bar{\mu})$&$4.0^{+1.4}_{-1.1}\times10^{-9}$&$5.4^{+2.5}_{-2.0}\times10^{-9}$\\
\hline $B_c\rightarrow D_{1}(2420)\mu \bar{\mu})$&$8.3^{+1.9}_{-1.5}\times10^{-10}$&$7.1^{+1.7}_{-1.7}\times10^{-10}$\\
\hline $B_c\rightarrow D_{1}(2430)\mu \bar{\mu})$&$1.2^{+0.5}_{-0.2}\times10^{-9}$&$9.7^{+4.5}_{-2.0}\times10^{-10}$\\
\hline $B_c\rightarrow D_{s1}(2460)\mu \bar{\mu})$&$4.7^{+1.2}_{-1.3}\times10^{-8}$&$4.5^{+1.1}_{-1.2}\times10^{-8}$\\
\hline $B_c\rightarrow D_{s1}(2536)\mu \bar{\mu})$&$ 3.7^{+0.4}_{-0.9}\times10^{-8}$&$3.4^{+0.5}_{-1.0}\times10^{-8}$\\
\hline $B_c\rightarrow D_{2}^*(2460)\mu \bar{\mu})$&$9.5^{+2.6}_{-2.1}\times10^{-10}$&$9.8^{+3.2}_{-2.7}\times10^{-10}$\\
\hline $B_c\rightarrow D_{s2}^*(2573)\mu \bar{\mu})$&$4.5^{+1.3}_{-1.0}\times10^{-8}$&$4.7^{+1.7}_{-1.4}\times10^{-8}$\\
\hline
\end{tabular} }
\end{center}\label{Table:Br}
\end{table}

\section{Discussions }

\subsection{Estimations of the Theoretical Uncertainties}
In the previous section, the numerical results of the $B_c\to D_{(s)J}^{(*)}\mu\bar{\mu}$ observables are discussed. In this part, we  discuss their theoretical uncertainties.

In this paper, we estimate the theoretical uncertainties of the observables including two aspects. First, the theoretical errors  from hadronic matrix elements are considered. Recall that our  hadronic currents are calculated in the BS method and the obtained form factors are dependent on the numerical values of the BS inputs. In order to estimate the according systematic uncertainties, we calculate the observables with changing the BS inputs by $\pm5\%$. Second, the systematic errors aroused by the factorization hypothesis are included. In the derivations of $\mathcal{M}_{ Ann, CS, CF}$, the factorization hypothesis~\cite{Factorization} is employed. In this method, in order to include the non-factorizable contributions, the number of colors $N_c$ in the expression $(C_1/N_c+C_2)$ or $(C_1+C_2/N_c)$ is treated as an adjustable parameter which should be determined by fitting the experimental data \cite{NF1,NF2,NF3,NF4}. But since that the present experimental data on  $B_c$ meson is still rare so that this parameter can not be obtained at the moment, we calculate the observables with $N_c=3$ but change the numerical values of $N_c$ within the region $[2,\infty]$ for estimating systematic uncertainties brought by factorization hypothesis.

Actually, in  recent years, several methods,  dealing with the non-factorizable contributions more systematically, have been devoted to investigating the $B_c$ decays, such as
perturbative QCD approach(PQCD)~\cite{Keum:2000ph,Lu:2000em} and QCD factorization (QCDF)
~\cite{Beneke:1999br}. However, the channels in which the PQCD and QCDF are workable must have energetic final particles. Moreover as to  $B_c\to D_{(s)J}^{(*)}l\bar{l}$, the finial mesons have small recoil momenta in the high $Q^2$ domain. Hence, in this paper, we choose to employ the factorization method~\cite{Factorization}. Similar situations can also be found in the calculations of $B_c\to D_{(s)}^{(*)}l\bar{l}$ \cite{
D.Ebert,AFaessler,Gengchaoqiang,Ho,TengWang,KAzizi1,KAzizi3,KAzizigamma,HYCheng,DDU,pakistan1,pakistan2,pakistan3} in which the factorization method has to be used extensively to account for the non-factorizable effects.

 Here we  stress that using the factorization assumption to deal with the non-factorizable effects is a temporary way in the early stage of investigating the rare $B_c$ decays. A more systematical method is important and necessary. Hence, more work in the future is required.

%
%
%
%

\subsection{Testing the Hadronic Matrix Elements }

In the previous subsection, by changing the BS inputs within $\pm5\%$, we  estimate the theoretical uncertainties from hadronic currents.
Strictly speaking, this only measures parts of the uncertainties, because the  systematic uncertainties from the approximations  made within the  BS method are not considered. Considering that  this kind of uncertainties are rather difficult to be systematically estimated, in fact, we do not control the hadronic uncertainties confidently. \footnote{To our knowledge, most (maybe all) of models, which are employed to calculate the hadronic matrix elements, suffer from this problem.}
Hence, testing whether the hadronic currents are properly evaluated is  important.

From Eq.~\eqref{eq:amplitudePB}, we see that within the transition amplitude $\mathcal{M}_{BP}$, the hadronic currents are multiplied by the Wilson coefficients $C^{\text{eff}}_{7 ,9},C_{10}$ which are sensitive to NP. This makes that  from the observables  of $B_c\to D_{(s)J}^{(*)}l\bar{l}$, it is quite involved to tell whether each hadronic current is correctly estimated. Hence, in order to test them, it is beneficial to analyze the channels in which the short distance interactions are not sensitive to NP and the hadronic matrix elements are similar or identical to the ones participating in $B_c\to D_{(s)J}^{(*)}l\bar{l}$.


First, we pay attentions to the decays $B_c\to D_{J}^{(*)}\mu\bar{\nu_{\mu}}$.
The processes $B_c\to D_{J}^{(*)}\mu\bar{\nu_{\mu}}$  are induced by the transitions $b\to u \mu\bar{\nu}_{\mu}$. From the experiences of $B$ decays, $b\to u \mu\bar{\nu}_{\mu}$ is dominated by the SM contributions~\cite{pdg}.
 In the SM, the according amplitude  reads $M(B_c\to D_{J}^{(*)}\mu\bar{\nu_{\mu}})=-iV^*_{ub}\frac{4G_f}{\sqrt{2}}\langle D_{J}^{(*)}| \bar{u}\gamma^{\alpha}(1-\gamma_5)b|B_c\rangle \bar{l}_{\mu}\gamma_{\alpha}(1-\gamma_5)l_{\nu}$. In light of the isospin symmetry of $u$ and $d$ quarks,  $\langle D_{J}^{(*)}| \bar{u}\gamma^{\alpha}(1-\gamma_5)b|B_c\rangle $s are almost identical to  $\langle D_{J}^{(*)}| \bar{d}\gamma^{\alpha}(1-\gamma_5)b|B_c\rangle $s. Hence, by means of investigating the $B_c\to D_{J}^{(*)}\mu\bar{\nu_{\mu}}$ observables experimentally, we can test the form factors of $\langle D_{J}^{(*)}| \bar{d}\gamma^{\alpha}(1-\gamma_5)b|B_c\rangle$.
In our previous paper~\cite{wangzhihuipwave}, the decays $B_c\to D_{J}^{(*)}\mu\bar{\nu_{\mu}}$ have been calculated.


\begin{figure}[htbp]
\centering
\subfigure[ ]{\includegraphics[width =
0.27\textwidth,height=0.1\textheight]{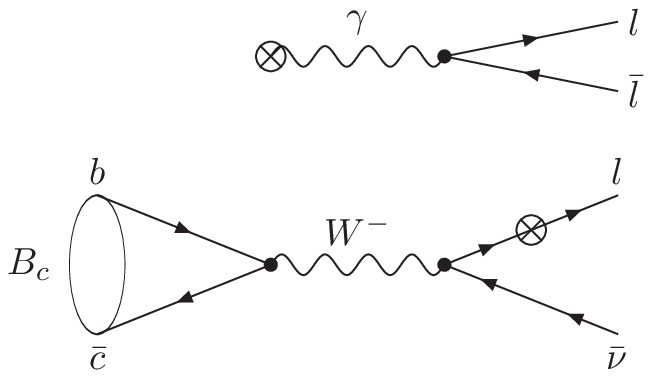}}
\subfigure[ ]{\includegraphics[width =
0.27\textwidth,height=0.1\textheight]{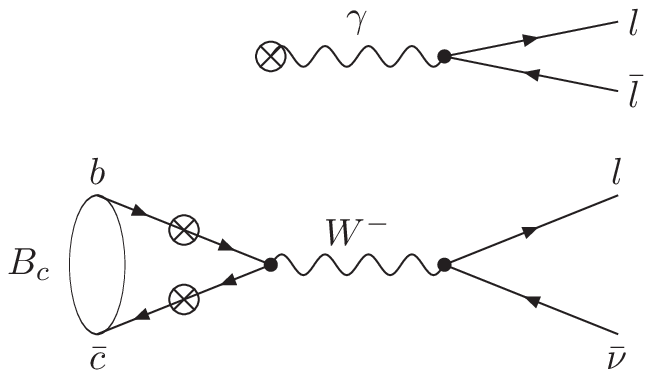}}
\subfigure[]{\includegraphics[width =
0.27\textwidth,height=0.1\textheight]{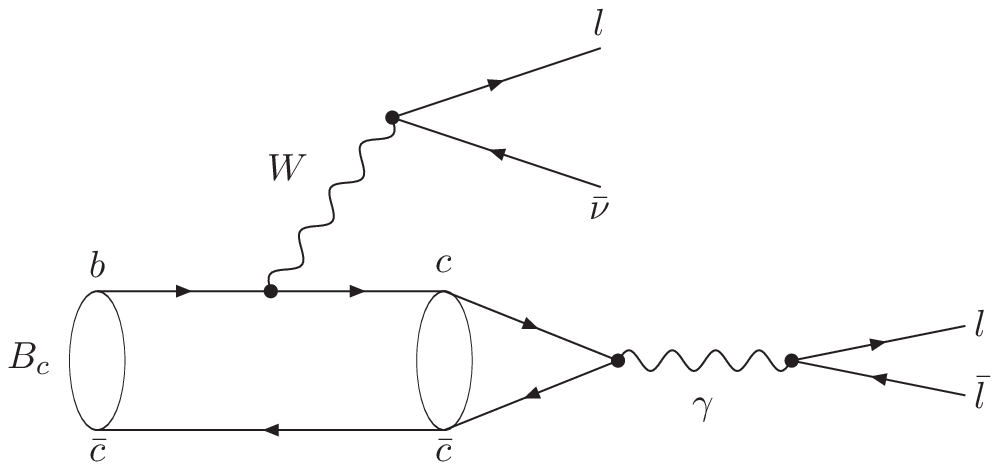}}
\caption{Typical Diagrams of $B_c\to l\bar{l} l\bar{\nu}$.}
\label{figure1-4}
\end{figure}

Second, we turn to investigating $B_c\to l_A\bar{l}_A l_B\bar{\nu}_{B}$, whose typical diagrams are illustrated in Fig.~\ref{figure1-4}.
For Fig.~\ref{figure1-4}~(a), the according hadronic matrix element is $\langle0|\bar{c}\gamma_{\mu}(1-\gamma_5)b|B_c\rangle$, which can be obtained from the future experimental data on pure leptonic decays $B_c\to l\bar{\nu}_l$. As to Fig.~\ref{figure1-4}~(b), the according hadronic matrix elements are the same as  $W_{1ann}+W_{2ann}$  in Eq.~\eqref{eq:formfactorAnn}, except the absence of $\langle f|\bar{s}(\bar{d})\gamma_{\nu}(1-\gamma_5)c|0\rangle$. Likewise, for Fig.~\ref{figure1-4}~(c),
 its hadronic current is similar to $W_{CF}$
 in Eq.~\eqref{eq:formfactorCF}, except lacking $\langle f|\bar{s}(\bar{d})\gamma_{\nu}(1-\gamma_5)c|0\rangle$. Hence, through experimentally detecting $B_c\to l_A\bar{l}_A l_B\bar{\nu}_{B}$, we can examine the hadronic currents $W_{1ann}+W_{2ann}$ and $W_{CF}$. (or, parts of $W_{1ann}+W_{2ann}$ and $W_{CF}$.)
Considering that in this paper we focus on the calculations of  $B_c\to D_{(s)J}^{(*)}l\bar{l}$, we do not show the results of $B_c\to l_A\bar{l}_A l_B\bar{\nu}_{B}$ here but put them into our future work.

However, for the other hadronic matrix elements $W_{T}$, $W_{3ann}$, $W_{4ann}$ and $W(B_c\to D_{sJ}^{(*)})$, the ideal channels  to examine them are difficult to find unless extra hypothesis is introduced. Hence, we attempt to test them in an indirect way: we use the same framework and the same set of inputs as the ones, which are used to calculate $W_{T}$, $W_{3ann}$, $W_{4ann}$ and $W(B_c\to D_{sJ}^{(*)})$, to investigate the processes $B_s\to D_{sJ}^{*}\mu\bar{\nu}$, $B\to D_{J}^{*}\mu\bar{\nu}$ and $B_c\to\chi_{cJ} \mu\bar{\nu}$. The reasons for choosing these channels are that 1) these channels are induced by $b\to c(u)\mu\bar{\nu}$ transitions, which are dominated by SM contributions from experiences of $B_{(s)}$ decays~\cite{pdg}; 2) unlike the non-leptonic decays, these semi-leptonic processes do not suffer from the theoretical uncertainties from the factorization problem. In our previous papers \cite{Yue:2013gxa,Jiang:2013vza}, the processes $B_s\to D_{sJ}^{*}\mu\bar{\nu}$, $B\to D_{J}^{*}\mu\bar{\nu}$ were calculated, while in Ref.~\cite{Wang:2011jt}, $B_c\to\chi_{cJ} \mu\bar{\nu}$ were analyzed.

In the paragraphs above, the channels $B_c\to D_{J}^{(*)}\mu\bar{\nu_{\mu}}$, $B_c\to l_A\bar{l}_A l_B\bar{\nu}_{B}$, $B_s\to D_{sJ}^{*}\mu\bar{\nu}$, $B\to D_{J}^{*}\mu\bar{\nu}$ and $B_c\to\chi_{cJ} \mu\bar{\nu}$ are recommended in order to test our hadronic matrix elements. At present, only the experimental results on $B\to D_{J}^{*}\mu\bar{\nu}$~\cite{pdg} are available and most of them are comparable with our theoretical results \cite{Yue:2013gxa,Jiang:2013vza} within the systemic errors.
If in the future more experimental results on the $B_{c,s}$ decays are reported, we can continue examining our hadronic matrix elements. Once the deviations appear between our predictions on $B_c\to D_{J}^{(*)}\mu\bar{\nu_{\mu}}$, $B_c\to l_A\bar{l}_A l_B\bar{\nu}_{B}$, $B_s\to D_{sJ}^{*}\mu\bar{\nu}$, $B\to D_{J}^{*}\mu\bar{\nu}$, $B_c\to\chi_{cJ} \mu\bar{\nu}$ and the future experimental observations, we need to check whether these deviations come from 1) the BS inputs or the approximations of the BS method; 2)  our assumption that $D_{(s)J}^{(*)}$ can be categorized as the conventional charmed(-strange) meson family.

 In order to examine the BS inputs and the approximations of the BS method, we should pay attentions to  the $B_{c,s,u,d}\to D^{(*)}_{s,d,u}(\eta_c, J/\psi) \mu\bar{\nu}$ decays whose finial mesons are of S-wave states. In our previous papers \cite{Fu:2011zzo,Zhang:2010ur}, the observables of the processes $B_{(s)}\to D^{(*)}_{(s)}\mu\bar{\nu}$ are estimated and the results are in good agreements with the experimental observations~\cite{pdg}. In Ref.~\cite{Chang:2014jca}, the $B_c\to J/\psi(\eta_c)\mu\bar{\nu}$ are analyzed and we expect that these channels can be tested by the future experimental data. If our results deviate from the future data, constraining our BS inputs or modifying BS method is required.

In this work, we take all the $D_{(s)J}^{(*)}$ mesons as the conventional charmed(-strange) mesons. However, there are still controversies on the natures of $D^*_{s0}(2317)$ and $D_{s1}(2460)$ mesons (A recent review on this problem can be found in Ref.~\cite{Song:2015nia}.)
For examining whether $D^*_{s0}(2317)$ and $D_{s1}(2460)$ mesons are pure $c\bar{s}$ states, we need to lay stress on their electromagnetic and strong decays. If the future data implies that this assumption is not suitable, we should modify our wave functions  describing $D_{(s)J}^{(*)}$ mesons.

\section{Conclusion}

%
In this paper, including the BP, Ann, CS and CF contributions,  we re-analyze the process $B_c\to D_{s0}^{*}(2317)\mu\bar{\mu}$ and first  calculate the decays $B_c\to D_{s1}(2460,2536)\mu\bar{\mu}$, $B_c\to D_{s2}^*(2573)\mu\bar{\mu}$ and $B_c\to D_{J}^{(*)}\mu\bar{\mu}$.  Their results are  illustrated  in Figs.~11-18. And our conclusions contain
\begin{enumerate}

\item If only BP effects are considered, our results on the $B_c\to D_{s0}^{*}(2317)\mu\bar{\mu}$  transition are agreeable with the ones in Ref.~\cite{BctoDs2317lightcone} but quite different from the ones in Ref.~\cite{BctoDs2317QCDsunrule}. Once Ann, CS and CF Feynman diagrams are contained, the $B_c\to D_{s0}^{*}(2317)\mu\bar{\mu}$  observables change considerably, as shown in Figs.~11 (a-d).
\item As plotted in Figs.~14, 18 (a-h),   the observables of the $B_c\to D_{s1}(2536)(D_{1}(2430))\mu\bar{\mu}$ processes behave quite sensitively to the Ann and CF influences. This makes that when these channels are analyzed, besides the BP and CS diagrams, it is necessary to include the  Ann and CF ones.
\item Unlike the case of $B_c\to D_{s1}(2536)(D_{1}(2430))\mu\bar{\mu}$,  the observables of the $B_c\to D_{s2}^*(2573)\mu\bar{\mu}$, $B_c\to D_2^*(2460)\mu\bar{\mu}$, $B_c\to D_{s1}(2536)\mu\bar{\mu}$ and $B_c\to D_1(2420)\mu\bar{\mu}$ processes are influenced by Ann and CF diagrams slightly. Hence, if only BP effects are interesting, theses channels offer  purer laboratories than the $B_c\to D_{s1}(2536)(D_{1}(2430))\mu\bar{\mu}$ processes.
\end{enumerate}

%
%

\section*{Acknowledgments}

Our gratitude are expressed to Ramesh Verma~(Punjabi University) and Hai-Yang Cheng~(Institute of Physics, Academia Sinica) with whom we had very important and helpful discussions on the mixing natures of the axial vector mesons. This work is supported in part by the National Natural Science Foundation of China (NSFC) under Grant Nos. 11175051, 11347193, 11175151 and 11235005, in part by the Fundamental Research Funds for the Central Universities, Program for Innovation Research of Science in Harbin Institute of Technology, and in part by  the Program for New Century Excellent Talents in University (NCET) by Ministry of Education of P. R. China (Grant No. NCET-13-0991).

\clearpage

\appendix
\appendixpage
\addcontentsline{toc}{section}{Appendices}\markboth{APPENDICES}{}

\begin{subappendices}%
\subsection{Definitions of $\mathcal{F}^{\alpha}_{V1-7}$ and $\mathcal{F}^{\alpha}_{A1-3}$}\label{AppP-Y}


Here we present the explicit expressions of $\mathcal{F}^{\alpha}_{V1-7}$ and $\mathcal{F}^{\alpha}_{A1-3}$.

\begin{equation}
\begin{split}
\mathcal{F}^{\alpha}_{V1}=&d_8 e_4 M_i^2 \left(-g^{\mu \nu }\right) \epsilon ^{\alpha P_fq_aq_b}+d_8 \epsilon ^{\alpha P_fP_iq_b} \left(2 e_4 \left(q_a^{\nu } P_i^{\mu }-q_a^{\mu } P_i^{\nu }\right)+e_2 M_i g^{\mu \nu }\right)+\epsilon ^{\alpha P_fP_iq_a} \left(d_6 e_3\right.\\
&\left.M_f M_i g^{\mu \nu }-2 d_8 e_4 \left(g^{\mu \nu } P_i\cdot q_b+q_b^{\mu } P_i^{\nu }-q_b^{\nu } P_i^{\mu }\right)\right)+\epsilon ^{\alpha P_iq_aq_b} \left(d_7 e_3 M_f M_i g^{\mu \nu }+\right.\\
&\left.2 d_8 e_4 \left(g^{\mu \nu } P_f\cdot P_i-P_f^{\nu } P_i^{\mu }+P_f^{\mu } P_i^{\nu }\right)\right).
\end{split}
\end{equation}

\begin{equation}
\begin{split}
\mathcal{F}^{\alpha}_{V2}=&-M_i \epsilon ^{\mu \alpha P_fq_a} \left(d_8 e_4 M_i q_b^{\nu }+d_6 e_3 M_f P_i^{\nu }\right)-d_8 M_i \left(e_4 M_i q_a^{\nu }+e_2 P_i^{\nu }\right) \epsilon ^{\mu \alpha P_fq_b}+M_i \epsilon ^{\mu \alpha q_aq_b}\\
&\left(d_7 e_3 M_f P_i^{\nu }-d_8 e_4 M_i P_f^{\nu }\right)+\epsilon ^{\mu \alpha P_fP_i} \left(-2 P_i^{\nu } \left(d_8 e_4 q_a\cdot q_b+d_6 M_f\right)+q_a^{\nu } \left(2 d_8 e_4 P_i\cdot q_b\right.\right.\\
&\left.\left.P_i\cdot q_b-d_6 e_3 M_f M_i\right)+d_8 e_2 M_i q_b^{\nu }\right)+\epsilon ^{\mu \alpha P_iq_a} \left(e_3 M_f M_i \left(d_7 q_b^{\nu }+d_6 P_f^{\nu }\right)-2 e_4 \left(d_8 \left(P_f^{\nu } P_i\cdot q_b\right.\right.\right.\\
&\left.\left.\left.-q_b^{\nu } P_f\cdot P_i\right)+d_5 M_f^2 P_i^{\nu }\right)\right)+\epsilon ^{\mu \alpha P_iq_b} \left(P_i^{\nu } \left(2 d_7 M_f-2 d_8 e_4 P_f\cdot q_a\right)+q_a^{\nu } \left(d_7 e_3 M_f M_i\right.\right.\\
&\left.\left.\left.+2 d_8 e_4 P_f\cdot P_i\right)+d_8 e_2 M_i P_f^{\nu }\right)\right).
\end{split}
\end{equation}

\begin{equation}
\begin{split}
\mathcal{F}^{\alpha}_{V3}=&M_i \epsilon ^{\mu P_fq_aq_b} \left(d_8 e_4 M_i g^{\alpha \nu }-d_4 e_3 q_b^{\alpha } P_i^{\nu }\right)+\epsilon ^{\mu P_fP_iq_a} \left(2 e_4 \left(q_b^{\alpha } \left(d_8-d_2 M_f\right) P_i^{\nu }+d_8 \left(g^{\alpha \nu } P_i\cdot q_b\right.\right.\right.\\
&\left.\left.\left.-q_b^{\nu } P_i^{\alpha }\right)\right)-e_3 M_i \left(d_4 q_b^{\alpha } q_b^{\nu }+d_6 M_f g^{\alpha \nu }\right)\right)+\epsilon ^{\mu P_iq_aq_b} \left(e_3 M_i \left(d_4 q_b^{\alpha } P_f^{\nu }-d_7 M_f g^{\alpha \nu }\right)-2 e_4 \left(d_3\right.\right.\\
&\left.\left.M_f q_b^{\alpha } P_i^{\nu }+d_8 \left(g^{\alpha \nu } P_f\cdot P_i-P_f^{\nu } P_i^{\alpha }\right)\right)\right)+\epsilon ^{\mu P_fP_iq_b} \left(-q_a^{\nu } \left(d_4 e_3 M_i \text{$\epsilon $1}\cdot q_b+2 d_8 e_4 \alpha \cdot P_i\right)\right.\\
&\left.-2 \left(d_4-d_8 e_4\right) q_a^{\alpha } P_i^{\nu }-d_8 e_2 M_i g^{\alpha \nu }\right).
\end{split}
\end{equation}

\begin{equation}
\begin{split}
\mathcal{F}^{\alpha}_{V4}=&M_i \epsilon ^{\nu \alpha P_fq_a} \left(d_8 e_4 M_i q_b^{\mu }+d_6 e_3 M_f P_i^{\mu }\right)+d_8 M_i \left(e_4 M_i q_a^{\mu }+e_2 P_i^{\mu }\right) \epsilon ^{\nu \alpha P_fq_b}+M_i \epsilon ^{\nu \alpha q_aq_b} \left(d_8 e_4 \right.\\
&\left.M_i P_f^{\mu }-d_7 e_3 M_f P_i^{\mu }\right)+\epsilon ^{\nu \alpha P_fP_i} \left(2 P_i^{\mu } \left(d_8 e_4 q_a\cdot q_b+d_6 M_f\right)+q_a^{\mu } \left(d_6 e_3 M_f M_i-2 d_8 e_4 P_i\cdot q_b\right)\right.\\
&\left.-d_8 e_2 M_i q_b^{\mu }\right)+\epsilon ^{\nu \alpha P_iq_a} \left(2 e_4 \left(d_8 \left(P_f^{\mu } P_i\cdot q_b-q_b^{\mu } P_f\cdot P_i\right)+d_5 M_f^2 P_i^{\mu }\right)-e_3 M_f M_i \left(d_7 q_b^{\mu }\right.\right.\\
&\left.\left.+d_6 P_f^{\mu }\right)\right)+\epsilon ^{\nu \alpha P_iq_b} \left(2 P_i^{\mu } \left(d_8 e_4 P_f\cdot q_a-d_7 M_f\right)-q_a^{\mu } \left(d_7 e_3 M_f M_i+2 d_8 e_4 P_f\cdot P_i\right)\right.\\
&\left.\left.-d_8 e_2 M_i P_f^{\mu }\right)\right).
\end{split}
\end{equation}

\begin{equation}
\begin{split}
\mathcal{F}^{\alpha}_{V5}=&\epsilon ^{\nu P_fP_iq_b} \left(M_i \left(d_4 e_3 q_a^{\mu } q_b^{\alpha }+d_8 e_2 g^{\alpha \mu }\right)+2 P_i^{\mu } \left(d_4 q_b^{\alpha }-d_8 e_4 q_a^{\alpha }\right)+2 d_8 e_4 q_a^{\mu } P_i^{\alpha }\right)+M_i \epsilon ^{\nu P_fq_aq_b}\\
&\left(d_4 e_3 q_b^{\alpha } P_i^{\mu }-d_8 e_4 M_i g^{\alpha \mu }\right)+\epsilon ^{\nu P_fP_iq_a} \left(2 e_4 \left(q_b^{\alpha } \left(d_2 M_f-d_8\right) P_i^{\mu }+d_8 \left(q_b^{\mu } P_i^{\alpha }-g^{\alpha \mu } P_i\cdot q_b\right)\right)\right.\\
&\left.+e_3 M_i \left(d_4 q_b^{\alpha } q_b^{\mu }+d_6 M_f g^{\alpha \mu }\right)\right)+\epsilon ^{\nu P_iq_aq_b} \left(e_3 M_i \left(d_7 M_f g^{\alpha \mu }-d_4 q_b^{\alpha } P_f^{\mu }\right)+2 e_4 \left(d_3 M_f q_b^{\alpha } P_i^{\mu }\right.\right.\\
&\left.\left.+d_8 \left(g^{\alpha \mu } P_f\cdot P_i-P_f^{\mu } P_i^{\alpha }\right)\right)\right).
\end{split}
\end{equation}

\begin{equation}
\begin{split}
\mathcal{F}^{\alpha}_{V6}=&M_i \epsilon ^{\mu \nu \alpha P_f} \left(M_i \left(d_8 e_4 q_a\cdot q_b+d_6 M_f\right)+d_8 e_2 P_i\cdot q_b\right)+\epsilon ^{\mu \nu \alpha P_i} \left(-2 \left(P_f\cdot P_i \left(d_8 e_4 q_a\cdot q_b\right.\right.\right.\\
&\left.\left.\left.+d_6 M_f\right)+d_7 M_f P_i\cdot q_b\right)+M_f M_i \left(d_5 e_2 M_f-e_3 \left(d_7 q_a\cdot q_b+d_6 P_f\cdot q_a\right)\right)+2 d_8 e_4 P_f\cdot q_a P_i\cdot q_b\right)\\
&-M_f M_i \epsilon ^{\mu \nu \alpha q_a} \left(e_3 \left(d_7 P_i\cdot q_b+d_6 P_f\cdot P_i\right)+d_5 e_4 M_f M_i\right)-M_i \epsilon ^{\mu \nu \alpha q_b} \left(M_i \left(d_8 e_4 P_f\cdot q_a\right.\right.\\
&\left.\left.-d_7 M_f\right)+d_8 e_2 P_f\cdot P_i\right).
\end{split}
\end{equation}

\begin{equation}
\begin{split}
\mathcal{F}^{\alpha}_{V7}=&M_i \epsilon ^{\mu \nu P_fq_b} \left(M_i \left(d_8 e_4 q_a^{\alpha }-d_4 q_b^{\alpha }\right)+d_8 e_2 P_i^{\alpha }\right)+\epsilon ^{\mu \nu P_fP_i} \left(2 P_i^{\alpha } \left(d_8 e_4 q_a\cdot q_b+d_6 M_f\right)\right.\\
&\left.+q_a^{\alpha } \left(d_6 e_3 M_f M_i-2 d_8 e_4 P_i\cdot q_b\right)+q_b^{\alpha } \left(M_i \left(d_4 e_3 q_a\cdot q_b+e_2 \left(d_2 M_f-d_8\right)\right)+2 d_4 P_i\cdot q_b\right)\right)\\
&+M_i \epsilon ^{\mu \nu P_fq_a} \left(q_b^{\alpha } \left(d_4 e_3 P_i\cdot q_b+e_4 M_i \left(d_8-d_2 M_f\right)\right)+d_6 e_3 M_f P_i^{\alpha }\right)+\epsilon ^{\mu \nu P_iq_a} \left(q_b^{\alpha } \left(2 e_4\right.\right.\\
&\left.\left. \left(d_3 M_f P_i\cdot q_b+\left(d_2 M_f-d_8\right) P_f\cdot P_i\right)-e_3 M_f M_i \left(d_1 M_f+d_7\right)\right)+2 d_5 e_4 M_f^2 P_i^{\alpha }\right)+\epsilon ^{\mu \nu P_iq_b}\\
&\left(q_b^{\alpha } \left(M_i \left(d_4 e_3 P_f\cdot q_a-d_3 e_2 M_f\right)+2 d_4 P_f\cdot P_i\right)+2 P_i^{\alpha } \left(d_8 e_4 P_f\cdot q_a-d_7 M_f\right)-q_a^{\alpha } \left(d_7 e_3 M_f M_i\right.\right.\\
&\left.\left.+2 d_8 e_4 P_f\cdot P_i\right)\right)+M_i \epsilon ^{\mu \nu q_aq_b} \left(q_b^{\alpha } \left(d_3 e_4 M_f M_i+d_4 e_3 P_f\cdot P_i\right)-d_7 e_3 M_f P_i^{\alpha }\right).
\end{split}
\end{equation}

\begin{equation}
\begin{split}
\mathcal{F}^{\alpha}_{A1}=&-d_7 M_f \left(-q_a^{\nu } g^{\alpha \mu } P_i\cdot q_b+q_a^{\mu } g^{\alpha \nu } P_i\cdot q_b-q_a^{\alpha } g^{\mu \nu } P_i\cdot q_b+g^{\alpha \mu } P_i^{\nu } q_a\cdot q_b-g^{\alpha \nu } P_i^{\mu } q_a\cdot q_b\right.\\
&\left.+g^{\mu \nu } P_i^{\alpha } q_a\cdot q_b+q_a^{\mu } q_b^{\alpha } P_i^{\nu }-q_a^{\nu } q_b^{\alpha } P_i^{\mu }-q_a^{\alpha } q_b^{\mu } P_i^{\nu }+q_a^{\nu } q_b^{\mu } P_i^{\alpha }+q_a^{\alpha } q_b^{\nu } P_i^{\mu }-q_a^{\mu } q_b^{\nu } P_i^{\alpha }\right)\\
&+d_4 q_b^{\alpha } \left(g^{\mu \nu } q_a\cdot q_b P_f\cdot P_i-g^{\mu \nu } P_f\cdot q_a P_i\cdot q_b+q_a^{\nu } q_b^{\mu } P_f\cdot P_i-q_a^{\mu } q_b^{\nu } P_f\cdot P_i-q_b^{\mu } P_i^{\nu } P_f\cdot q_a\right.\\
&\left.+q_b^{\nu } P_i^{\mu } P_f\cdot q_a+q_a^{\mu } P_f^{\nu } P_i\cdot q_b-q_a^{\nu } P_f^{\mu } P_i\cdot q_b-P_f^{\nu } P_i^{\mu } q_a\cdot q_b+P_f^{\mu } P_i^{\nu } q_a\cdot q_b\right)+d_1 M_f^2 q_b^{\alpha }\\
&\left(-\left(q_a^{\mu } P_i^{\nu }-q_a^{\nu } P_i^{\mu }\right)\right)-d_6 M_f \left(-q_a^{\alpha } g^{\mu \nu } P_f\cdot P_i-q_a^{\nu } g^{\alpha \mu } P_f\cdot P_i+q_a^{\mu } g^{\alpha \nu } P_f\cdot P_i\right.\\
&\left.+g^{\alpha \mu } P_i^{\nu } P_f\cdot q_a-g^{\alpha \nu } P_i^{\mu } P_f\cdot q_a+g^{\mu \nu } P_i^{\alpha } P_f\cdot q_a+q_a^{\alpha } P_f^{\nu } P_i^{\mu }-q_a^{\alpha } P_f^{\mu } P_i^{\nu }-q_a^{\mu } P_f^{\nu } P_i^{\alpha }+q_a^{\nu } P_f^{\mu } P_i^{\alpha }\right).
\end{split}
\end{equation}

\begin{equation}
\begin{split}
\mathcal{F}^{\alpha}_{A2}=&d_2 M_f q_b^{\alpha } \left(-g^{\mu \nu } P_f\cdot P_i+P_f^{\nu } P_i^{\mu }-P_f^{\mu } P_i^{\nu }\right)-d_3 M_f q_b^{\alpha } \left(g^{\mu \nu } P_i\cdot q_b+q_b^{\mu } P_i^{\nu }-q_b^{\nu } P_i^{\mu }\right)-d_8\\
&\left(-q_b^{\alpha } g^{\mu \nu } P_f\cdot P_i-q_b^{\nu } g^{\alpha \mu } P_f\cdot P_i+q_b^{\mu } g^{\alpha \nu } P_f\cdot P_i+P_f^{\nu } g^{\alpha \mu } P_i\cdot q_b-P_f^{\mu } g^{\alpha \nu } P_i\cdot q_b+q_b^{\alpha } P_f^{\nu } P_i^{\mu }\right.\\
&\left.-q_b^{\alpha } P_f^{\mu } P_i^{\nu }-q_b^{\mu } P_f^{\nu } P_i^{\alpha }+q_b^{\nu } P_f^{\mu } P_i^{\alpha }\right)+d_1 M_f^2 \left(-q_b^{\alpha }\right) g^{\mu \nu }-d_7 M_f \left(q_b^{\nu } g^{\alpha \mu }-q_b^{\mu } g^{\alpha \nu }+q_b^{\alpha } g^{\mu \nu }\right)\\
&-d_4 q_b^{\alpha } \left(q_b^{\mu } P_f^{\nu }-q_b^{\nu } P_f^{\mu }\right)-d_5 M_f^2 \left(g^{\alpha \mu } P_i^{\nu }-g^{\alpha \nu } P_i^{\mu }+g^{\mu \nu } P_i^{\alpha }\right)-d_6 M_f \left(P_f^{\nu } g^{\alpha \mu }-P_f^{\mu } g^{\alpha \nu }\right).
\end{split}
\end{equation}

\begin{equation}
\begin{split}
\mathcal{F}^{\alpha}_{A3}=&d_2 M_f q_b^{\alpha } \left(-g^{\mu \nu } P_f\cdot q_a+q_a^{\mu } P_f^{\nu }-q_a^{\nu } P_f^{\mu }\right)-d_3 M_f q_b^{\alpha } \left(g^{\mu \nu } q_a\cdot q_b+q_a^{\nu } q_b^{\mu }-q_a^{\mu } q_b^{\nu }\right)\\
&-d_8 \left(-q_b^{\alpha } g^{\mu \nu } P_f\cdot q_a-q_b^{\nu } g^{\alpha \mu } P_f\cdot q_a+q_b^{\mu } g^{\alpha \nu } P_f\cdot q_a+P_f^{\nu } g^{\alpha \mu } q_a\cdot q_b-P_f^{\mu } g^{\alpha \nu } q_a\cdot q_b\right.\\
&\left.+q_a^{\mu } q_b^{\alpha } P_f^{\nu }-q_a^{\nu } q_b^{\alpha } P_f^{\mu }-q_a^{\alpha } q_b^{\mu } P_f^{\nu }+q_a^{\alpha } q_b^{\nu } P_f^{\mu }\right)+d_5 M_f^2 \left(-\left(q_a^{\nu } g^{\alpha \mu }-q_a^{\mu } g^{\alpha \nu }+q_a^{\alpha } g^{\mu \nu }\right)\right).
\end{split}
\end{equation}

\subsection{Definitions of $P_i$, $P_f$, $\epsilon_A$, $\epsilon_{T}$ and $\epsilon^{\mu}_ {H}$ }\label{AppObser}
During calculating the physical observables, we must specify the $P_i$, $P_f$, $\epsilon_A$, $\epsilon_{T}$ and $\epsilon^{\mu}_ {H}$.
In the initial meson rest frame, we have $
P_{i}^{\alpha}=(M_i,0,0,0)$ and $P_{f}^{\alpha}=(E_f,0,0,P_f^3)$.
The polarization vectors $\epsilon_A^{\alpha}$ are chosen as
$\epsilon_A^{\alpha}(\pm1)=\frac{1}{\sqrt{2}}(0,\pm1
,+i,0)$ and $\epsilon_A^{\alpha}(0)=\frac{1}{M_f}(-P_f^3,0,0,-E_f)$.
The polarization tensors $\epsilon_T^{\alpha\beta}$ can be constructed in terms of the polarization vectors $\epsilon_A^{\alpha}$, which
are written as
\begin{equation}\begin{split}
\epsilon&_{T}^{\alpha\beta}(\pm2)=\epsilon_A(\pm1)^{\alpha}\epsilon_A(\pm1)^{\beta},\\
\epsilon&_{T}^{\alpha\beta}(\pm1)=\sqrt{\frac{1}{2}}\left\{ \epsilon _{A}(\pm1)^{\alpha}\epsilon_A(0)^{\beta}+\epsilon _{A}(0)^{\alpha}\epsilon_A(\pm1)^{\beta}\right\},\\
\epsilon&_{T}^{\alpha\beta}=\sqrt{\frac{1}{6}}\left\{\epsilon_A(+1)^{\alpha}\epsilon_A(-1)^{\beta}+\epsilon_A(-1)^{\alpha}\epsilon_A(+1)^{\beta}\right\}
+\sqrt{\frac{2}{3}}\epsilon_A(0)^{\alpha}\epsilon_A(0)^{\beta}.
\end{split}\label{eq:specifyepsilonT}\end{equation}
Besides, we define the helicity
amplitudes as~\cite{AFaessler}
\begin{equation}\begin{split}
\epsilon&^{\mu}_ {H} (t)=\frac{1}{\sqrt{Q^2}}(M_i-E_f,0,0,-P_f^3),\\
\epsilon&^{\mu}_H (\pm1)=\frac{1}{\sqrt{2}}(0,\mp1,+i,0),\\
\epsilon&^{\mu}_H (0)=\frac{1}{\sqrt{Q^2}}(-P_f^3,0,0,M_i-E_f).
\end{split}\end{equation}

\end{subappendices}

\begin{figure}[htbp]
\centering \subfigure[Form-Factors of $W^{\mu}$ and $W^{\mu}_T$
induced by penguin and box diagrams]{\includegraphics[width =
0.49\textwidth,height=0.4\textheight]{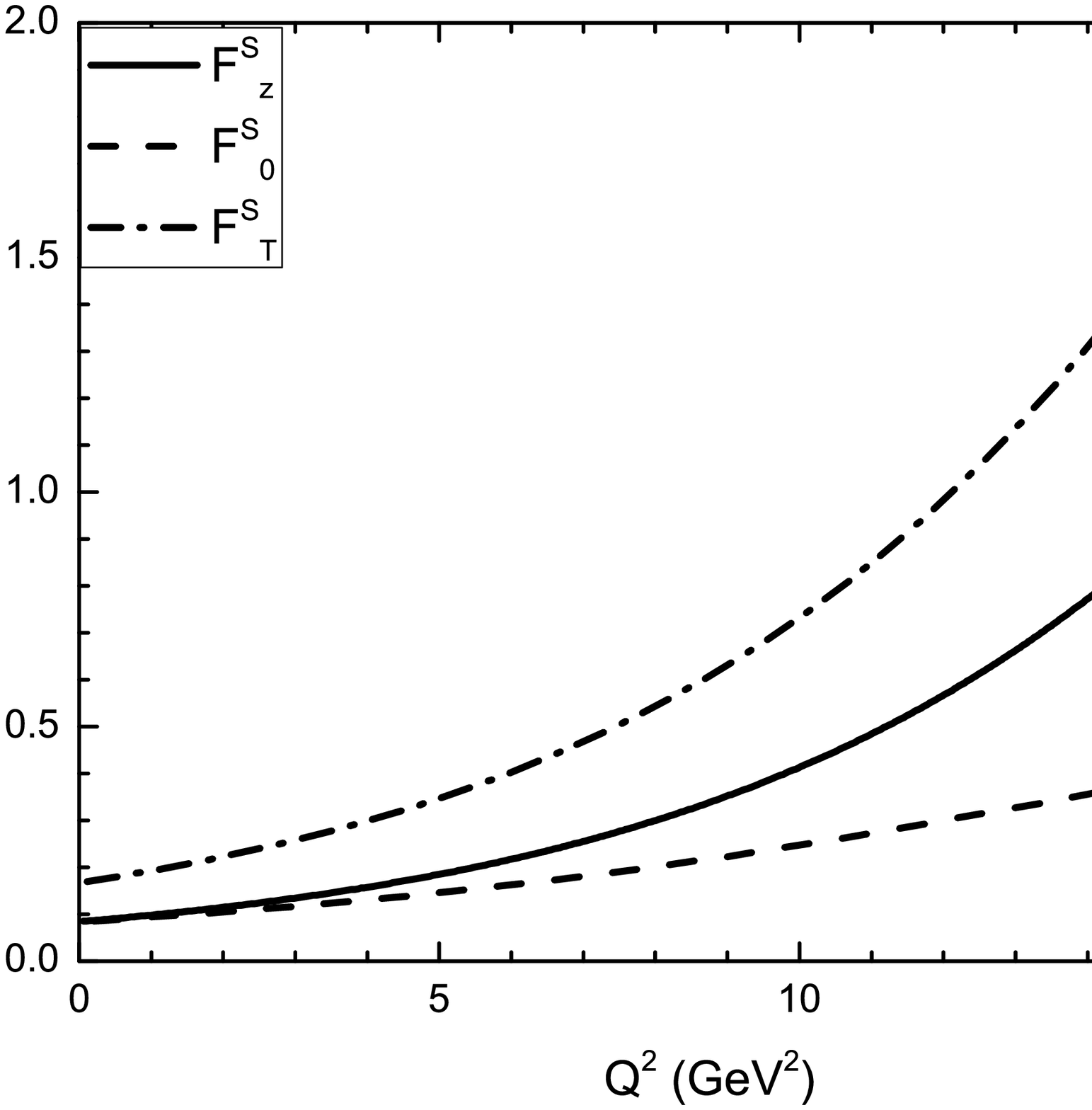}}
\subfigure[Form-Factors of $W^{\mu}_{ann}$ induced by annihilation
diagrams]{\includegraphics[width =
0.49\textwidth,height=0.4\textheight]{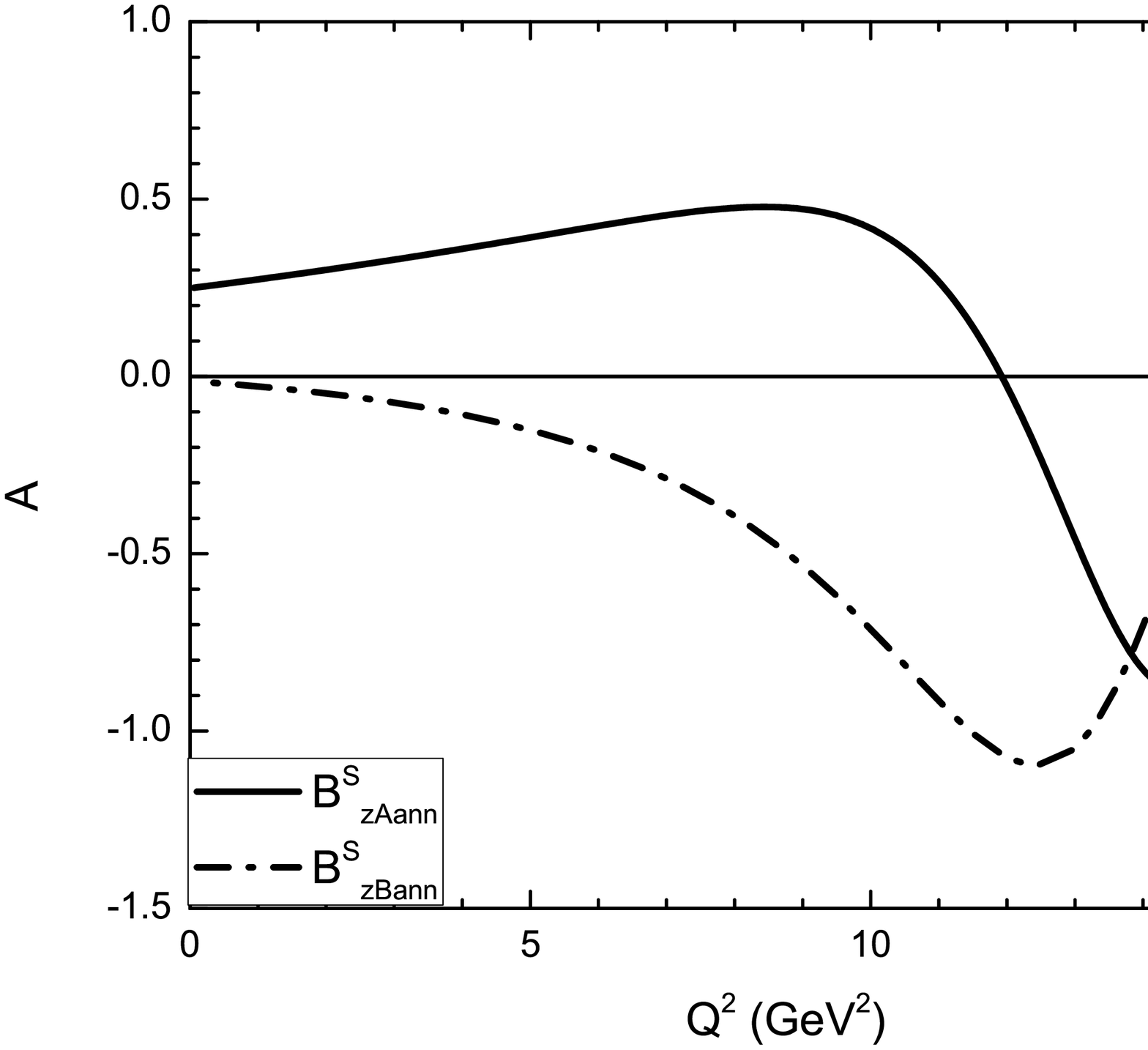}}
\caption{Form-Factors of $B_c\rightarrow D_{s0}^*(2317) l\bar{l}$, where $B^S_{zAann}$  stands for $\text{Re}[B^S_{zann}]$, while $B^S_{zBann}$ denotes $\text{Im}[B^S_{zann}]$.}
\end{figure}

\begin{figure}[htbp]
\centering \subfigure[Form-factors of $W^{\mu}$ induced by $Z^0$
penguin and box diagrams]{\includegraphics[width =
0.49\textwidth,height=0.4\textheight]{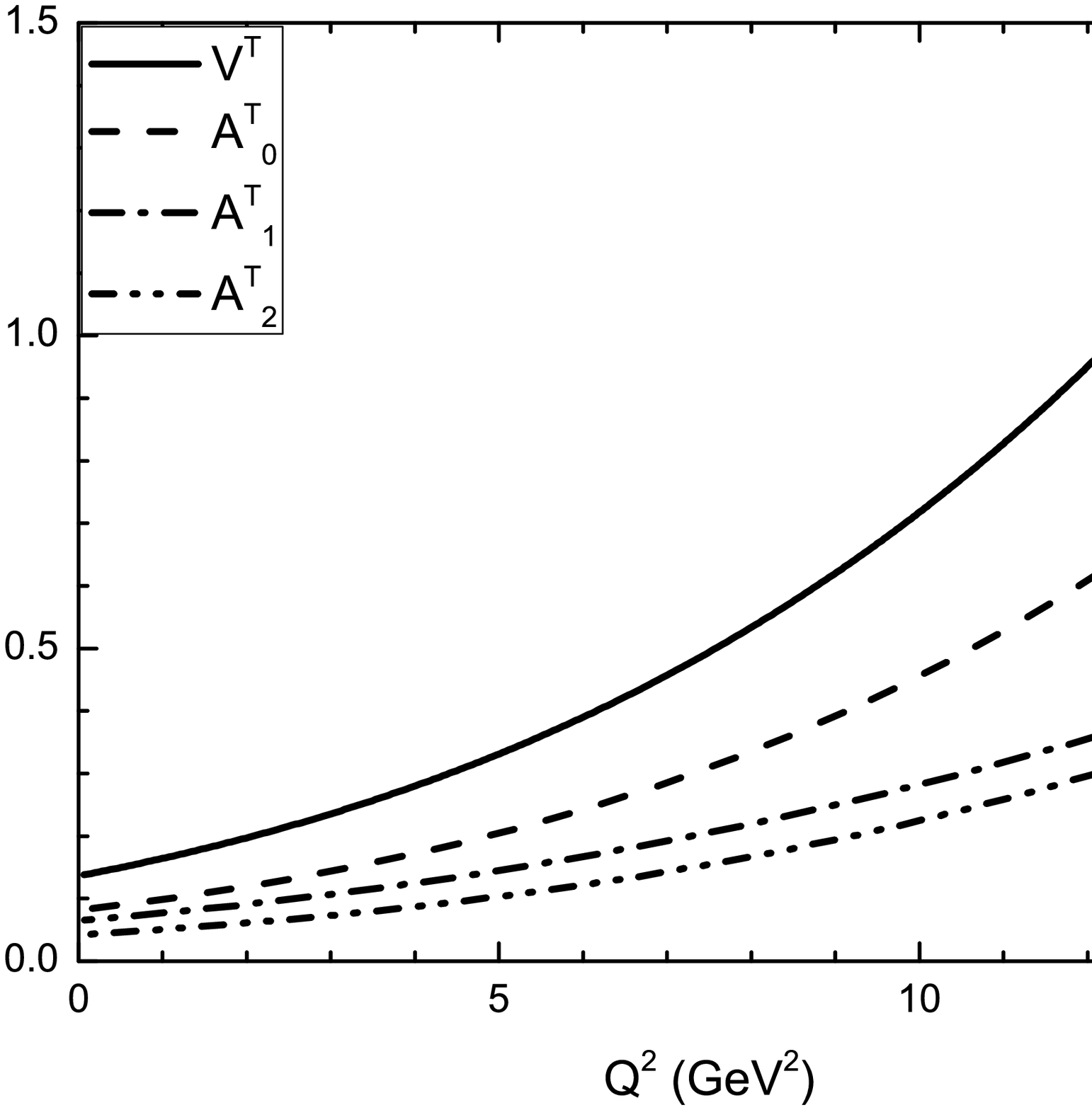}}
\subfigure[Form-factors of $W^{\mu}_T$ induced by $\gamma$ penguin diagrams]{\includegraphics[width = 0.49\textwidth,height=0.4\textheight]{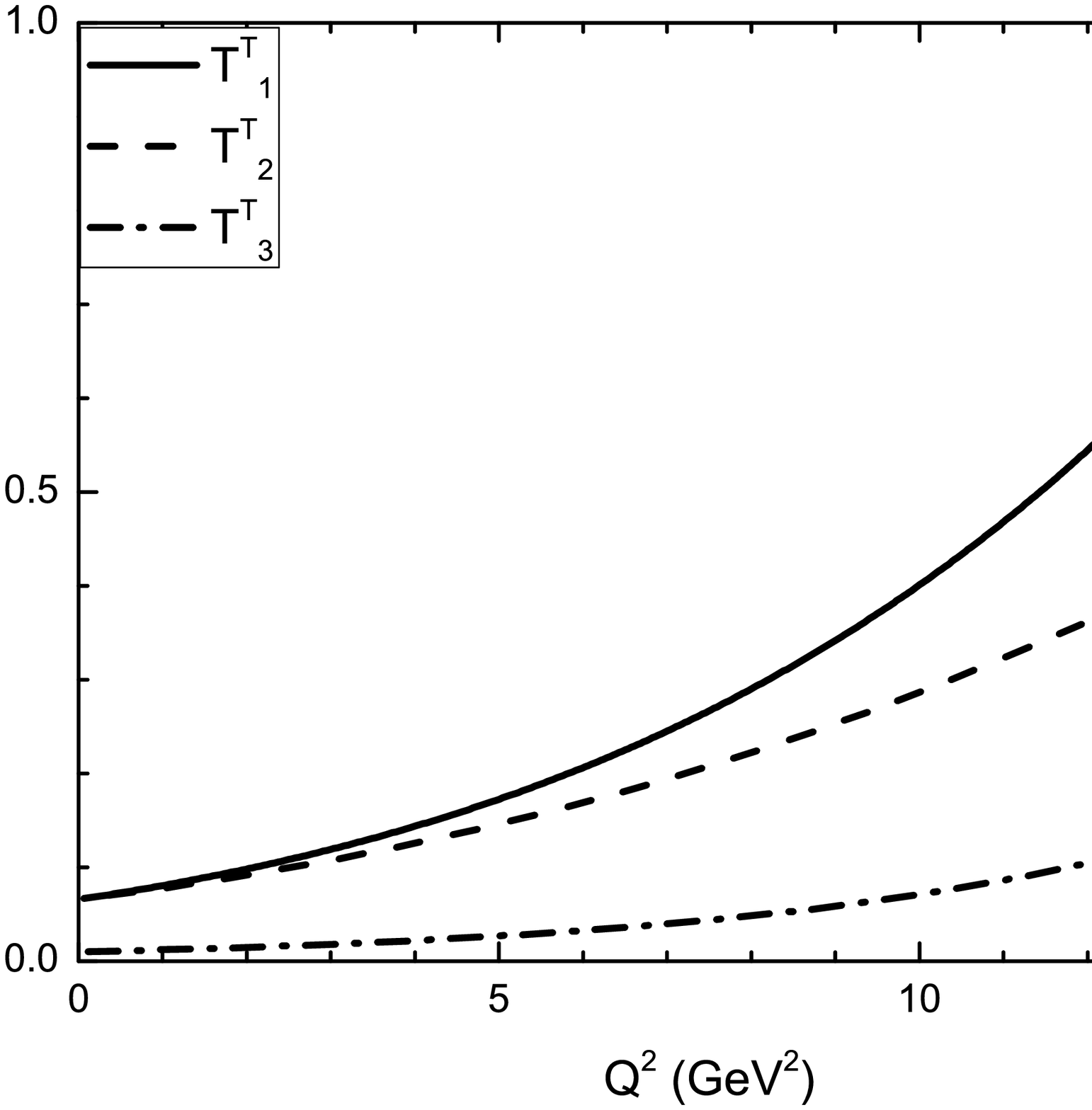}}\\
\subfigure[Real parts of $W^{\mu}_{ann}$ Form-Factors induced by
annihilation diagrams]{\includegraphics[width =
0.49\textwidth,height=0.4\textheight]{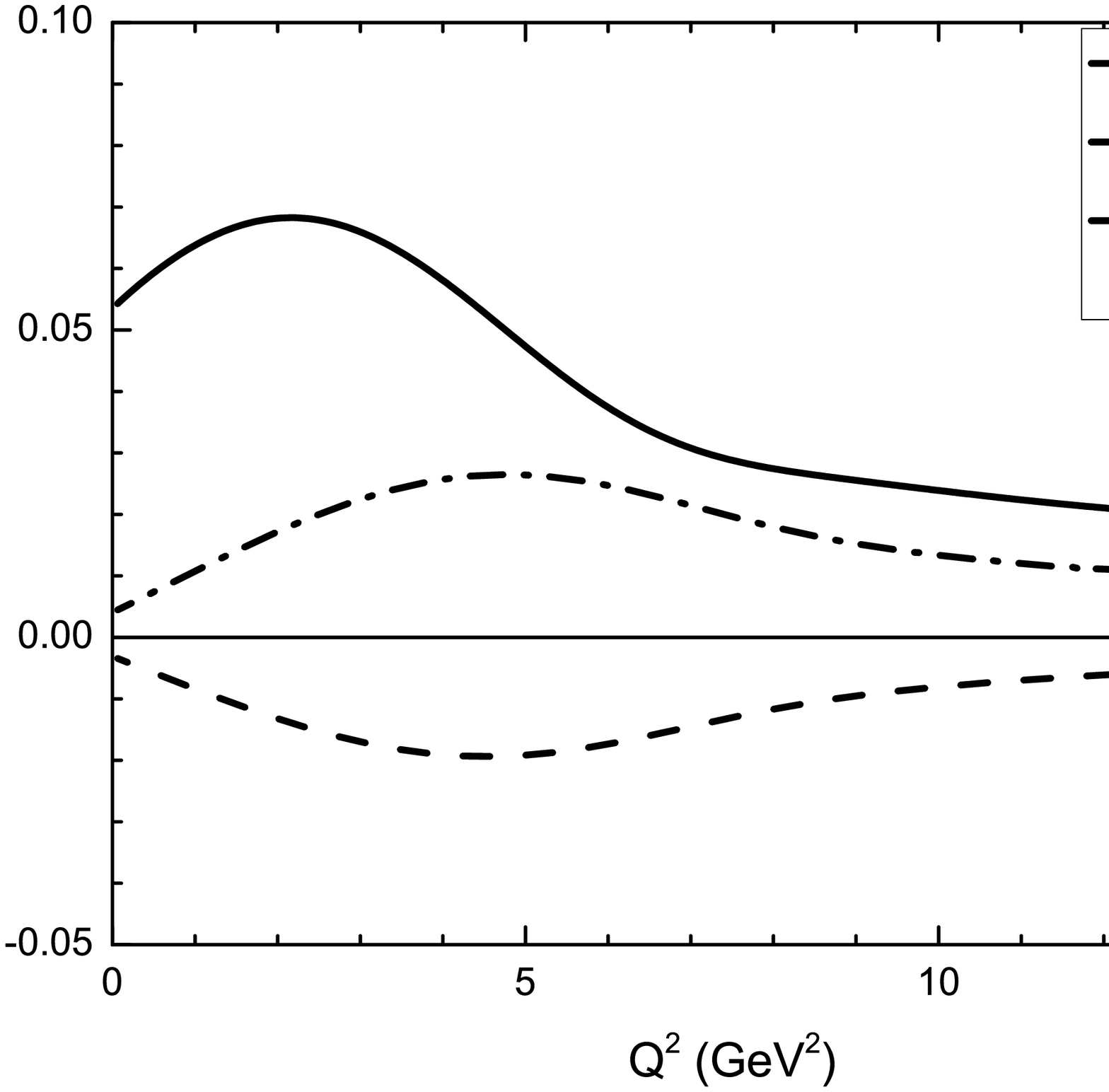}}
\subfigure[Imaginary parts of $W^{\mu}_{ann}$ Form-Factors induced
by annihilation
diagrams]{\includegraphics[width = 0.49\textwidth,height=0.4\textheight]{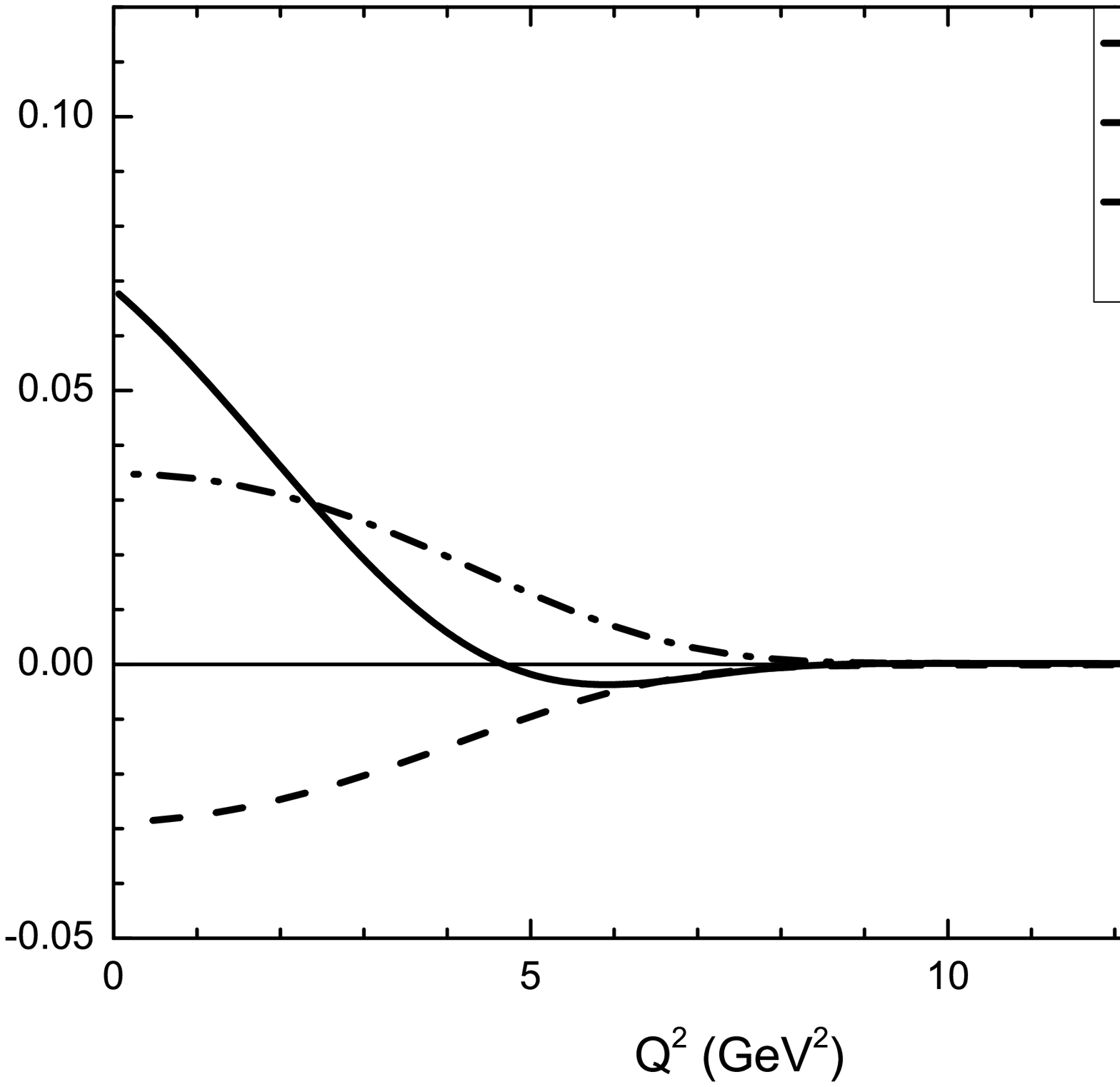}}\\
\caption{Form-factors of $B_c\rightarrow D_{s2}^*(2573)l\bar{l}$, where $V^T_{Aann}$ and $T^T_{1,zAann}$ stand for $\text{Re}[V^T_{ann}]$ and $\text{Re}[T^T_{1,zann}]$, respectively, while $V^T_{Bann}$ and $T^T_{1,zBann}$ denote $\text{Im}[V^T_{ann}]$ and $\text{Im}[T^T_{1,zann}]$.}
\end{figure}

\begin{figure}[htbp]
\centering \subfigure[Form-factors of $W^{\mu}$ induced by $Z^0$
penguin and box diagrams]{\includegraphics[width =
0.49\textwidth,height=0.4\textheight]{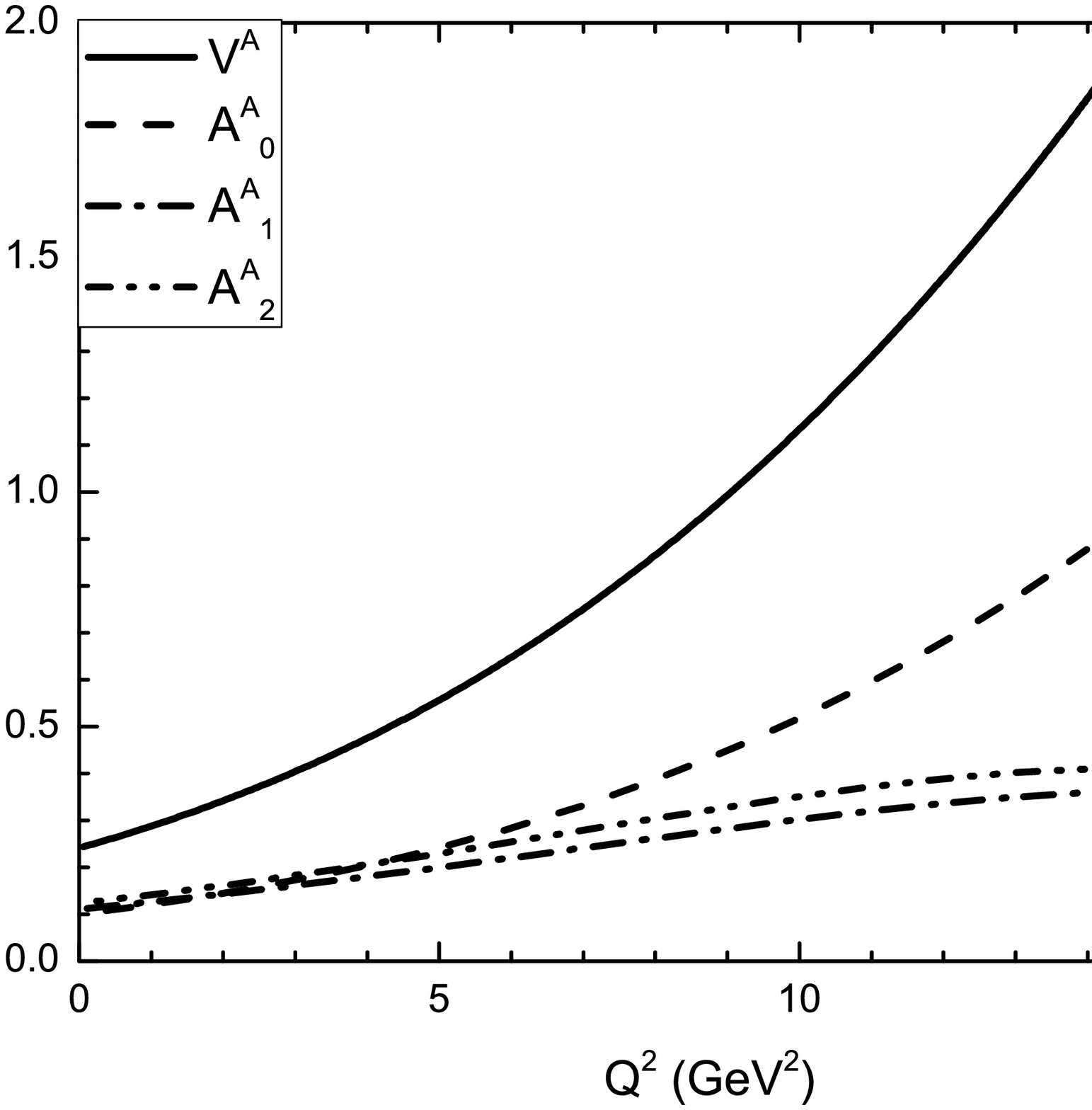}}
\subfigure[Form-factors of $W^{\mu}_T$ induced by $\gamma$ penguin diagrams]{\includegraphics[width = 0.49\textwidth,height=0.4\textheight]{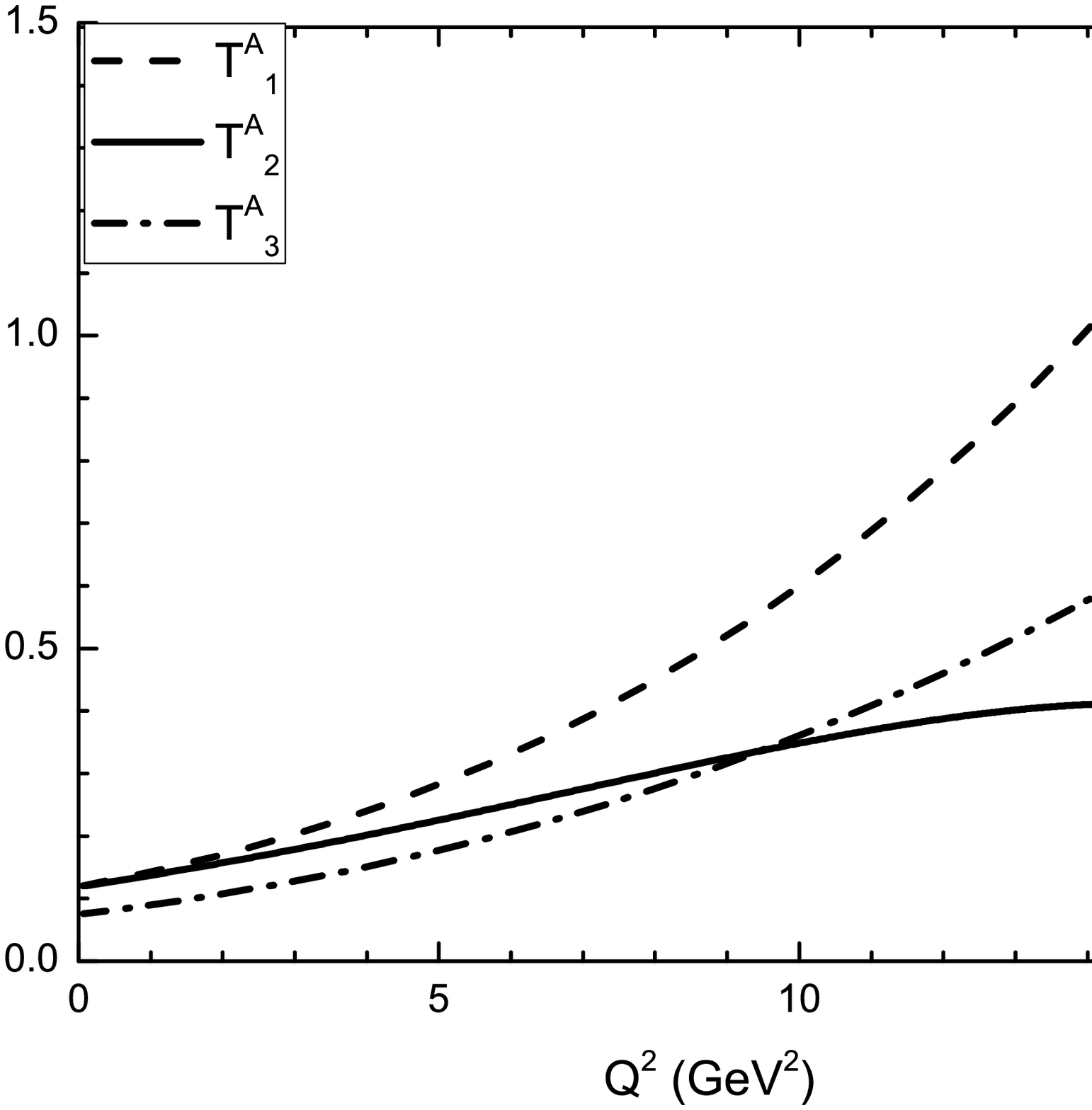}}\\
\subfigure[Real parts of $W^{\mu}_{ann}$ Form-Factors induced by
annihilation diagrams]{\includegraphics[width =
0.49\textwidth,height=0.4\textheight]{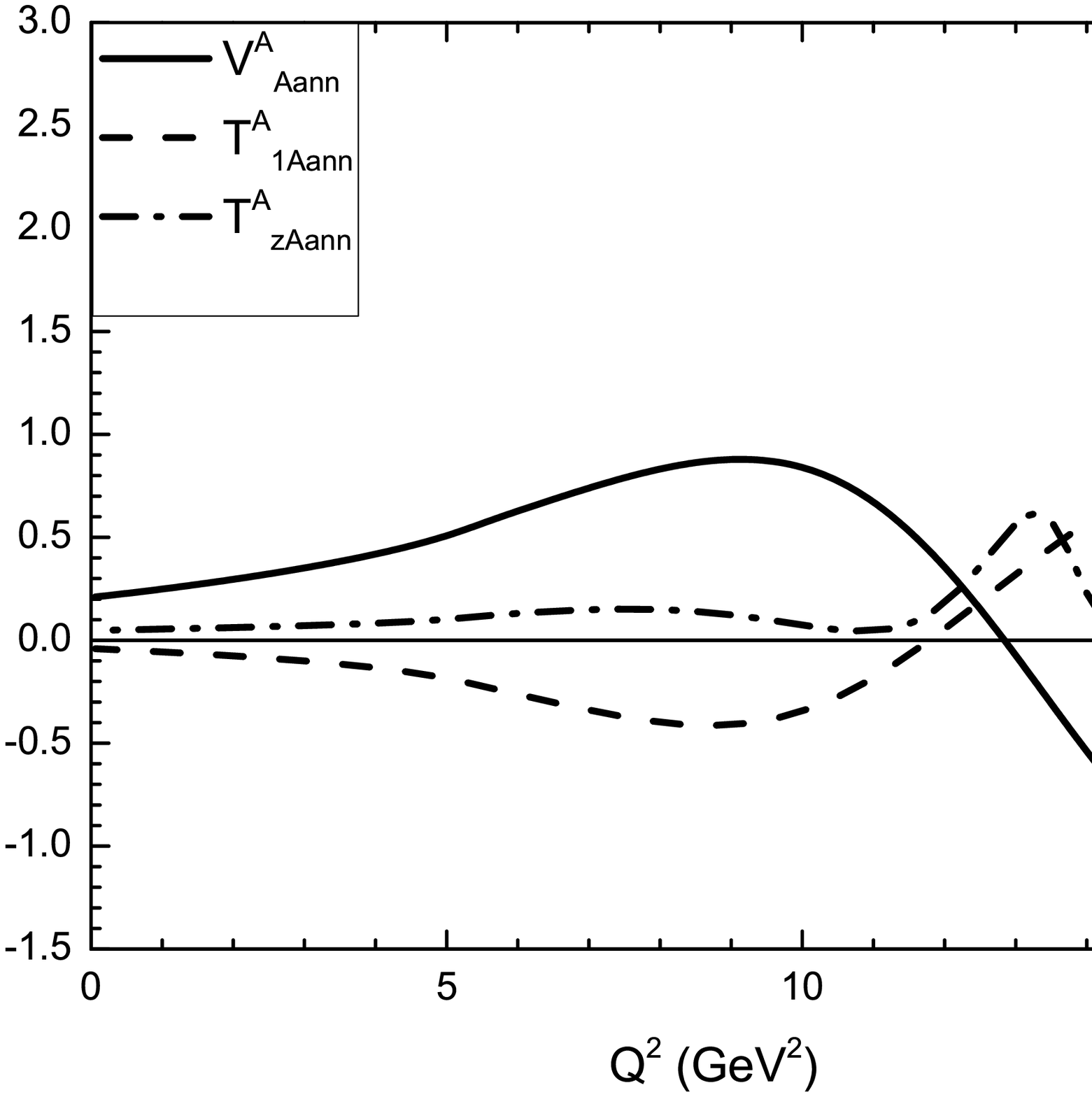}}
\subfigure[Imaginary parts of $W^{\mu}_{ann}$ Form-Factors induced
by annihilation
diagrams]{\includegraphics[width = 0.49\textwidth,height=0.4\textheight]{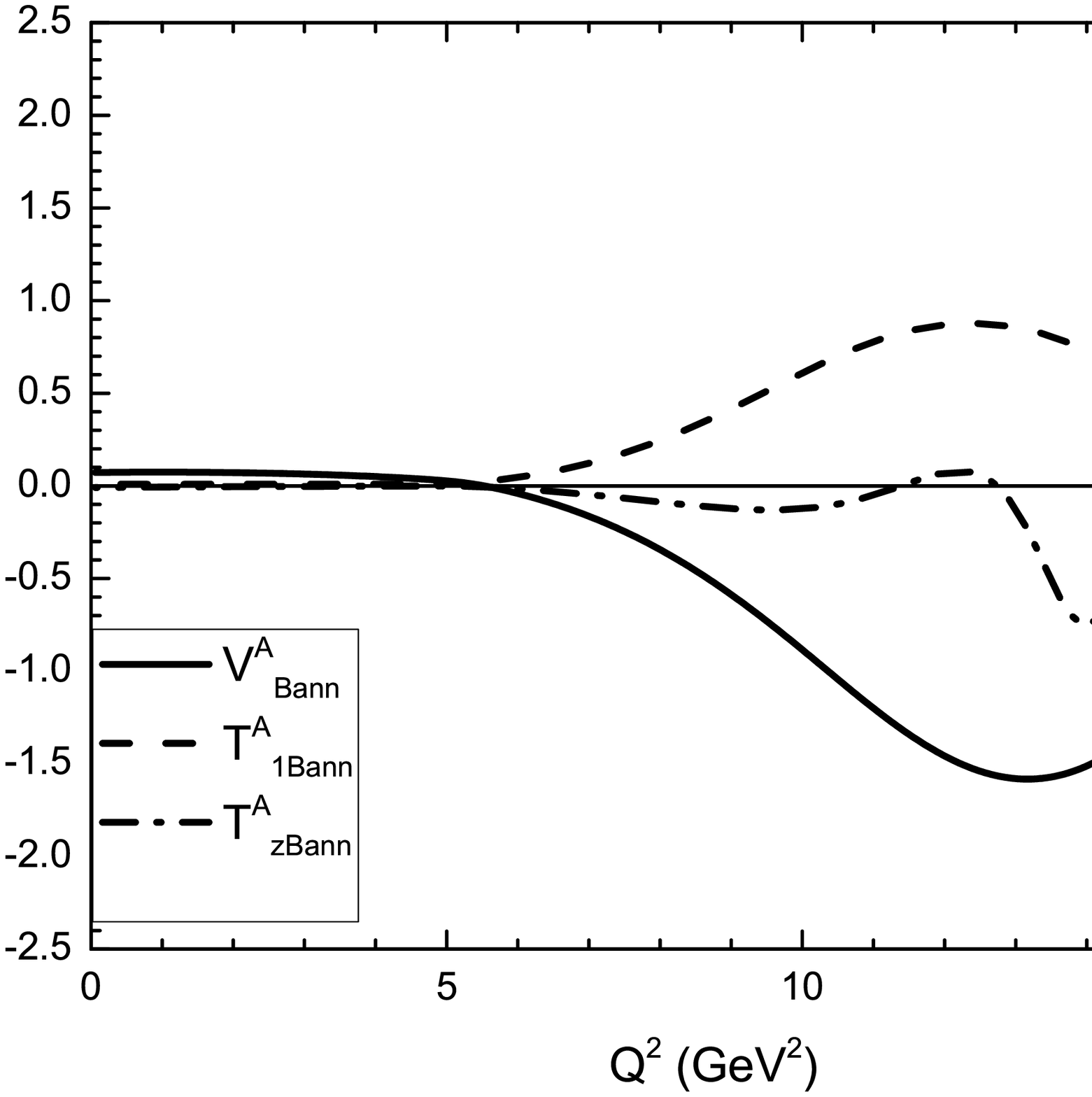}}\\
\caption{Form-factors of $B_c\rightarrow D_{s1}(2460)l\bar{l}$, where $V^A_{Aann}$ and $T^A_{1,zAann}$ stand for $\text{Re}[V^A_{ann}]$ and $\text{Re}[T^A_{1,zann}]$, respectively, while $V^A_{Bann}$ and $T^A_{1,zBann}$ denote $\text{Im}[V^A_{ann}]$ and $\text{Im}[T^A_{1,zann}]$.}
\end{figure}

\begin{figure}[htbp]
\centering \subfigure[Form-factors of $W^{\mu}$ induced by $Z^0$
penguin and box diagrams]{\includegraphics[width =
0.49\textwidth,height=0.4\textheight]{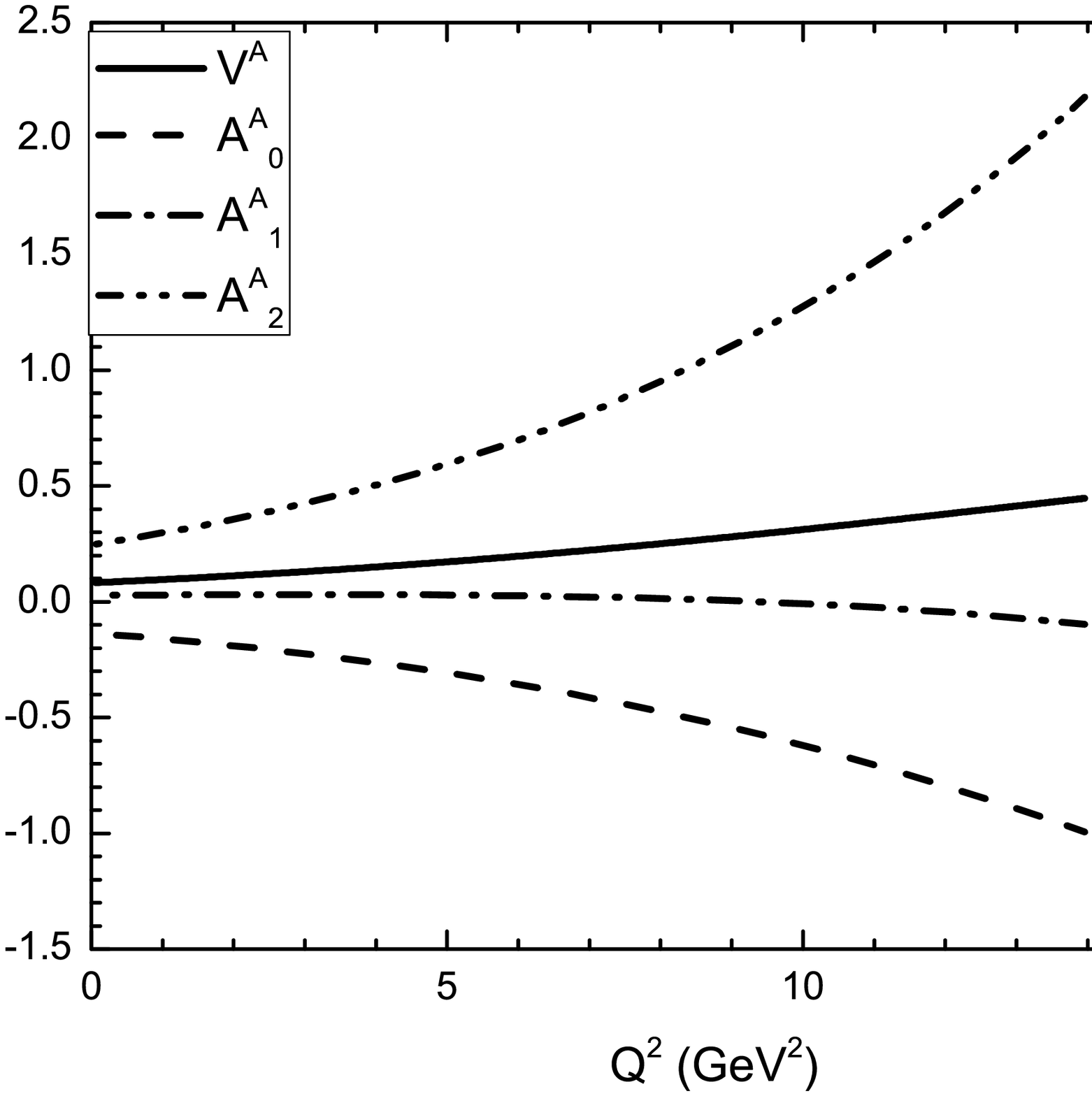}}
\subfigure[Form-factors of $W^{\mu}_T$ induced by $\gamma$ penguin diagrams]{\includegraphics[width = 0.49\textwidth,height=0.4\textheight]{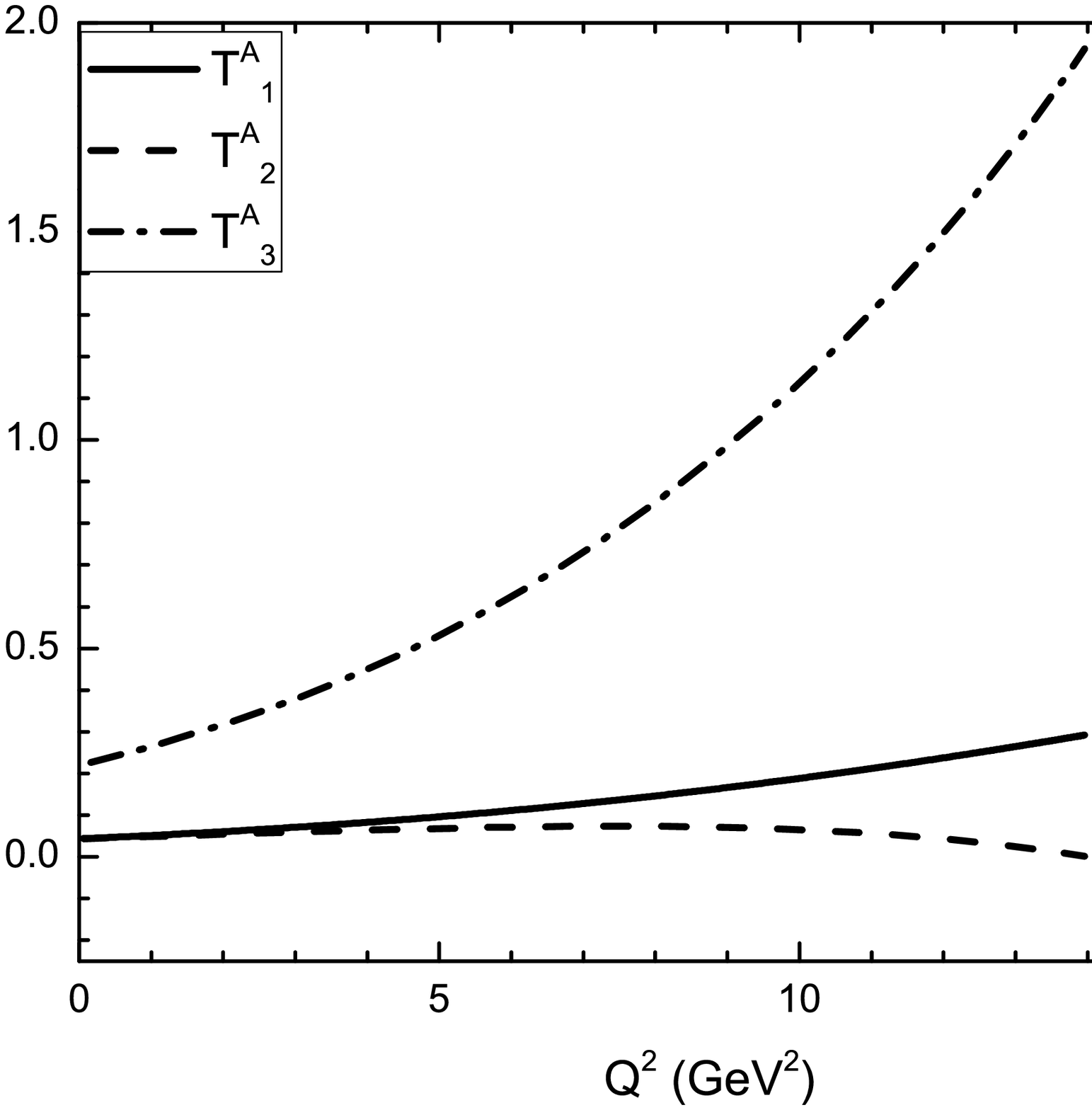}}\\
\subfigure[Real parts of $W^{\mu}_{ann}$ Form-Factors induced by
annihilation diagrams]{\includegraphics[width =
0.49\textwidth,height=0.4\textheight]{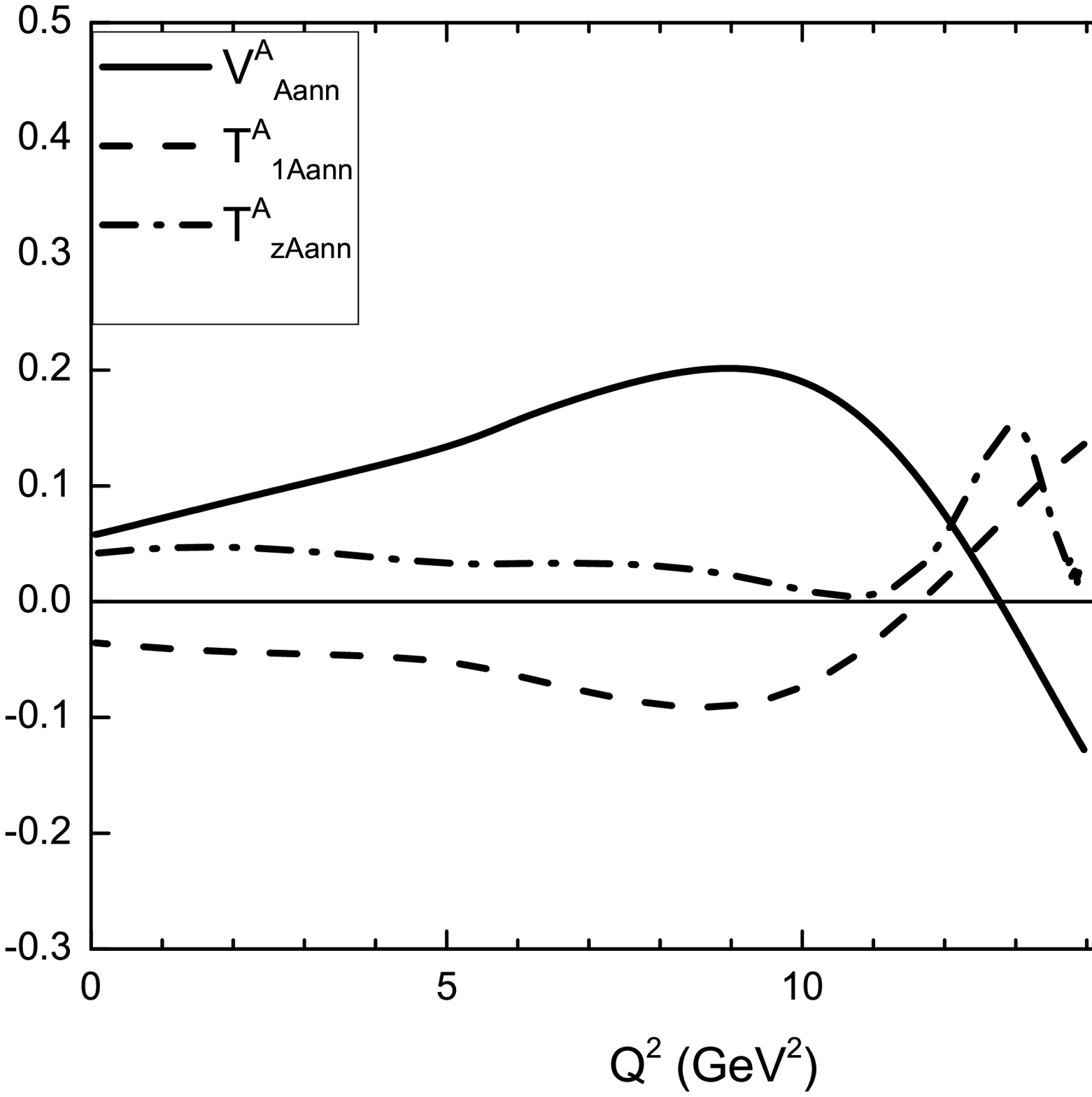}}
\subfigure[Imaginary parts of $W^{\mu}_{ann}$ Form-Factors induced
by annihilation
diagrams]{\includegraphics[width = 0.49\textwidth,height=0.4\textheight]{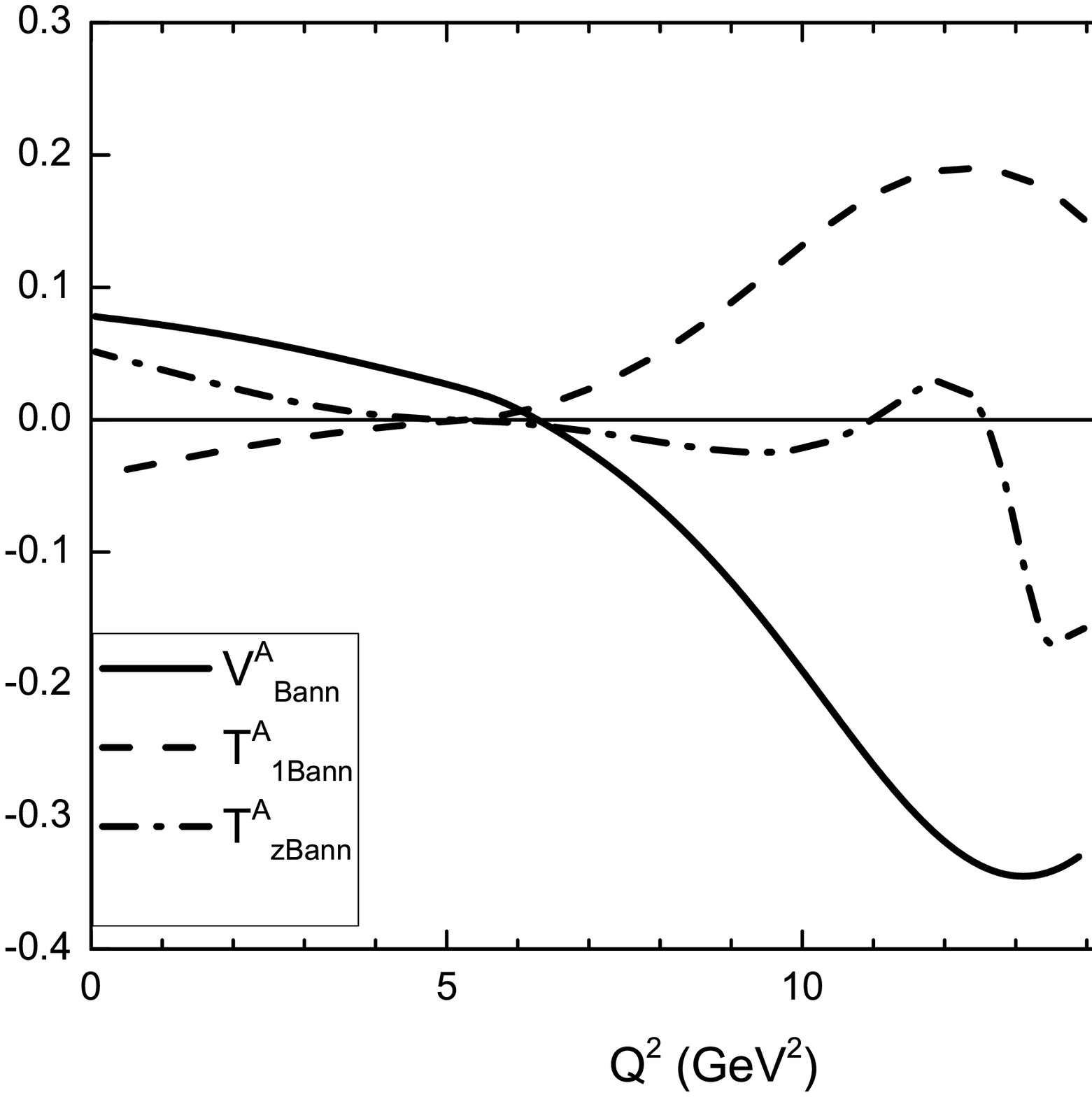}}\\
\caption{Form-factors of $B_c\rightarrow D_{s1}(2536)l\bar{l}$, where $V^A_{Aann}$ and $T^A_{1,zAann}$ stand for $\text{Re}[V^A_{ann}]$ and $\text{Re}[T^A_{1,zann}]$, respectively, while $V^A_{Bann}$ and $T^A_{1,zBann}$ denote $\text{Im}[V^A_{ann}]$ and $\text{Im}[T^A_{1,zann}]$.}
\end{figure}

\begin{figure}[htbp]
\centering \subfigure[Form-Factors of $W^{\mu}$ and $W^{\mu}_T$
induced by penguin and box diagrams]{\includegraphics[width =
0.49\textwidth,height=0.4\textheight]{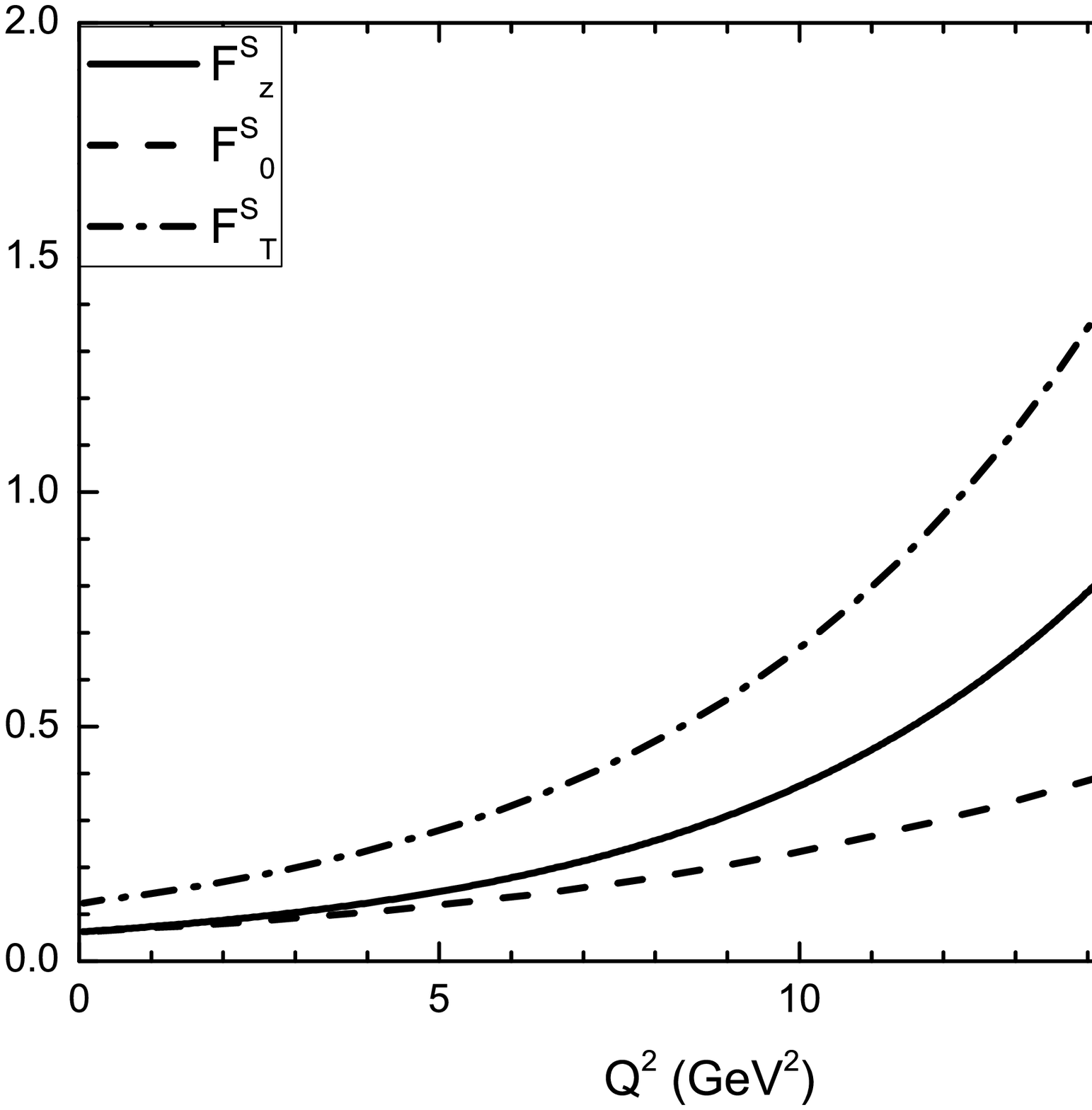}}
\subfigure[Form-Factors of $W^{\mu}_{ann}$ induced by annihilation
diagrams]{\includegraphics[width =
0.49\textwidth,height=0.4\textheight]{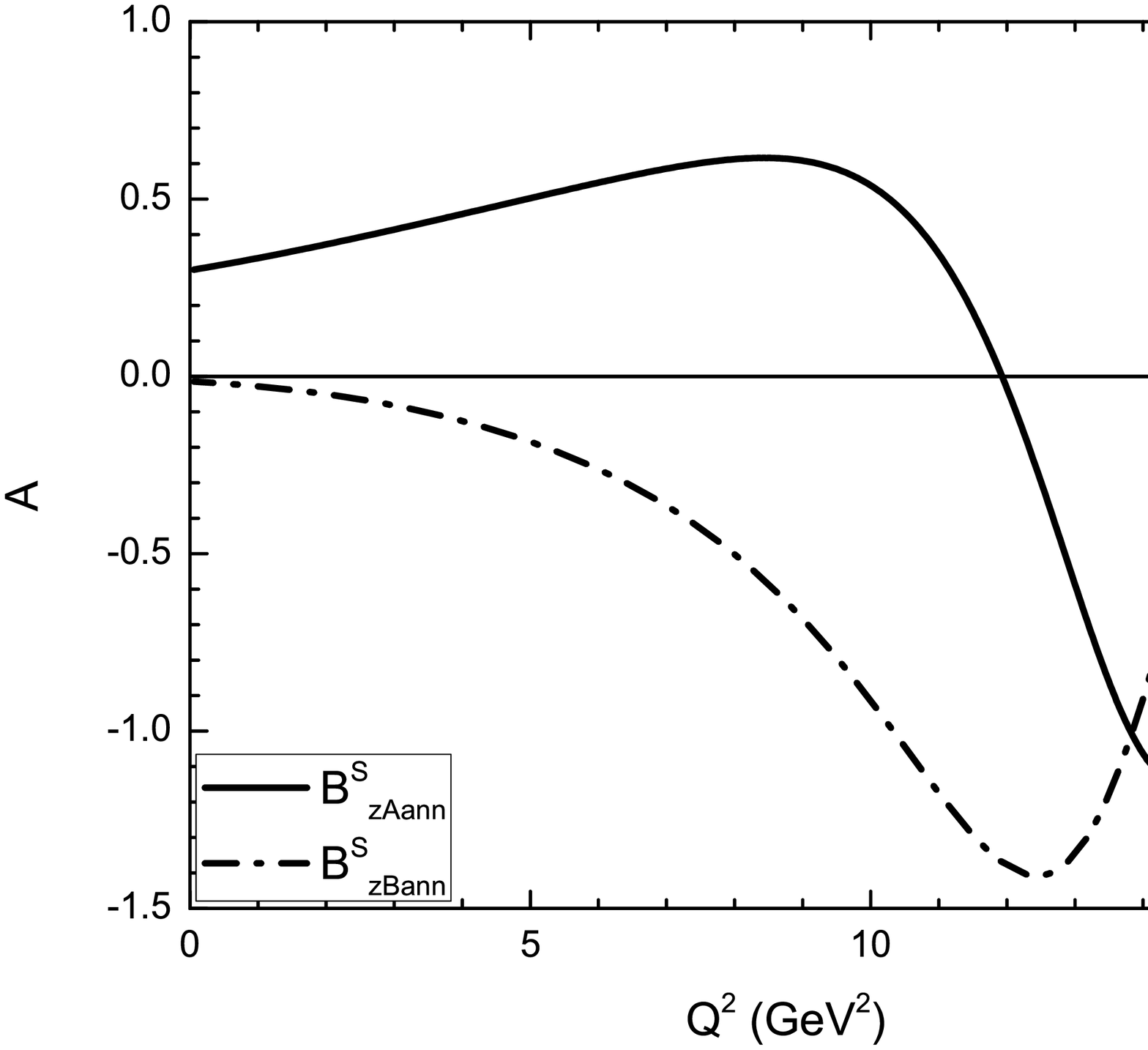}}
\caption{Form-Factors of $B_c\rightarrow D_0^*(2400) l\bar{l}$, where $B^S_{zAann}$  stands for $\text{Re}[B^S_{zann}]$, while $B^S_{zBann}$ denotes $\text{Im}[B^S_{zann}]$.}
\end{figure}
\begin{figure}[htbp]
\centering \subfigure[Form-factors of $W^{\mu}$ induced by $Z^0$
penguin and box diagrams]{\includegraphics[width =
0.49\textwidth,height=0.4\textheight]{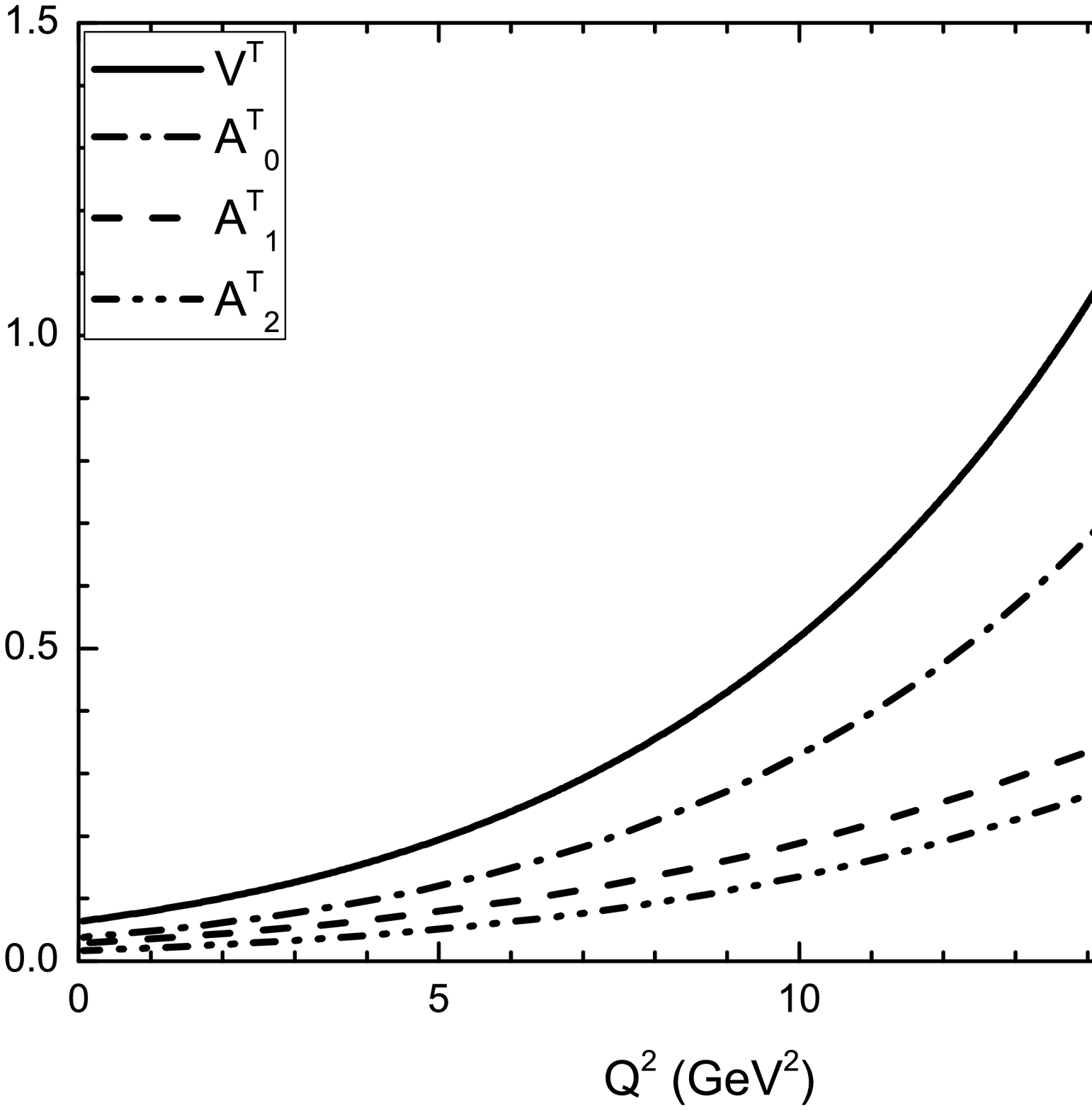}}
\subfigure[Form-factors of $W^{\mu}_T$ induced by $\gamma$ penguin diagrams]{\includegraphics[width = 0.49\textwidth,height=0.4\textheight]{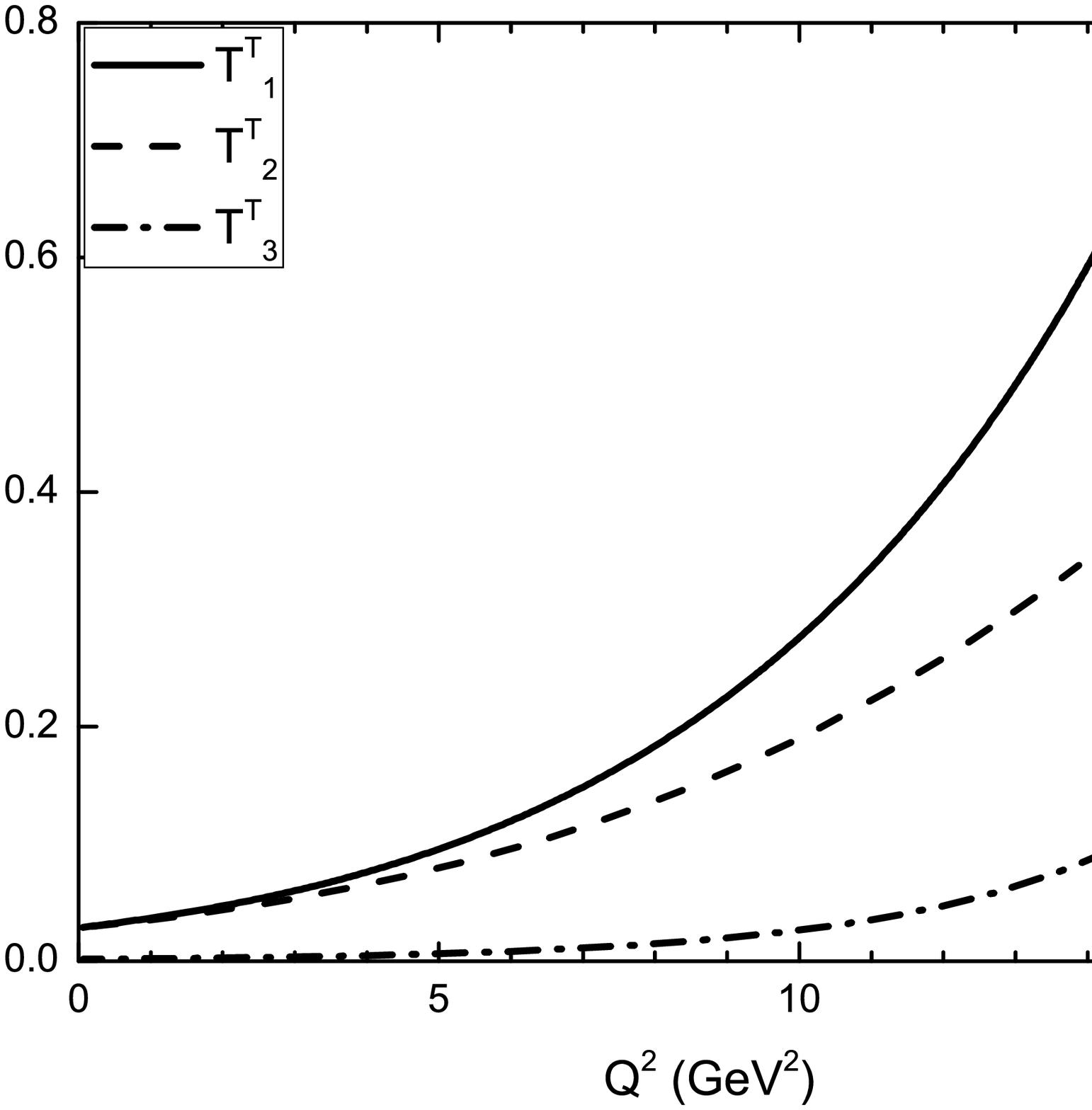}}\\
\subfigure[Real parts of $W^{\mu}_{ann}$ Form-Factors induced by
annihilation diagrams]{\includegraphics[width =
0.49\textwidth,height=0.4\textheight]{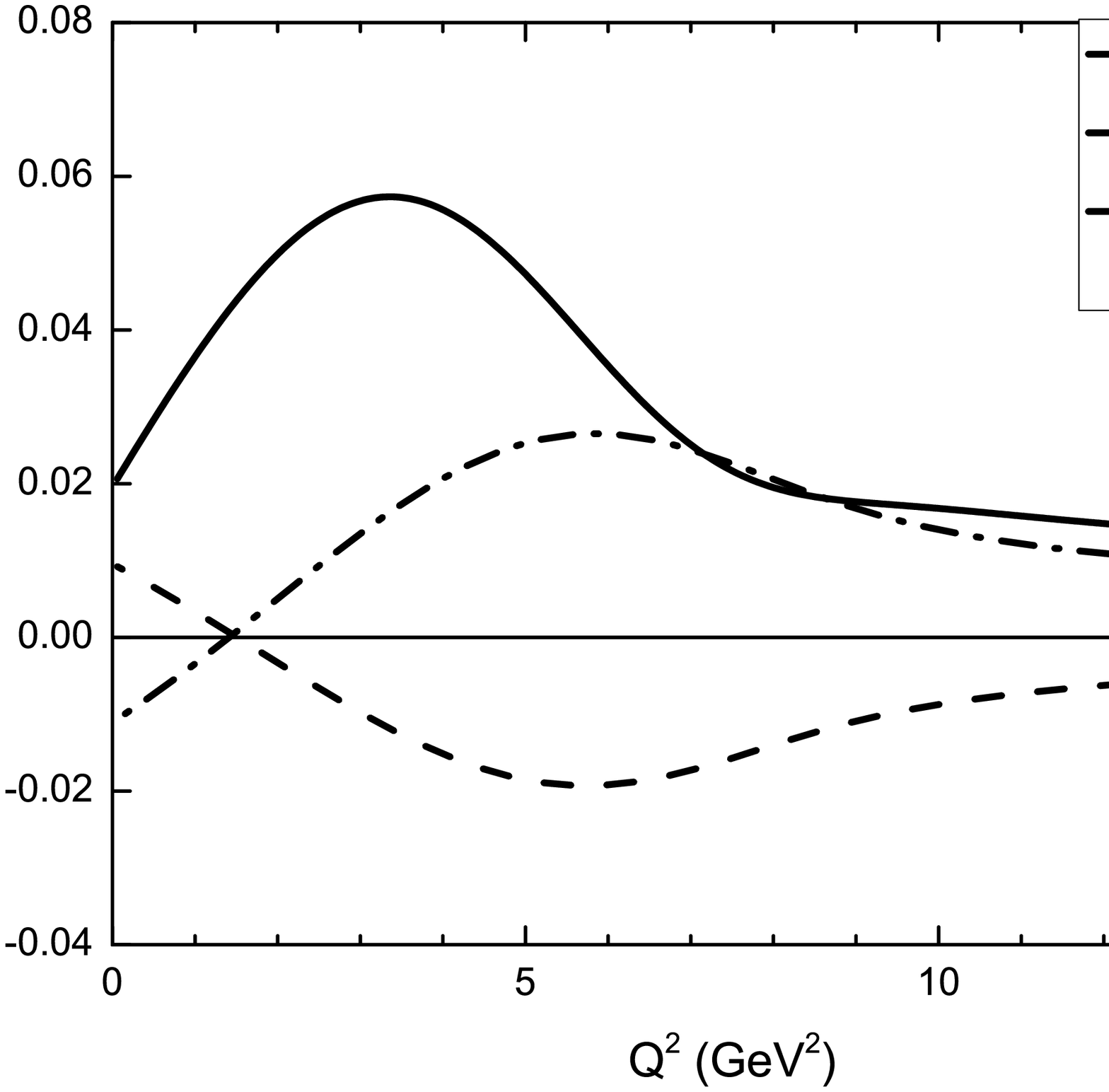}}
\subfigure[Imaginary parts of $W^{\mu}_{ann}$ Form-Factors induced
by annihilation
diagrams]{\includegraphics[width = 0.49\textwidth,height=0.4\textheight]{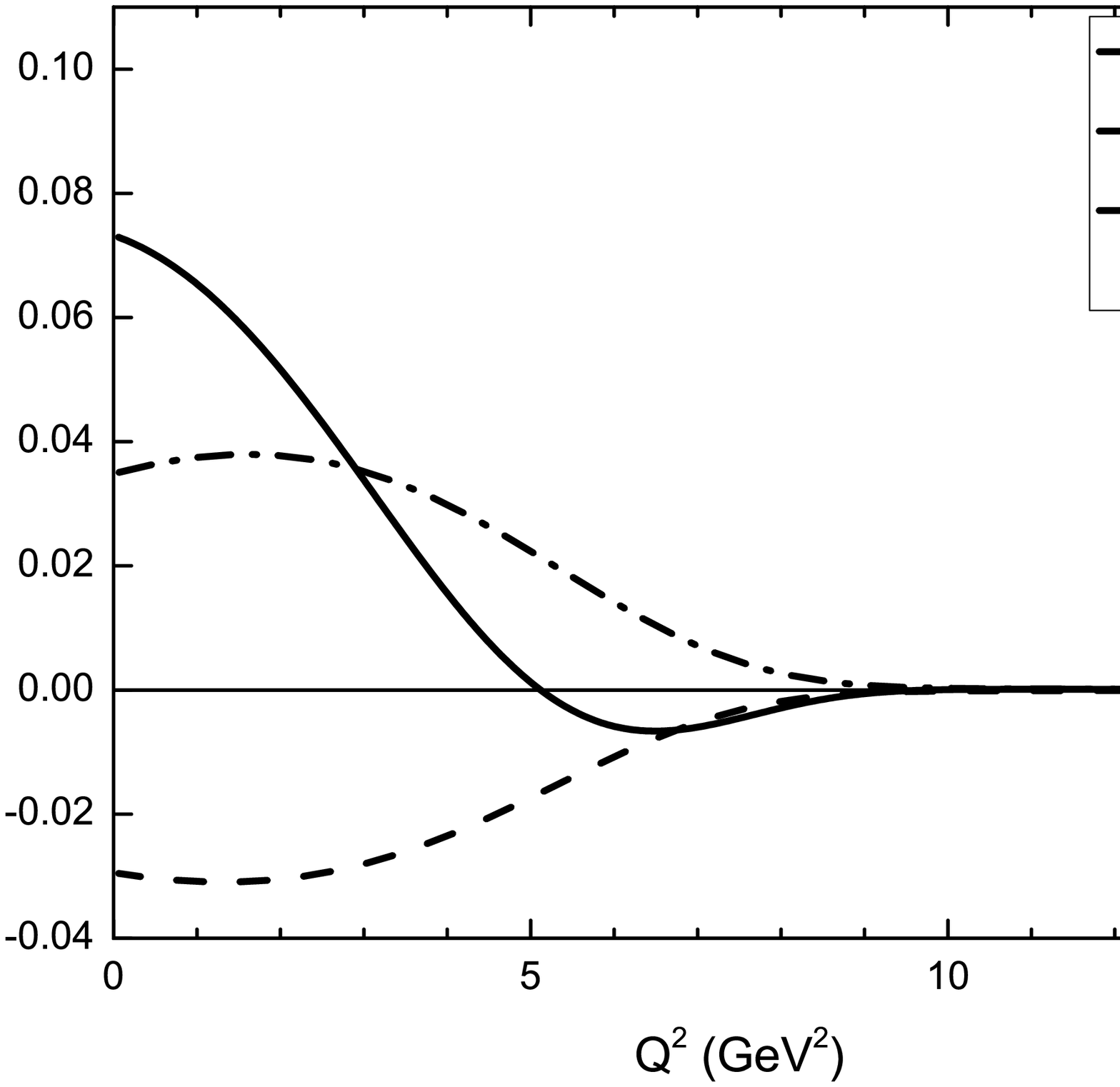}}\\
\caption{Form-factors of $B_c\rightarrow D_2^*(2460)l\bar{l}$, where $V^T_{Aann}$ and $T^T_{1,zAann}$ stand for $\text{Re}[V^T_{ann}]$ and $\text{Re}[T^T_{1,zann}]$, respectively, while $V^T_{Bann}$ and $T^T_{1,zBann}$ denote $\text{Im}[V^T_{ann}]$ and $\text{Im}[T^T_{1,zann}]$.}
\end{figure}

\begin{figure}[htbp]
\centering \subfigure[Form-factors of $W^{\mu}$ induced by $Z^0$
penguin and box diagrams]{\includegraphics[width =
0.49\textwidth,height=0.4\textheight]{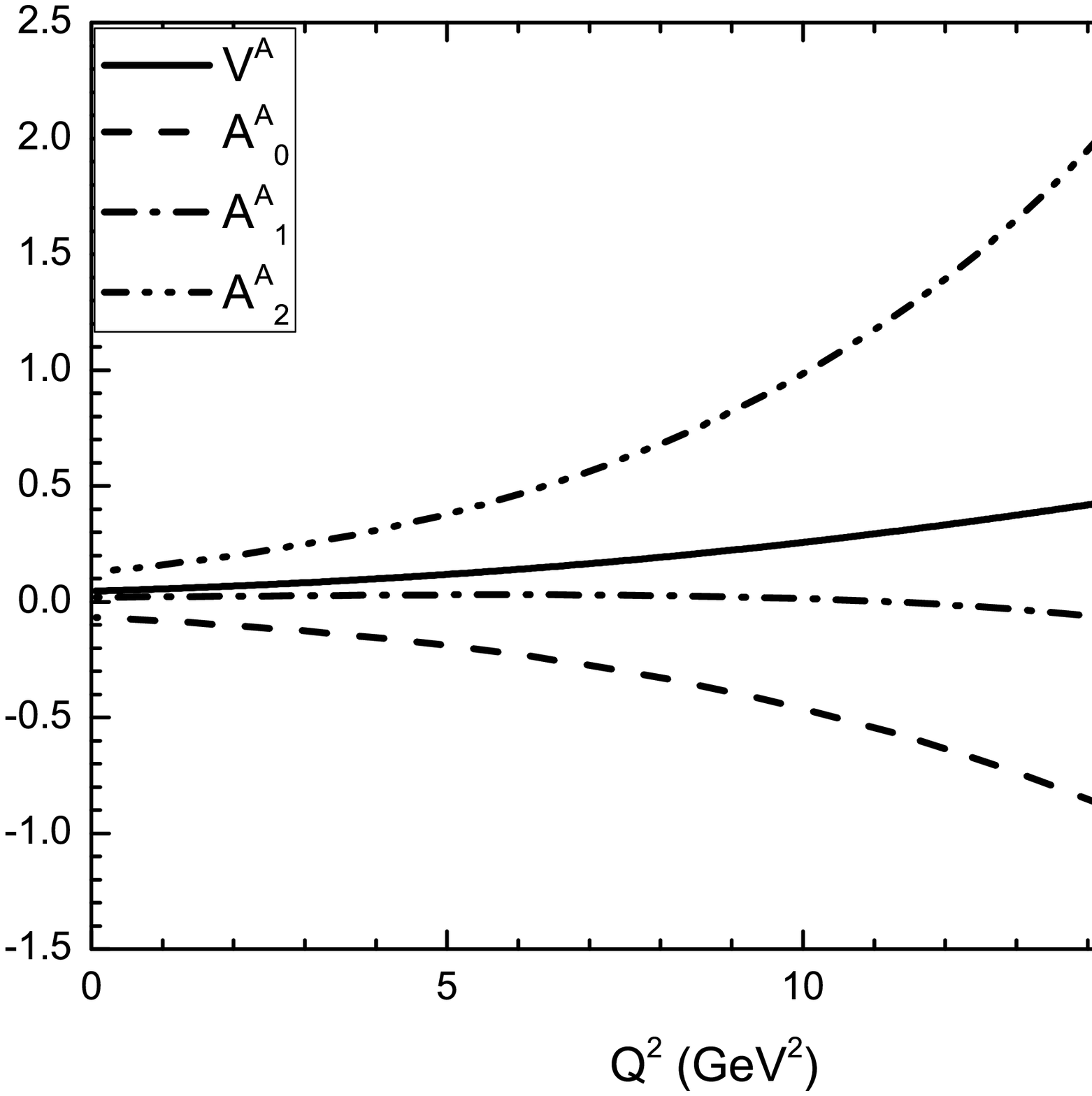}}
\subfigure[Form-factors of $W^{\mu}_T$ induced by $\gamma$ penguin diagrams]{\includegraphics[width = 0.49\textwidth,height=0.4\textheight]{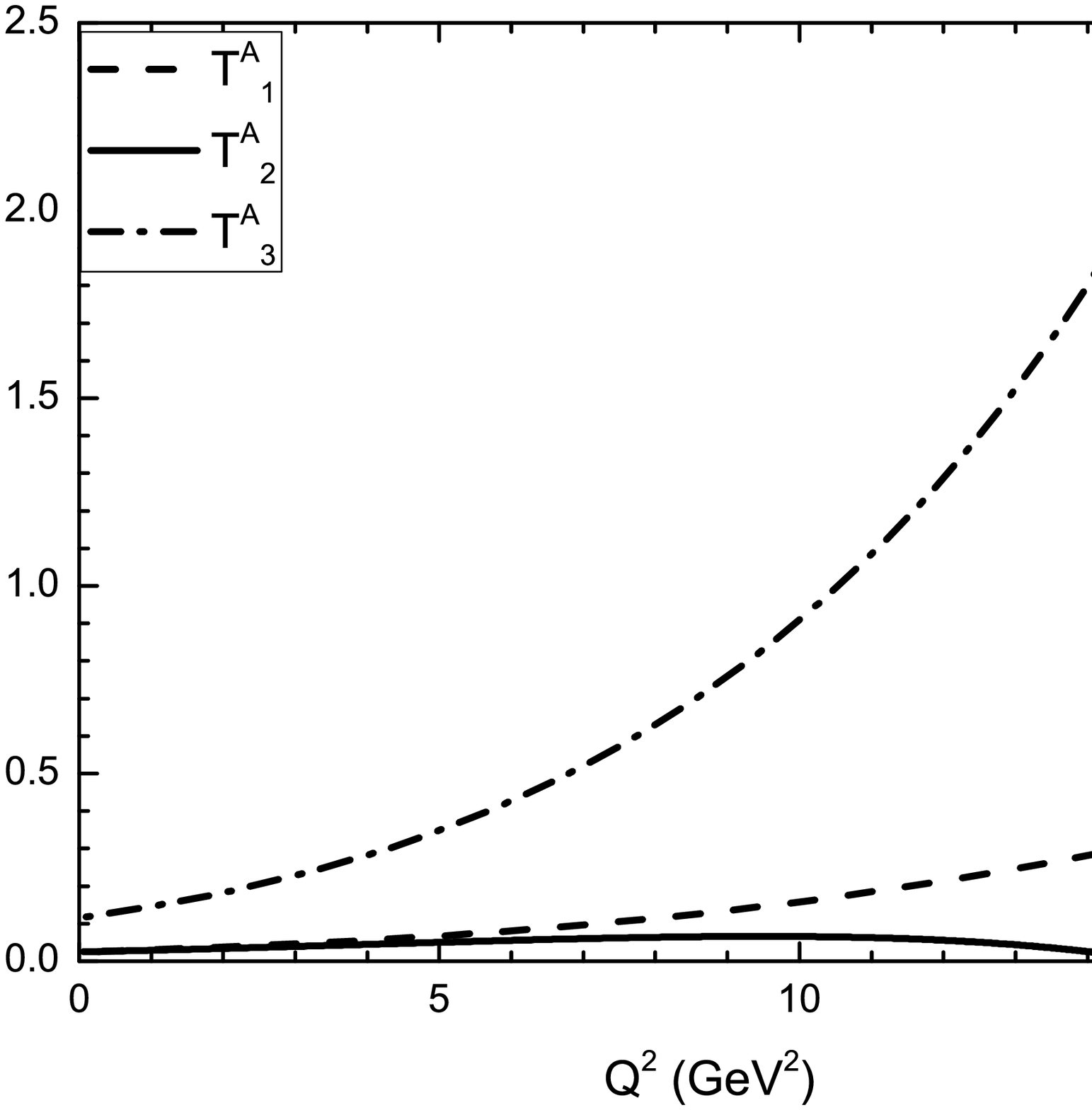}}\\
\subfigure[Real parts of $W^{\mu}_{ann}$ Form-Factors induced by
annihilation diagrams]{\includegraphics[width =
0.49\textwidth,height=0.4\textheight]{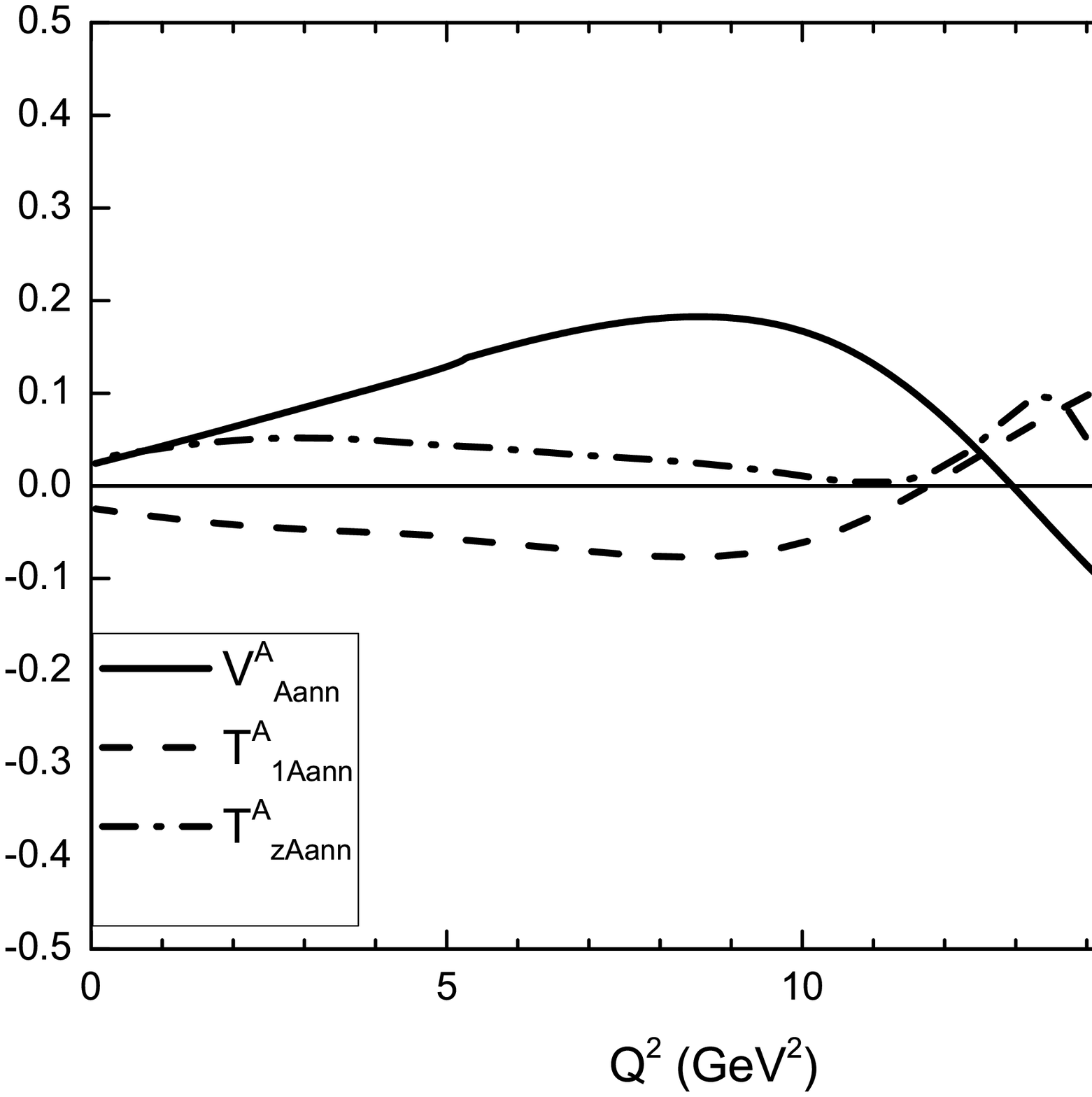}}
\subfigure[Imaginary parts of $W^{\mu}_{ann}$ Form-Factors induced
by annihilation
diagrams]{\includegraphics[width = 0.49\textwidth,height=0.4\textheight]{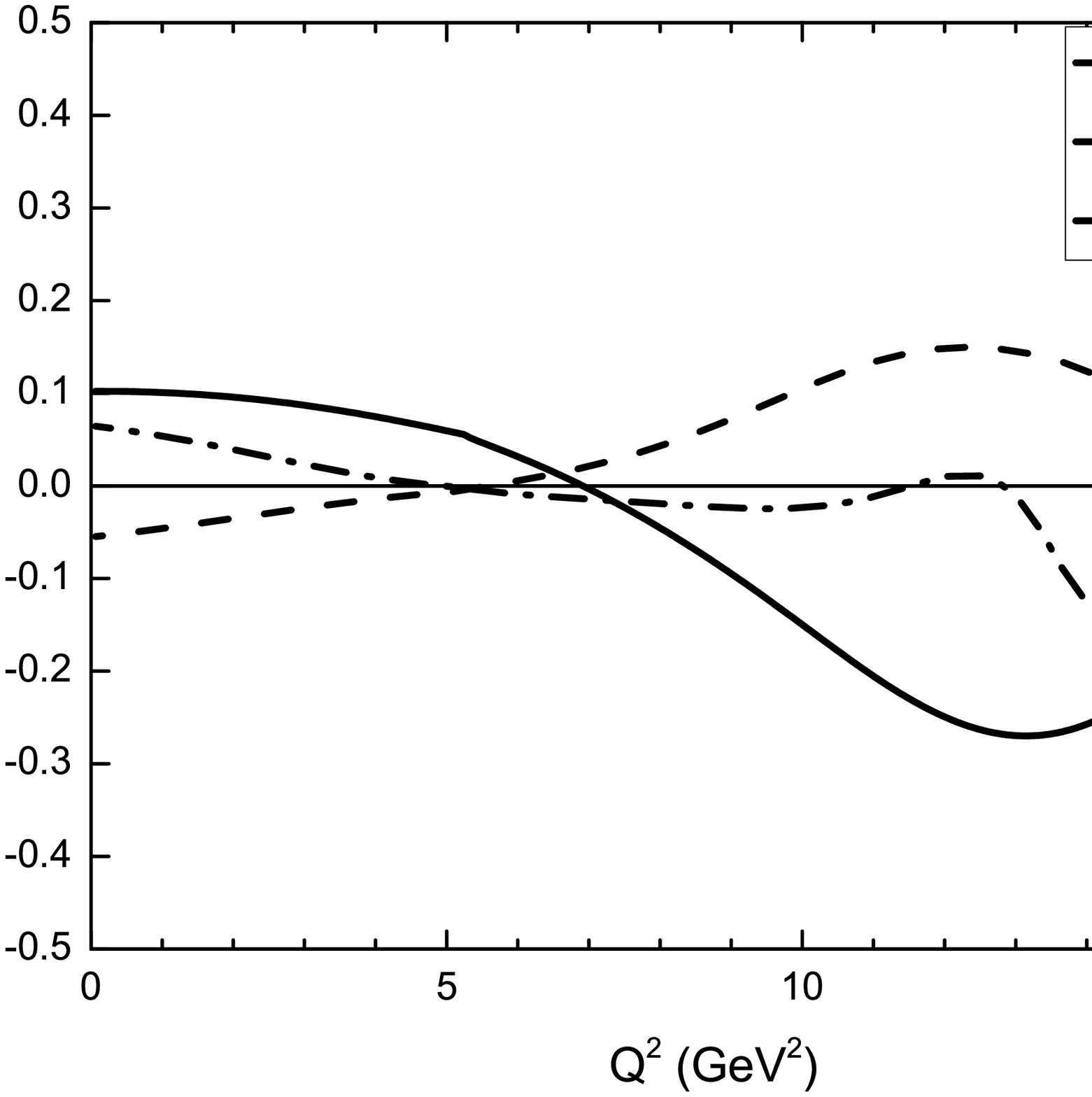}}\\
\caption{Form-factors of $B_c\rightarrow D_{1}(2420)l\bar{l}$, where $V^A_{Aann}$ and $T^A_{1,zAann}$ stand for $\text{Re}[V^A_{ann}]$ and $\text{Re}[T^A_{1,zann}]$, respectively, while $V^A_{Bann}$ and $T^A_{1,zBann}$ denote $\text{Im}[V^A_{ann}]$ and $\text{Im}[T^A_{1,zann}]$.}
\end{figure}

\begin{figure}[htbp]
\centering \subfigure[Form-factors of $W^{\mu}$ induced by $Z^0$
penguin and box diagrams]{\includegraphics[width =
0.49\textwidth,height=0.4\textheight]{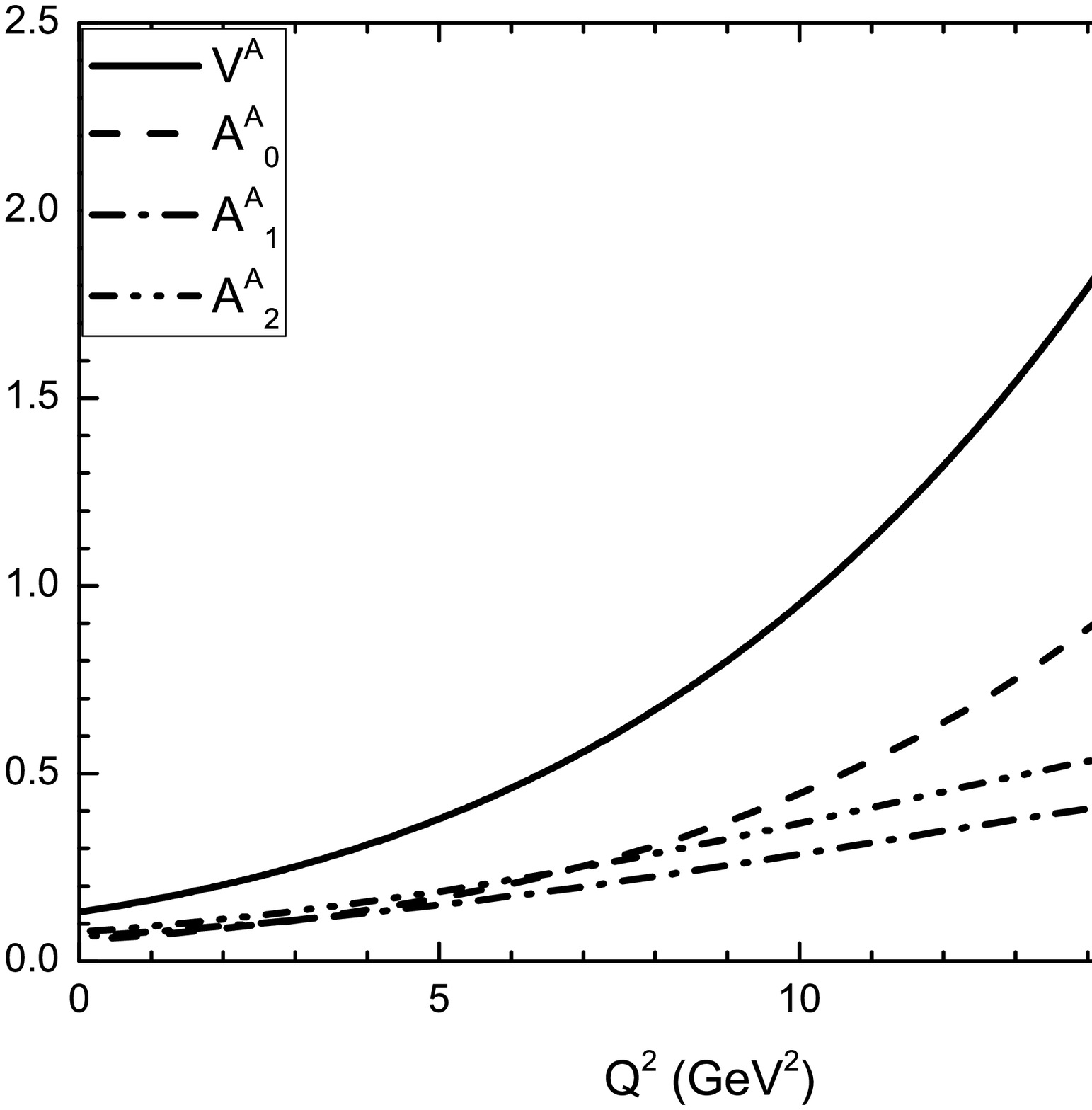}}
\subfigure[Form-factors of $W^{\mu}_T$ induced by $\gamma$ penguin diagrams]{\includegraphics[width = 0.49\textwidth,height=0.4\textheight]{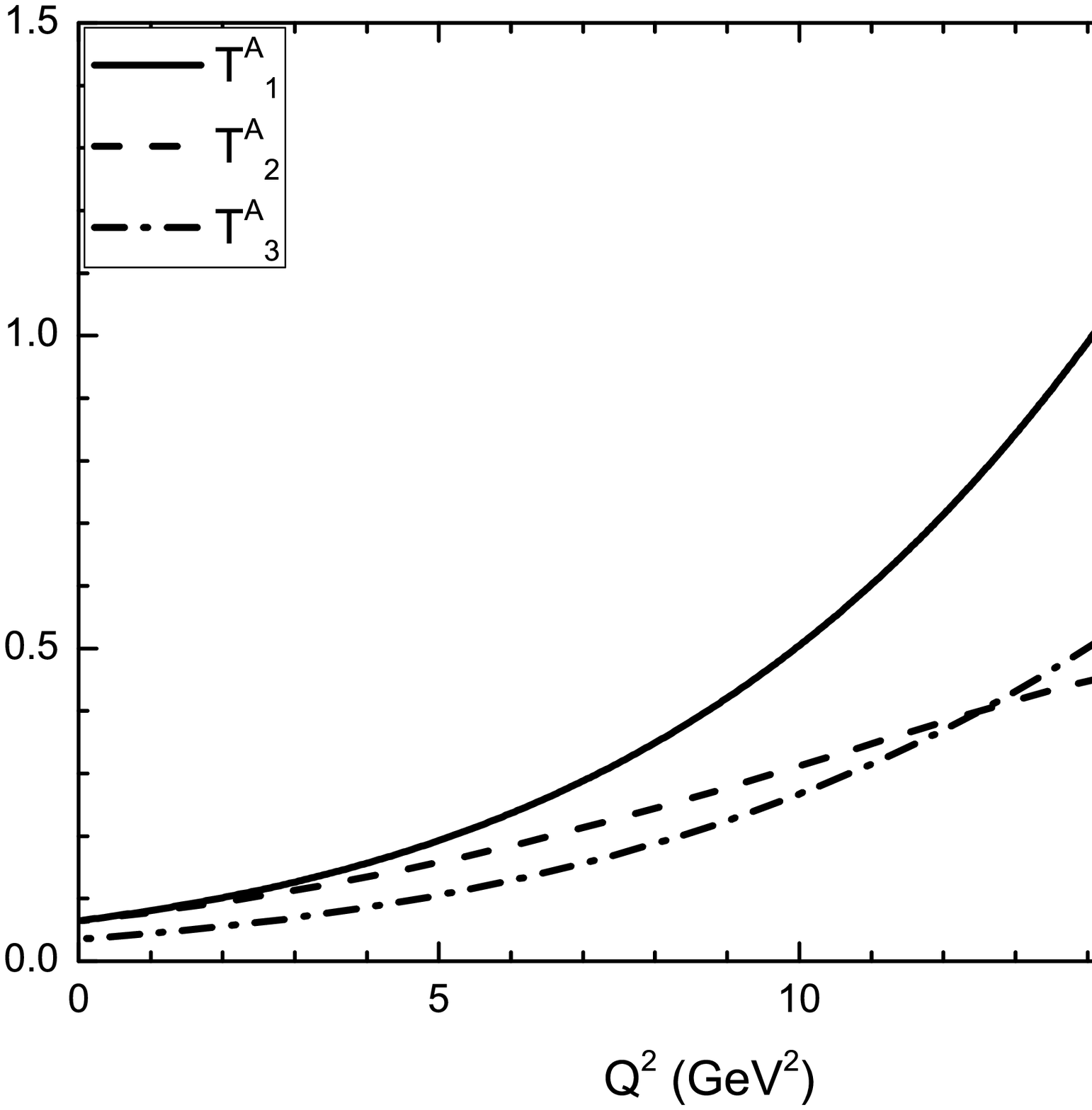}}\\
\subfigure[Real parts of $W^{\mu}_{ann}$ Form-Factors induced by
annihilation diagrams]{\includegraphics[width =
0.49\textwidth,height=0.4\textheight]{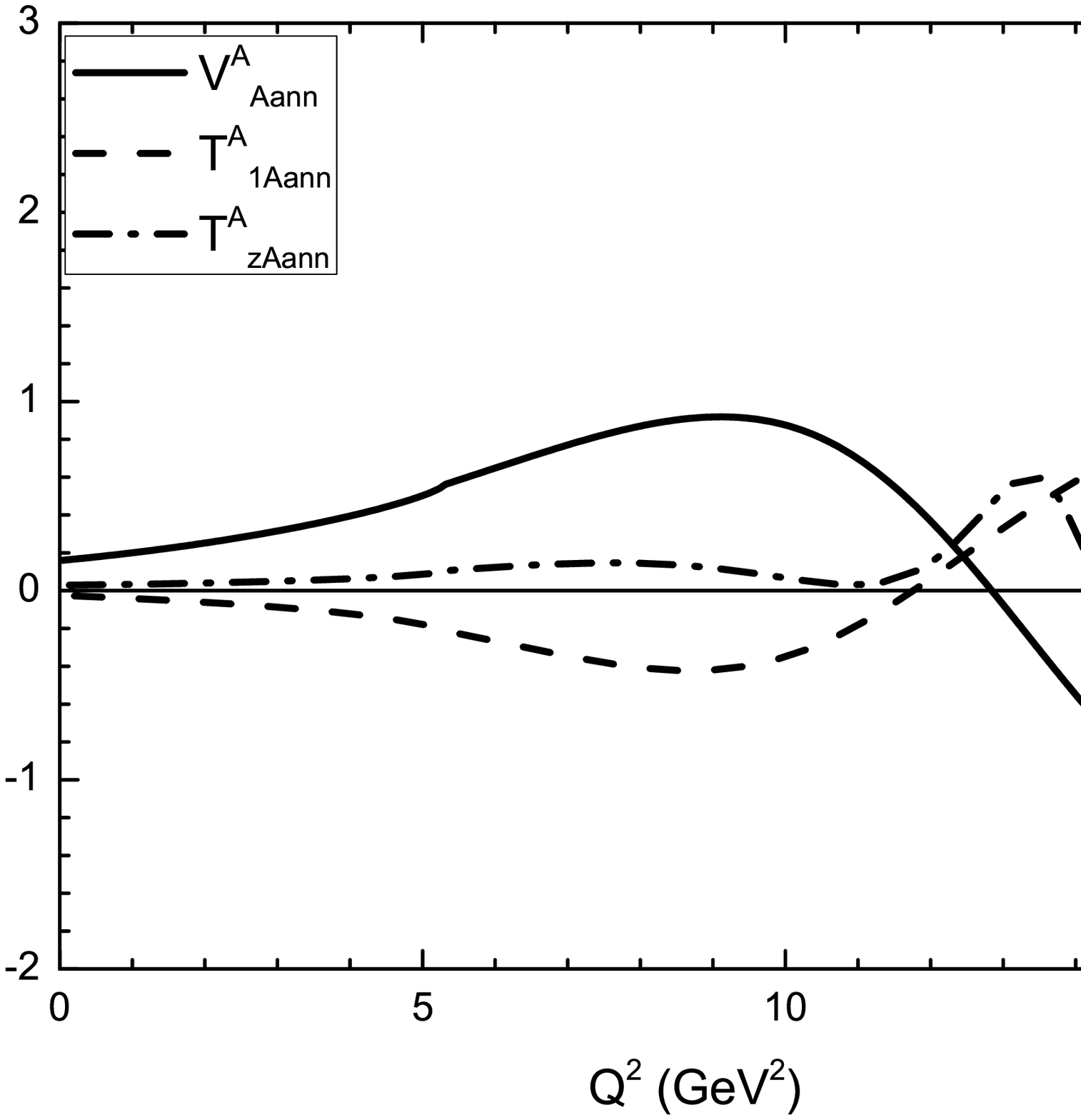}}
\subfigure[Imaginary parts of $W^{\mu}_{ann}$ Form-Factors induced
by annihilation
diagrams]{\includegraphics[width = 0.49\textwidth,height=0.4\textheight]{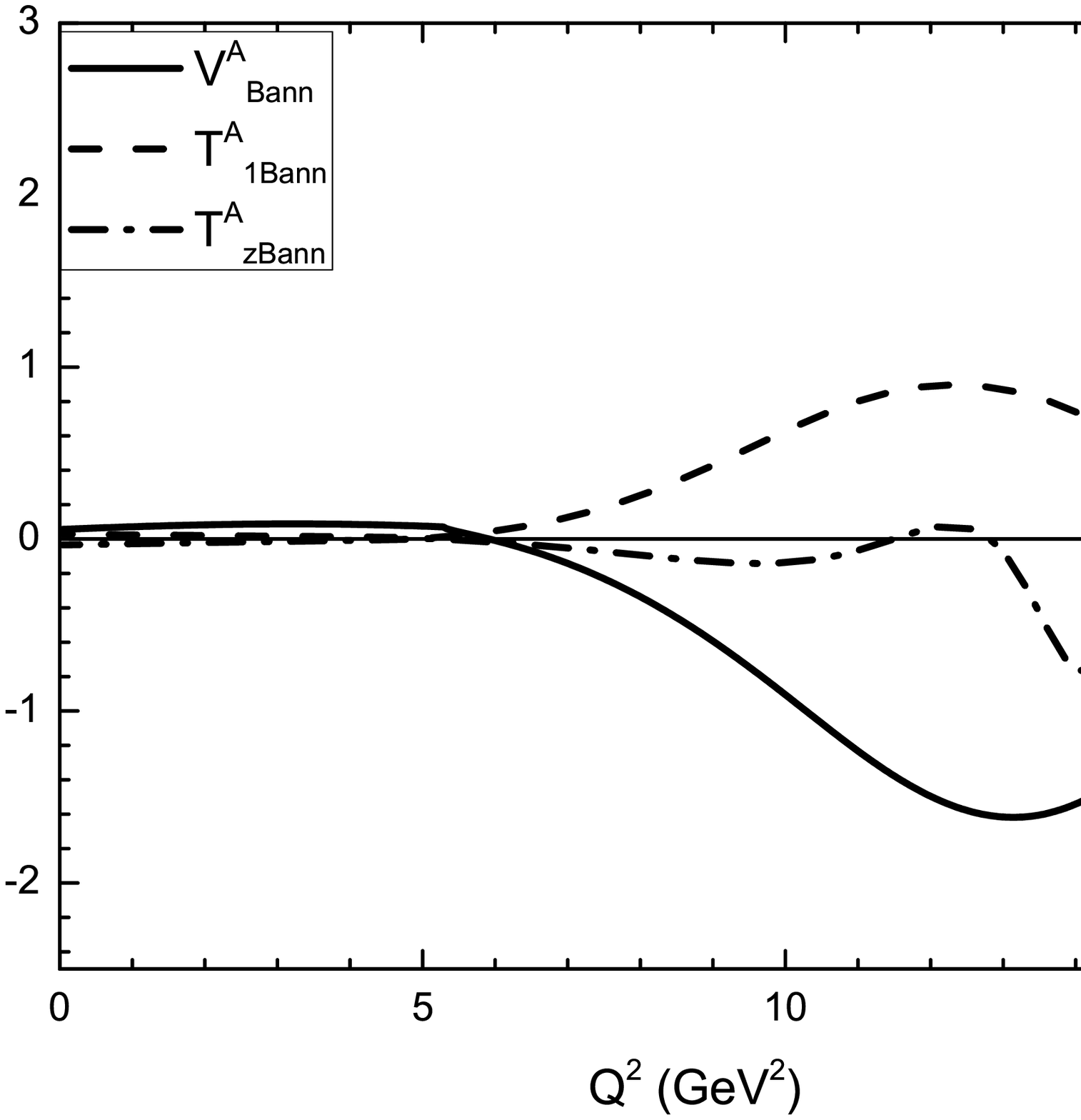}}\\
\caption{Form-factors of $B_c\rightarrow D_{1}(2430)l\bar{l}$, where $V^A_{Aann}$ and $T^A_{1,zAann}$ stand for $\text{Re}[V^A_{ann}]$ and $\text{Re}[T^A_{1,zann}]$, respectively, while $V^A_{Bann}$ and $T^A_{1,zBann}$ denote $\text{Im}[V^A_{ann}]$ and $\text{Im}[T^A_{1,zann}]$.}
\end{figure}

\begin{figure}[htbp]
\centering
\subfigure[Differential branching fraction]{\includegraphics[width = 0.49\textwidth,height=0.4\textheight]{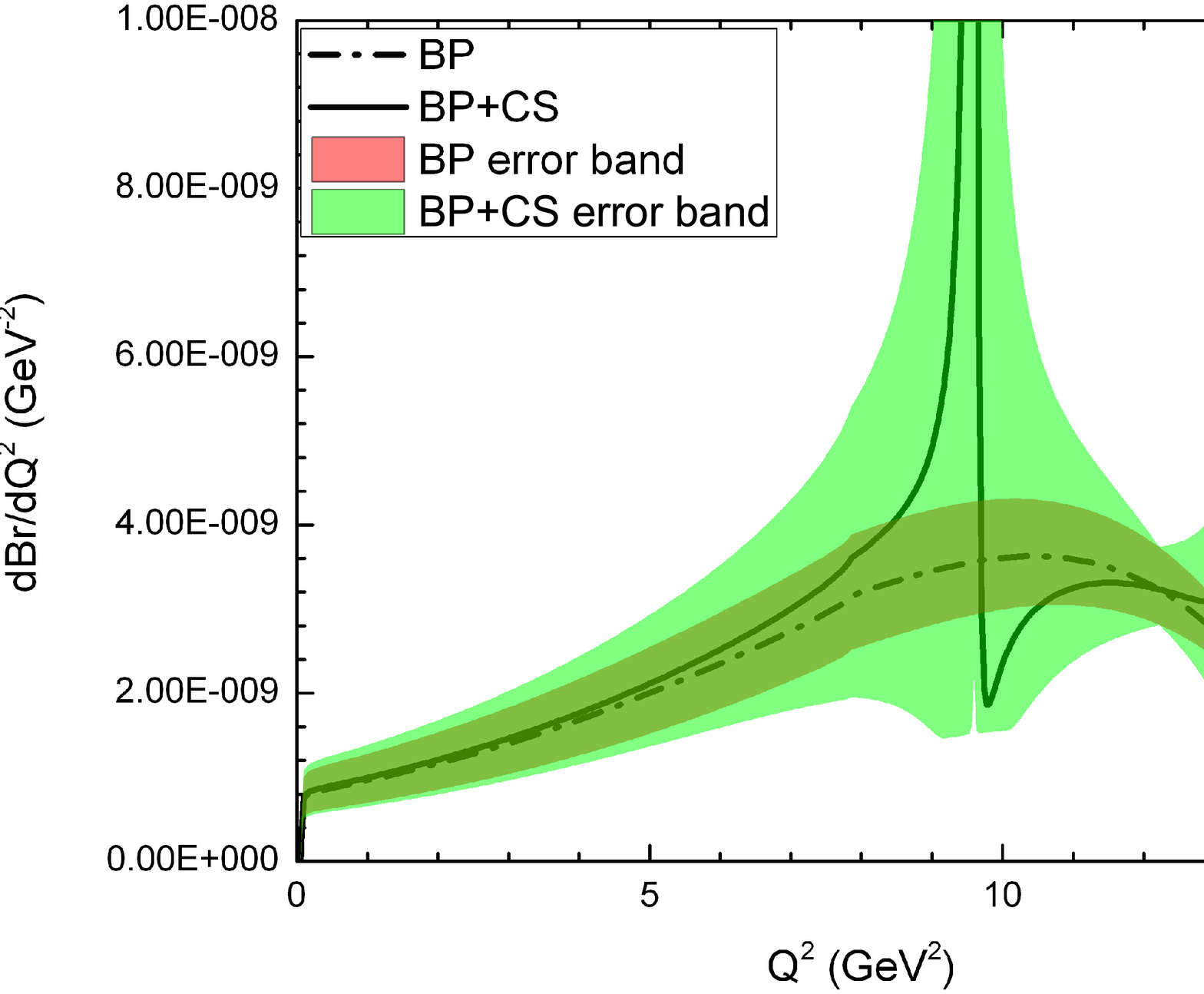}}
\subfigure[Differential branching fraction]{\includegraphics[width = 0.49\textwidth,height=0.4\textheight]{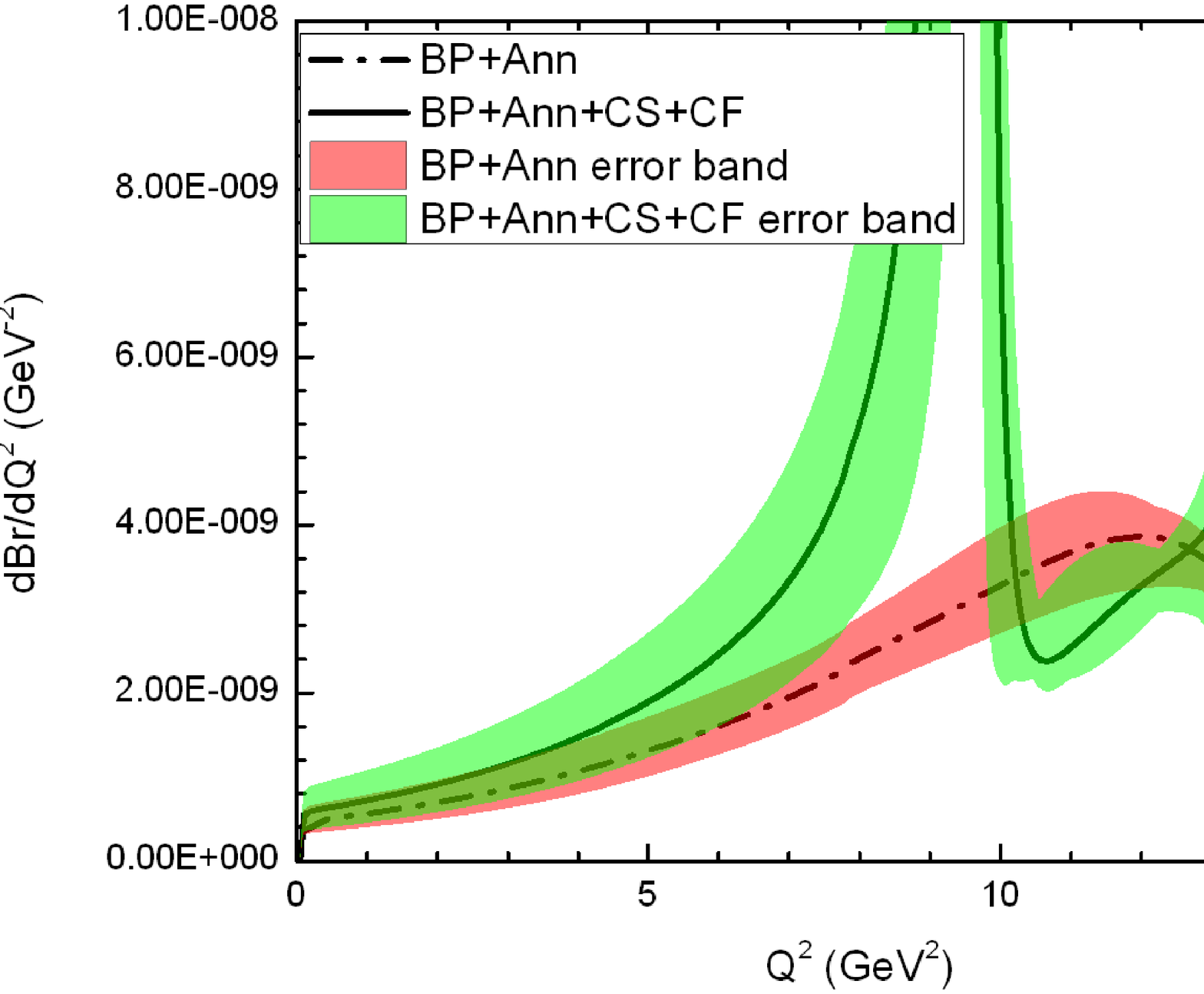}}\\
\subfigure[Leptonic longitudinal polarization asymmetry]{\includegraphics[width = 0.49\textwidth,height=0.4\textheight]{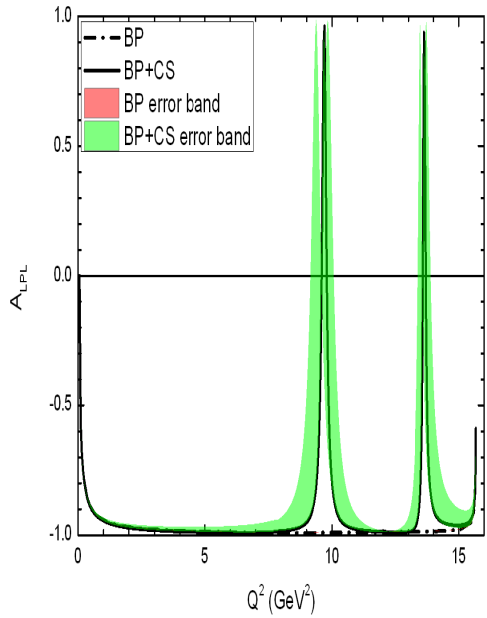}}
\subfigure[Leptonic longitudinal polarization asymmetry]{\includegraphics[width = 0.49\textwidth,height=0.4\textheight]{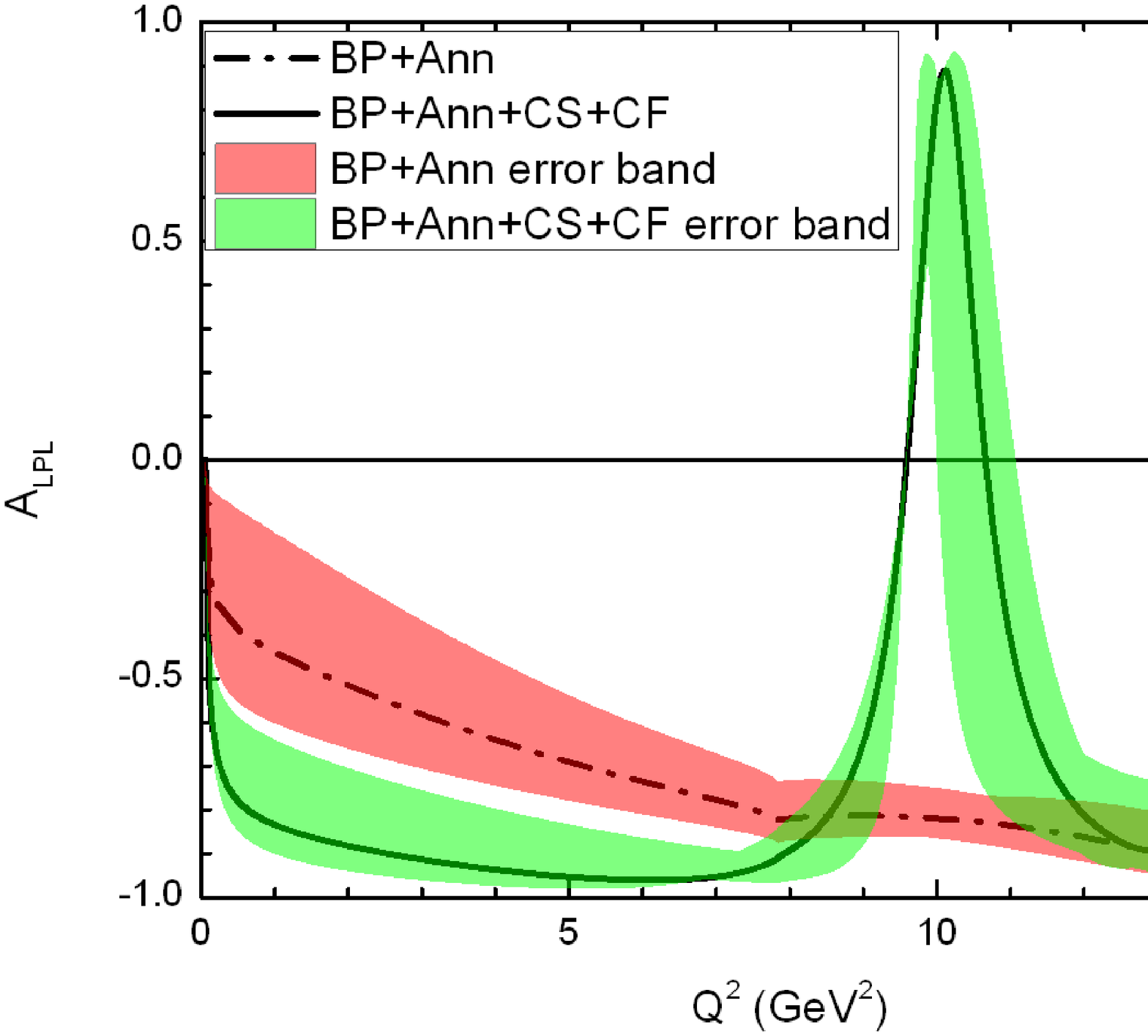}}
\caption{Observables of $B_c\rightarrow D_{s0}^*(2317) \mu\bar{\mu}$. }
\end{figure}

\begin{figure}[htbp]
\centering
\subfigure[Differential branching fraction]{\includegraphics[width = 0.49\textwidth,height=0.2\textheight]{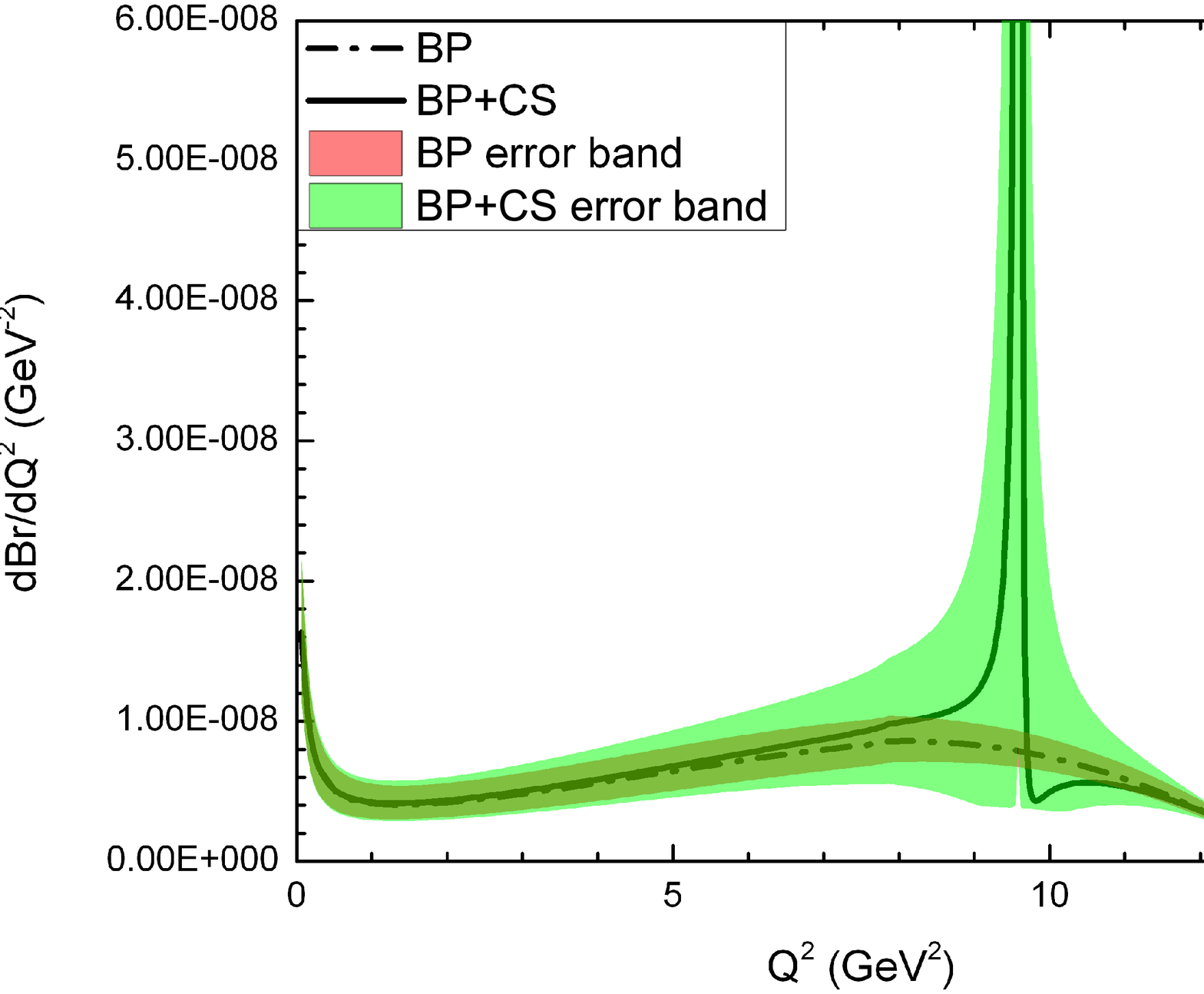}}
\subfigure[Differential branching fraction]{\includegraphics[width = 0.49\textwidth,height=0.2\textheight]{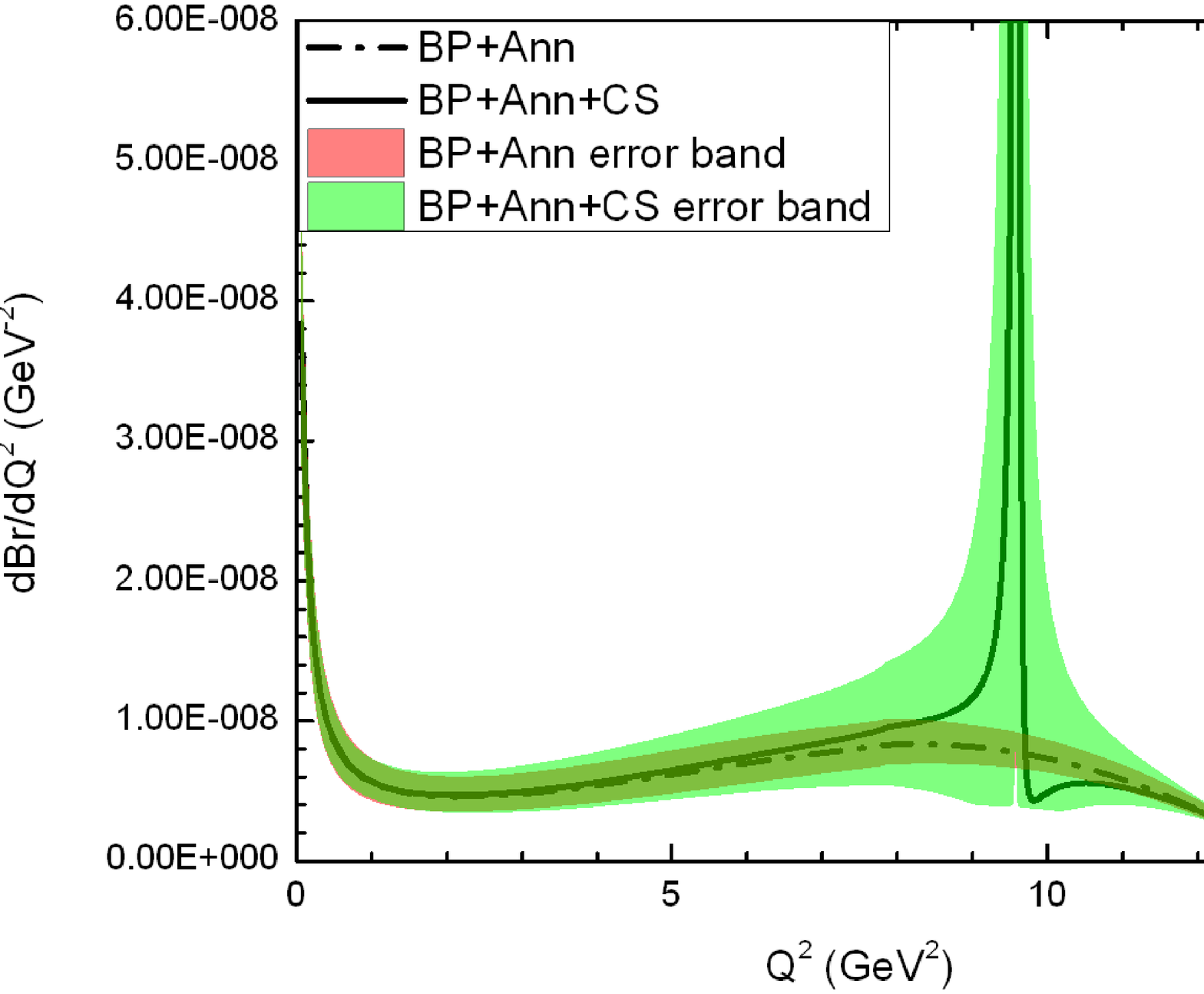}}\\
\subfigure[Leptonic longitudinal polarization asymmetry]{\includegraphics[width = 0.49\textwidth,height=0.2\textheight]{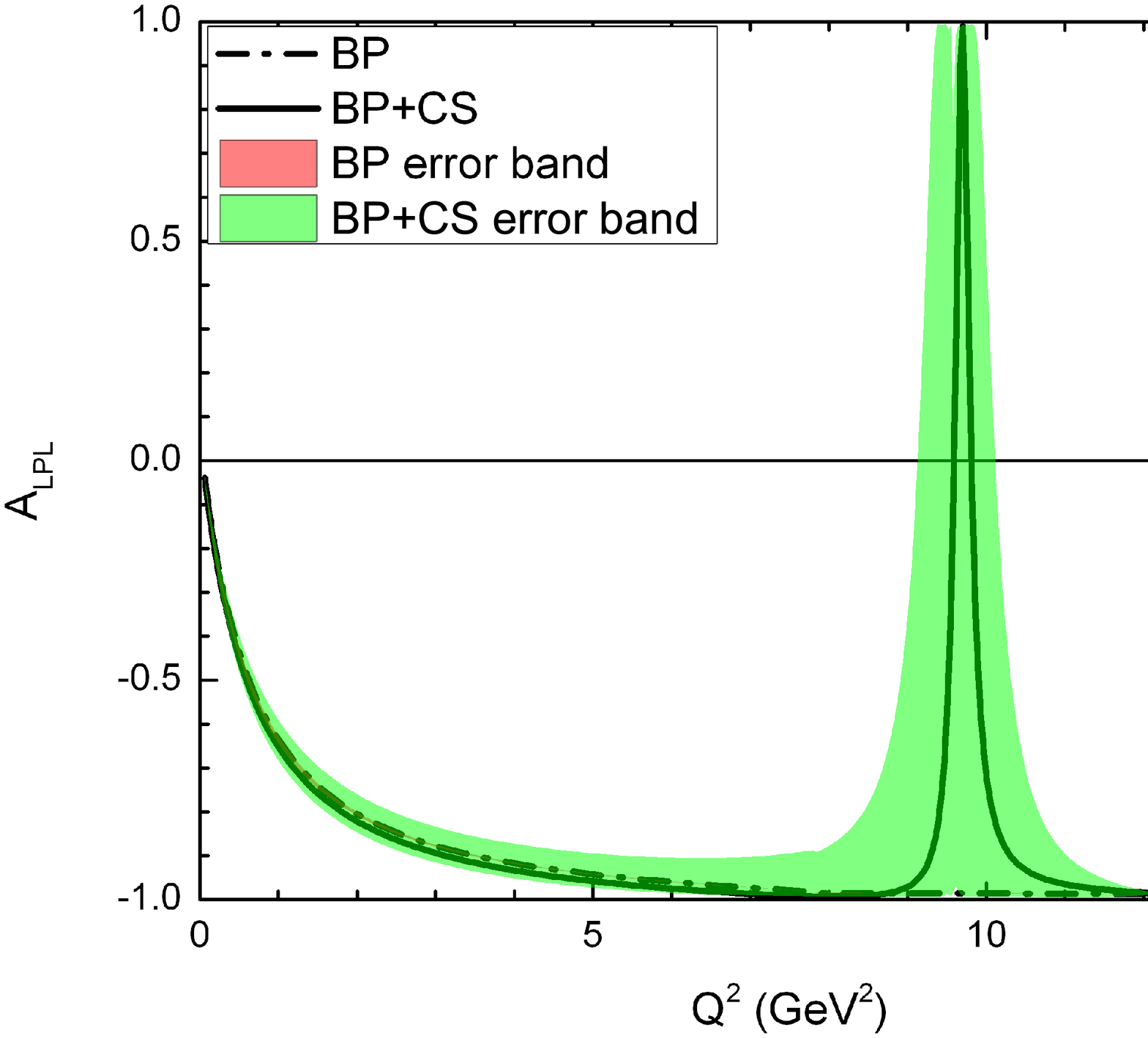}}
\subfigure[Leptonic longitudinal polarization asymmetry]{\includegraphics[width = 0.49\textwidth,height=0.2\textheight]{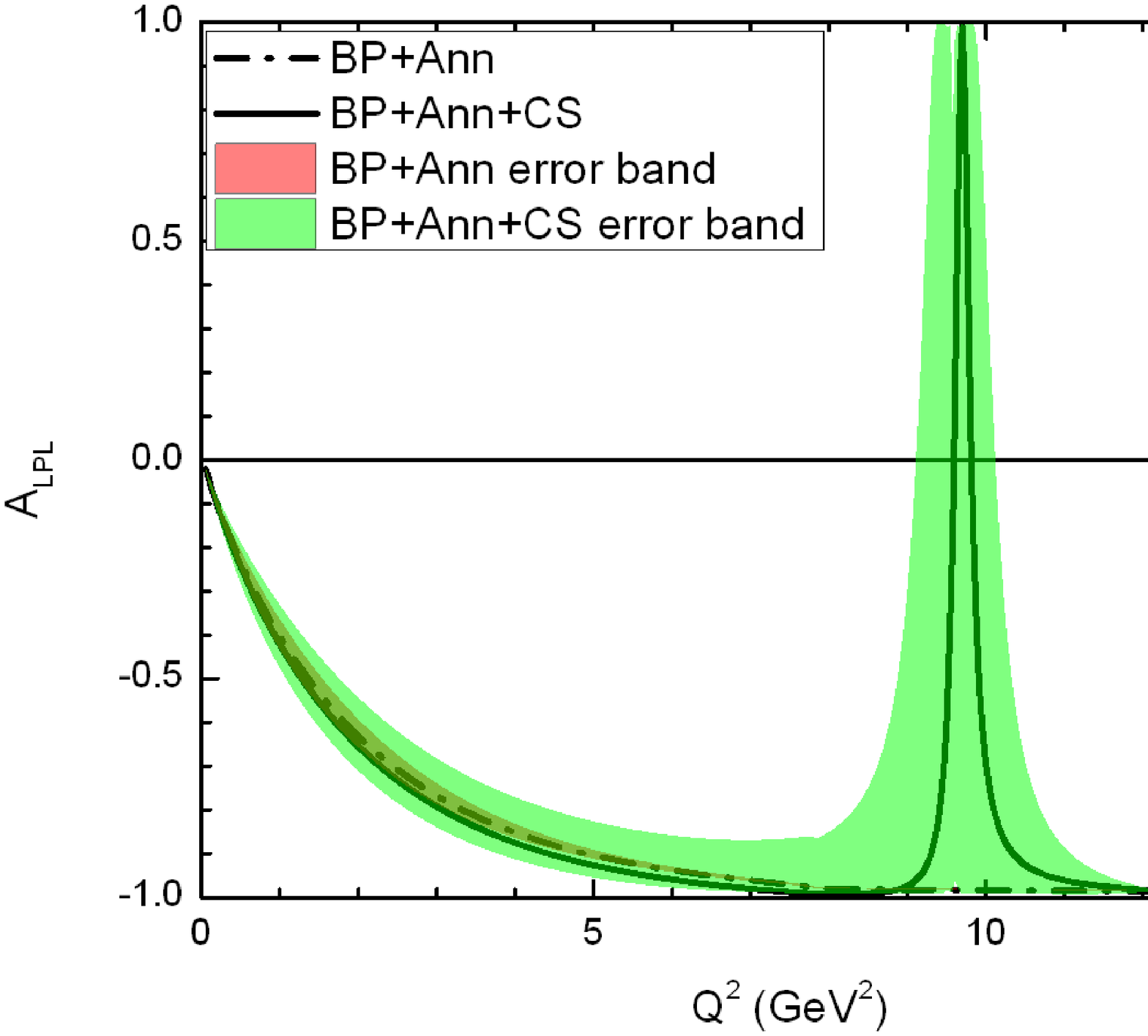}}\\
\subfigure[Forward-backward asymmetry]{\includegraphics[width = 0.49\textwidth,height=0.2\textheight]{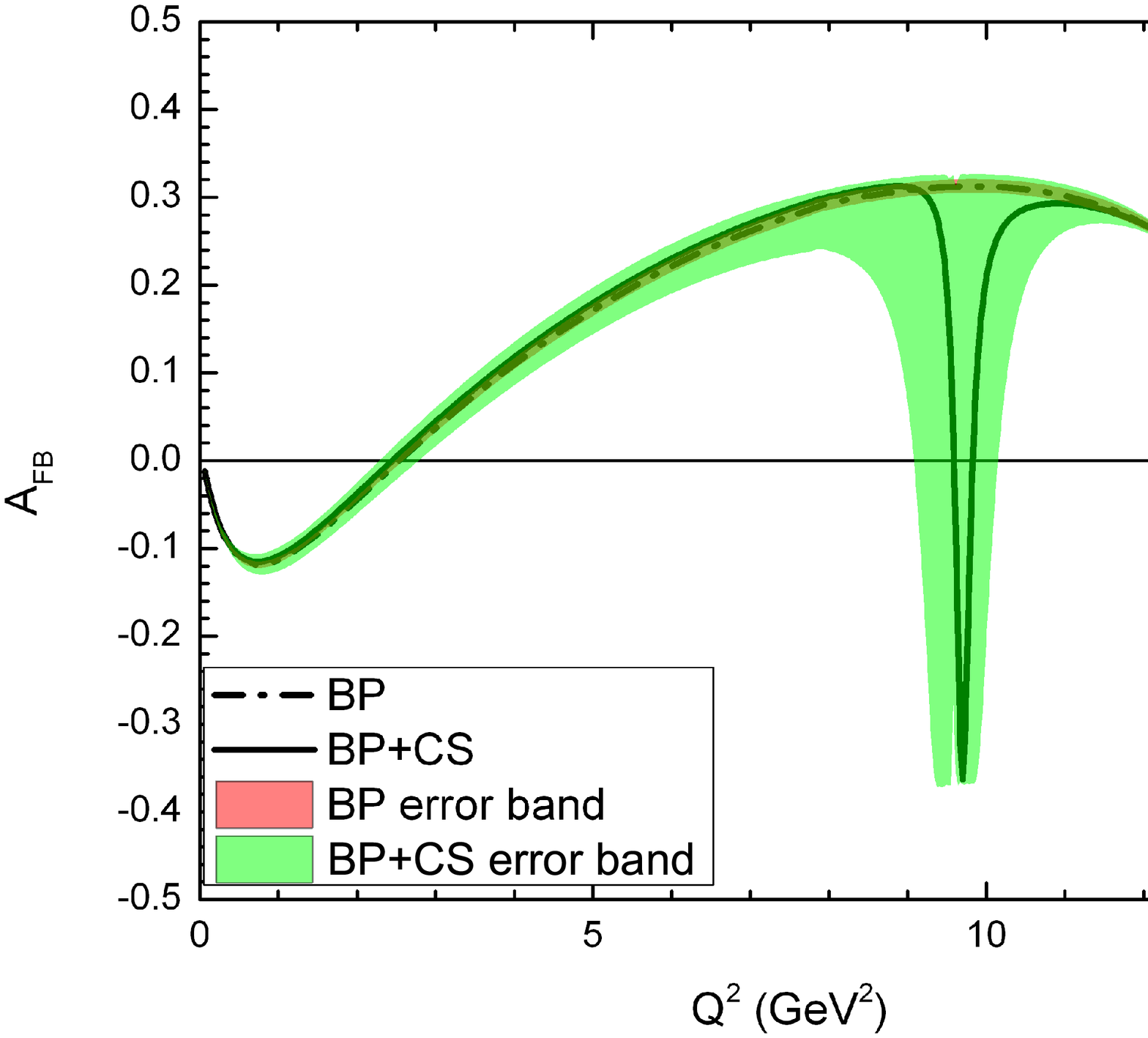}}
\subfigure[Forward-backward asymmetry]{\includegraphics[width = 0.49\textwidth,height=0.2\textheight]{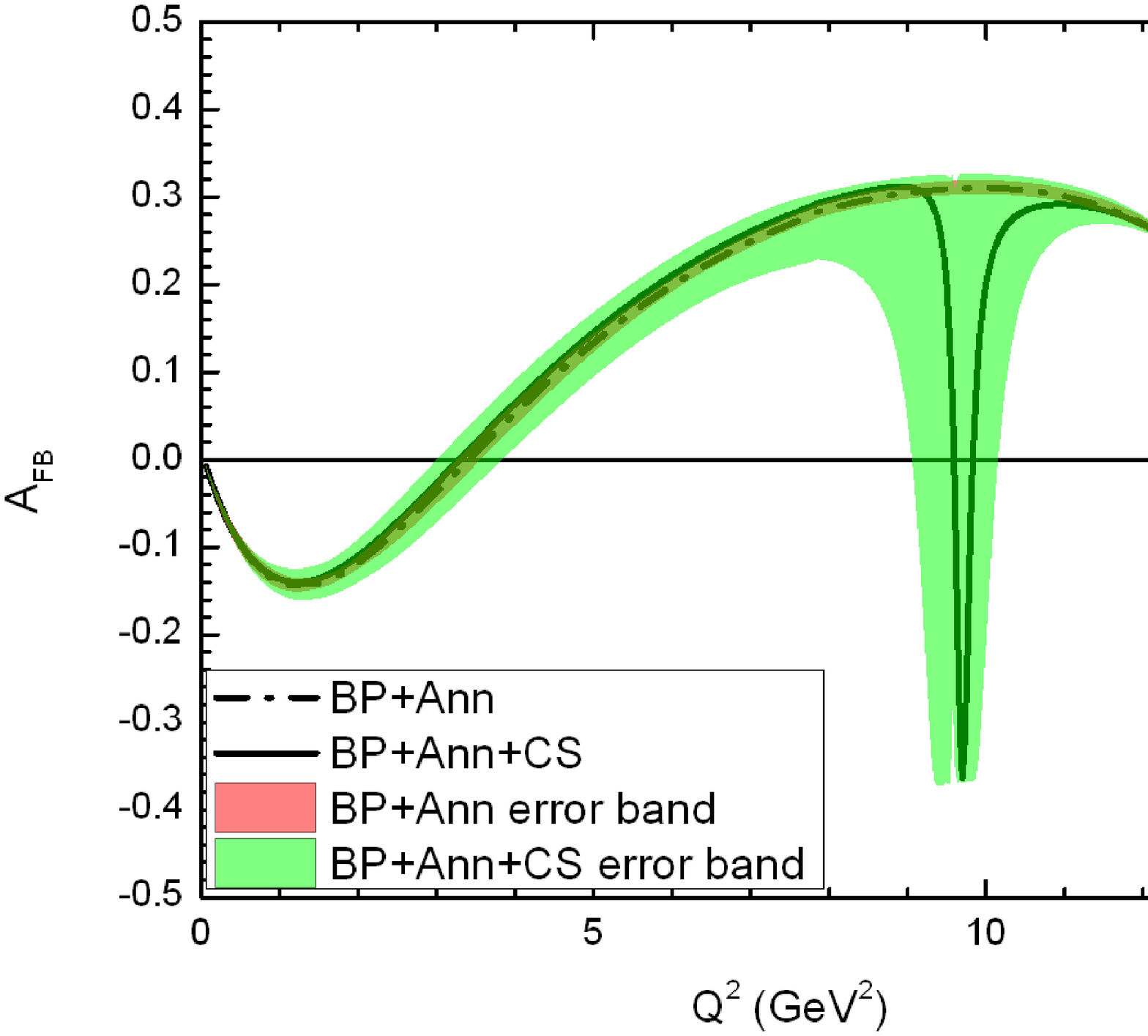}}\\
\subfigure[Longitudinal polarization of the
meson]{\includegraphics[width =
0.49\textwidth,height=0.2\textheight]{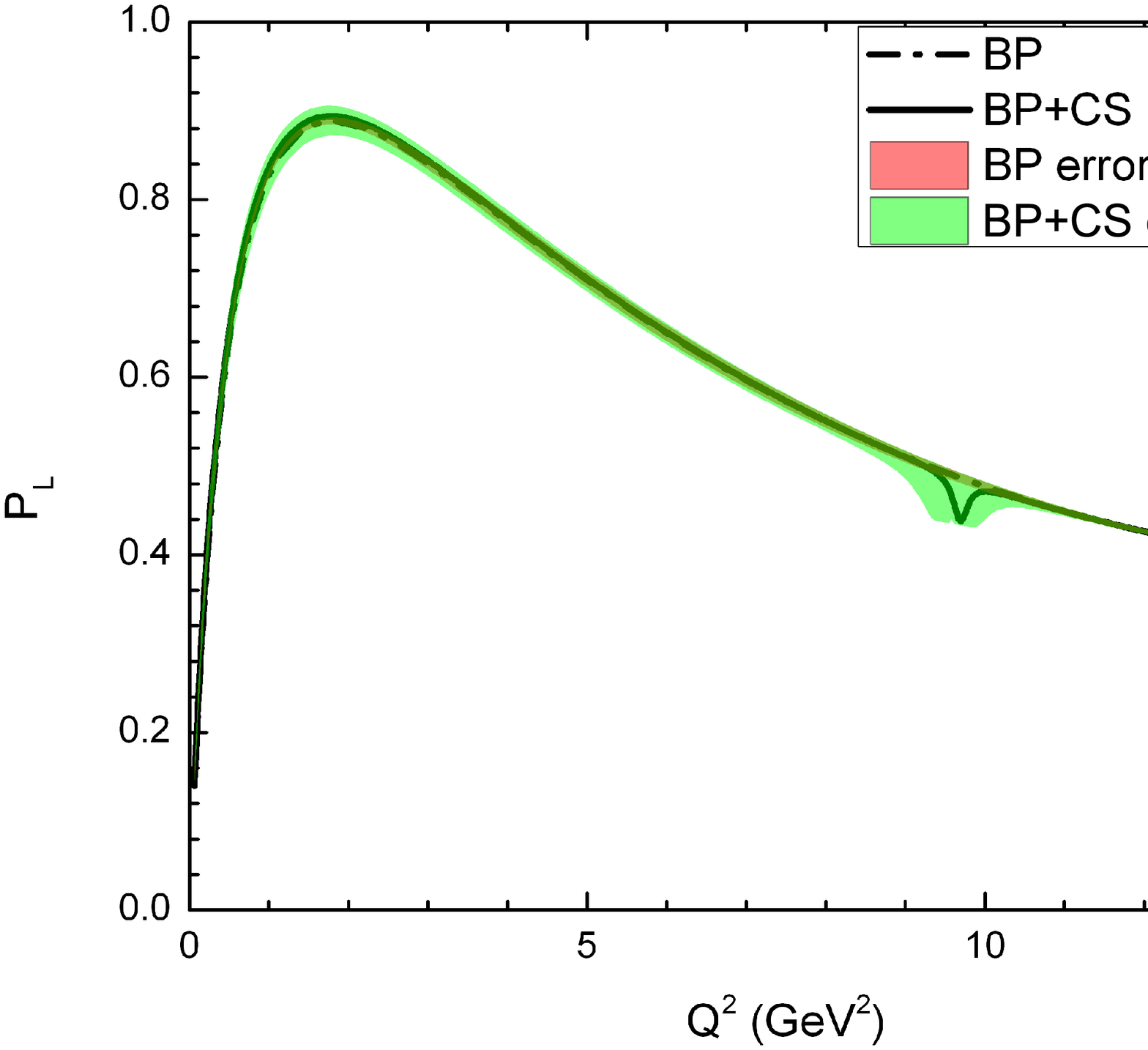}}
\subfigure[Longitudinal polarization of the final meson]{\includegraphics[width = 0.49\textwidth,height=0.2\textheight]{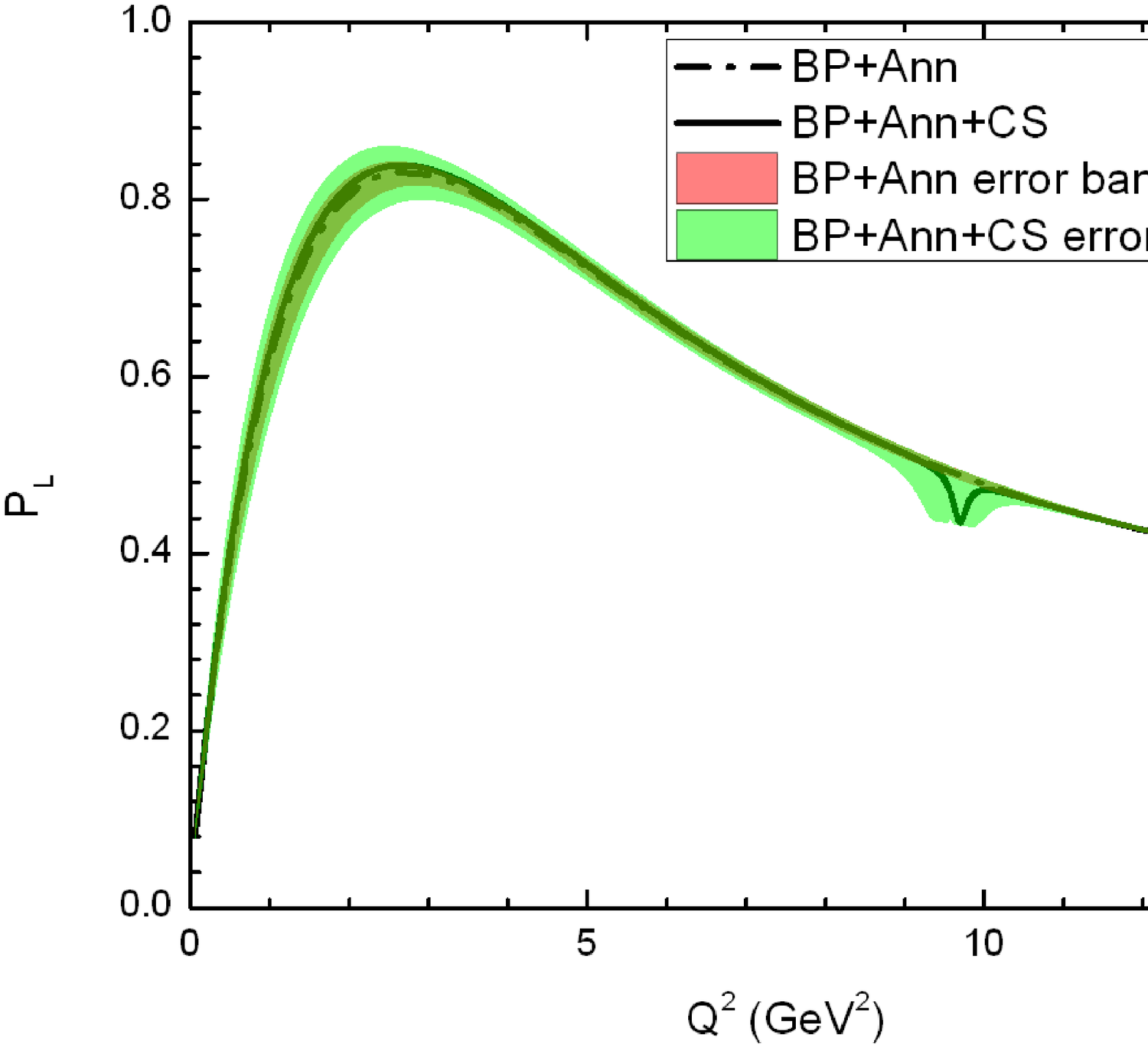}}\\
\caption{Observables of $B_c\rightarrow D_{s2}^*(2573) \mu\bar{\mu}$. }
\end{figure}

\begin{figure}[htbp]
\centering
\subfigure[Differential branching fraction]{\includegraphics[width = 0.49\textwidth,height=0.2\textheight]{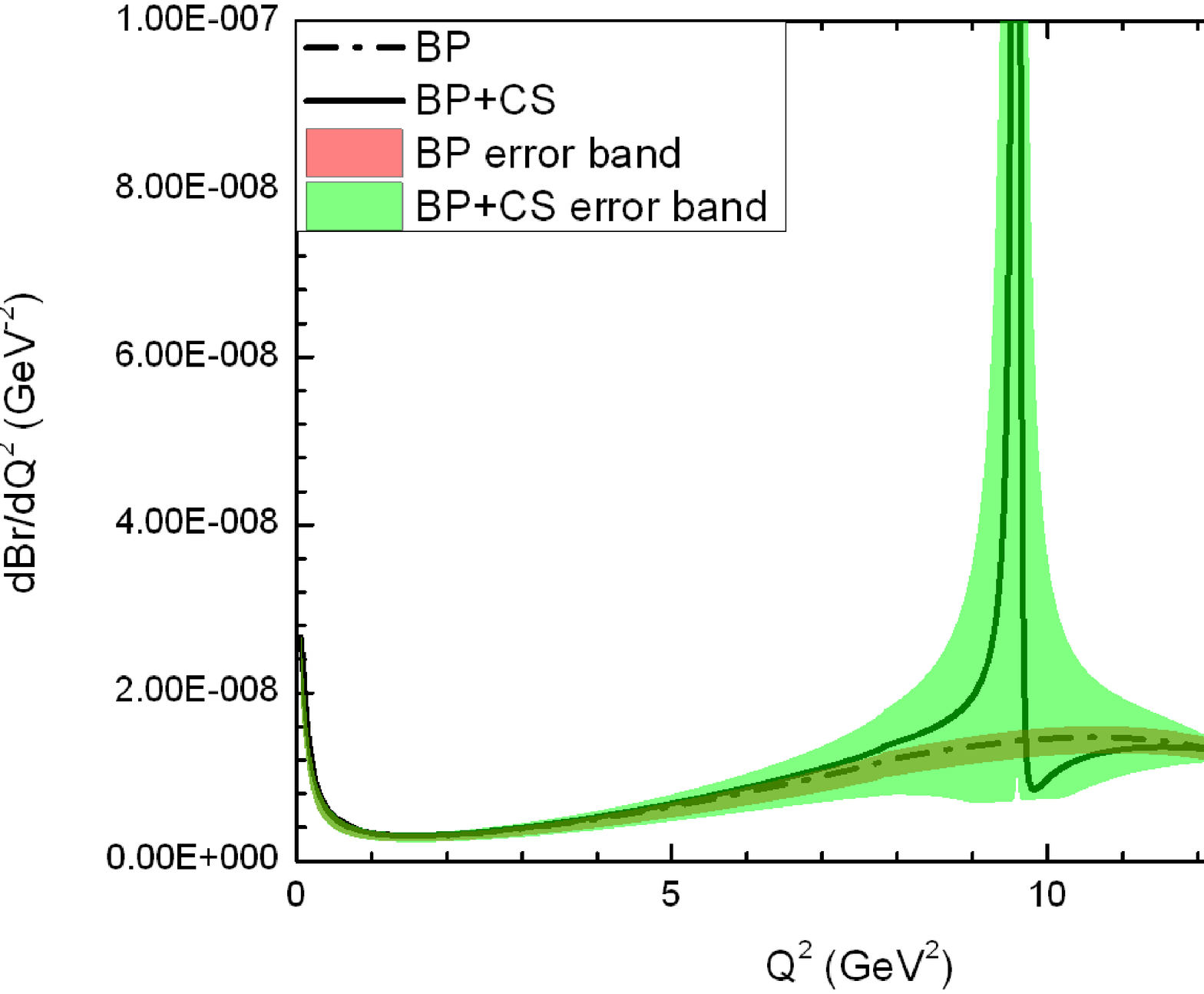}}
\subfigure[Differential branching fraction]{\includegraphics[width = 0.49\textwidth,height=0.2\textheight]{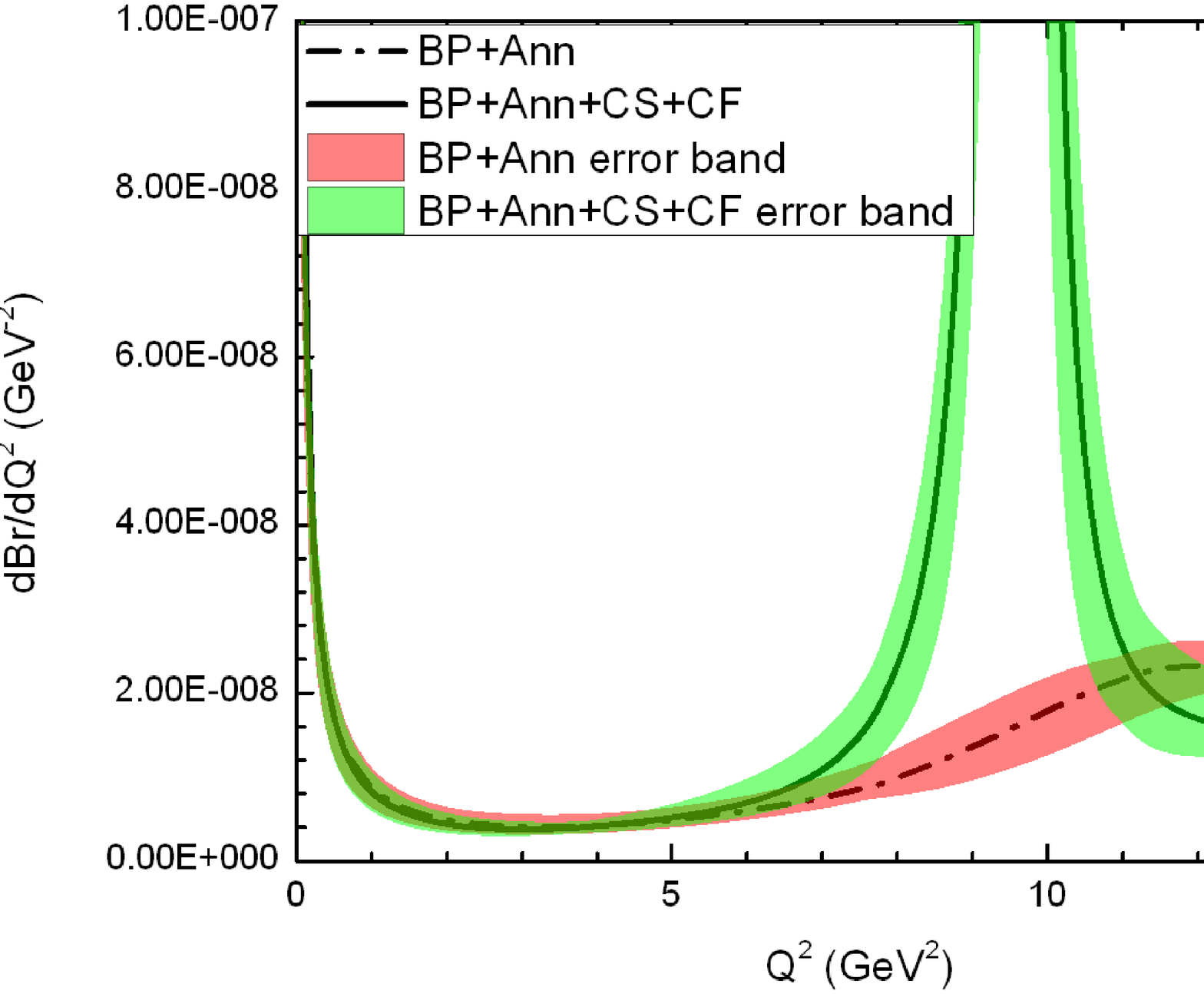}}\\
\subfigure[Leptonic longitudinal polarization asymmetry]{\includegraphics[width = 0.49\textwidth,height=0.2\textheight]{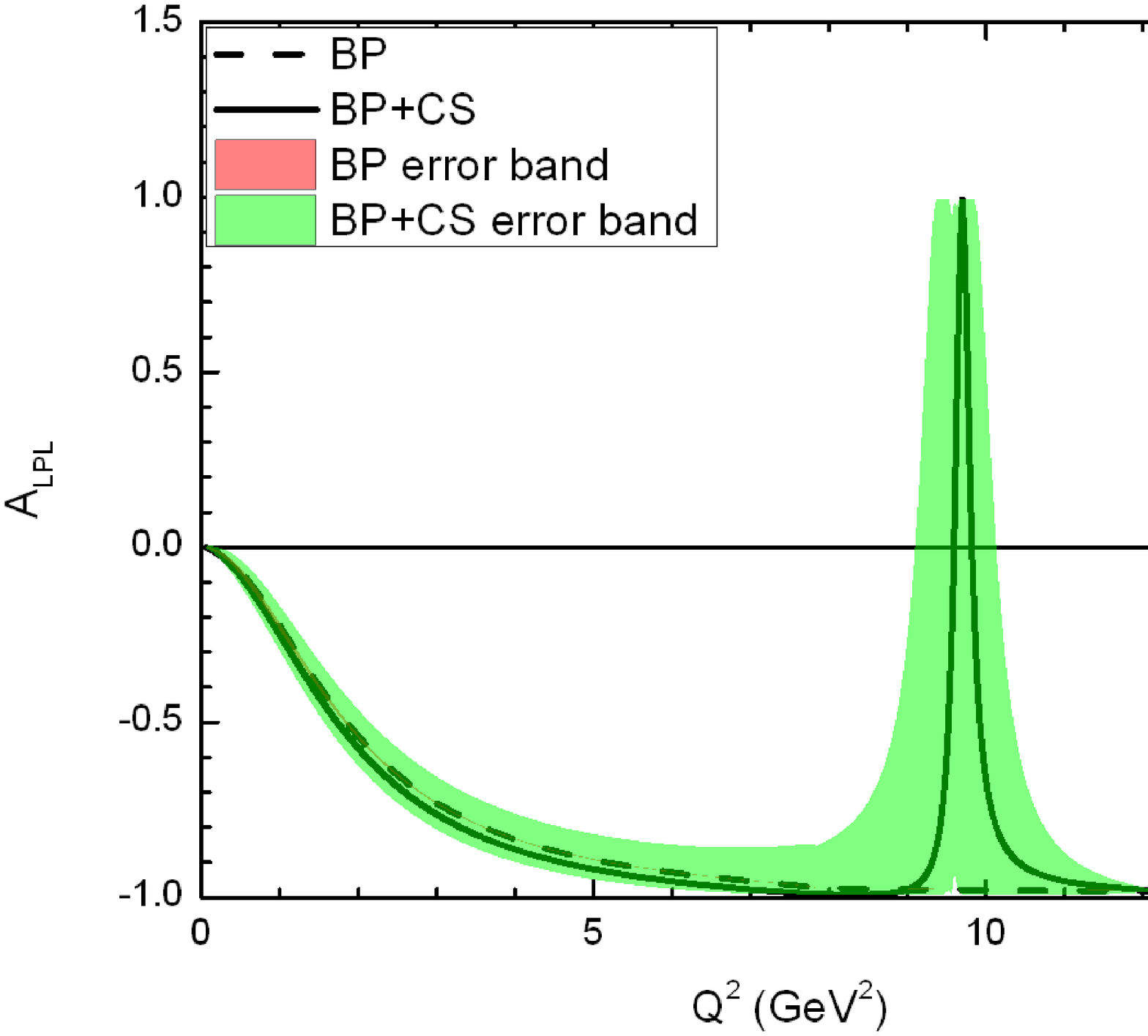}}
\subfigure[Leptonic longitudinal polarization asymmetry]{\includegraphics[width = 0.49\textwidth,height=0.2\textheight]{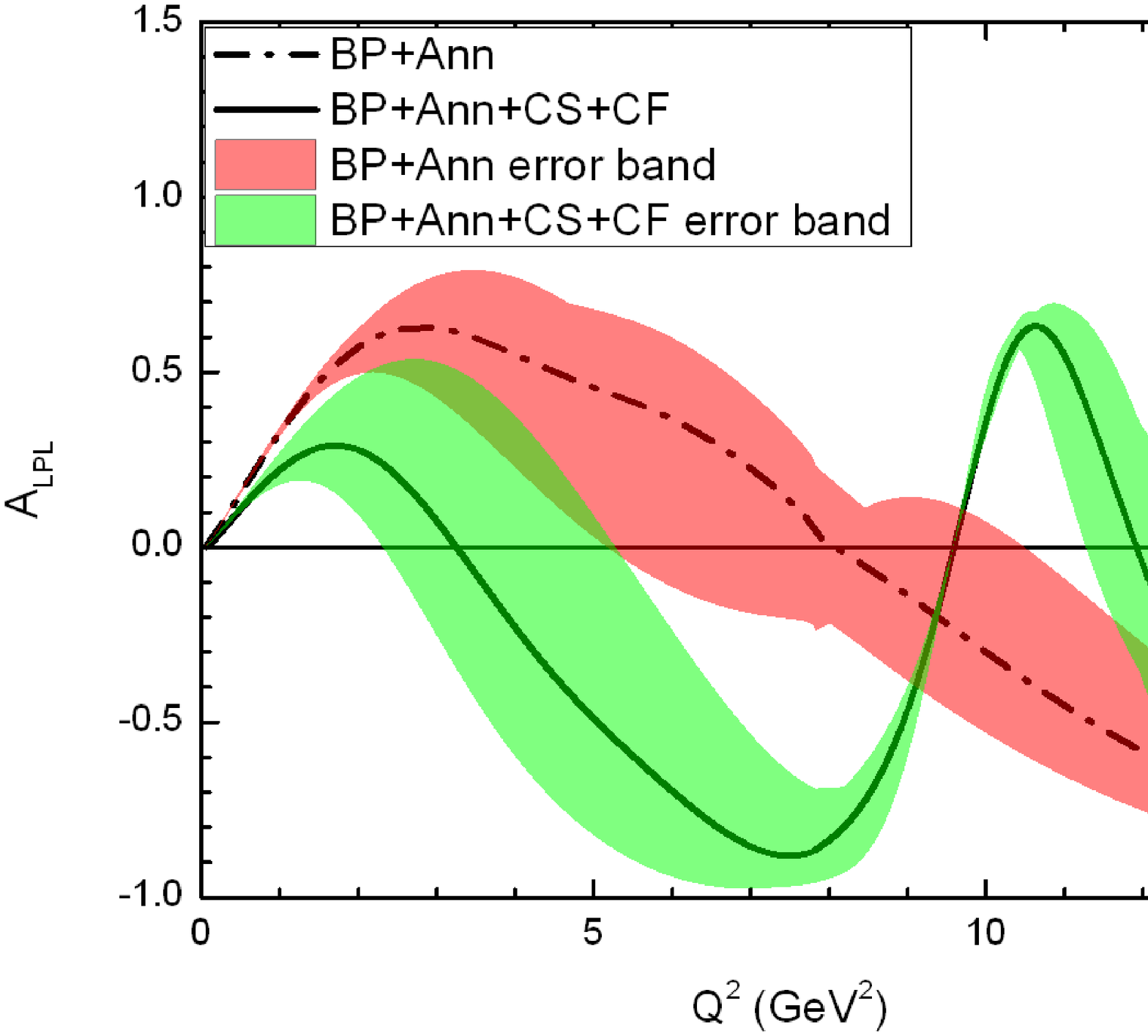}}\\
\subfigure[Forward-backward asymmetry]{\includegraphics[width = 0.49\textwidth,height=0.2\textheight]{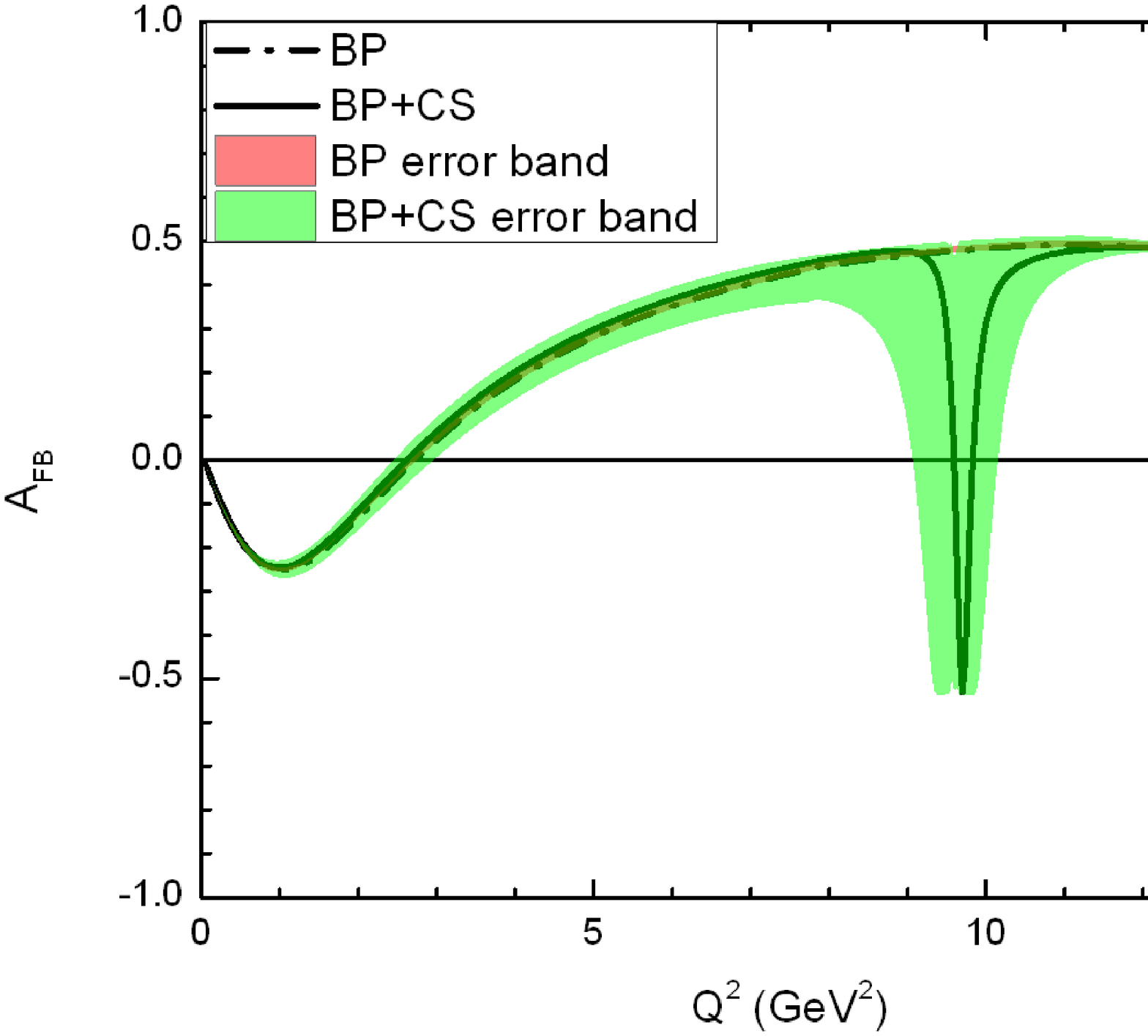}}
\subfigure[Forward-backward asymmetry]{\includegraphics[width = 0.49\textwidth,height=0.2\textheight]{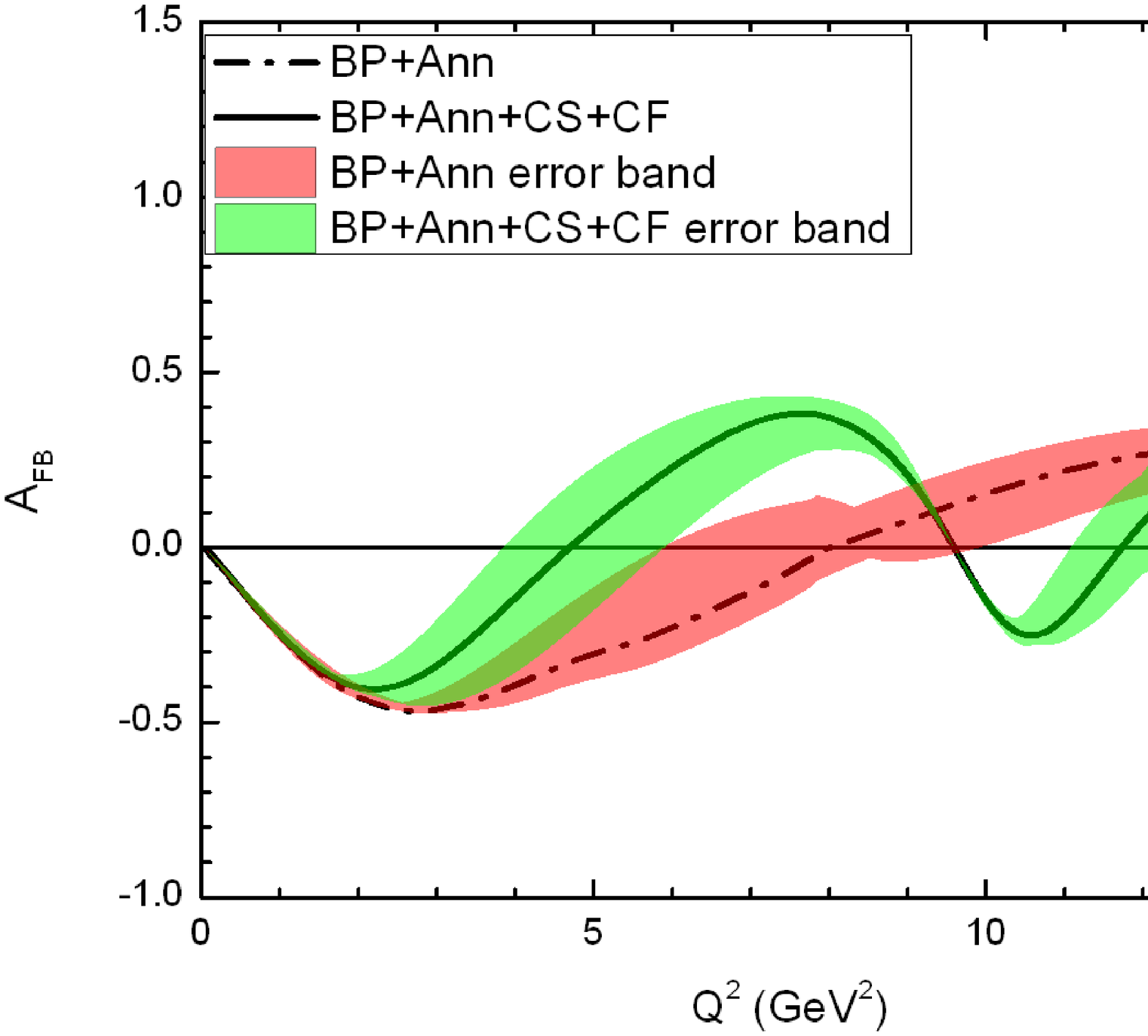}}\\
\subfigure[Longitudinal polarization of the
meson]{\includegraphics[width =
0.49\textwidth,height=0.2\textheight]{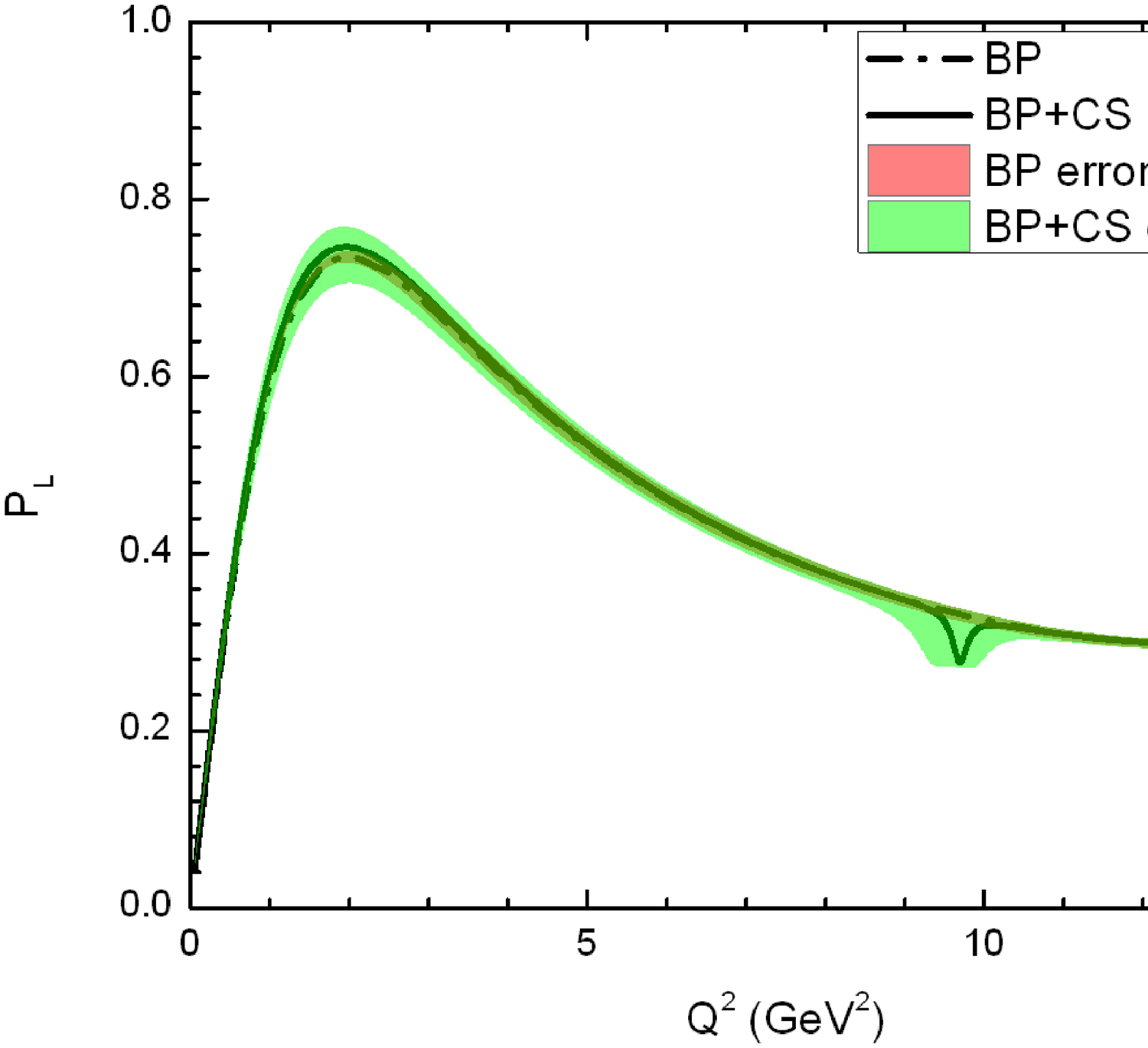}}
\subfigure[Longitudinal polarization of the final meson]{\includegraphics[width = 0.49\textwidth,height=0.2\textheight]{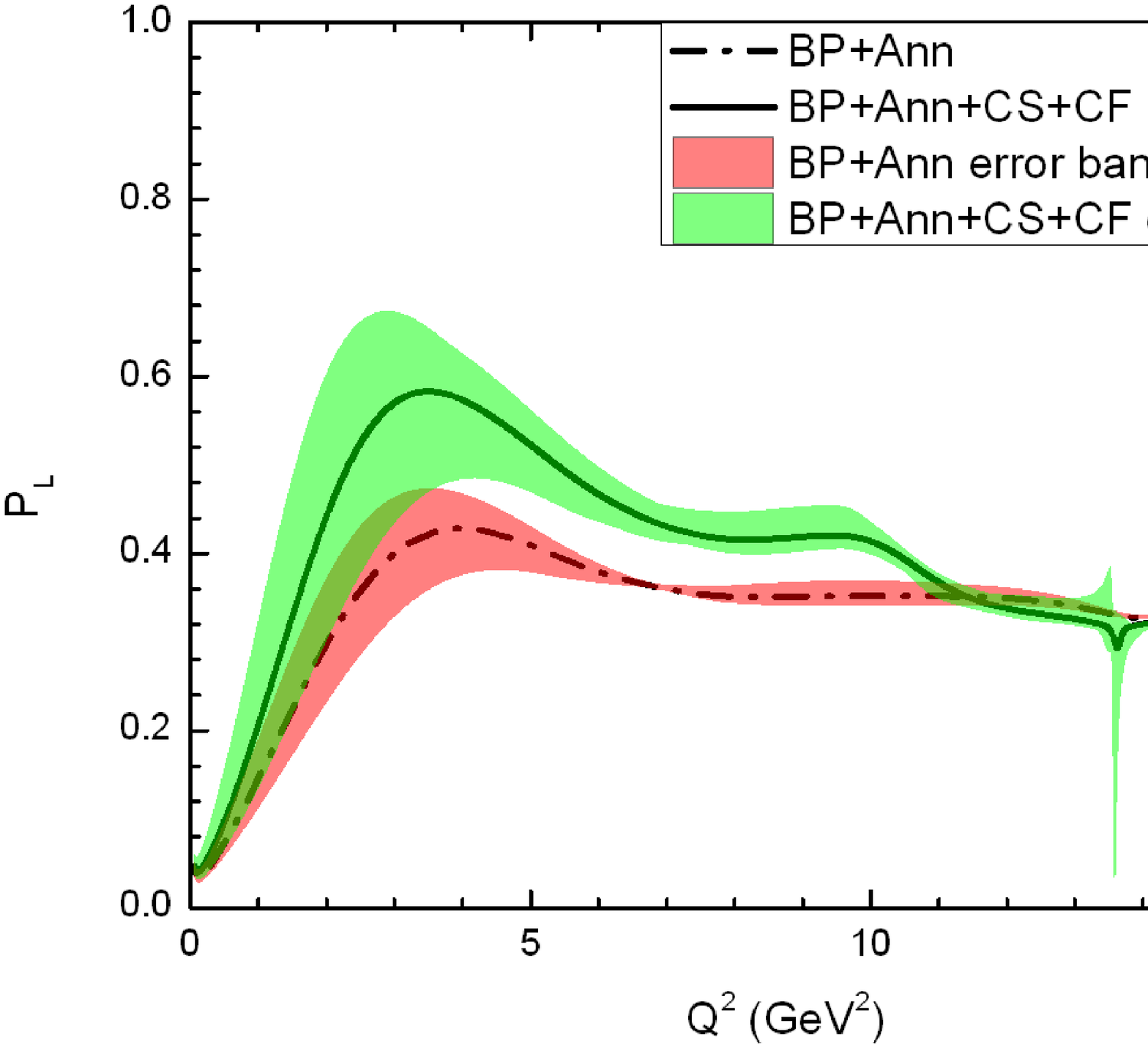}}\\

\caption{Observables of $B_c\rightarrow D_{s1}(2460) \mu\bar{\mu}$. }
\end{figure}

\clearpage

\begin{figure}[htbp]
\centering
\subfigure[Differential branching fraction]{\includegraphics[width = 0.49\textwidth,height=0.2\textheight]{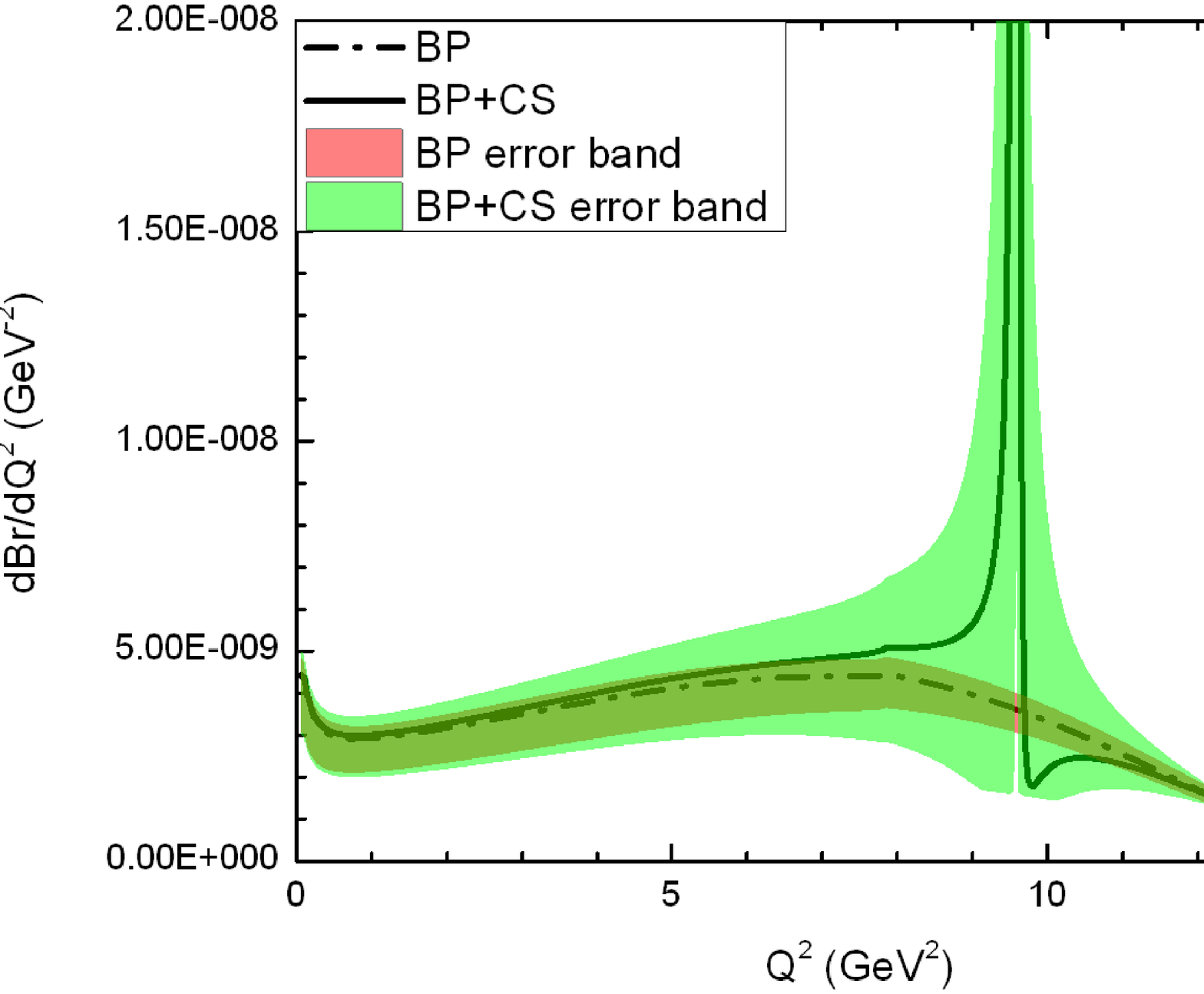}}
\subfigure[Differential branching fraction]{\includegraphics[width = 0.49\textwidth,height=0.2\textheight]{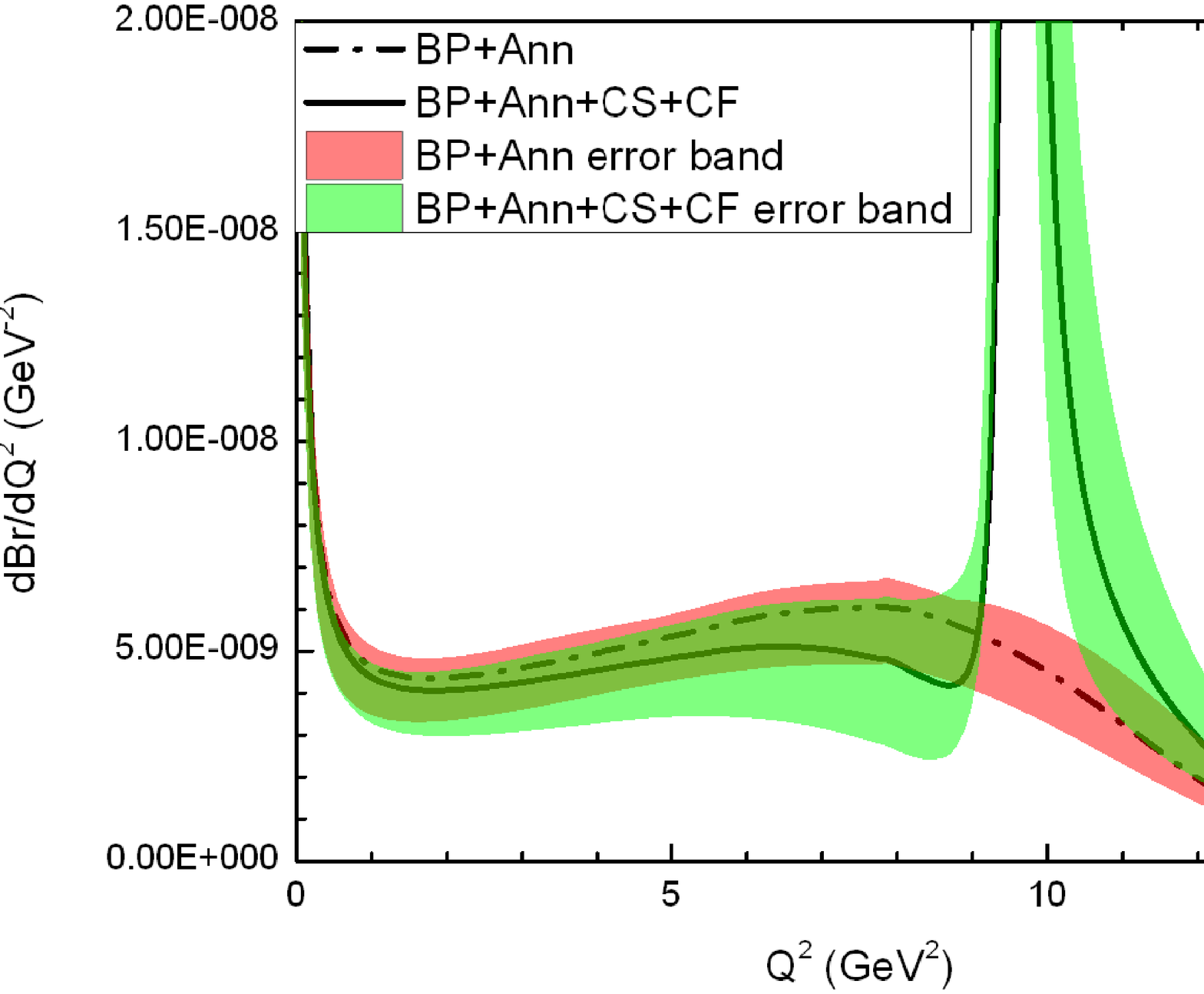}}\\
\subfigure[Leptonic longitudinal polarization asymmetry]{\includegraphics[width = 0.49\textwidth,height=0.2\textheight]{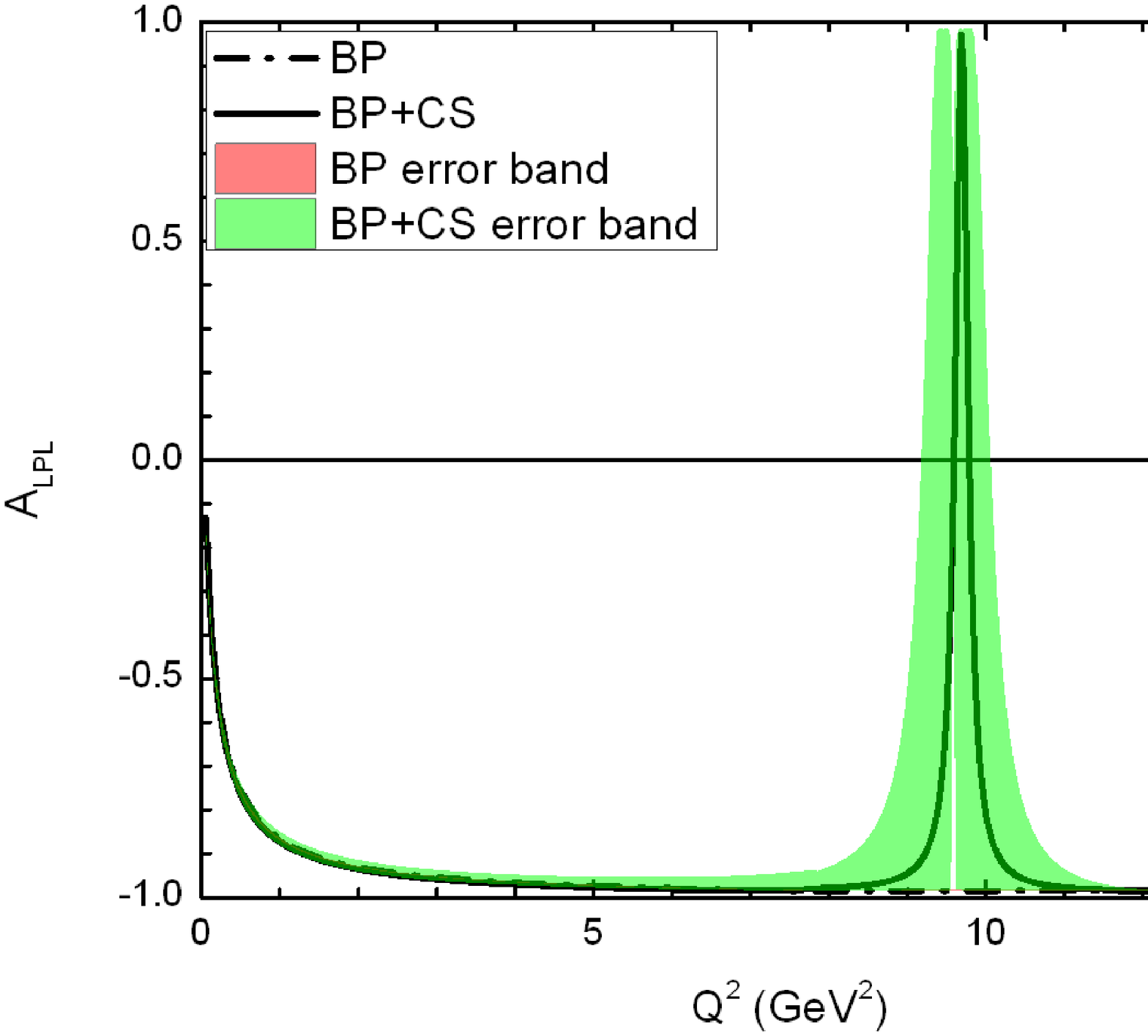}}
\subfigure[Leptonic longitudinal polarization asymmetry]{\includegraphics[width = 0.49\textwidth,height=0.2\textheight]{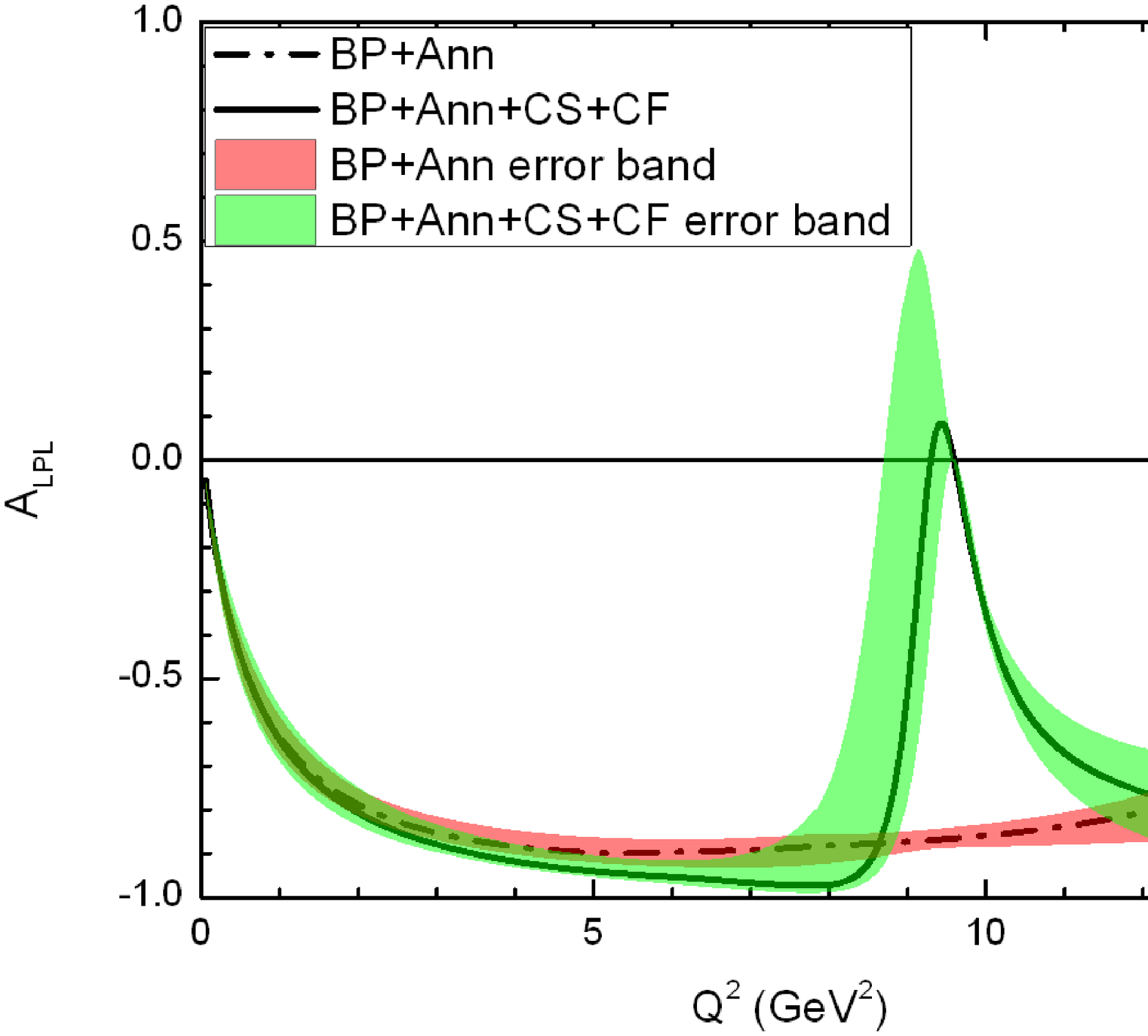}}\\
\subfigure[Forward-backward asymmetry]{\includegraphics[width = 0.49\textwidth,height=0.2\textheight]{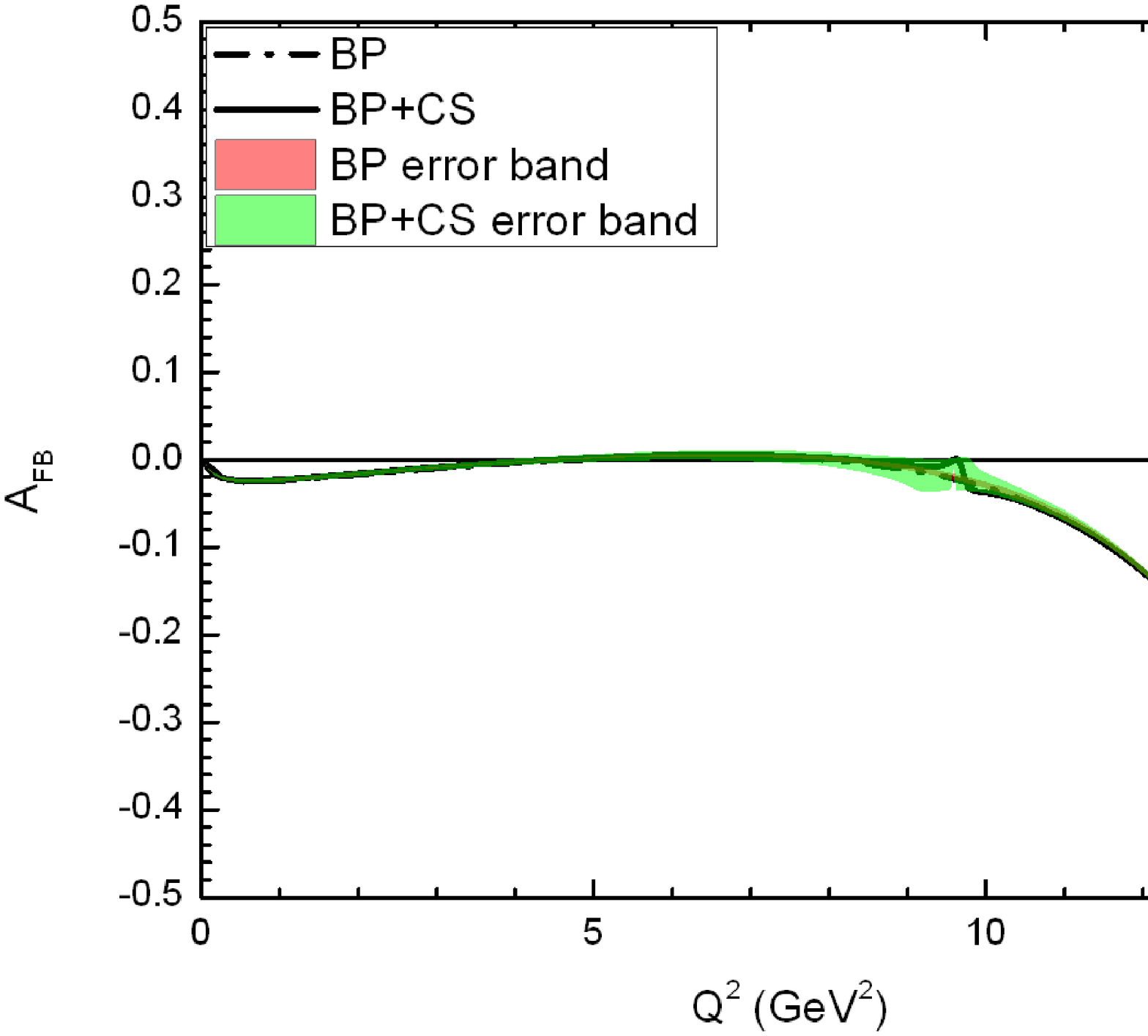}}
\subfigure[Forward-backward asymmetry]{\includegraphics[width = 0.49\textwidth,height=0.2\textheight]{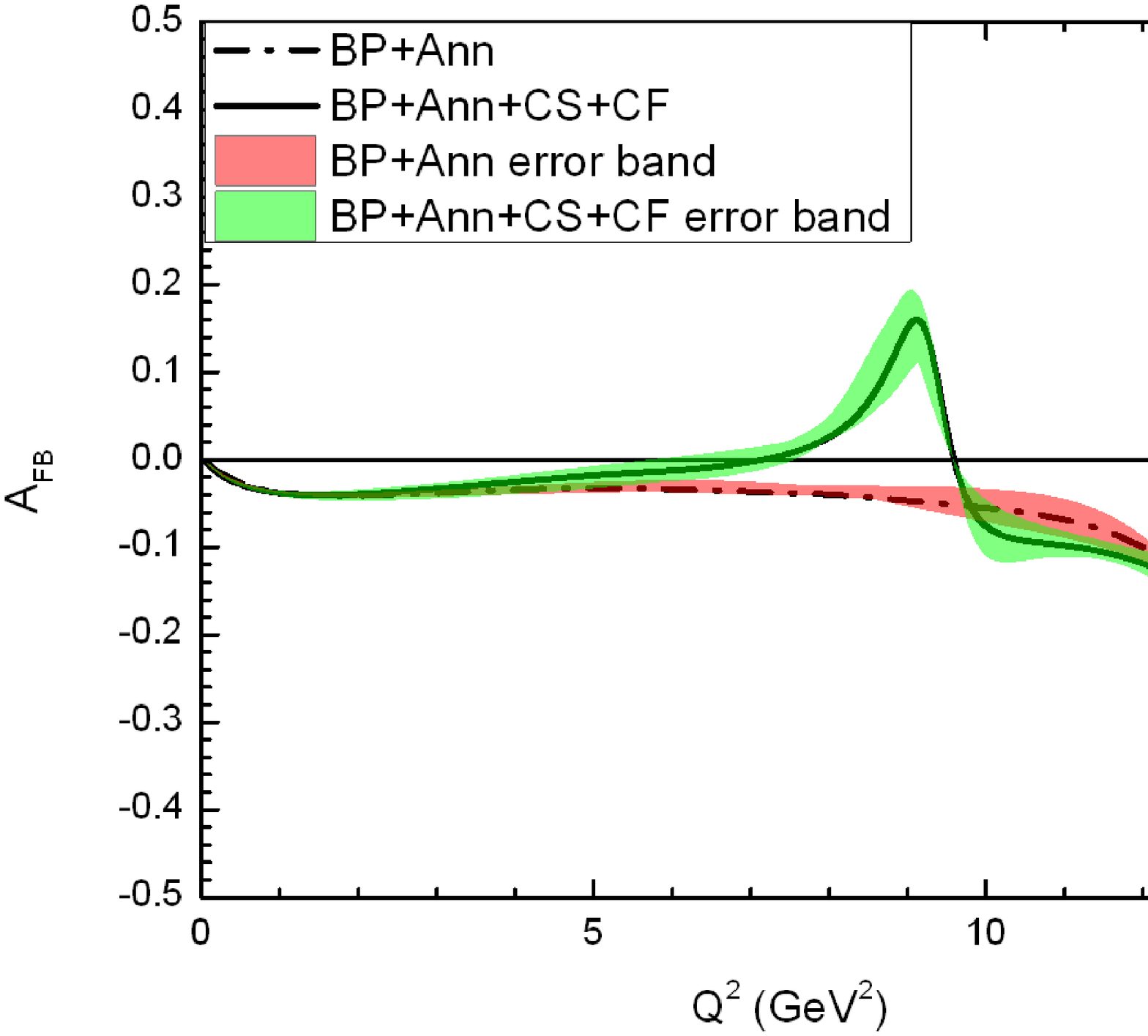}}\\
\subfigure[Longitudinal polarization of the
meson]{\includegraphics[width =
0.49\textwidth,height=0.2\textheight]{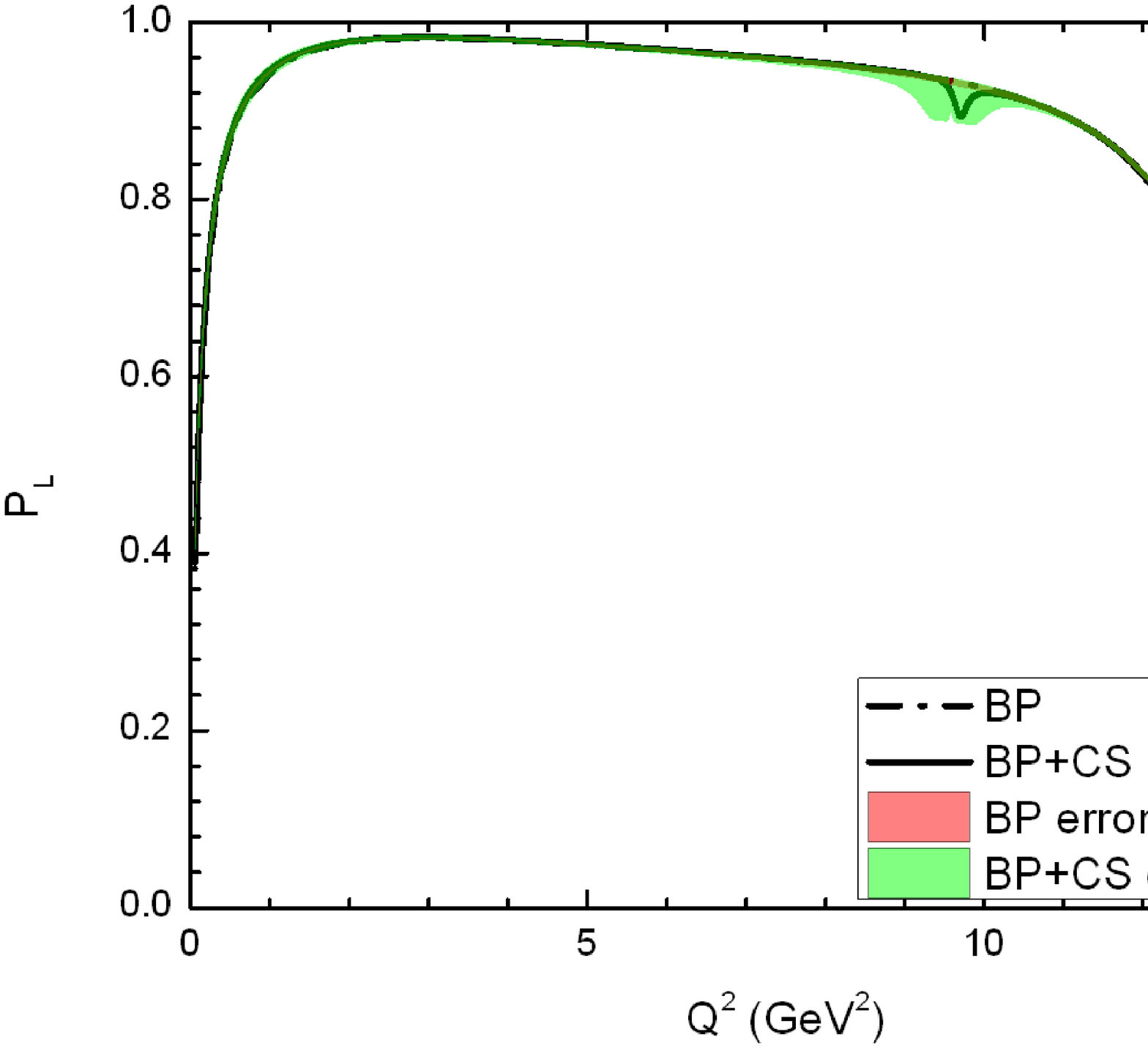}}
\subfigure[Longitudinal polarization of the final meson]{\includegraphics[width = 0.49\textwidth,height=0.2\textheight]{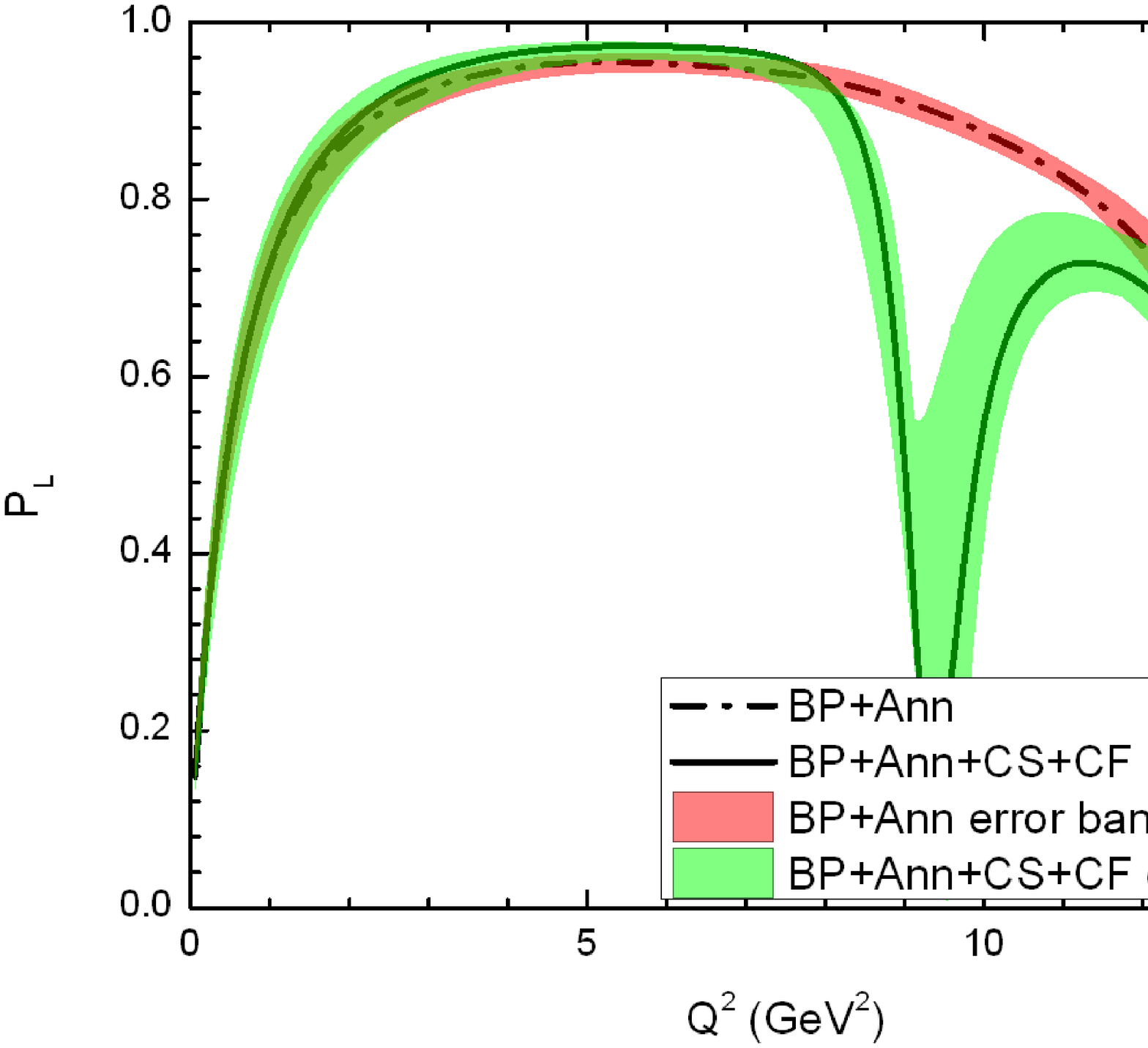}}\\

\caption{Observables of $B_c\rightarrow D_{s1}(2536) \mu\bar{\mu}$. }
\end{figure}

\begin{figure}[htbp]
\centering
\subfigure[Differential branching fraction]{\includegraphics[width = 0.49\textwidth,height=0.4\textheight]{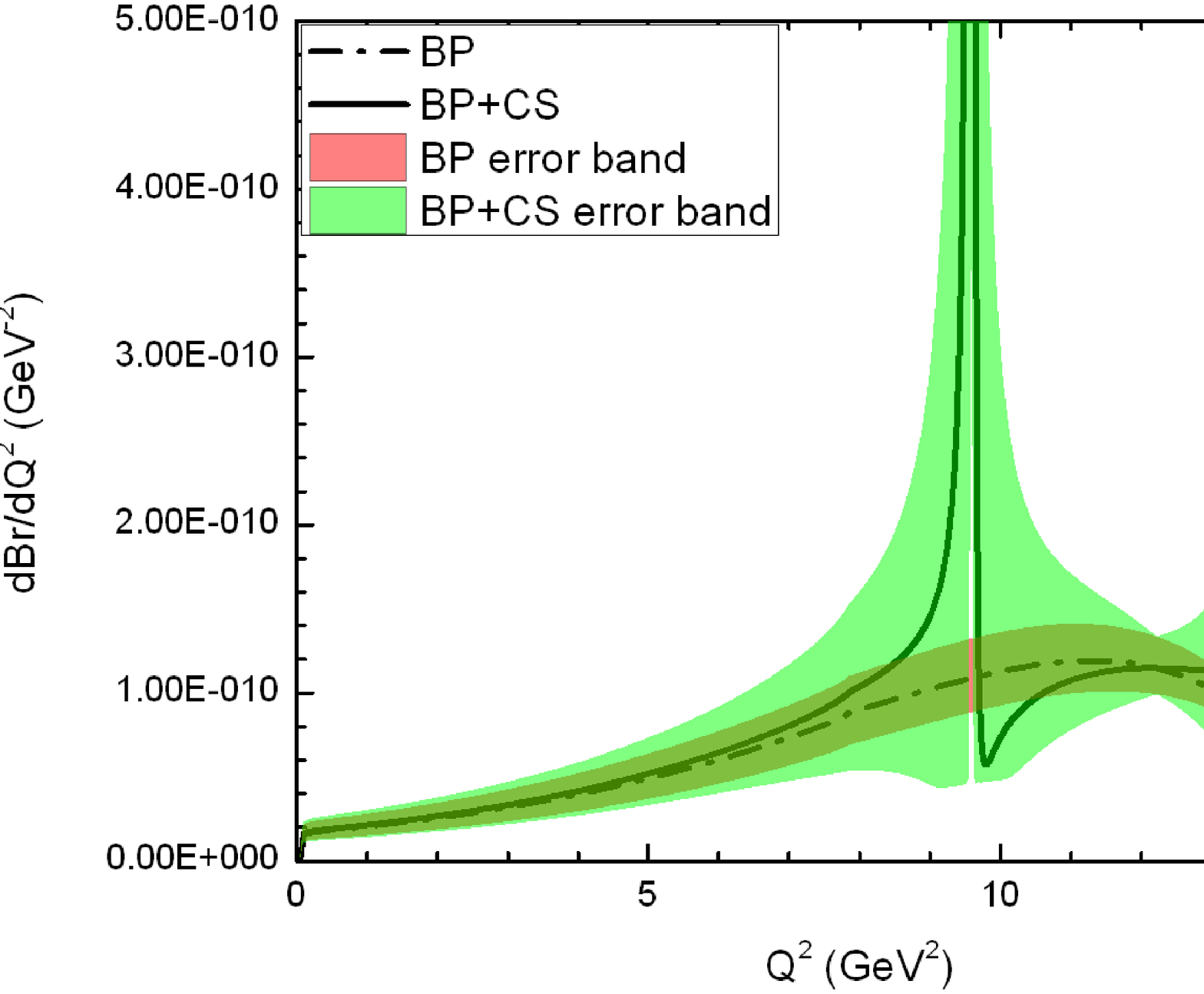}}
\subfigure[Differential branching fraction]{\includegraphics[width = 0.49\textwidth,height=0.4\textheight]{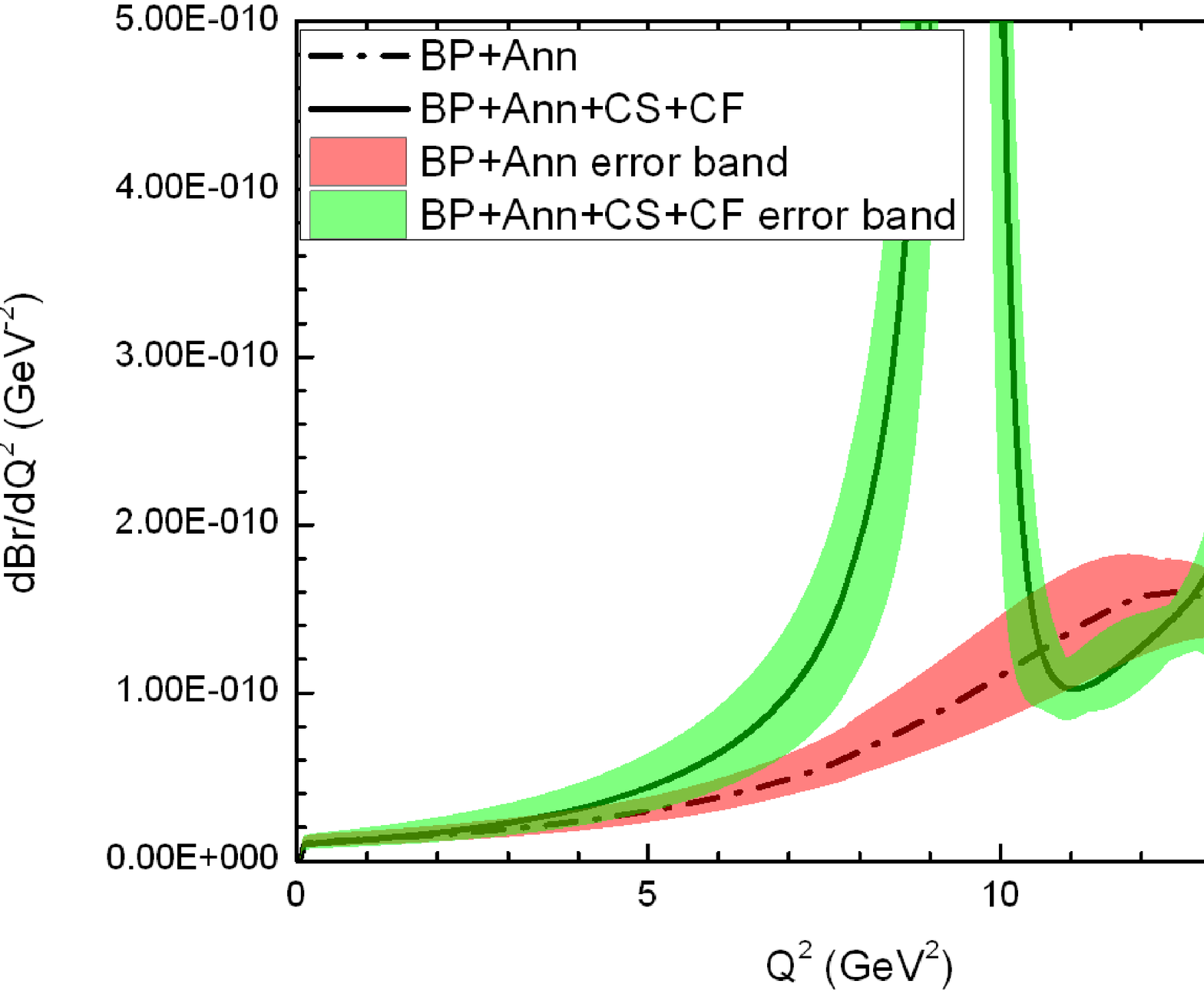}}\\
\subfigure[Leptonic longitudinal polarization asymmetry]{\includegraphics[width = 0.49\textwidth,height=0.4\textheight]{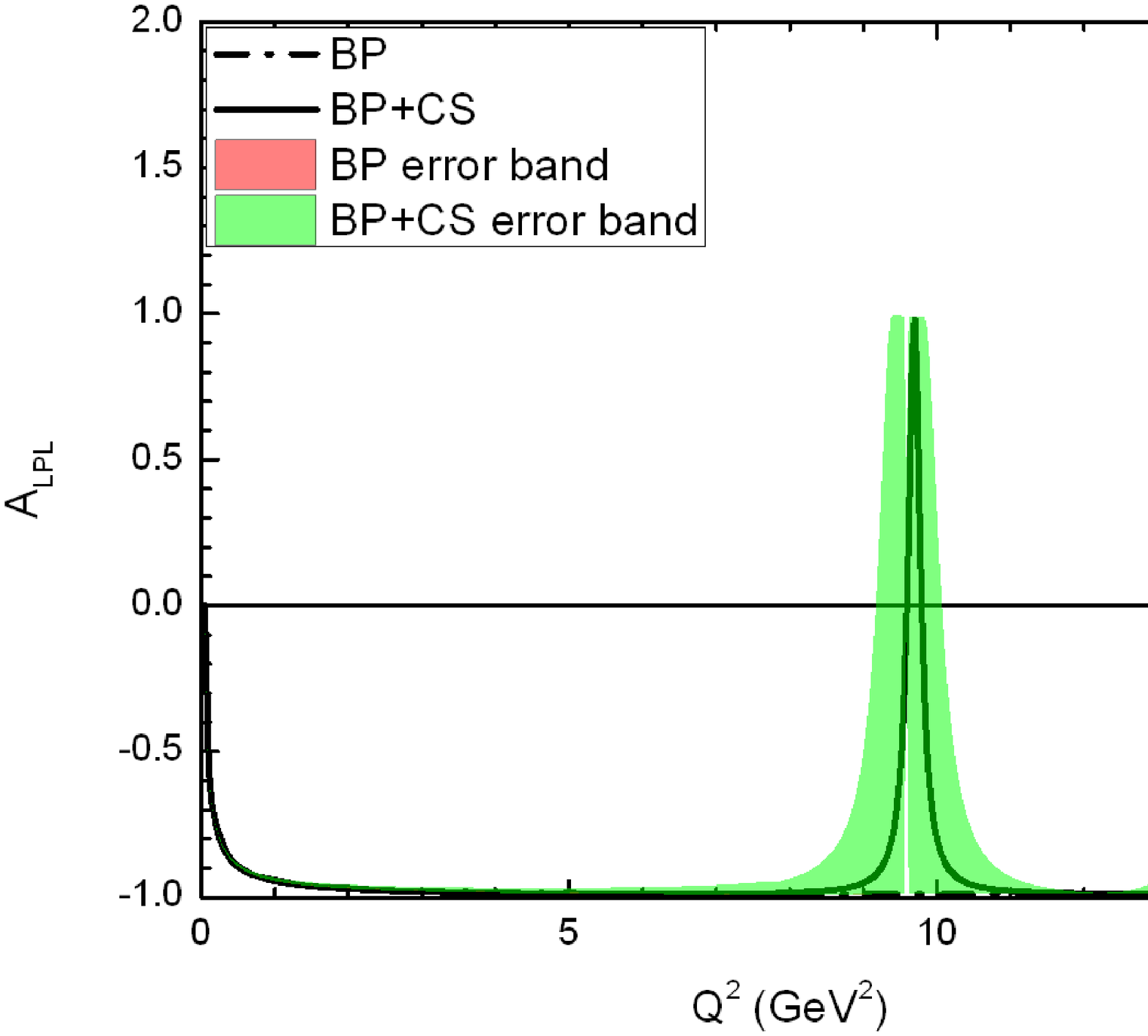}}
\subfigure[Leptonic longitudinal polarization asymmetry]{\includegraphics[width = 0.49\textwidth,height=0.4\textheight]{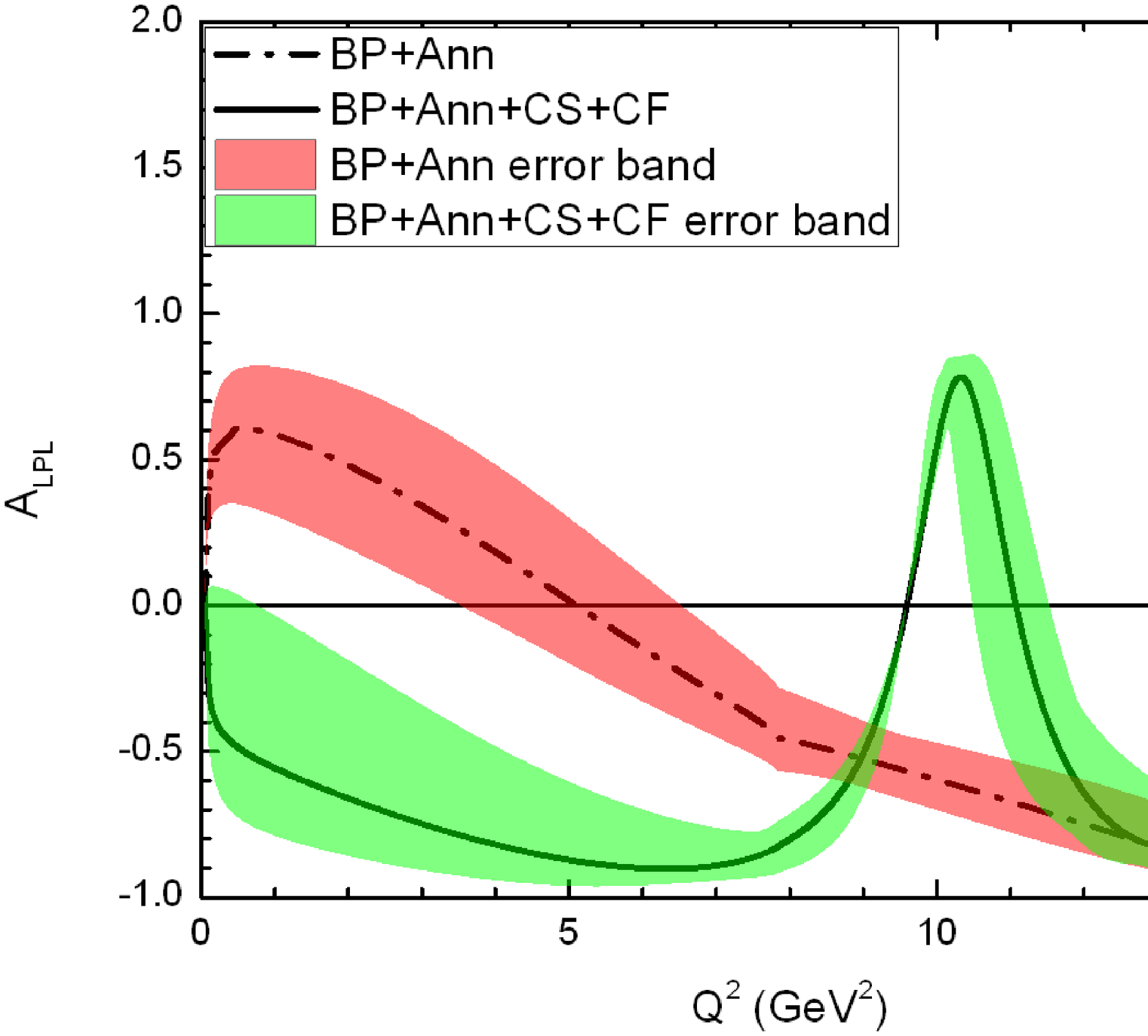}}
\caption{Observables of $B_c\rightarrow D_0^*(2400) \mu\bar{\mu}$. }
\end{figure}

\begin{figure}[htbp]
\centering
\subfigure[Differential branching fraction]{\includegraphics[width = 0.49\textwidth,height=0.2\textheight]{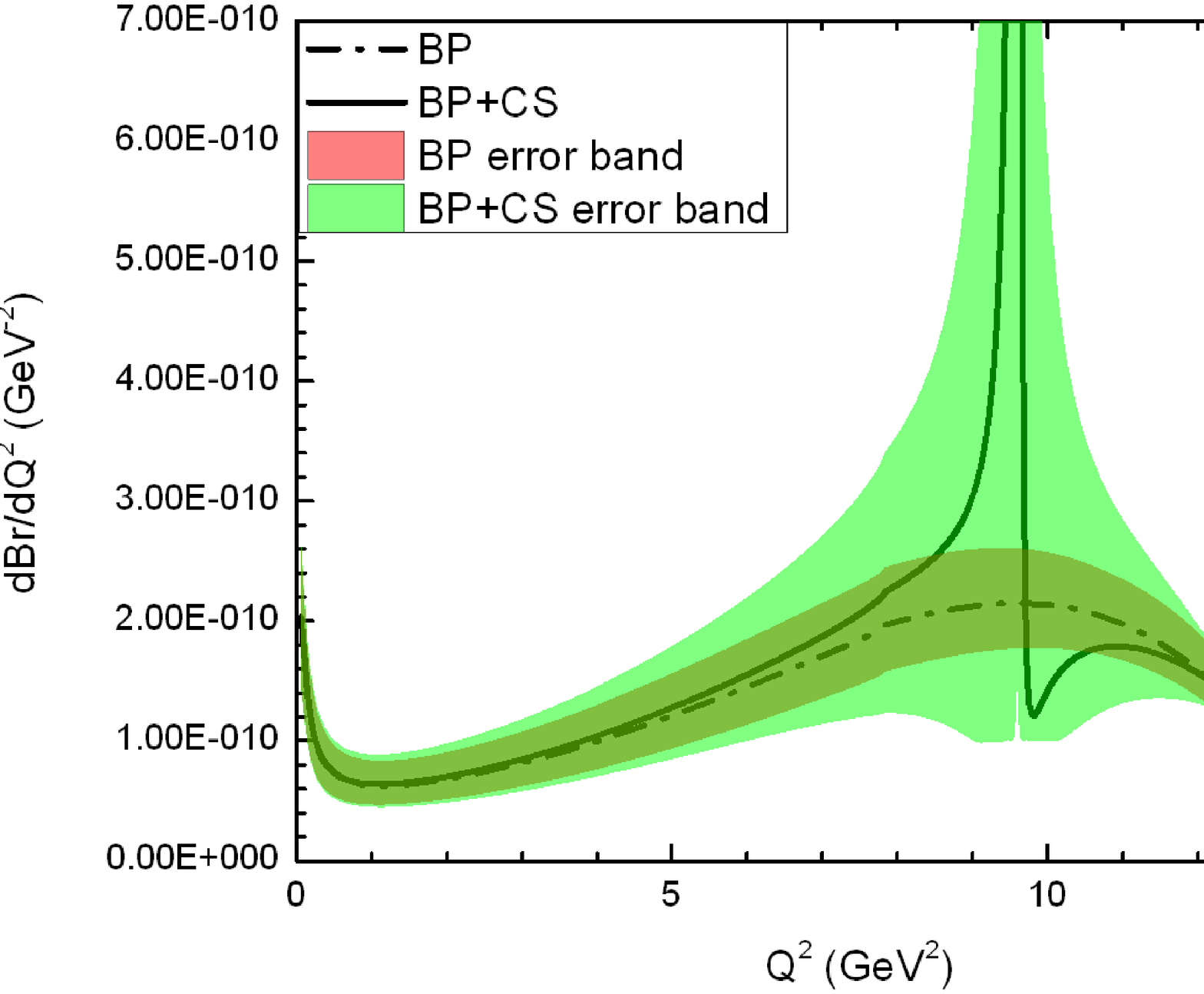}}
\subfigure[Differential branching fraction]{\includegraphics[width = 0.49\textwidth,height=0.2\textheight]{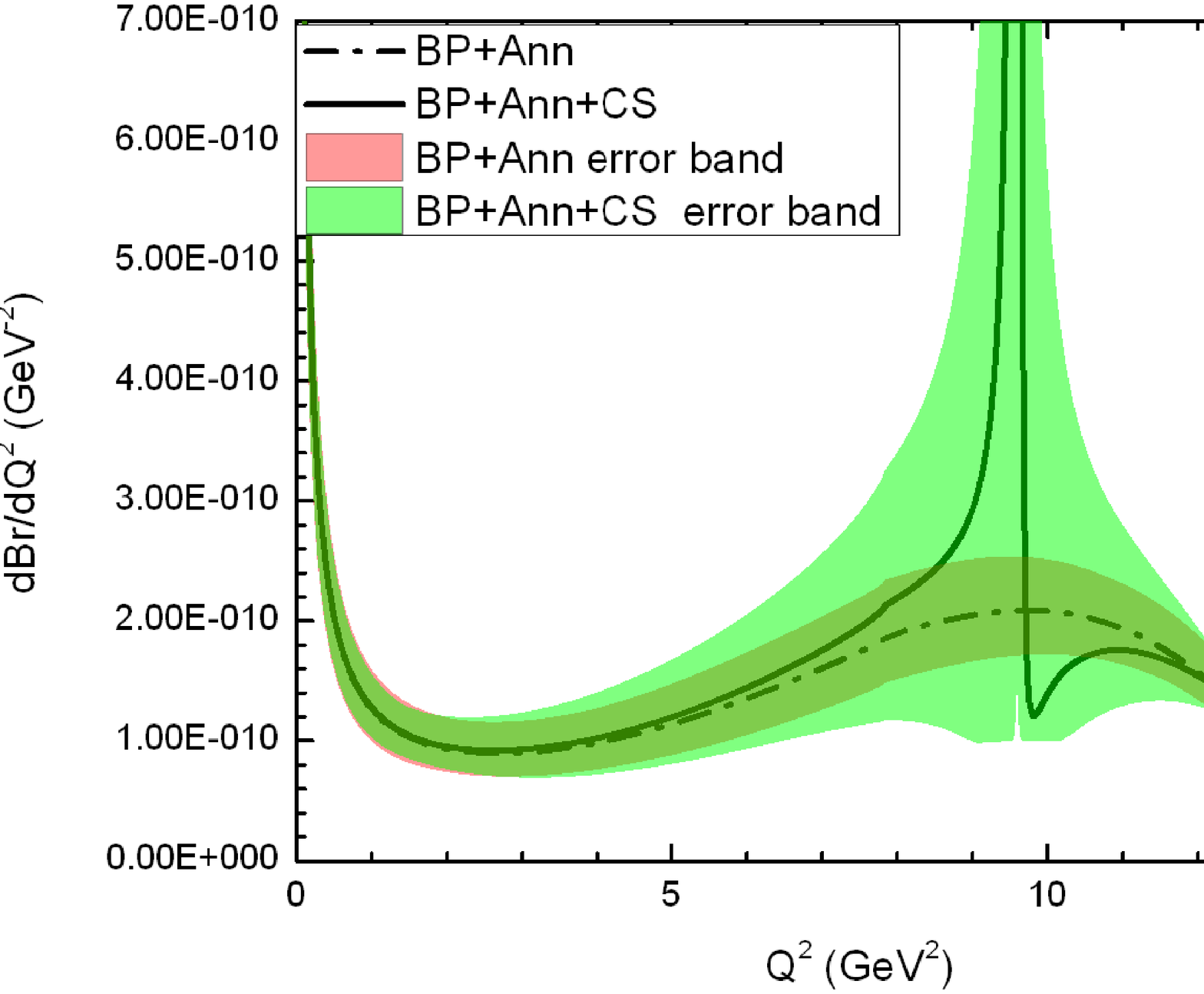}}\\
\subfigure[Leptonic longitudinal polarization asymmetry]{\includegraphics[width = 0.49\textwidth,height=0.2\textheight]{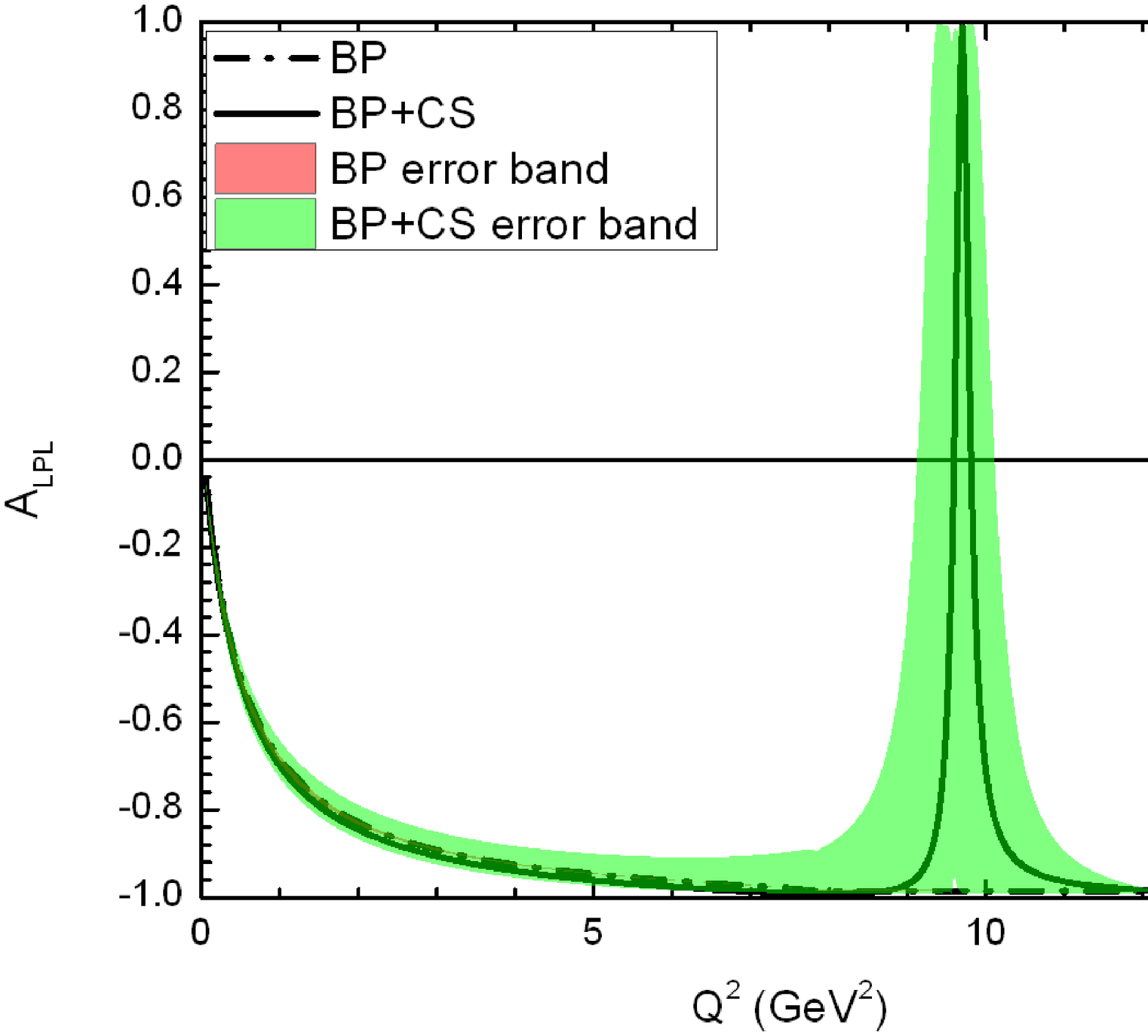}}
\subfigure[Leptonic longitudinal polarization asymmetry]{\includegraphics[width = 0.49\textwidth,height=0.2\textheight]{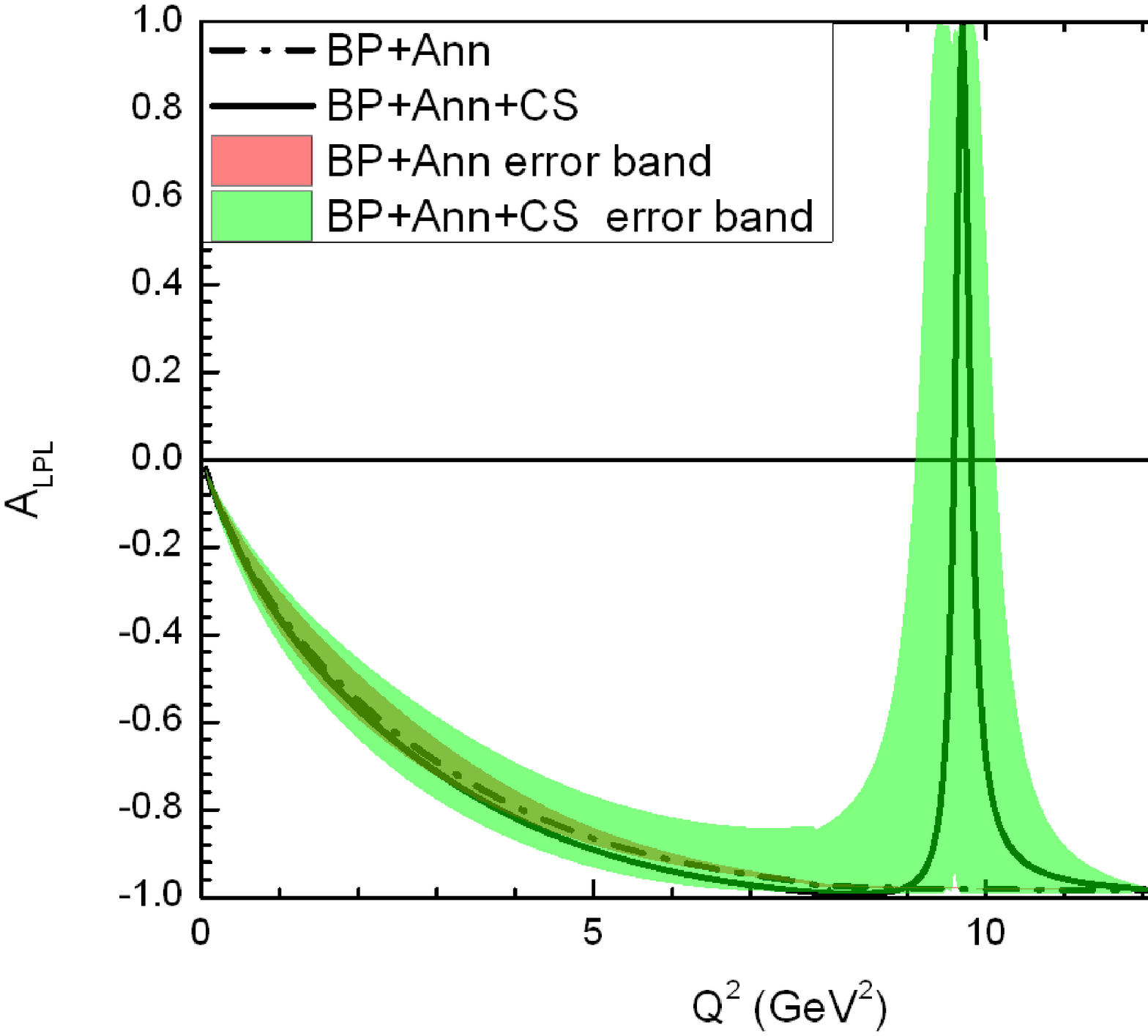}}\\
\subfigure[Forward-backward asymmetry]{\includegraphics[width = 0.49\textwidth,height=0.2\textheight]{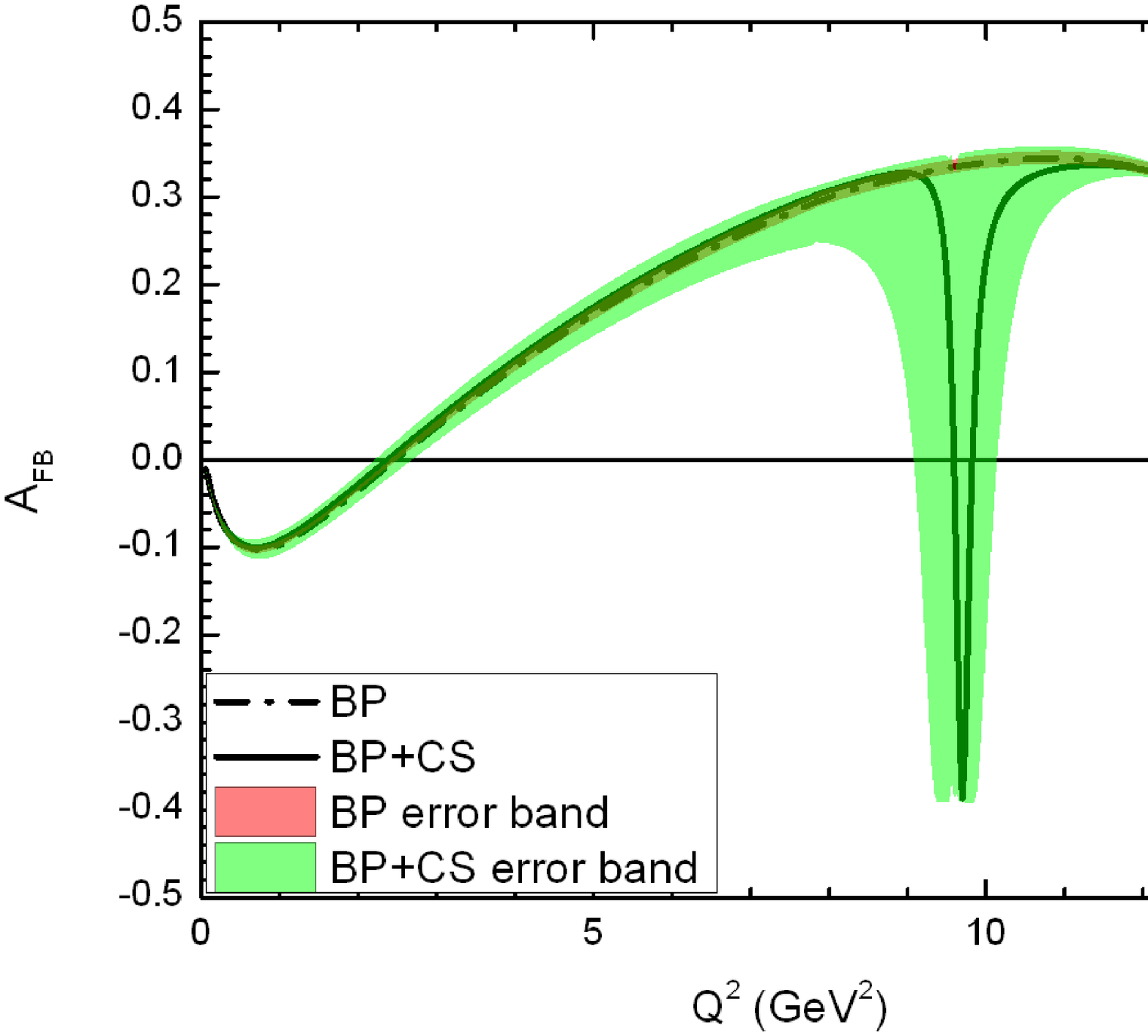}}
\subfigure[Forward-backward asymmetry]{\includegraphics[width = 0.49\textwidth,height=0.2\textheight]{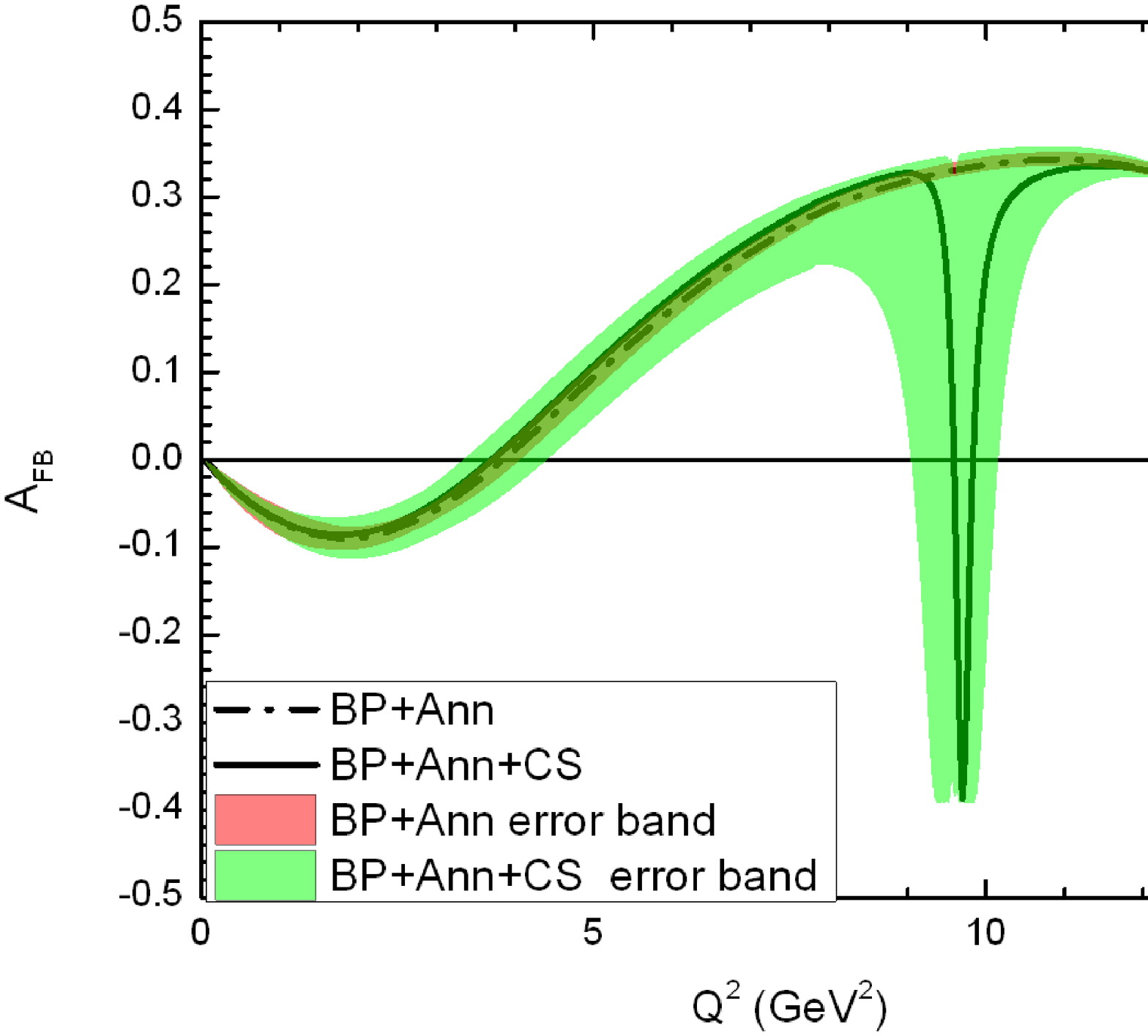}}\\
\subfigure[Longitudinal polarization of the
meson]{\includegraphics[width =
0.49\textwidth,height=0.2\textheight]{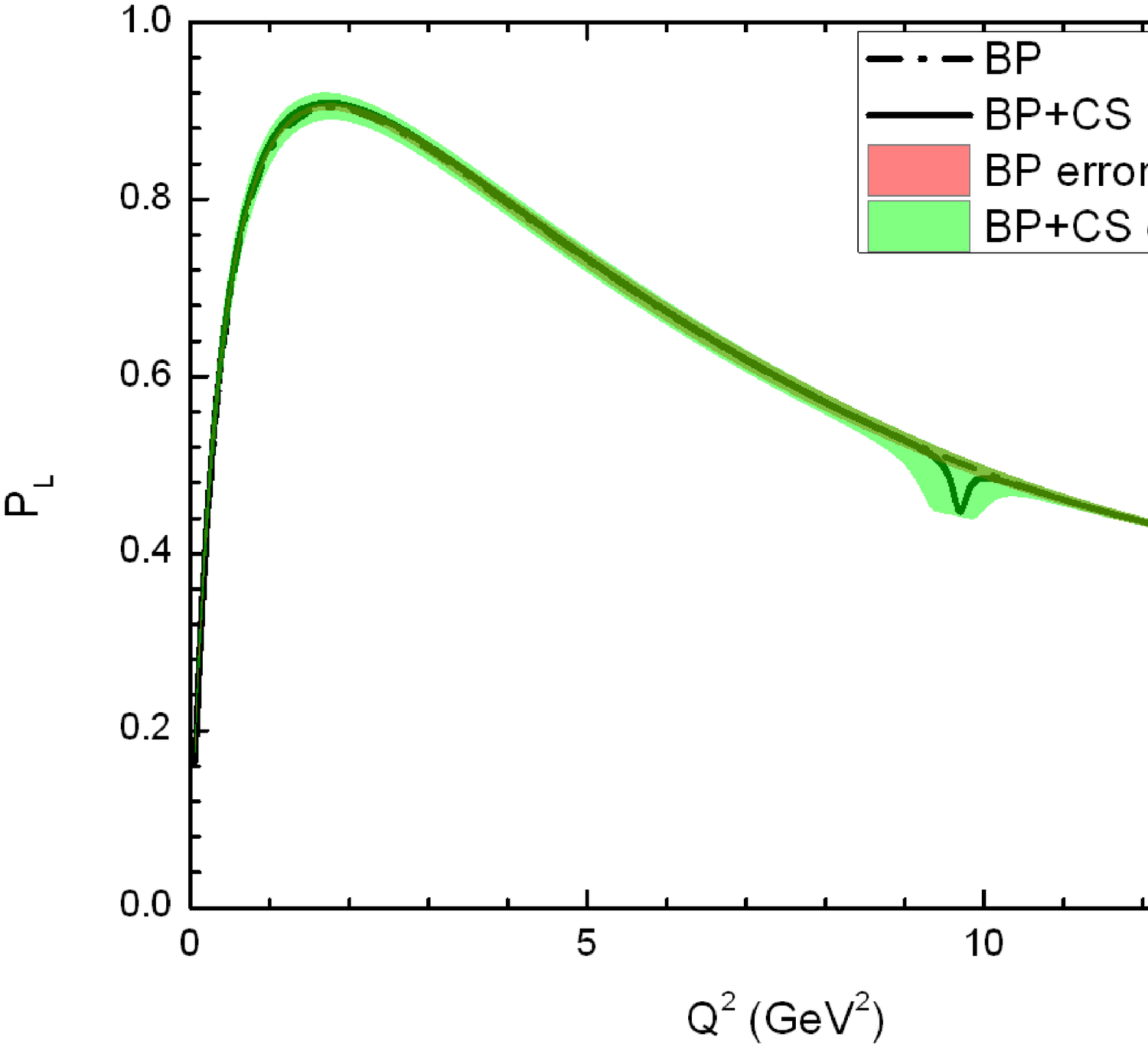}}
\subfigure[Longitudinal polarization of the final meson]{\includegraphics[width = 0.49\textwidth,height=0.2\textheight]{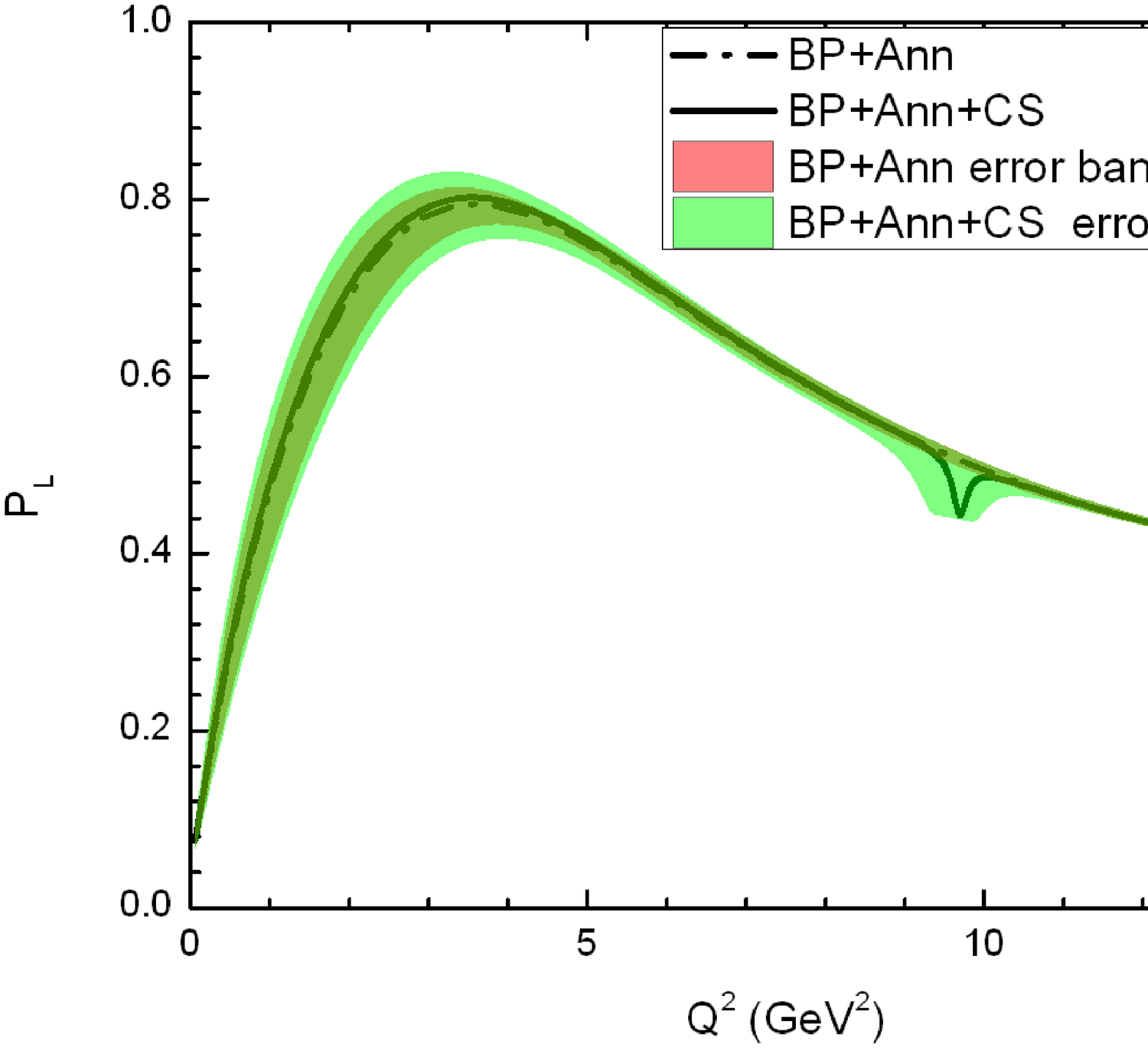}}\\

\caption{Observables of $B_c\rightarrow D_2^*(2460) \mu\bar{\mu}$. }
\end{figure}

\begin{figure}[htbp]
\centering
\subfigure[Differential branching fraction]{\includegraphics[width = 0.49\textwidth,height=0.2\textheight]{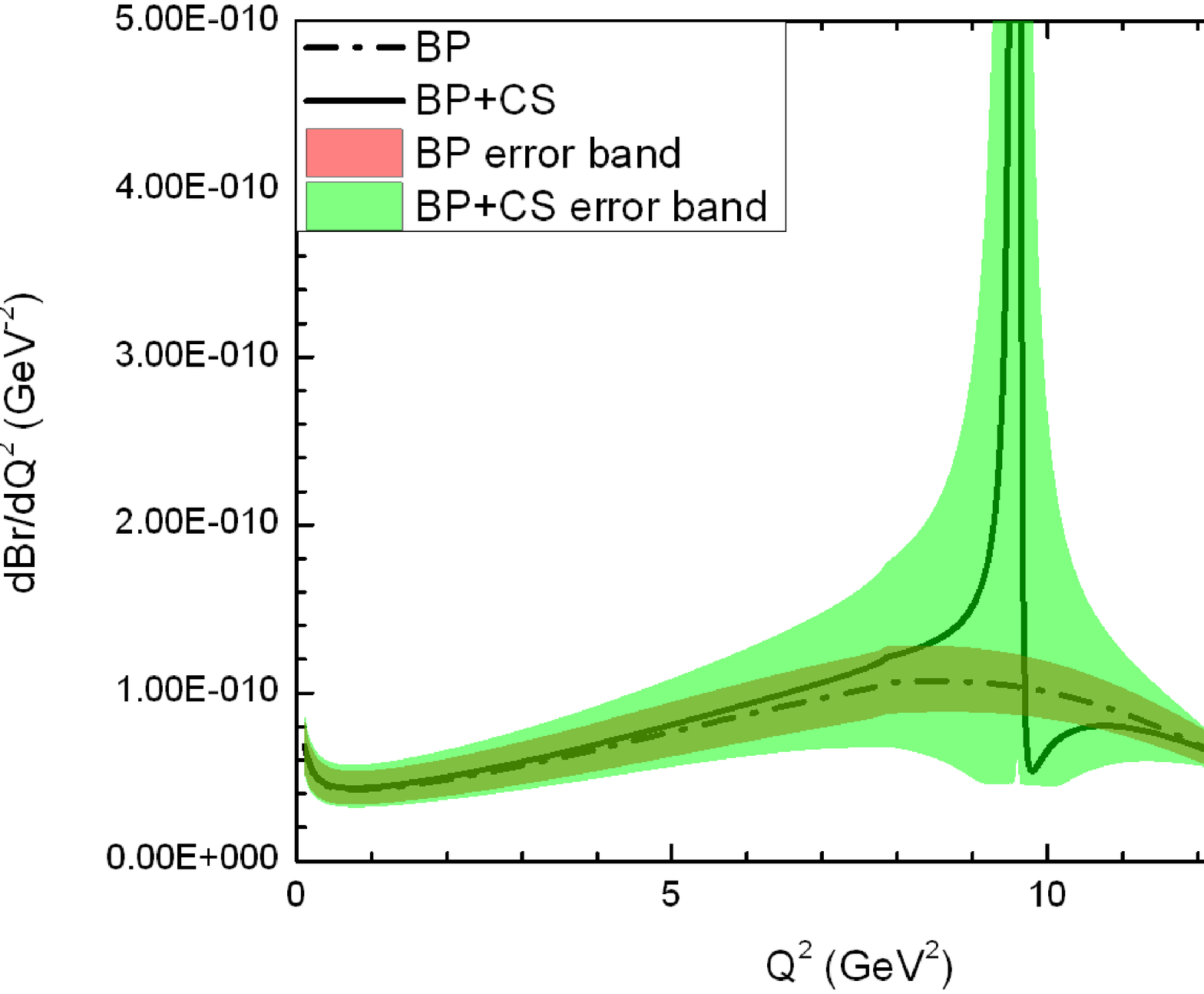}}
\subfigure[Differential branching fraction]{\includegraphics[width = 0.49\textwidth,height=0.2\textheight]{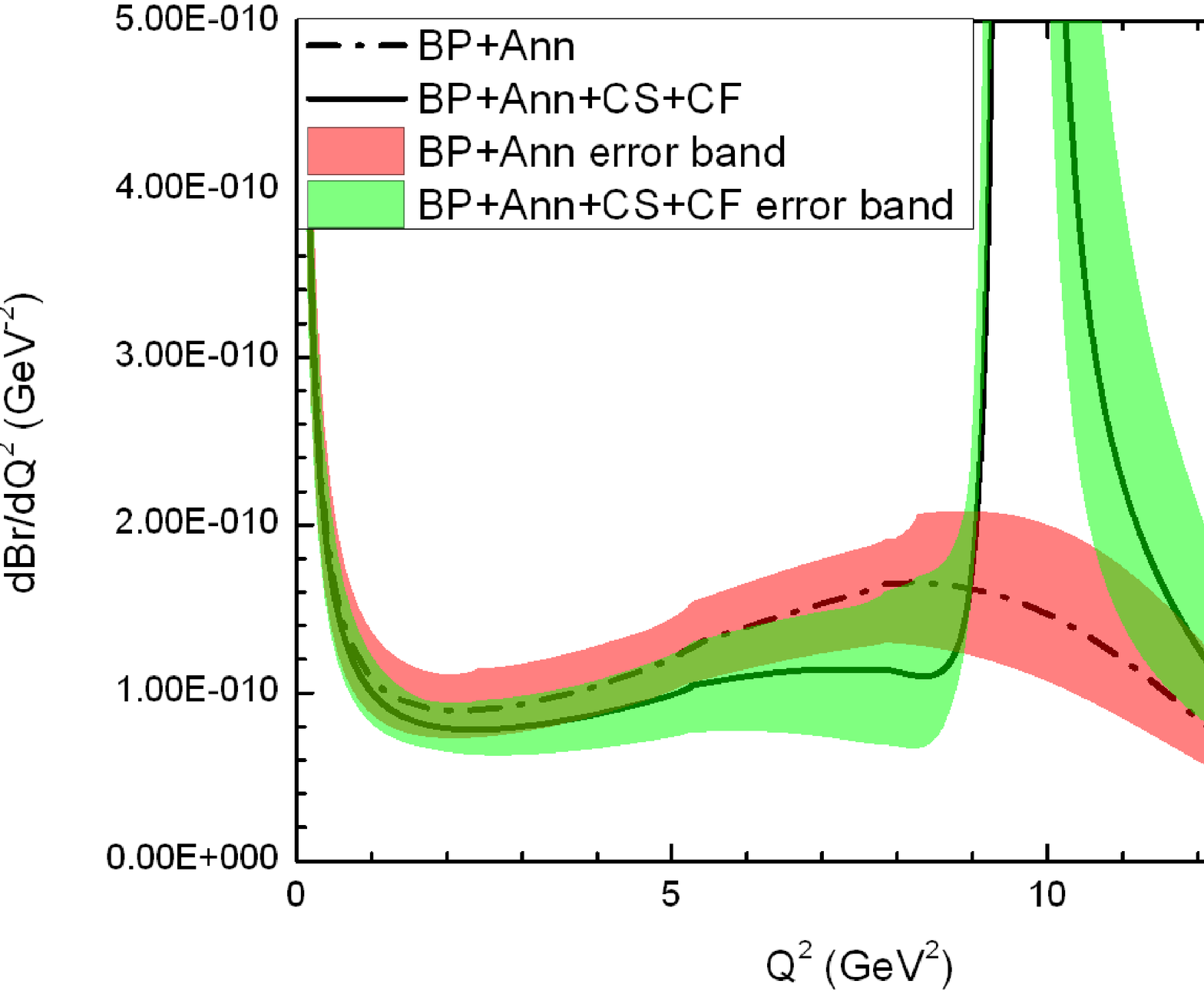}}\\
\subfigure[Leptonic longitudinal polarization asymmetry]{\includegraphics[width = 0.49\textwidth,height=0.2\textheight]{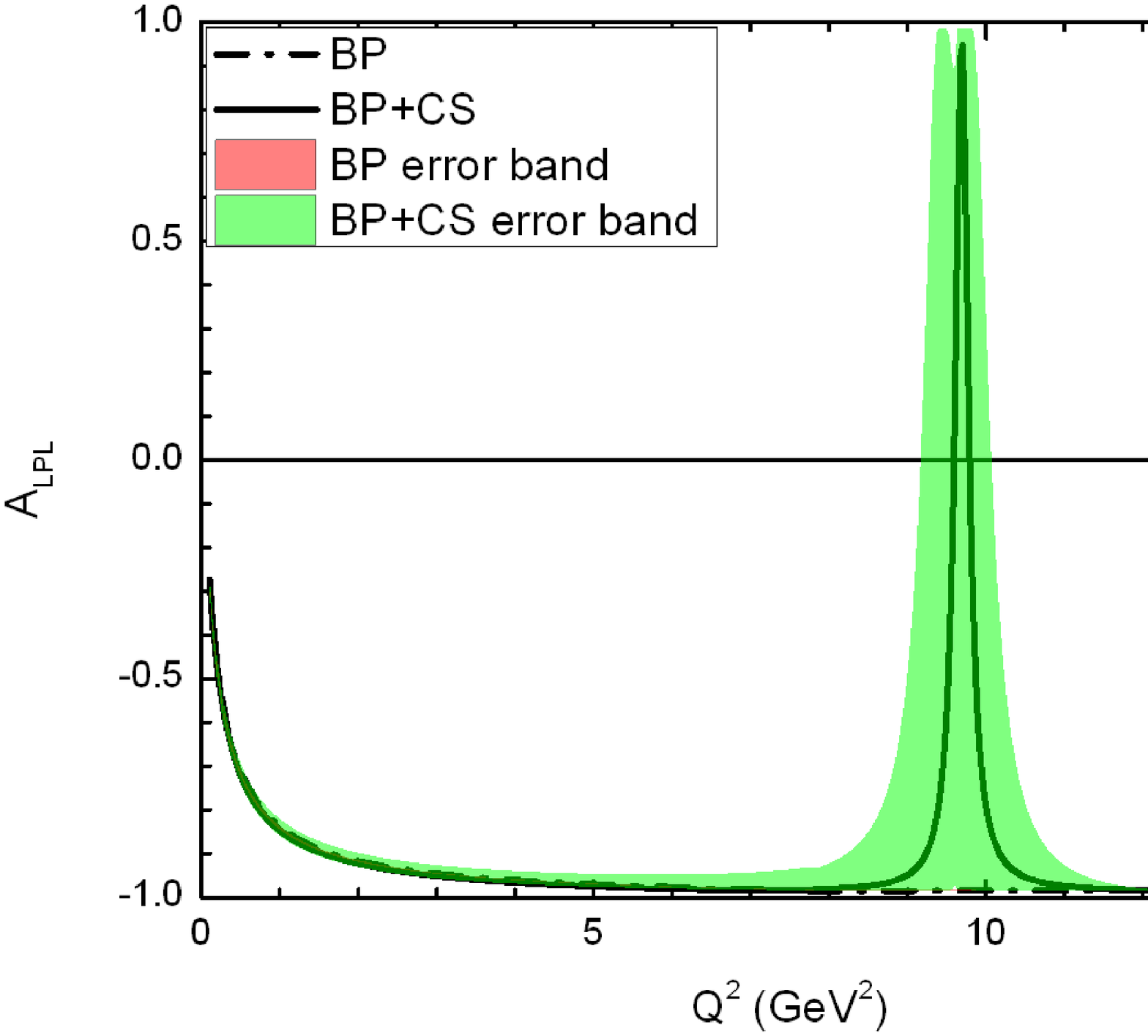}}
\subfigure[Leptonic longitudinal polarization asymmetry]{\includegraphics[width = 0.49\textwidth,height=0.2\textheight]{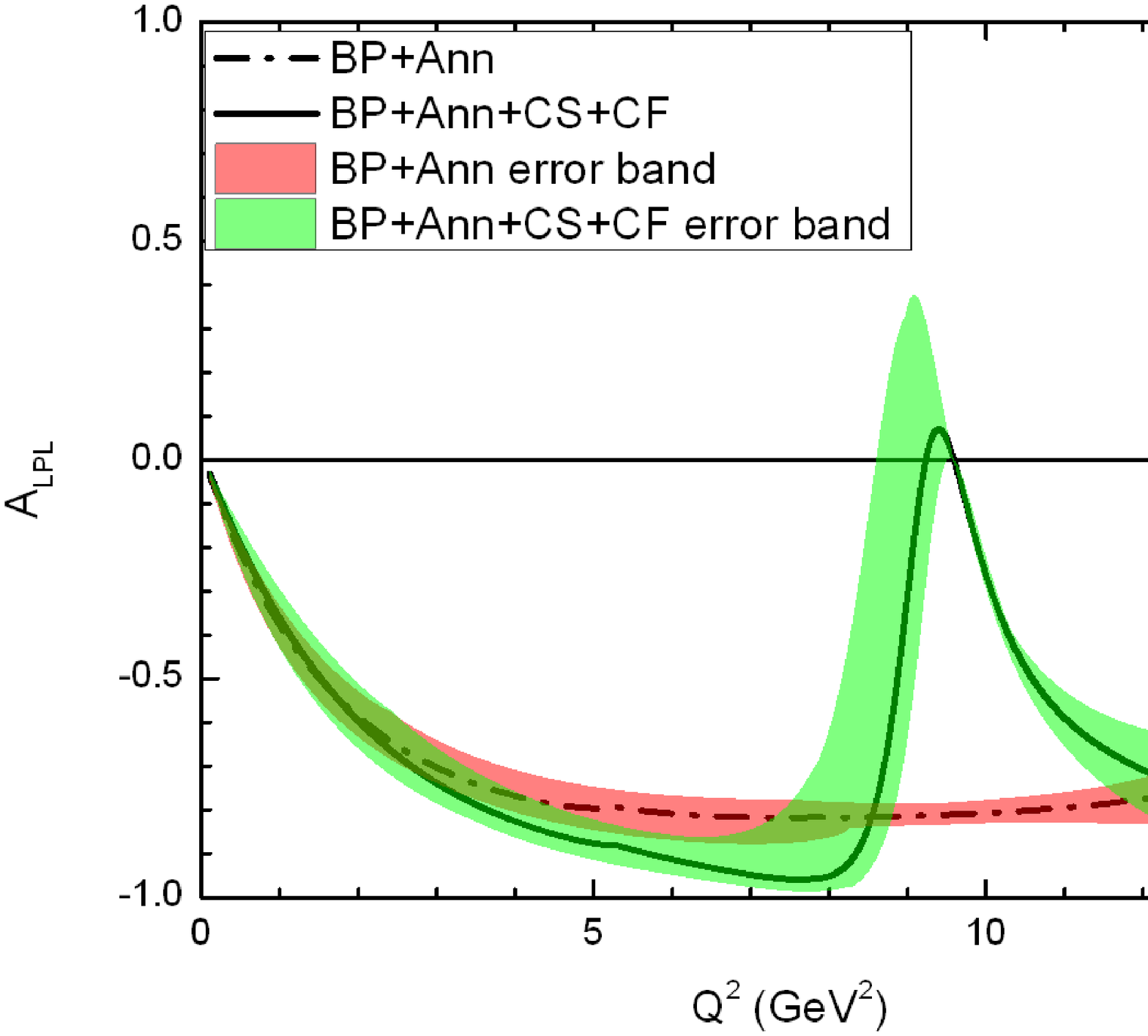}}\\
\subfigure[Forward-backward asymmetry]{\includegraphics[width = 0.49\textwidth,height=0.2\textheight]{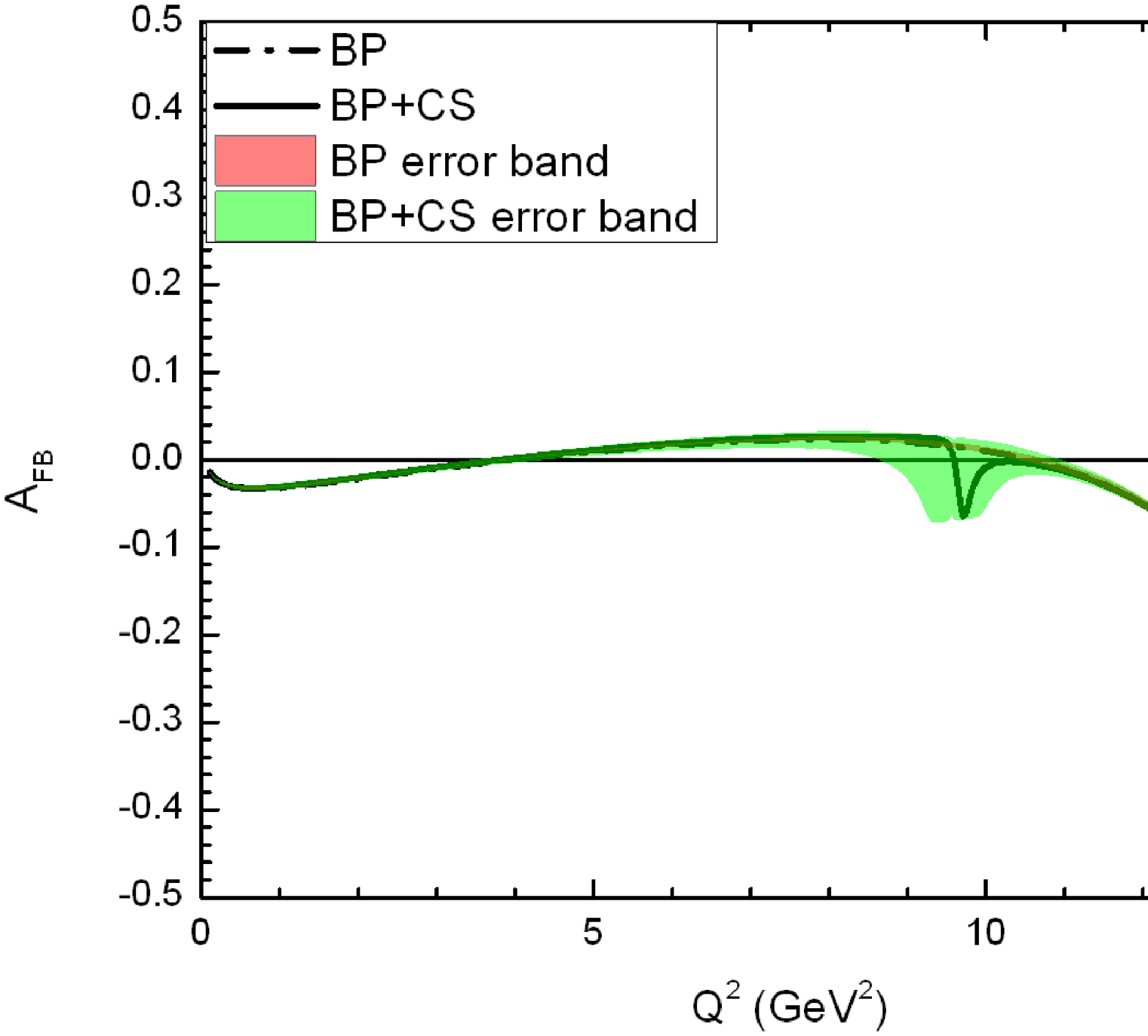}}
\subfigure[Forward-backward asymmetry]{\includegraphics[width = 0.49\textwidth,height=0.2\textheight]{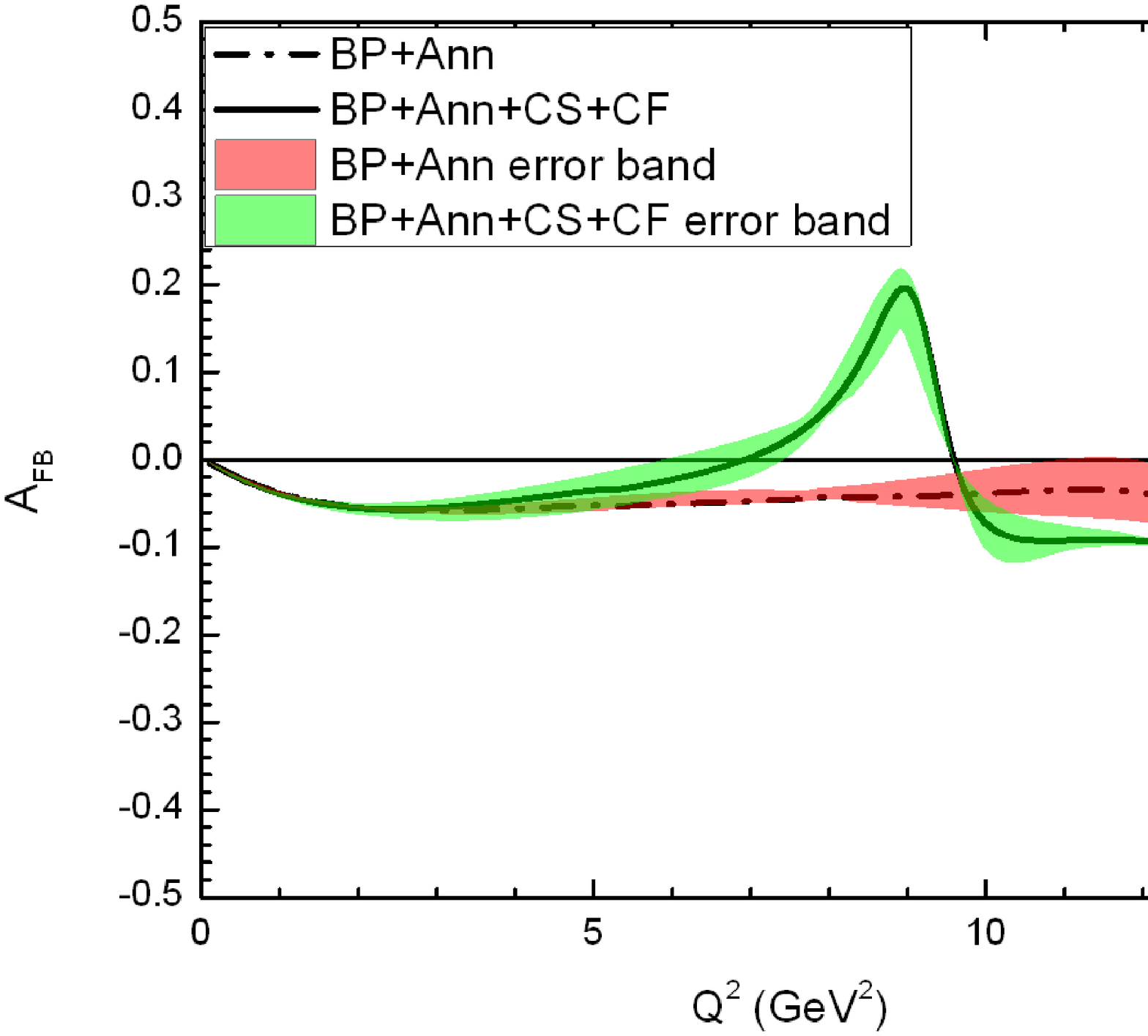}}\\
\subfigure[Longitudinal polarization of the
meson]{\includegraphics[width =
0.49\textwidth,height=0.2\textheight]{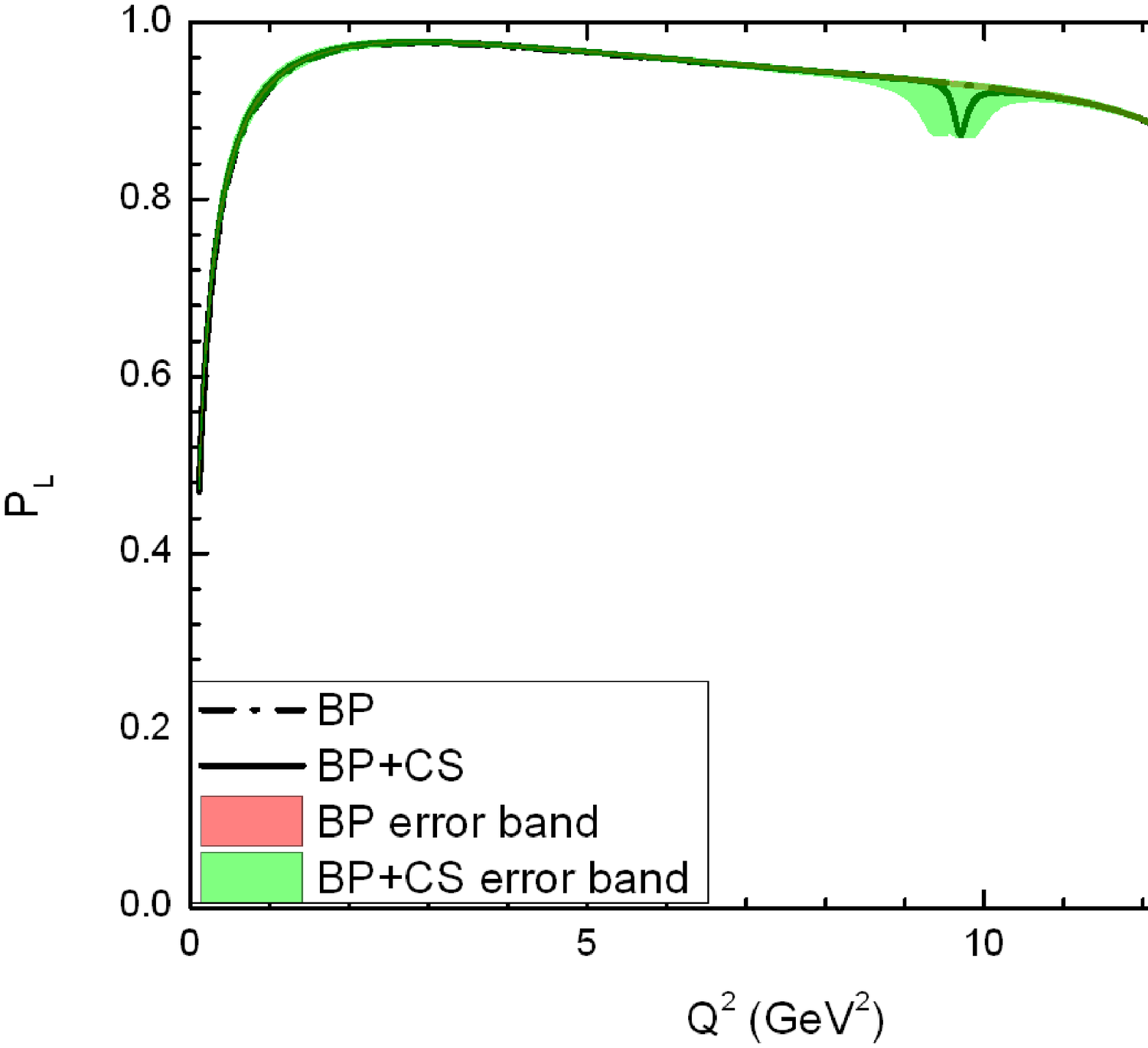}}
\subfigure[Longitudinal polarization of the final meson]{\includegraphics[width = 0.49\textwidth,height=0.2\textheight]{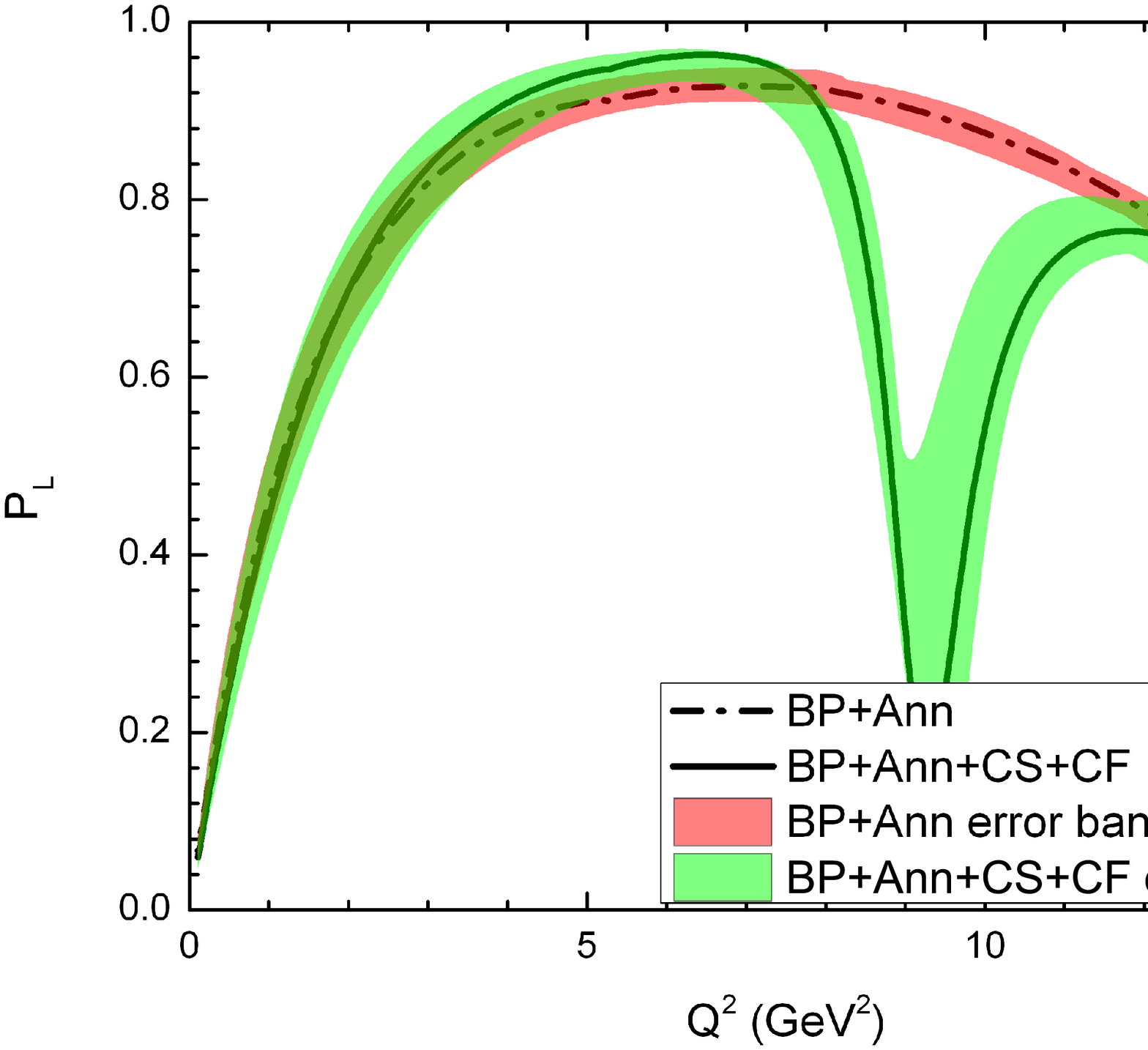}}\\

\caption{Observables of $B_c\rightarrow D_{1}(2420) \mu\bar{\mu}$. }
\end{figure}

\begin{figure}[htbp]
\centering
\subfigure[Differential branching fraction]{\includegraphics[width = 0.49\textwidth,height=0.2\textheight]{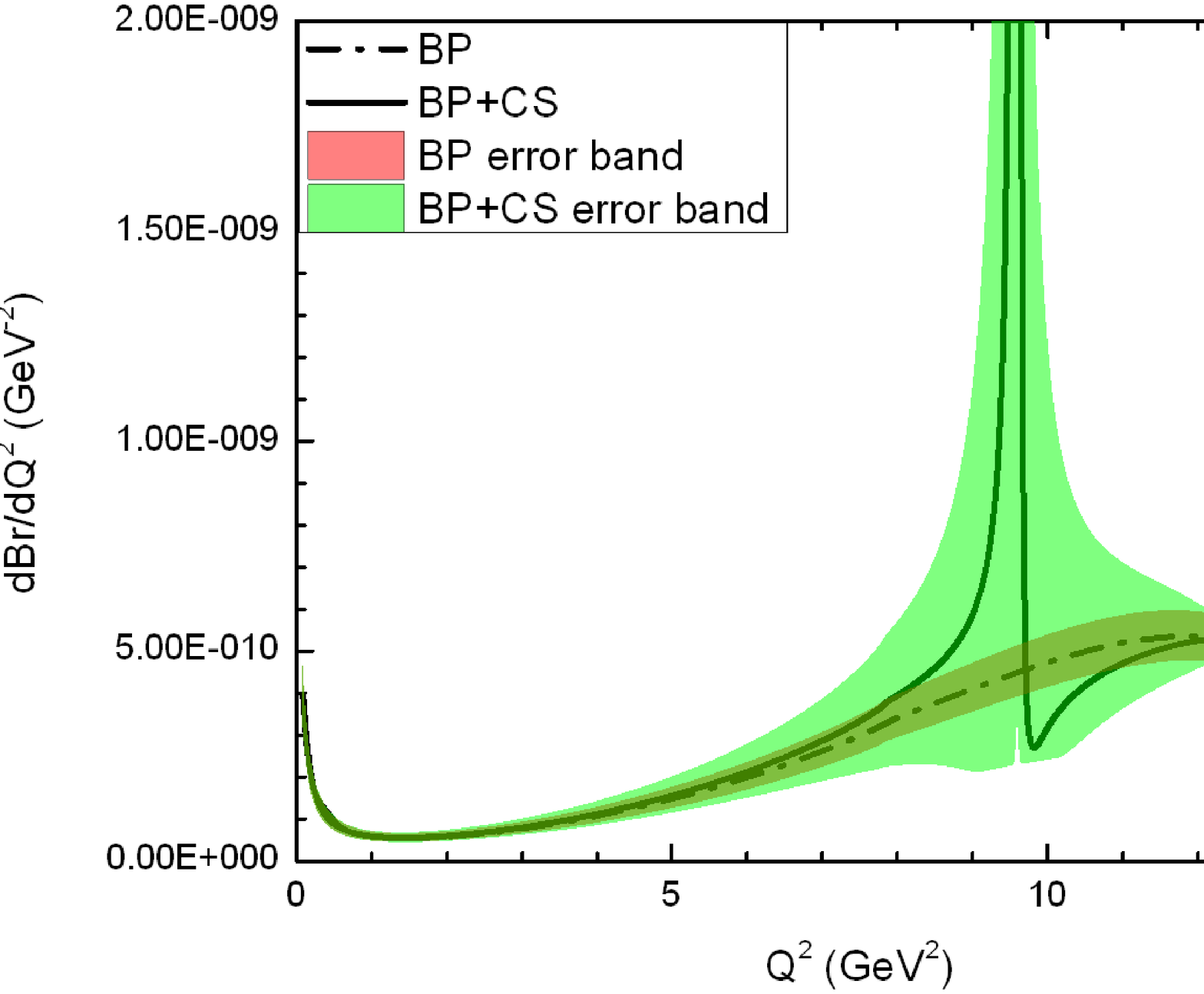}}
\subfigure[Differential branching fraction]{\includegraphics[width = 0.49\textwidth,height=0.2\textheight]{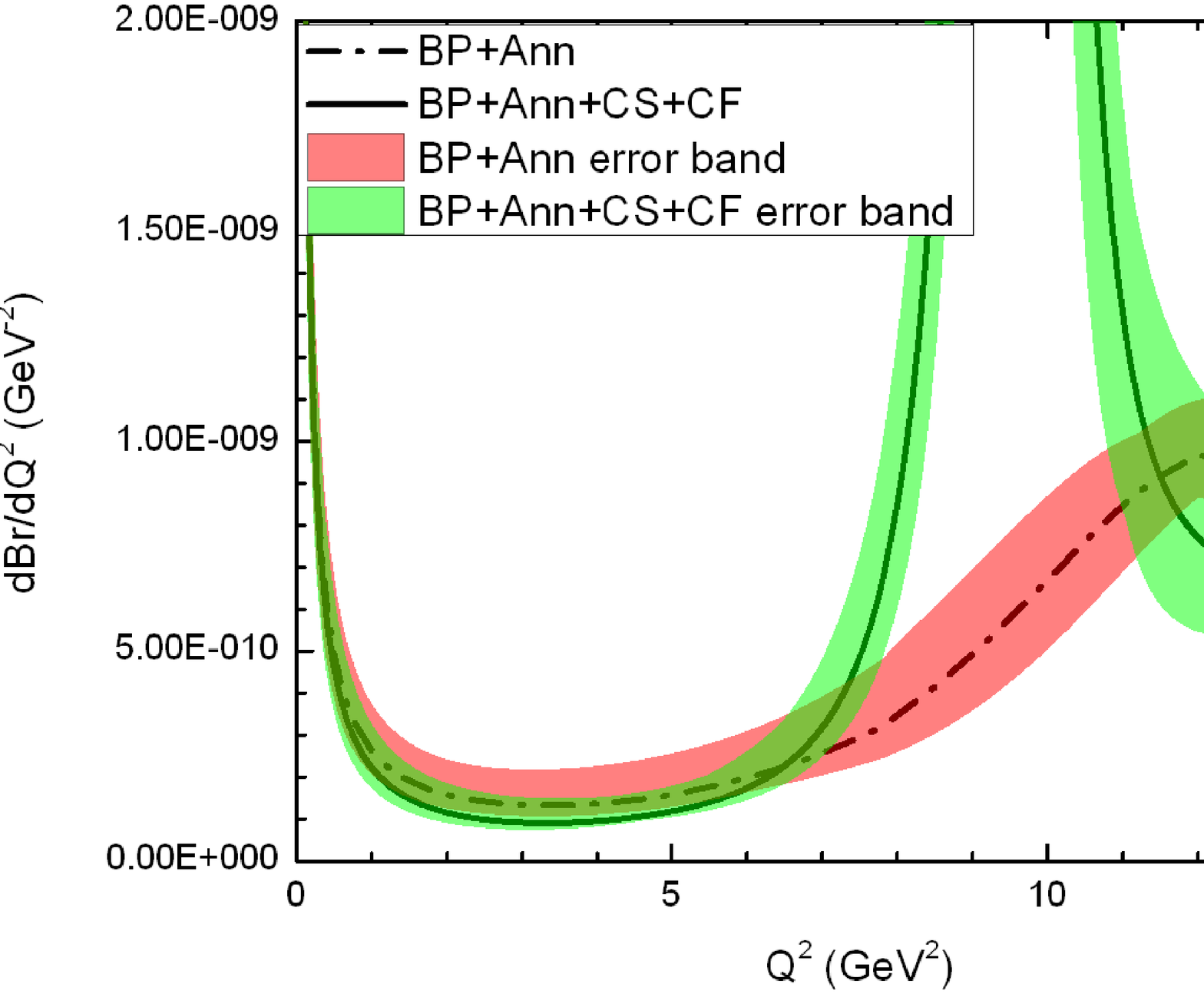}}\\
\subfigure[Leptonic longitudinal polarization asymmetry]{\includegraphics[width = 0.49\textwidth,height=0.2\textheight]{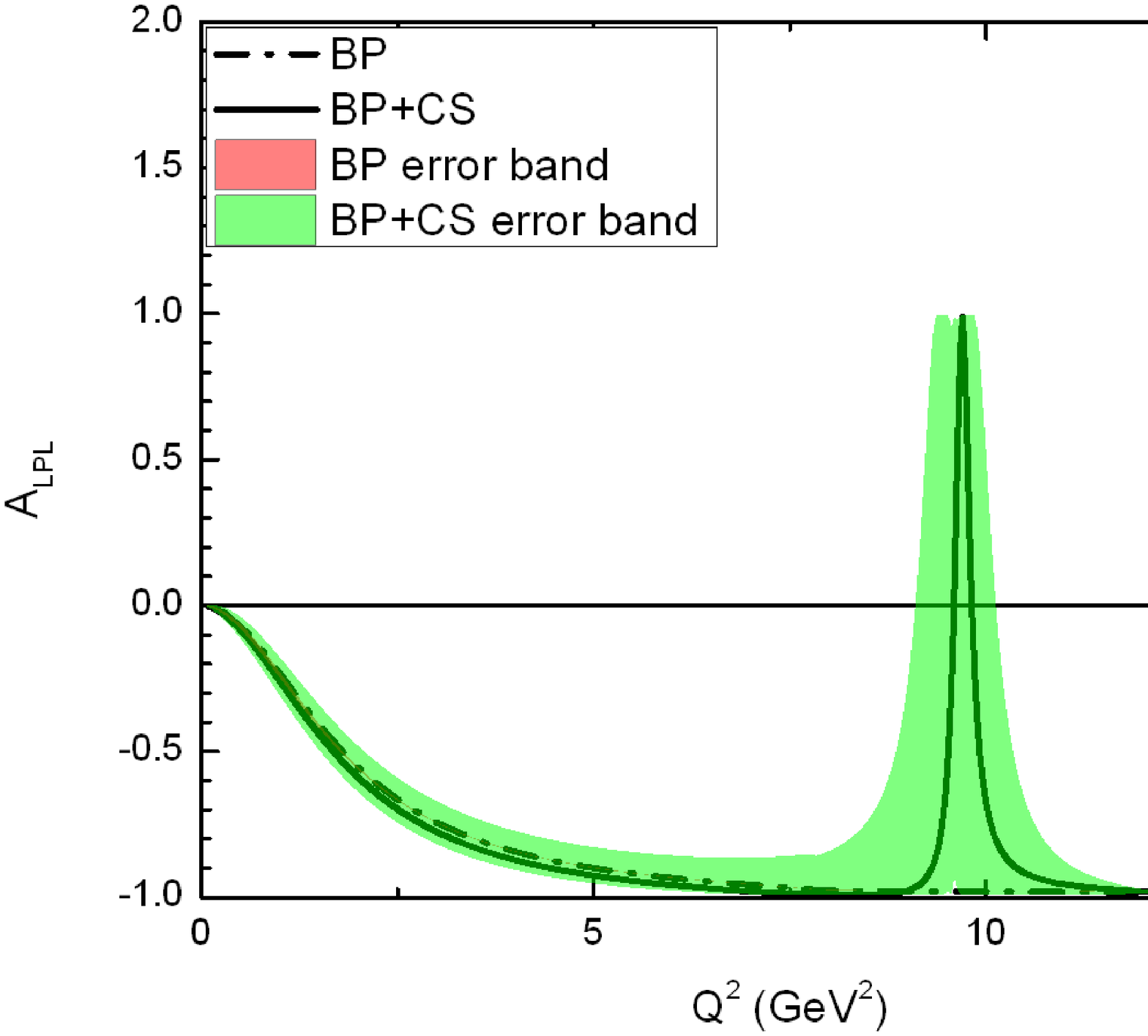}}
\subfigure[Leptonic longitudinal polarization asymmetry]{\includegraphics[width = 0.49\textwidth,height=0.2\textheight]{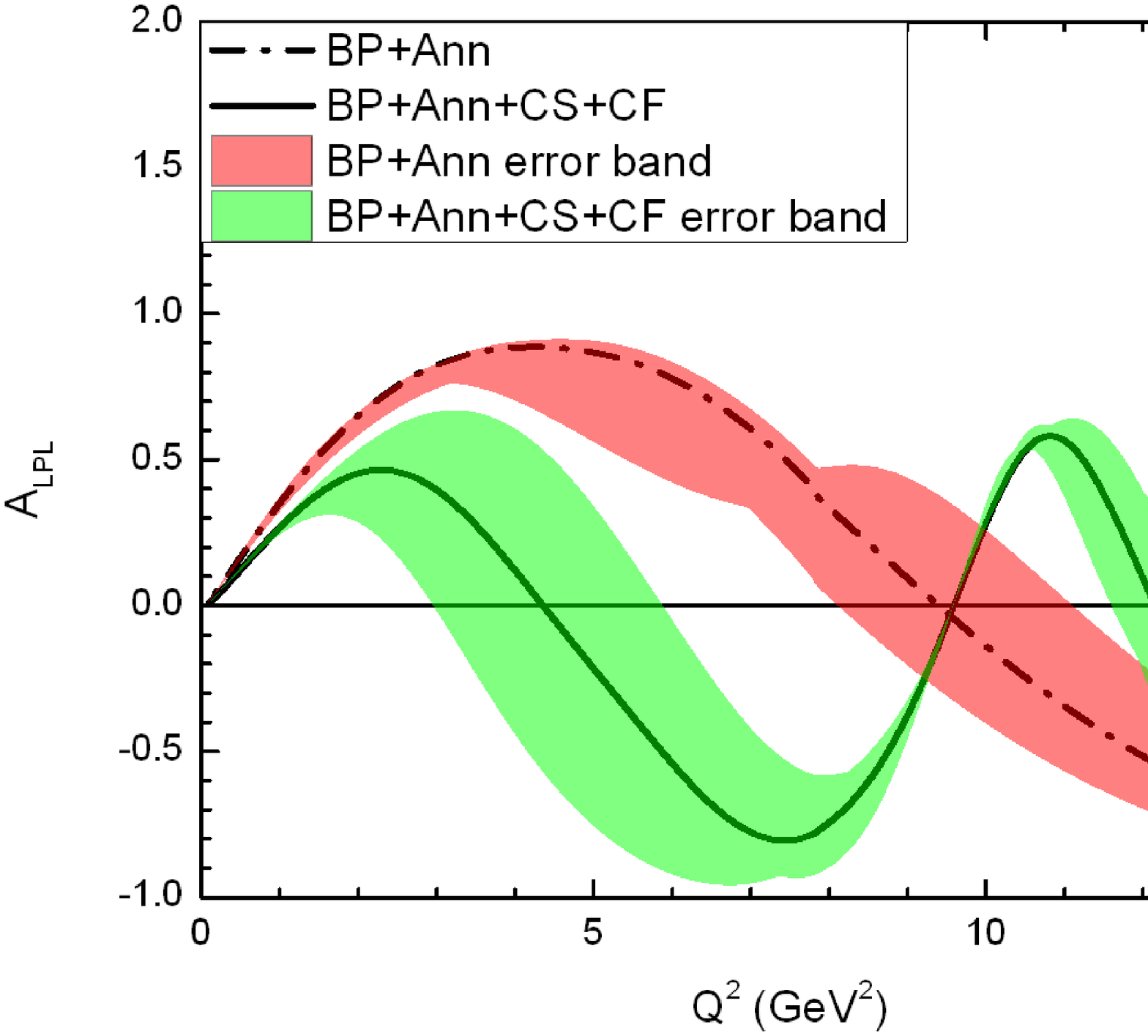}}\\
\subfigure[Forward-backward asymmetry]{\includegraphics[width = 0.49\textwidth,height=0.2\textheight]{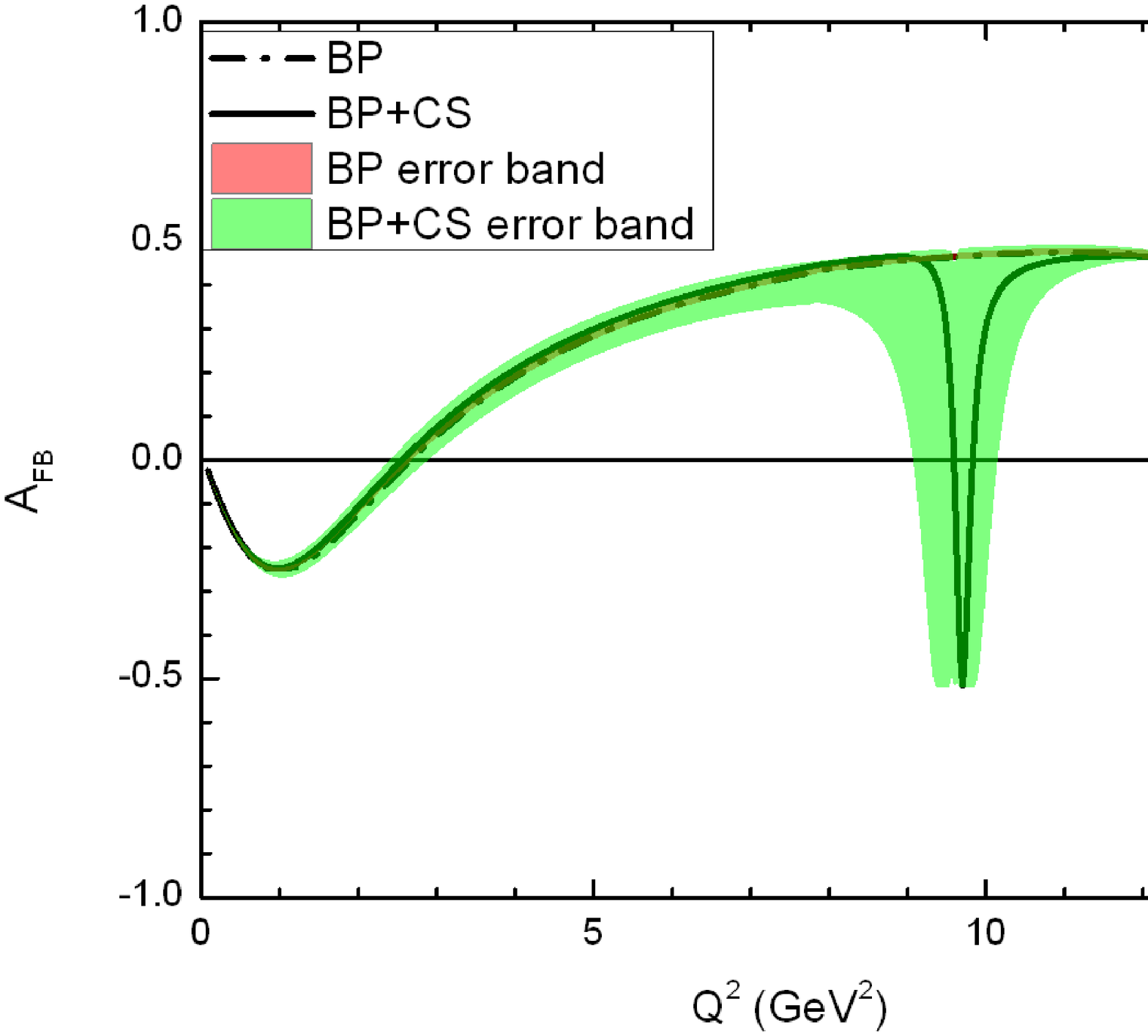}}
\subfigure[Forward-backward asymmetry]{\includegraphics[width = 0.49\textwidth,height=0.2\textheight]{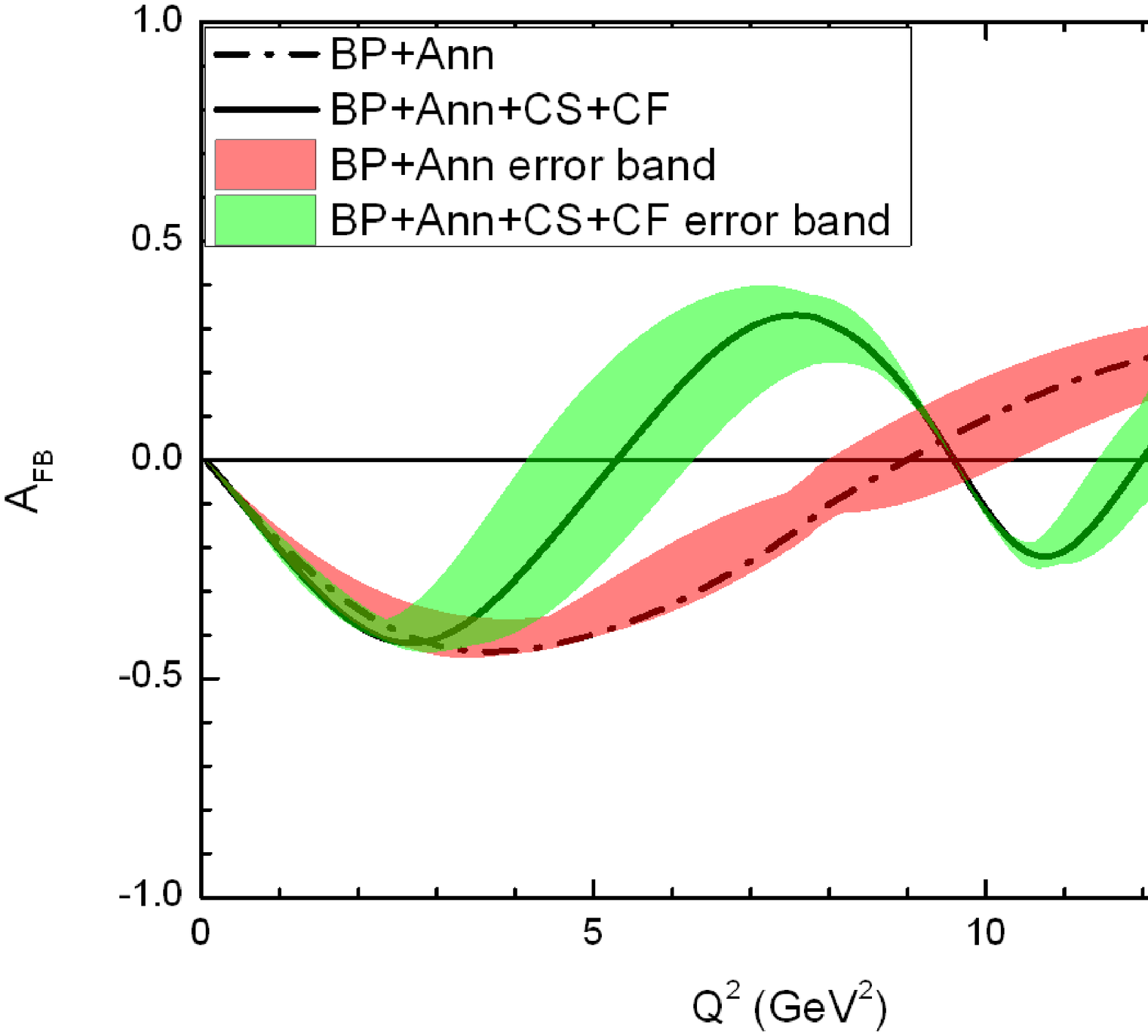}}\\
\subfigure[Longitudinal polarization of the
meson]{\includegraphics[width =
0.49\textwidth,height=0.2\textheight]{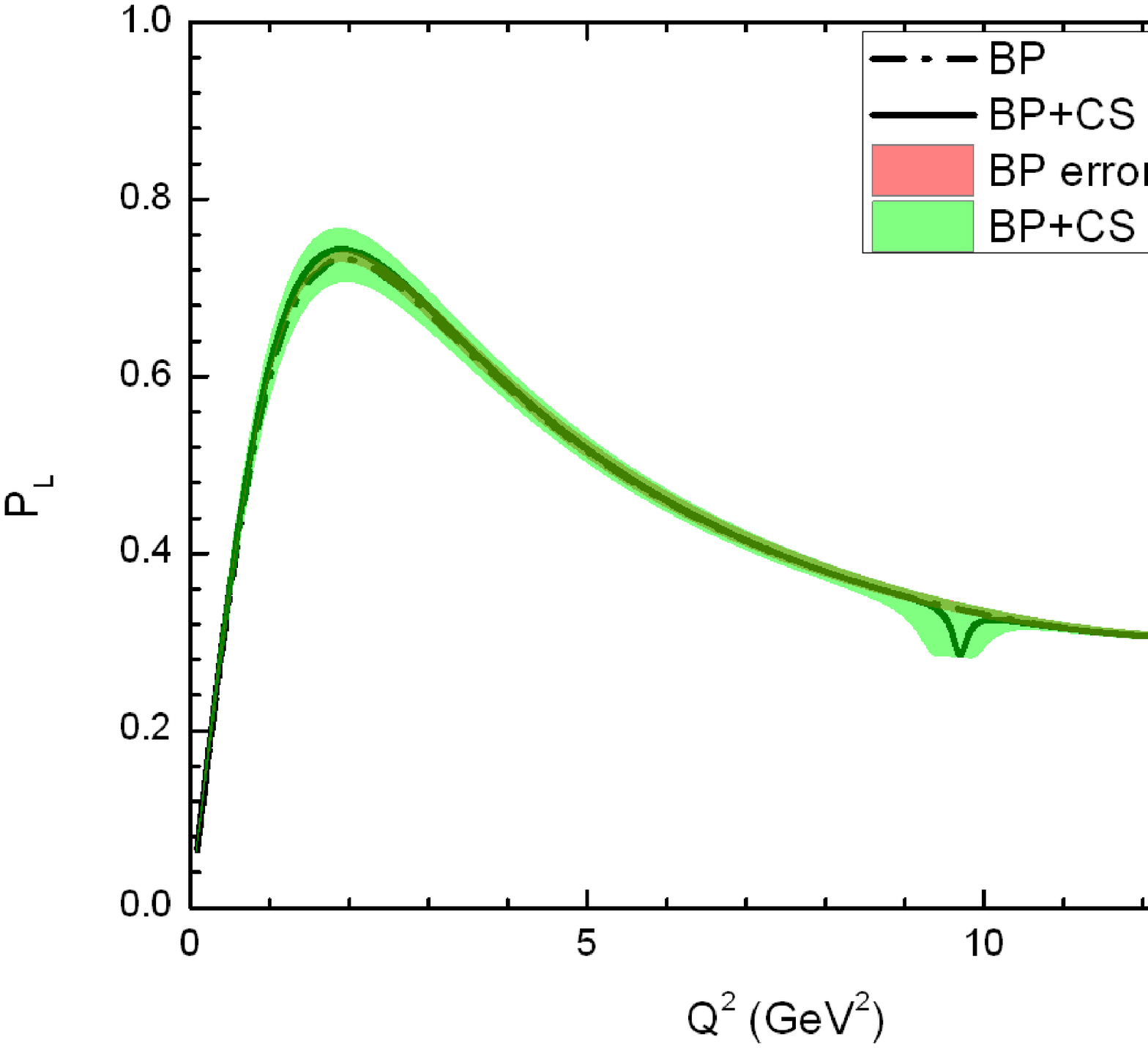}}
\subfigure[Longitudinal polarization of the final meson]{\includegraphics[width = 0.49\textwidth,height=0.2\textheight]{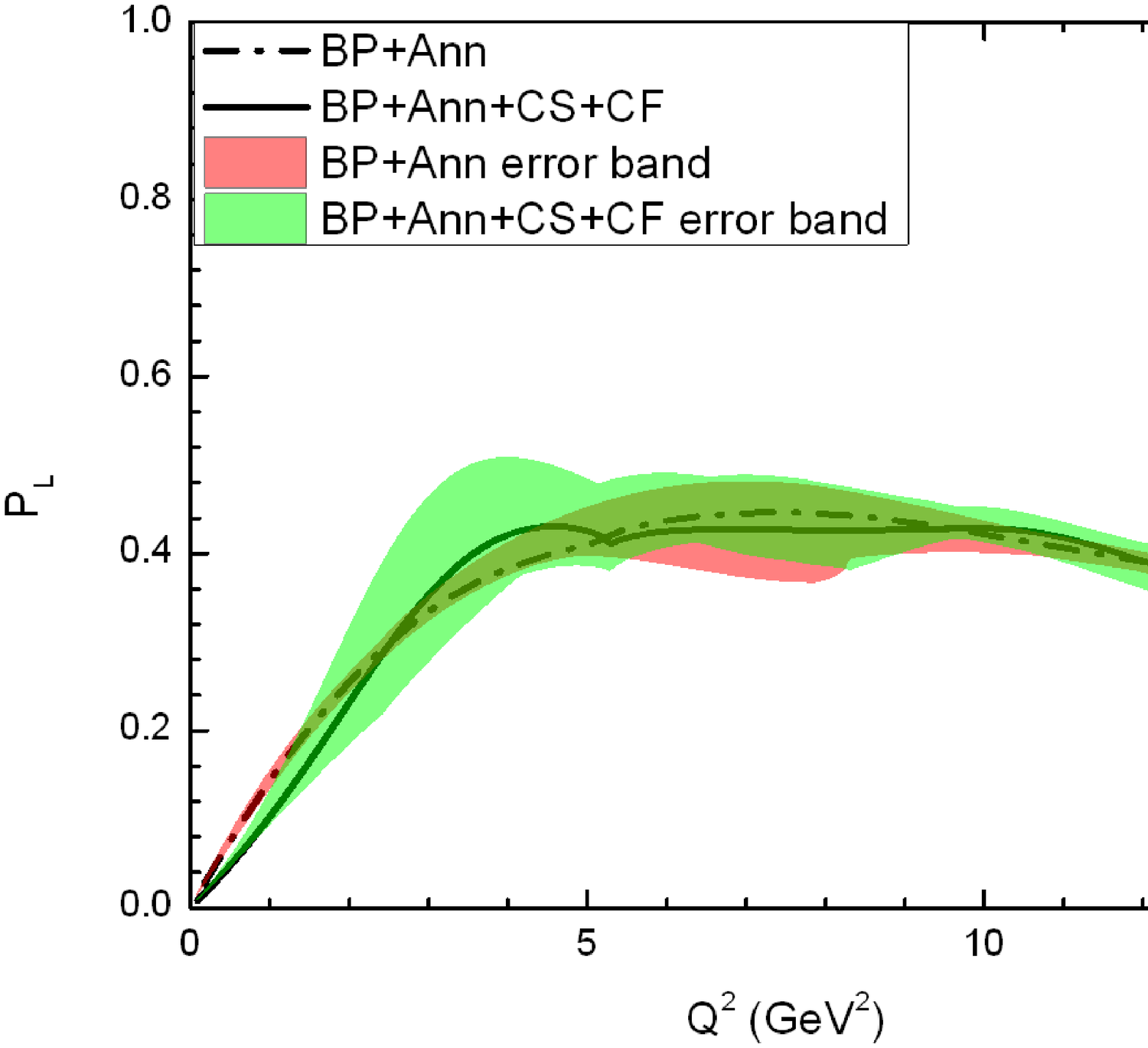}}\\

\caption{Observables of $B_c\rightarrow D_{1}(2430) \mu\bar{\mu}$.}
\end{figure}

\end{document}